\documentclass[journal=cgdefu,manuscript=article]{achemso}
\usepackage{graphicx}
\usepackage{amsmath}
\usepackage{wrapfig}
\usepackage{physics}
\usepackage{caption}
\usepackage{textcomp}
\usepackage{subfiles}
\usepackage{longtable}
\usepackage[pdftex]{hyperref}
\usepackage{array}
\usepackage{ragged2e}
\newcolumntype{P}[1]{>{\RaggedRight\hspace{0pt}}p{#1}}

\usepackage{multicol}
\usepackage{setspace}

\graphicspath{{../Figures/}}

\title{Adding anisotropy to the standard quasi-harmonic approximation still fails in several ways to capture organic crystal thermodynamics}
\author{Nathan S. Abraham}
\affiliation{Department of Chemical and Biological Engineering, University of Colorado Boulder, Boulder, CO 80309, USA}
\author{Michael R. Shirts}
\affiliation{Department of Chemical and Biological Engineering, University of Colorado Boulder, Boulder, CO 80309, USA}
\email{michael.shirts@colorado.edu}

\begin{document}
\begin{multicols}{2}
\begin{singlespace}

\maketitle
\bibliographystyle{plain}

\section{Abstract}
We evaluate the accuracy of varying thermal expansion models for the 
quasi-harmonic approximation (QHA) relative to molecular dynamics (MD) for 10 
sets of enantiotropic organic polymorphs. Relative to experiment we find that MD, 
using an off-the-shelf point
charge potential gets the sign of the enthalpic contributions correct for 6
of the 10 pairs of polymorphs and the sign of the entropic contributions
correct for all pairs. 
We find that anisotropic QHA provides little improvement to the error in 
free energy differences from MD relative to isotropic QHA, but does a better job 
capturing the thermal expansion of the crystals. A form of entropy--enthalpy compensation
allows the free energy differences of QHA to deviate less than 0.1 kcal/mol from 
MD for most polymorphic pairs, despite errors up to 0.4 kcal/mol in the entropy and 
enthalpy. Deviations in the free energy of QHA and MD do not clearly correlate
with molecular flexibility, clarifying a previously published finding. Much of the 
error previously found between QHA and MD for these flexible molecules is 
reduced when QHA is run from a lattice minimum consistent with the same basin 
as MD, rather than the energy-minimized experimental crystal structure. 
Specifically, performing anisotropic QHA on lattice minimum quenched from 
low-temperature replica exchange simulations reduced the error previously 
found by 0.2 kcal/mol on average. However, these conformationally flexible 
molecules can have many low-temperature conformational minima, and the choice 
of an inconsistent minima causes free energies estimated from QHA to deviate 
from MD at temperatures as low as 10 K. We also find finite 
size errors in the polymorph free energy differences using anisotropic 
QHA, with free energy differences as large as 0.5 kcal/mol between unit and supercells
loosely correlated with differences in anisotropic thermal expansion. These 
larger system sizes are computationally more accessible because our cheaper 
1D variant of anisotropic QHA, which gives free energies within within 
0.02 kcal/mol of the fully anisotropic approach at all temperature studied.
The errors between MD and experiment are 1--2 orders of magnitude larger than 
those seen between QHA and MD, so the quality of the force field used is still 
of primary concern, but this study illustrates a number of other important factors 
that must be considered to obtain quantitative organic crystal thermodynamics.

\section{Introduction}
    Solid organics can pack stably in multiple forms, each of which can have 
significantly different chemical and materials properties from each other. The ability of a 
molecule to arrange in multiple solid forms, or polymorphs, can change 
crystal solubility,~\cite{cheuk_solid_2015,maher_solubility_2012} charge 
mobility,~\cite{valle_organic_2004,stevens_temperature-mediated_2015} 
hardness,~\cite{singhal_drug_2004} and 
reactivity.~\cite{ghosh_understanding_2016,dreger_phase_2016,van_der_heijden_crystallization_2004} Relative 
stability of crystals can change with 
temperature~\cite{grzesiak_comparison_2003,yoshino_contribution_1999,vemavarapu_crystal_2002,badea_fusion_2007,yu_measuring_2005,cherukuvada_pyrazinamide_2010,torrisi_solid_2008,stolar_solid-state_2016,maher_solubility_2012,alcobe_temperature-dependent_1994,sacchetti_thermodynamic_2000,cesaro_thermodynamic_1980} 
and 
pressure~\cite{seryotkin_high-pressure_2013,boldyreva_effect_2002,cansell_phase_1993,gajda_pressure-promoted_2011}, 
so we must consider entropic effects and crystal lattice changes when modeling 
the crystal thermodynamics.

    Methods such as molecular dynamics (MD) and the quasi-harmonic approximation 
(QHA) are important for understanding the relative stability of crystals. QHA 
estimates the thermodynamics as a function of temperature and pressure by 
computing free energy due to harmonic vibrations around each crystal lattice 
minimum, and then determining the lattice geometry that minimizes the Gibbs free 
energy at each temperature as the crystal is expanded from the initial crystal 
lattice minimum. It is common to model thermal expansion isotropically, where 
the lattice vectors remain proportional and angles fixed. MD provides a 
more accurate description of the entropy and therefore free energy 
by generating the full conformational ensemble of the crystal, rather than 
neglecting anharmonic motions. Additionally, QHA assumes a single lattice is 
representative of the crystal at low temperatures whereas MD samples an 
ensemble of lattice configurations. 

    Our previous studies show that an isotropic QHA model can often
yield free energy differences between polymorphs that fall within numerical 
error of the polymorph free energy differences computed by MD for small and 
rigid molecules. However, free energies from QHA deviates from those 
generated by MD by $>$ 0.1 kcal/mol at 300 K for dynamically disordered crystals or 
crystals with greater conformational flexibility.~\cite{dybeck_capturing_2017} 
The small-molecule crystals examined in our previous study all showed some 
level of anisotropic expansion, and the error in the QHA free energy 
qualitatively increased with the degree of anisotropy. We concluded 
that the deviations of the isotropic QHA from MD were either due to 1) an 
inaccurate thermal expansion model or 2) anharmonic motions in the crystal 
lattice, or 3) some combination of the two. Without modeling QHA anisotropically, 
we could not definitively determine the source of the deviations of QHA from MD.

    We recently developed a method to efficiently determine the true anisotropic 
free energy minimum within the quasi-harmonic framework.~\cite{abraham_thermal_2018} 
This method relies on determining gradient of the lattice parameters with respect 
to temperature. Our gradient approach identifies high temperature free energy 
minimum that are 0.01--0.23 kcal/mol lower than the minimum identified with an 
isotropic model, altering the polymorph free energy differences of piracetam and 
resorcinol by 0.02--0.12 kcal/mol for unit cells of 4--8 
molecules.~\cite{abraham_thermal_2018} Our fully anisotropic approach and even 
a simplified 1D-anisotropic approach outperformed the accuracy of current 
quasi-anisotropic methods that only consider the anisotropy due to the potential 
energy as a function of crystal lattice 
vectors.~\cite{erba_how_2015,erba_assessing_2015,erba_structural_2015,erba_combining_2014,maul_thermal_2016}

    In this paper we re-evaluate the 10 of the 12 polymorph pairs previously 
studied with MD and isotropic QHA to determine whether an anisotropic thermal 
expansion model can improve the performance of QHA relative to 
MD.~\cite{abraham_thermal_2018} Specifically, we:
  \begin{itemize}
    \item evaluate the use of an off-the-shelf point charge potential in modeling the entropy, enthalpy, stability, and thermal expansion of organic polymorphs;
    \item determine how well our 1D-QHA variant compares to QHA using full anisotropic expansion over a larger data set
      than our initial anisotropic methods paper~\cite{abraham_thermal_2018};
    \item determine if finite size effects significantly contribute to the free energy ranking of polymorphs;
    \item discuss the appropriate ways to compare QHA and MD;
    \item determine if anisotropic QHA can reduce errors with respect to MD found previously with isotropic QHA;
    \item evaluate if entropy--enthalpy compensation plays a significant role in differences between QHA and MD; and	
    \item highlight crystal behaviors found in MD that QHA cannot account for.
  \end{itemize}

\section{Methods}
\begin{figure*}
      \begin{center}
      \includegraphics[width=16cm]{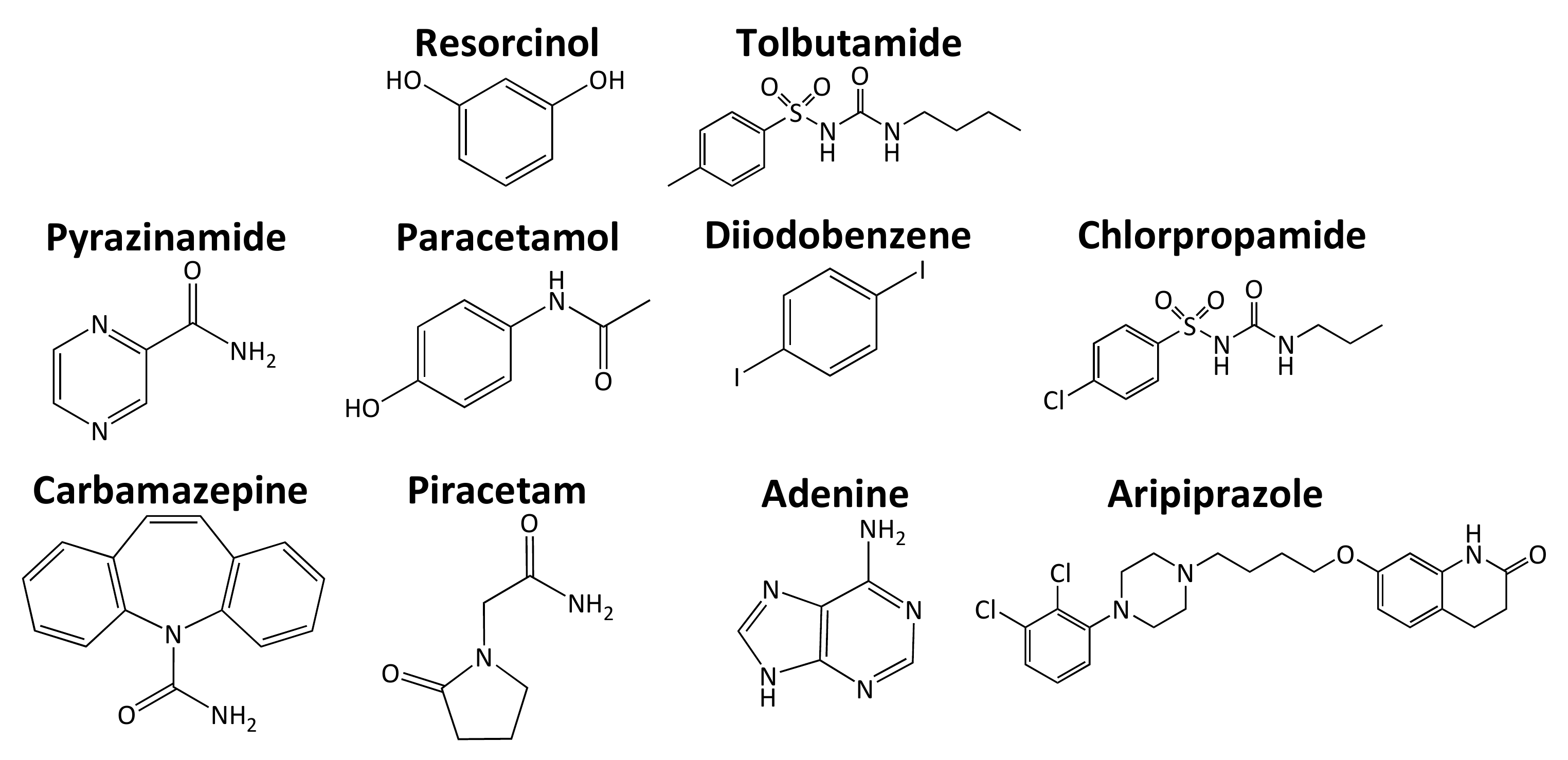}
      \end{center}
      \caption{Molecules examined in this study.
      \label{figure:molecule}}
    \end{figure*}

    In this study we examine the experimentally known polymorphs of the 10 
molecules shown in Figure \ref{figure:molecule} using QHA and MD. All 10 
of these molecules were examined in our previous study.~\cite{dybeck_capturing_2017}
    Most MD results in this study are from our previous 
publication~\cite{dybeck_capturing_2017}, and a methodological discussion of MD 
can be found there.~\cite{dybeck_effects_2016,dybeck_capturing_2017} 
Modifications of the MD procedure and where these modified simulations are used 
will be discussed in the context of the paper. All isotropic QHA results have 
been updated from that previous study and we describe those changes made here. 
In our previous study we examined 12 enantiotropic pairs of polymorphs, but have excluded
two molecules due to the experimentally known plastic phases with dynamic disorder 
at each molecular site. The dynamic disorder was confirmed with MD, both by 
others and ourselves, for the rotation of 
cyclopentane~\cite{torrisi_solid_2008,dybeck_capturing_2017} and trans/gauche 
isomerization of succinonitrile~\cite{Hore2009,dybeck_capturing_2017} in the 
disordered crystals. This behavior is inherently anharmonic, meaning there is no chance that QHA can properly 
model them, and hence they are left out of this study. Furthermore, both plastic phase 
crystals had their lattices constrained for stability in 
MD in previous studies~\cite{dybeck_capturing_2017}, so they were not modeling fully anisotropic
thermal expansion.

\subsection{Quasi-Harmonic Approximation}
    In the quasi-harmonic approximation (QHA) the Gibbs free energy is computed
by determining the lattice geometry that minimizes the sum of the potential and
the energy of the static, harmonic lattice vibrations, as shown in 
eq \ref{eq:QHA}. 
    \begin{eqnarray}
      G(T) &=& \min_{y} f(y,T) \label{eq:QHA_min}  \\
      f(y,T) &=& \min_{\boldsymbol{x}} \left(U(y,\boldsymbol{x})\right)  \nonumber \\ 
             & & + A_{v}(y,T) + PV \label{eq:QHA} \\
      y &=& {V, \boldsymbol{C}, \lambda} \nonumber	  
    \end{eqnarray}
where $T$ is the temperature, $P$ is the pressure, $\min_{\boldsymbol{x}} 
\left(U(y,\boldsymbol{x})\right)$ is the lattice energy of the minimum energy 
cell geometry $y$ and geometry optimized lattice coordinates $\boldsymbol{x}$, and 
$A_{v}$ is the Helmholtz vibrational free energy of a harmonic oscillator. We define
the cell geometry as $y$, which for isotropic expansion is the volume $V$, 
anisotropically with the six dimensional lattice tensor $\boldsymbol{C}$, or using 
our modified 1D-anisotropic expansion defined by $\lambda$.

    The Helmholtz free energy ($A_{v}$) of the lattice harmonic vibrations 
contribute to the entropic portion of QHA. We use the classical version of 
$A_{v}$ for comparison to MD, shown in eq \ref{eq:classical}.
    \begin{eqnarray}
      A_{v}(V,T) = \sum_{k} \beta^{-1} \ln{\left(\beta \hbar \omega_{k}(V)\right)}
	  \label{eq:classical}
    \end{eqnarray}
where $\beta$ is $(k_{B} T)^{-1}$, $\hbar$ is the reduced Planck constant, 
and $\omega_{k}$ is the frequency of the phonon of the $k^{th}$ vibrational mode 
of the crystal lattice.

\subsubsection{Gr\"{u}neisen Parameters}
    There are two common methods within the QHA framework to estimate the phonon frequencies of 
the crystal lattice at each lattice geometry of interest. The first way is to 
compute and diagonalize the mass-weighted Hessian of every crystal structure, 
which is computationally demanding. The second way is to use the Gr\"{u}neisen 
parameter approach assumes that the changes in the frequencies of a particular 
phonon are constant as the crystal is strained in a particular direction. Use 
of the Gr\"{u}neisen parameter to calculate the thermodynamics of organic and 
inorganic material is common.~\cite{heit_predicting_2015,ramirez_quasi-harmonic_2012,nath_high-throughput_2016,huang_efficient_2016,dybeck_capturing_2017}
This assumption can introduce errors, but we have shown for polymorph free 
energy differences that those errors are generally less than 0.02 
kcal/mol.~\cite{abraham_thermal_2018} This is smaller than the experimental 
error in most cases, so we will exclusively use the Gr\"{u}neisen parameter 
approach for QHA calculations in this paper when calculating polymorph free 
energy differences.

    The standard definition of the Gr\"{u}neisen due to a volume change is in 
eq \ref{eq:iso_Gru}, which is directly applicable only for constant strains, 
such as isotropic expansion, is:
    \begin{eqnarray}
	  \gamma_k = - \frac{V}{\omega_k} \frac{\partial \omega_k}{\partial V}
	  \label{eq:iso_Gru}
	\end{eqnarray}
where $\gamma_{k}$ is the Gr\"{u}neisen parameter for the $k^{th}$ vibrational 
frequency. Eq~\ref{eq:iso_Gru} is solved numerically, which requires 
diagonalization of the mass-weighted Hessian at two volumes to produce reference 
frequencies. The corresponding Gr\"{u}neisen parameters can be used to solve the 
frequencies at all isotropic volumes relative to a reference point, generally the 
lattice minimum structure.

    The Gr\"{u}neisen parameter for the volume can be extended to a crystal 
placed under any strain,~\cite{choy_thermal_1984,abraham_thermal_2018} which we 
will use for anisotropic expansion. Eq~\ref{eq:aniso_Gru} is the anisotropic version of the isotropic equation(eq~\ref{eq:iso_Gru}).
    \begin{eqnarray}
	  \gamma_{k,i} = - \left.\frac{1}{\omega_k} \frac{\partial \omega_k}{\partial \eta_i}\right|_{\eta_j \ne \eta_i}
	  \label{eq:aniso_Gru}
	\end{eqnarray}
Here, the Gr\"{u}neisen parameter for the $k^{th}$ vibrational mode due to the 
$i^{th}$ strain ($\eta_{i}$) applied to the crystal is $\gamma_{k,i}$. The 
symmetric 3$\times$3 strain matrix $\boldsymbol{\eta}$ allows us to compute 
6 sets of Gr\"{u}neisen parameters numerically. With the Gr\"{u}neisen 
parameters and reference frequencies, we can compute the frequencies of the lattice 
parameters at any cell geometry of interest, by integrating 
eq~\ref{eq:aniso_Gru}. A full discussion of the use of both the isotropic and 
anisotropic Gr\"{u}neisen parameters can be found in previous 
work.~\cite{abraham_thermal_2018}

\subsubsection{Thermal Expansion}
    We previously implemented a method to determine the crystals thermal 
expansion,~\cite{abraham_thermal_2018} which can be used to solve bi-optimization 
problems,~\cite{gould_differentiating_2016} such as QHA. Eq~\ref{eq:general_Grad}
gives the general formulation for determining the crystal thermal expansion. If we 
have a reference structure, the 0 K minimum (i.e. classical lattice minimum), we can 
compute the thermal expansion and numerically integrate with temperature.
    \begin{eqnarray}
	  \frac{\partial y}{\partial T} &=& \left(\frac{\partial^{2} G}{\partial y^{2}}\right)^{-1} \frac{\partial S}{\partial y} \label{eq:general_Grad} \\
	  y &=& V, \lambda, \boldsymbol{C} \nonumber
	\end{eqnarray}
In eq~\ref{eq:general_Grad}, $T$ is the temperature, $S$ is the entropy, and 
$G$ is the Gibbs free energy. By computing the numerical derivatives of the entropy 
and Gibbs free energy we can determine the gradient of $y$, the cell geometry, with 
respect to temperature.
\\
\textit{Isotropic Expansion} \\
    Isotropic expansion assumes that the lattice vectors remain proportional to 
one another and the lattice angles remain fixed. Due to these constraints the 
solution to QHA in eq~\ref{eq:QHA} requires minimization of a single variable, 
the volume ($y = V$). We compute the rate of isotropic thermal expansion using 
eq~\ref{eq:general_Grad} and replacing $y$ with the crystal volume ($V$).
\\
\textit{Anisotropic Expansion} \\
    Anisotropic expansion allows the crystal lattice to relax to the harmonic 
free energy minimum structure by removing all constraints in isotropic 
expansion. In eq~\ref{eq:QHA} we minimize the free energy as a function of all 
six crystal lattice parameters ($y = \boldsymbol{C}$), which makes the problem 
of minimizing the Gibbs free energy for QHA more complex. We compute the 
thermal expansion for all six parameters by using eq~\ref{eq:general_Grad}
and replacing $y$ with a array of the six lattice parameters ($\boldsymbol{C}$).
\\
\textit{1D-Anisotropic Expansion} \\
    Computing eq~\ref{eq:general_Grad} for anisotropic expansion 
($\boldsymbol{C}$) requires 73 lattice optimizations to determine a single 
six-dimensional gradient, which becomes increasingly expensive to compute for 
the entire temperature range of interest. In our previous 
work,~\cite{abraham_thermal_2018} we found if the ratio of anisotropic 
expansion was kept constant at all temperatures we could achieve the same free 
energy differences within 0.005 kcal/mol of full anisotropic expansion. In that 
paper, we presented eq~\ref{eq:1D_expansion},
    \begin{eqnarray}
	  C_i(\lambda) = C_i(\lambda=0) + \kappa_i \lambda(T)
	  \label{eq:1D_expansion}
	\end{eqnarray}
where the lattice parameter $C_i$ is a function of the variable $\lambda$ and 
$\kappa_i$. In this case, $\lambda$ is a single parameter describing 
the expansion in eq~\ref{eq:QHA}, which is zero at the 0 K lattice minimum 
value of $C_i$. $\kappa_i$ is the gradient computed in eq~\ref{eq:general_Grad} 
for $y=\boldsymbol{C}$ at 0 K.

\subsection{Molecular Dynamics}
    A more complete approach to compute the thermodynamics of the crystals, 
including the free energy difference between polymorphs, is molecular dynamics 
(MD), which generates (in theory) the full configurational ensemble at a given temperature 
of interest. For our method, there are two necessary steps for computing the 
free energy differences of polymorphs using MD: 1) determine $\Delta G$ between polymorphs at 
a reference temperature with simulations of a series of non-physical intermediate states and 2) determine $\Delta G$ as a function of temperature for 
each polymorph using simulations at range of temperatures.  
For the first step, we can drive each crystal along a reversible 
thermodynamic path to an ideal gas state by turning off intermolecular energies to determine the reference free energy 
differences. Then, with simulations at intermediate temperatures we can 
relate each temperature point back to the reference energy difference between 
polymorphs using the Multistate Bennett Acceptance Ratio (MBAR). A full 
discussion of this approach can be found in our previous 
work.~\cite{dybeck_effects_2016,dybeck_capturing_2017}

\subsection{Computational Details}
    We have provided as supporting information a .zip file including all 
computational details to re-run simulations for both MD and QHA. For 
specifications on how we chose certain parameters we refer the reader back to 
our previous publications that present the methods for 
MD~\cite{dybeck_effects_2016,dybeck_capturing_2017} and QHA.~\cite{abraham_thermal_2018}
\\
\textit{Molecular Dynamics} \\
    For a number of crystals, we found that the lattice parameters from the 
previous study~\cite{dybeck_capturing_2017} did not continuously change with 
temperature at low $T$ for MD, so we re-ran the crystals with temperature replica exchange (REMD)
to better escape from metastable states and thus allow for better convergence to
the 0 K lattice energy minimum. Replica exchange parameters were chosen
to give an average exchange probability of $\sim0.3$ between temperature 
samples from 10 K up to 350 K. Temperature replica exchange allows the 
simulations at different temperatures to exchange with one another, which
causes the crystal to be heated up and reannealed throughout simulation while maintaining the proper ensemble distribution at all temperatures.~\cite{zhang_convergence_2005}
Temperature replica exchange was required 
to compute the low-temperature ensembles of
for tolbutamide, chlorpropamide, and 
aripiprazole. We provide input files for both MD and QHA all within the 
input\_files.zip, which contains all parameters for individual crystals and 
simulation settings for GROMACS 2018.
\\
\textit{Quasi-Harmonic Approximation} \\
    All QHA calculations were performed using our Python based lattice dynamics 
code available on GitHub at \url{http://github.com/shirtsgroup/Lattice\_dynamics}. The 
code currently wraps around a number of molecular modeling packages. In this 
paper, all lattice vibrations, energy evaluations, and optimizations were 
performed using Tinker 8.1 molecular modeling 
distributed TINKER code to correct for errors in the Hessian calculation. Those 
modifications are discussed in the main paper and supporting information of our 
previous work and have since been updated in Tinker 8.7.~\cite{abraham_thermal_2018}

    Lattice structures were retrieved from the Cambridge Crystallographic Data 
Center (\url{https://www.ccdc.cam.ac.uk/structures/}) and a unit and supercell were 
lattice optimized. Supercells were created assuring that the three lattice 
vectors were twice the van der Waals cutoff, (8 \r{A}). Specifications on cell 
size can be found in the Table S\ref{table:systems}. Each molecule was 
parameterized using the OPLS-AA~\cite{robertson_improved_2015,jorgensen_development_1996} 
classical fixed point charge potential: 
    \begin{eqnarray}
	  U_{total} &=& U_{bond} + U_{angle} + U_{dihedral} \nonumber \\
                  & &  + U_{vdw} + U_{coulombic}
    \end{eqnarray}
where the exact forms of the each of the energy terms are defined in the references.

    The crystal structures were geometry and lattice optimized to the lattice minimum structure. 
Optimization of the crystal structure was performed with Tinker's {\tt xtalmin} 
executable to an RMS gradient/atom of $10^{-5}$. The molecules 
centers of mass, keeping the coordinates relative to the center of mass constant 
constant. The crystal was then geometry optimized using Tinker's {\tt minimize} 
executable to an RMS gradient/atom of $10^{-4}$. These values were chosen to 
maximize convergence and numerical stability for the gradient method, 
and this choice is discussed fully in the supporting information of our previous 
work.~\cite{abraham_thermal_2018}

    To determine $y(T)$ across the entire temperature range we use a 
4$^{\mathrm{th}}$ order Runge-Kutta integrator using the thermal gradient 
approach to satisfy eq~\ref{eq:QHA_min}. For most crystals, we found that 3 
steps of 100 K each up to 300 K ensures that the crystal was at a locally 
metastable free energy minimum at all temperatures. If the structure is not at a 
free energy minimum with respect to box geometry variables at 300 K, the 
structure was re-run with 6 Runge-Kutta steps. For the unit cells only, if 6 
steps could not maintain the crystal at a free energy minimum then 20 steps were 
run. All results shown are for the maximum temperature step that could be 
achieved while assuring that it remained at a free energy minimum with respect 
to box geometry under expansion (temperatures given in Table 
S\ref{table:1D_6D_dG}). Results for the graphed intermediate points between Runge-Kutta 
steps were calculated with a third order spline that was fit to the lattice 
parameters with temperature.~\cite{hairer_solving_2008,abraham_thermal_2018}

\subsection{Details of experimental data used for comparison}
    We have exhaustively collected experimental results reported in literature 
to compare thermal expansion and thermodynamic stability to for MD. In previous 
work we compared entropic and enthalpic contributions to polymorph relative 
stabilities, but have found some inconsistencies in those reported results~\cite{dybeck_capturing_2017} and 
reevaluate these calculations here. Provided with 
our supporting information are two files, experimental\_expansion.csv and 
experimental\_stability.csv, containing the literature results we used for 
comparison. In the expansion results file we provide the reported experimental 
lattice parameters, citation references and links, reported temperatures, and 
reference codes if the structure was submitted to the CCDC. For the stability 
results we report the values seen in Figure~\ref{figure:experimental_stability} 
along with the reported temperatures and literature references.

\section{Results and Discussion}
    The approaches presented above allow us to evaluate the effectiveness of 
QHA methods with different treatments of thermal expansion relative to MD for 
polymorph thermodynamics, including free energy differences between polymorphs, 
and lattice expansion. The differences between approaches for estimating the 
temperature dependence are greatest at high temperatures, so all comparisons 
are shown at the maximum temperature at which QHA is stable.

    For QHA, we found that discontinuous readjustment of the molecules within 
the lattice occurs frequently during expansion, causing the method to expand to a structure that 
no longer corresponds with free energy minimum of the chosen expansion variable.
As the crystal expands, the free volume around the molecules increases, allowing 
molecules to readjust into alternate minima more favorable to the lattice 
energy than the minima found by continuous deformation of lower temperature minima. 
These structural disruptions are problematic for the numerical stability of the 
gradient approach and generally cause the crystal to expand to a structure that 
is not a free energy minimum. For example, upon expansion 
of tolbutamide, at a certain point the alkyl tail can move into a newly created 
free volume, causing the quasi-harmonic free energy to become discontinuous 
with temperature.

    We determine if the crystal is at a free energy minimum with respect to 
the geometry parameter $y$ by checking if:
  \begin{eqnarray}
    \left(\frac{\partial G}{\partial y}\right)_{Backward} < 0 < \left(\frac{\partial G}{\partial y}\right)_{Forward}
    \label{eq:defining_Gmin}
  \end{eqnarray}
If our crystal is at or near a free energy minimum, then the forward and 
backward numerical solution to eq~\ref{eq:defining_Gmin} must be positive and 
negative respectively. If both are the same sign, then we must no longer be at 
a minimum.

\begin{figure*}
  \begin{center}
  \includegraphics[width=12cm]{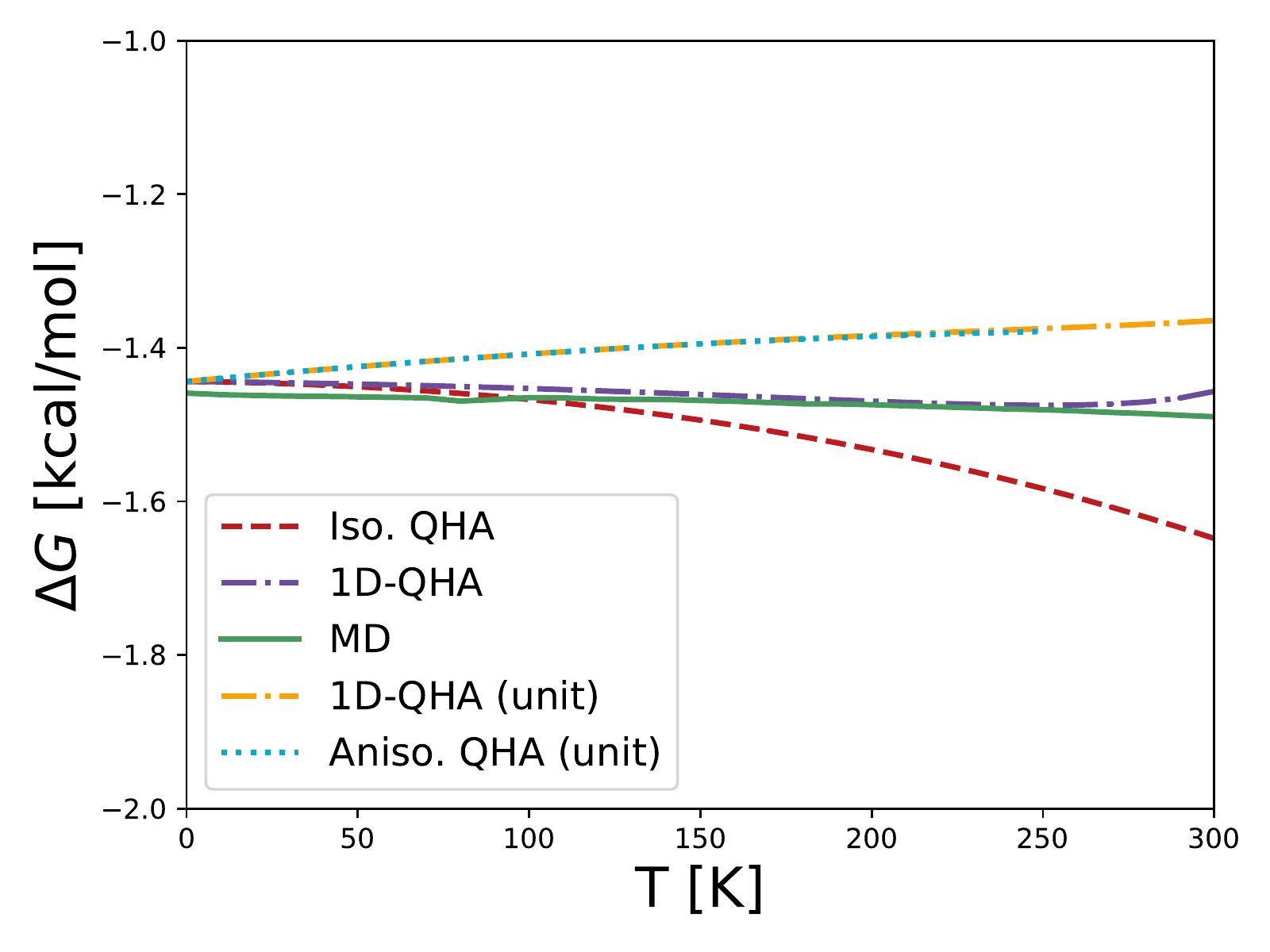}
  \end{center}
  \caption{Polymorph free energy differences of piracetam form I relative to form III
  \label{figure:bismev_dG}}
\end{figure*}

\begin{figure*}
  \begin{center}
  \includegraphics[width=16cm]{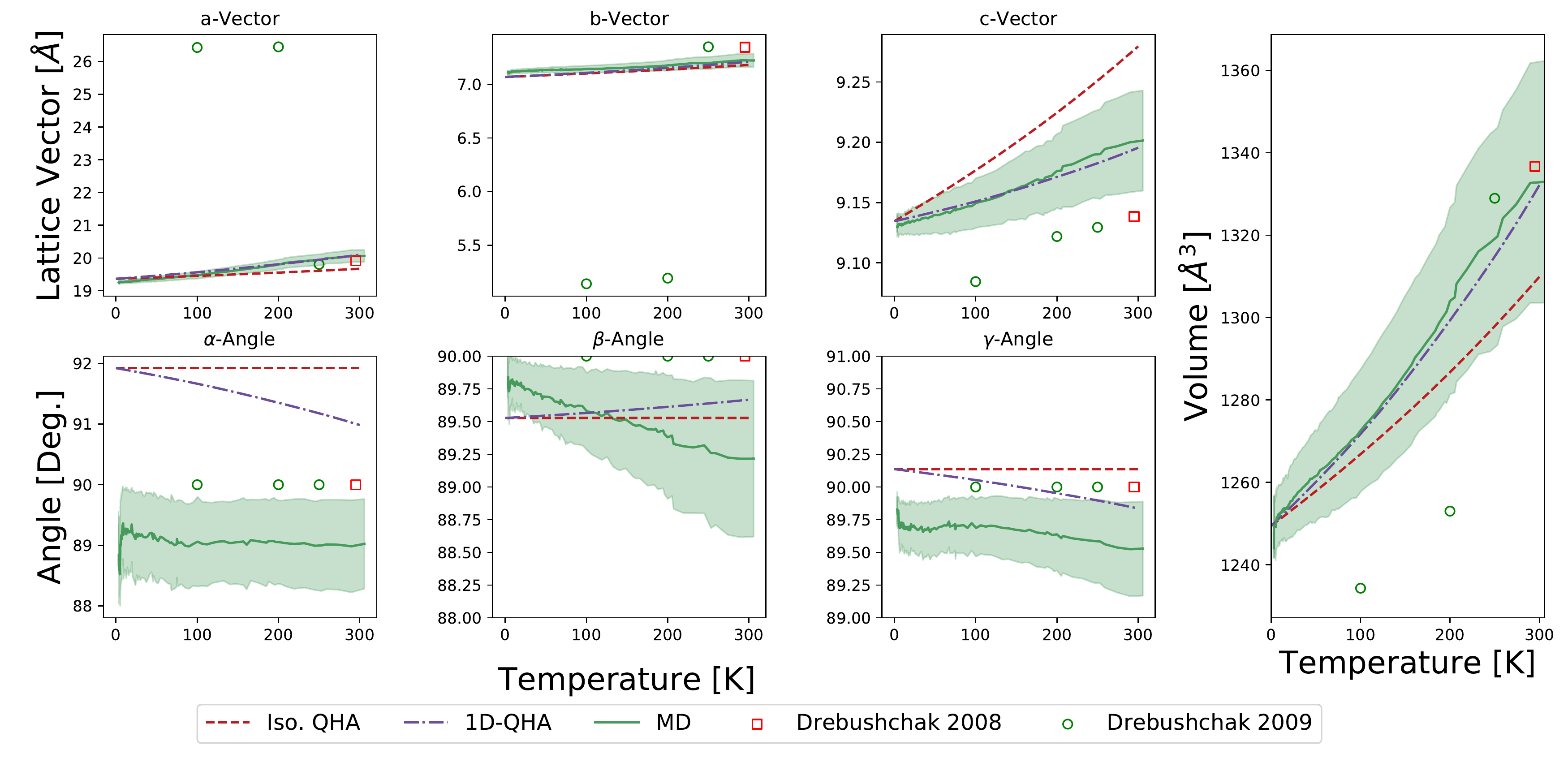}
  \end{center}
  \caption{Chlorpropamide form V-restructured percent expansion from 0 K with experimental results.~\cite{drebushchak_two_2008,drebushchak_conformational_2009}
   \label{figure:dh_bedmig7}}
\end{figure*}

    All results using QHA are shown at the maximum temperature ($T_{max}$), 
which is the highest temperature to satisfy eq~\ref{eq:defining_Gmin}.
The values of $T_{max}$ are reported in Table S\ref{table:1D_6D_dG} and 
Table S\ref{table:UvS_dG}. Plots of the polymorph free energy differences and 
lattice expansions versus temperature for the 10 molecules studied are provided 
(Figures S\ref{figure:bismev_dG} -- S\ref{figure:dh_bedmig7}). Further 
quantification of the numerical and structural instability is discussed in the 
work where we initially presented the gradient 
method.~\cite{abraham_thermal_2018}

\subsection{Comparisons to Experimental Results}
\begin{figure*}
      \begin{center}
      \includegraphics[width=8cm]{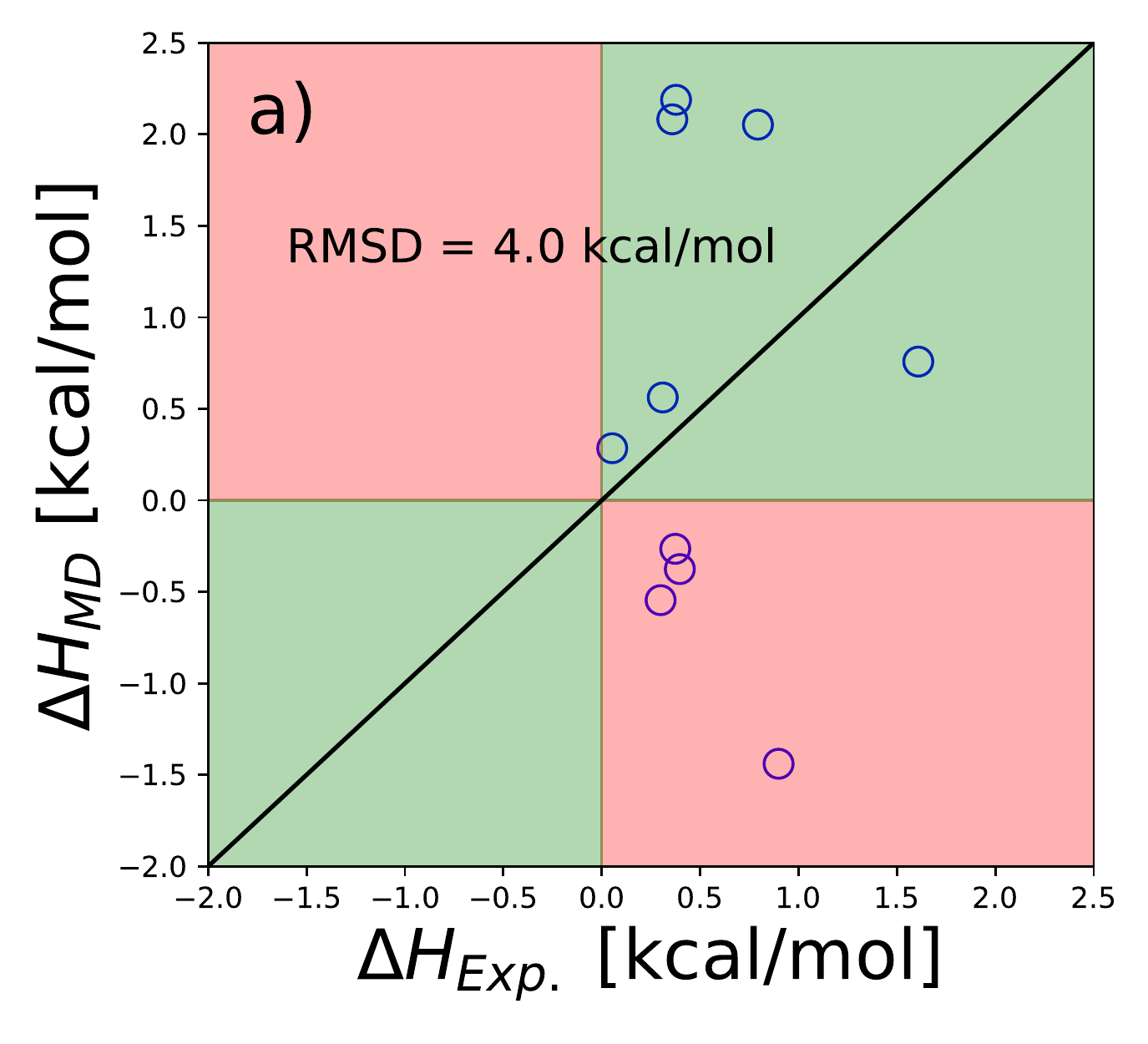}
      \includegraphics[width=8cm]{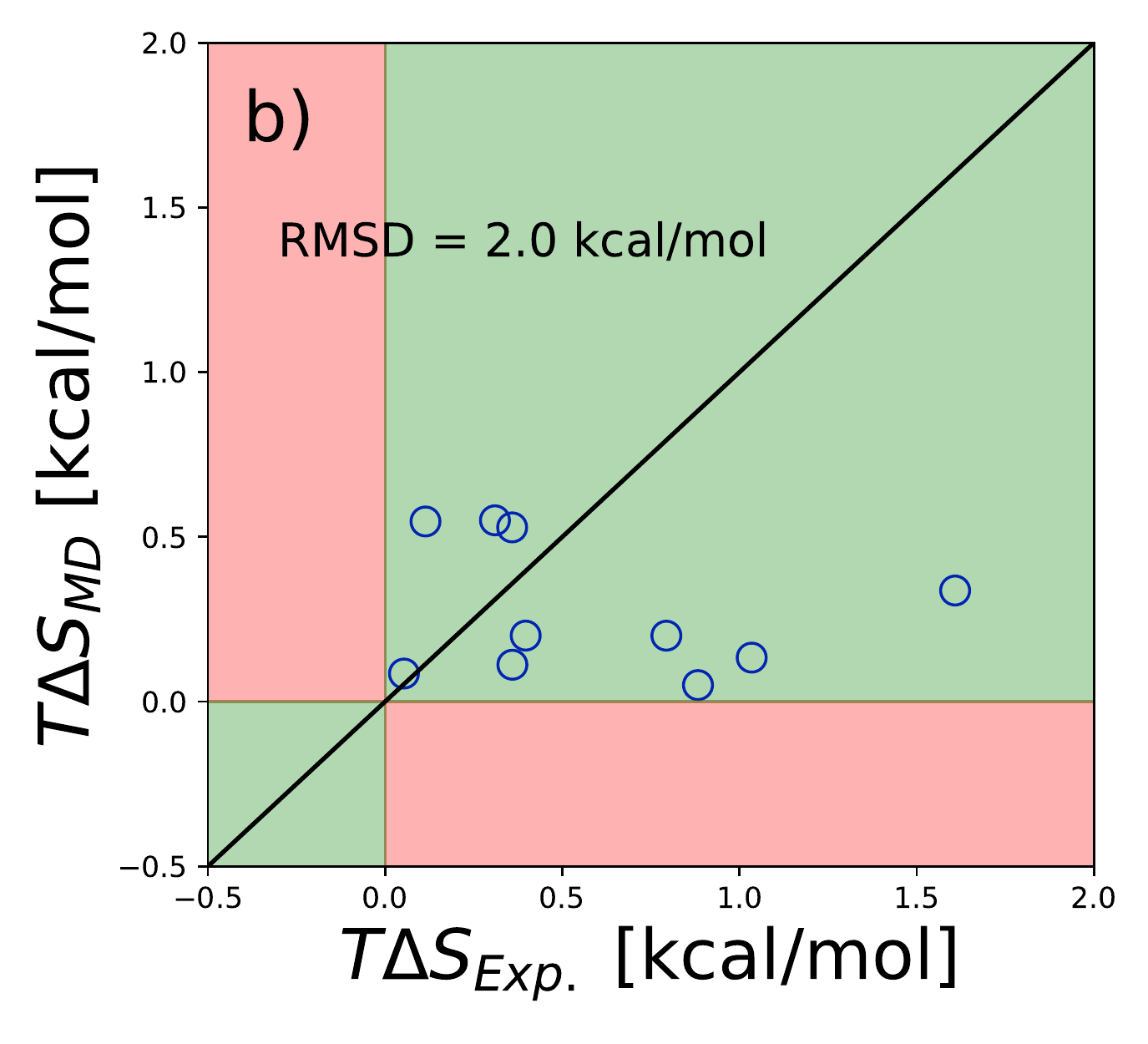}
      \end{center}
      \caption{300 K polymorph differences from MD for a) enthalpy and
      b) entropy versus experimental values taken from 
      literature.~\cite{kimura_characterization_1999,yoshino_contribution_1999,cherukuvada_pyrazinamide_2010,braun_conformational_2009,stolar_solid-state_2016,sacchetti_thermodynamic_2000,grzesiak_comparison_2003,maher_solubility_2012,van_miltenburg_low-temperature_2001,drebushchak_transitions_2008}
      The sign of the enthalpy is only correct for MD for 6 of the 10 molecules, with
      no clear distinguishing molecular or crystallographic features separating the crystals with
      the incorrect sign. This contrasts with the entropy, which is correct in sign for all 10 molecules.
      If the polymorph pairs fall in the green quadrants than MD has the same sign as experiment, and
      the wrong sign in the red quadrants.
      \label{figure:experimental_stability}}
    \end{figure*}

    We found that the OPLS-AA point charge potential is a poor predictor of the 
sign of enthalpic differences for polymorphs relative to experiment, as is 
expected for the simplicity of the energy function, but does accurately 
estimate the sign of the entropic differences. In 
Figure~\ref{figure:experimental_stability}a and 
\ref{figure:experimental_stability}b the enthalpic and entropic contributions 
to the polymorph free energy differences at 300 K for the supercells using MD 
are shown relative to their experimental values, which are reported at varying
temperatures that generally corresponding with the polymorph transition 
temperature.~\cite{kimura_characterization_1999,yoshino_contribution_1999,cherukuvada_pyrazinamide_2010,braun_conformational_2009,stolar_solid-state_2016,sacchetti_thermodynamic_2000,grzesiak_comparison_2003,maher_solubility_2012,van_miltenburg_low-temperature_2001,drebushchak_transitions_2008}
The sign of the enthalpic contributions is only correct for 6 of the 10 
polymorph pairs. The 4 molecules where the simulation has the wrong enthalpic 
ranking are piracetam, pyrazinamide, resorcinol, and paracetamol, and none of 
the enthalpic errors in these molecules have obvious correlations to chemical 
groups present, flexibility, crystal Z and Z' values, or relative packing. 
This contrasts with the polymorph entropy differences, where the sign is 
correct for MD for all 10 molecules. The RMSD between theory and experiment are 
4.0 and 2.0 kcal/mol for $\Delta H$ and $T \Delta S$, respectively. The most 
likely large sources of deviation from experiment could be due to 1) the experimental
values coming from the polymorph transition temperature, not 300 K and 2)
the use of a point charge potential to model the crystal energetics.

\begin{figure*}
      \begin{center}
      \includegraphics[width=16cm]{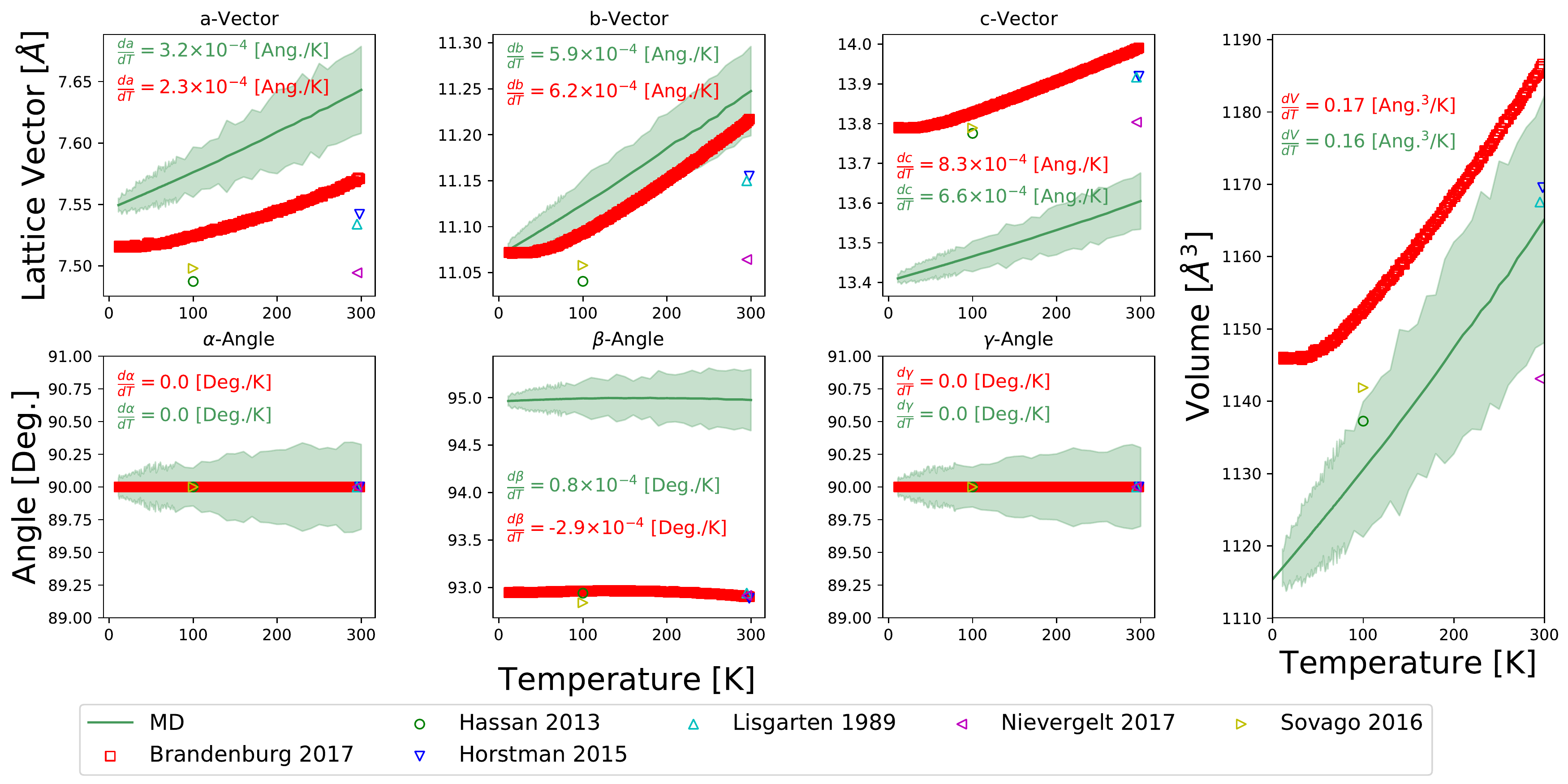}
      \end{center}
      \caption{Thermal expansion of carbamazepine form III comparing MD to experimental
      results provided in literature.~\cite{brandenburg_thermal_2017,horstman_crystallization_2015,lisgarten_crystal_1989,nievergelt_growing_2016,sovago_electron_2016,el_hassan_electron_2013}
      While the lattice parameters differ between theory and experiment, the thermal expansion
      is similar between MD and experiment for all lattice parameters and volume. At
      low temperatures we find the largest deviation between experiment and MD. The shaded green area is the standard deviation in the lattice parameter during the MD simulation.
      \label{figure:dh_cbmzpn_p3}}
    \end{figure*}

\begin{figure*}
  \begin{center}
  \includegraphics[width=16cm]{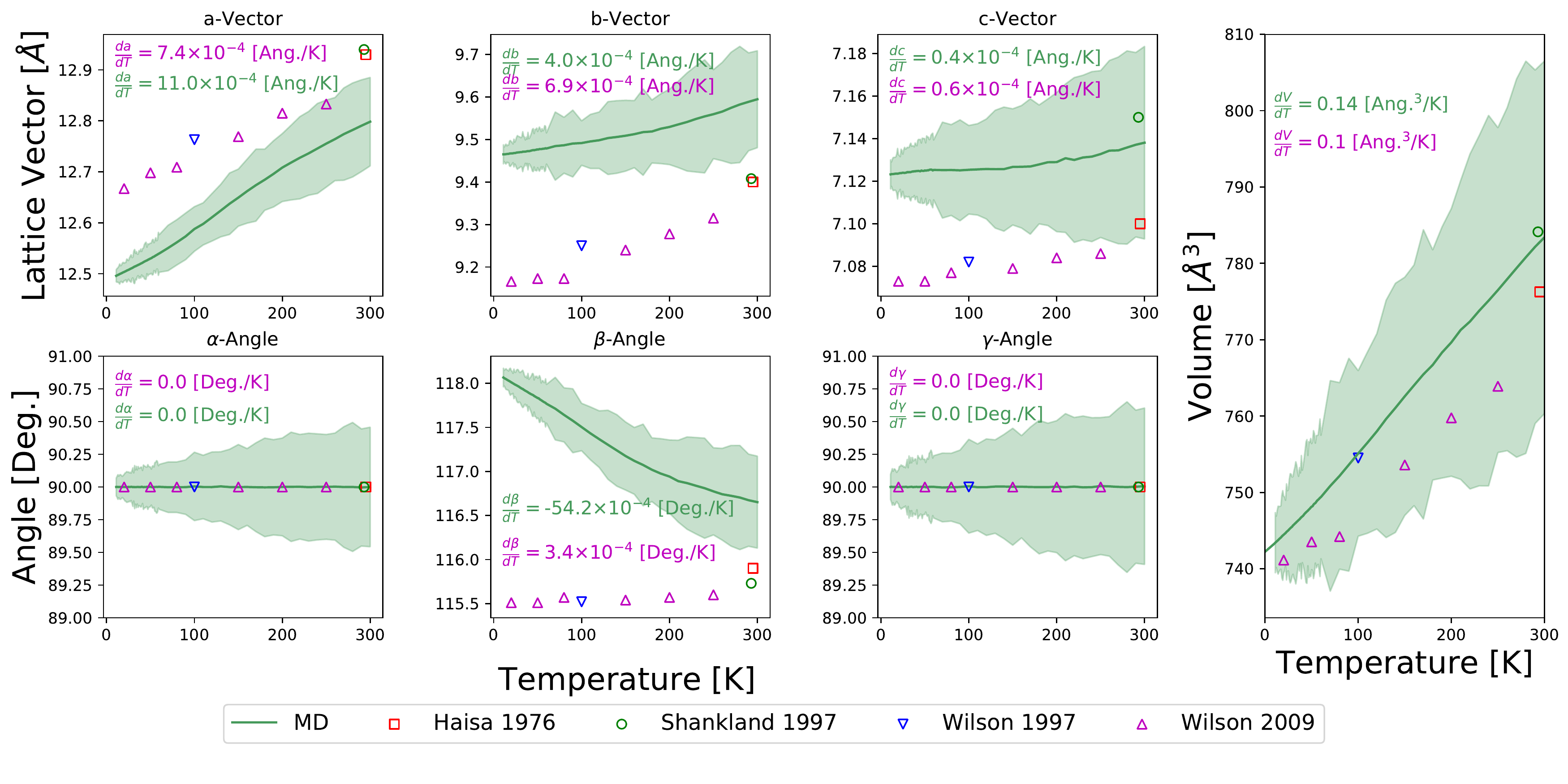}
  \end{center}
  \caption{Thermal expansion of paracetamol form I comparing MD to experimental
      results provided in literature.~\cite{haisa_monoclinic_1976,wilson_variable_2009,wilson_single-crystal_1997,wilson_neutron_1997}
      While the lattice parameters differ between theory and experiment, the thermal expansion
      is similar between MD and experiment for all lattice parameters and volume,
      except for the $\beta$ angle. The shaded green area is the standard deviation in the lattice parameter during the MD simulation.
  \label{figure:dh_hxacan_p1}}
\end{figure*}

    The lattice geometries for MD differ moderately from experiment, though the 
volumes match experiment more closely. We can quantitatively compare to 
experiment using exhaustive experimental results from Brandenburg and Wilson 
for carbamazepine form III and paracetamol form I, 
respectively.~\cite{brandenburg_thermal_2017,wilson_variable_2009} The 
expansion of the lattice vectors, angles, and volume are shown for 
carbamazepine form III (Figure~\ref{figure:dh_cbmzpn_p3}) and paracetamol form 
I (Figure~\ref{figure:dh_hxacan_p1}). Prediction accuracy of lattice vectors 
varies both with crystal and lattice vector. We find that if the lattice 
vectors are orthogonal for experiment, MD always gets the corresponding 
angle correct or within error as seen with $\alpha$ and $\beta$ angles for 
carbamazepine and paracetamol. Despite errors in the individual lattice 
parameters, the crystal volume is similar between MD and experimental values. 
In the case of paracetamol (Figure~\ref{figure:dh_hxacan_p1}) the experimental 
volume is within error of MD.

    The change in the lattice vectors and orthogonal angles in MD generally 
agrees with experiment despite errors in the actual geometry. In 
Figures~\ref{figure:dh_cbmzpn_p3} and~\ref{figure:dh_hxacan_p1} we provide the 
thermal expansion of each parameter by fitting a linear fit to MD and 
experimental data at all temperatures. In carbamazepine the change in the slope of 
experimental lattice parameters at low temperatures could be indicative of 
quantum behavior from the zero point energy, so for this crystal we perform the linear fit on values above 
100 K. For all of the lattice vectors we see that the thermal expansion between 
MD and experiment are the same order of magnitude and sign. For the orthogonal 
angles there is no expansion. However, the thermal expansion of the 
$\beta$ angle in both polymorphs has the wrong sign and in the case of paracetamol
form I is an order of magnitude different. For carbamazepine, we see larger divergence between 
experiment and MD at low temperatures ($T <$ 100 K), which would be indicative
of quantum zero-point energy effects. For both polymorphs there is also good agreement with 
experiment for the volumetric thermal expansion. Accurately 
modeled thermal expansion helps to explain why we also get the correct sign of 
the entropic differences for all 10 polymorph pairs in 
Figure~\ref{figure:experimental_stability}b.

\subsection{Testing the Validity of 1D-Anisotropic QHA}
    Anisotropic expansion is relatively expensive compared to isotropic 
expansion, though still 2 orders of magnitude cheaper than free energy 
estimation with MD. Our previous work shows that a 1D-variant of 
anisotropic expansion can be a sufficient substitute to speed up the method 
with little effect on the accuracy of computed free energy 
differences.~\cite{abraham_thermal_2018} Fully anisotropic expansion requires 73 
structure optimizations every time the thermal gradient is computed. This 
contrasts with the 1D-approach, which requires the initial 73 optimizations at 
0 K followed by 3 optimizations at all subsequent points for numerical 
integration. When we first presented the gradient approach we found that the 
1D-anisotropic QHA approach computed polymorph free energy differences within 
0.01 kcal/mol of full anisotropic expansion for the two sets of polymorphs 
tested. For this study, we ran both 1D-QHA and anisotropic QHA on unit cells of 
the quenched experimental structures (independent of MD) to evaluate the 
effectiveness of the 1D constraint. 

    The constraints applied to 1D-anisotropic approach provide numerical 
stability, allowing eq~\ref{eq:defining_Gmin} to be satisfied at 300 K more
frequently. At each integration step for expansion our program checks if
eq~\ref{eq:defining_Gmin} is satisfied, verifying if the crystal is at a free 
energy minimum. We will only report result up to the temperature ($T_{max}$)
where eq~\ref{eq:defining_Gmin} is satisfied. When using fully anisotropic
expansion, only 11 of the 20 crystals were able to satisfy eq~\ref{eq:defining_Gmin}
at all temperatures ($T_{max} = $ 300 K). This contrasts with the 1D-approach, 
where 19 of the 20 crystals remained at a free energy minimum up to 300 K.
The only molecules where the fully anisotropic approach remained at a free energy
minimum for both polymorphs were the relatively rigid resorcinol, adenine, and 
carbamazepine, suggesting that rigid molecules are less likely to experience 
numerical disruptions in QHA. 

\begin{figure*}
  \begin{center}
  \includegraphics[width=8cm]{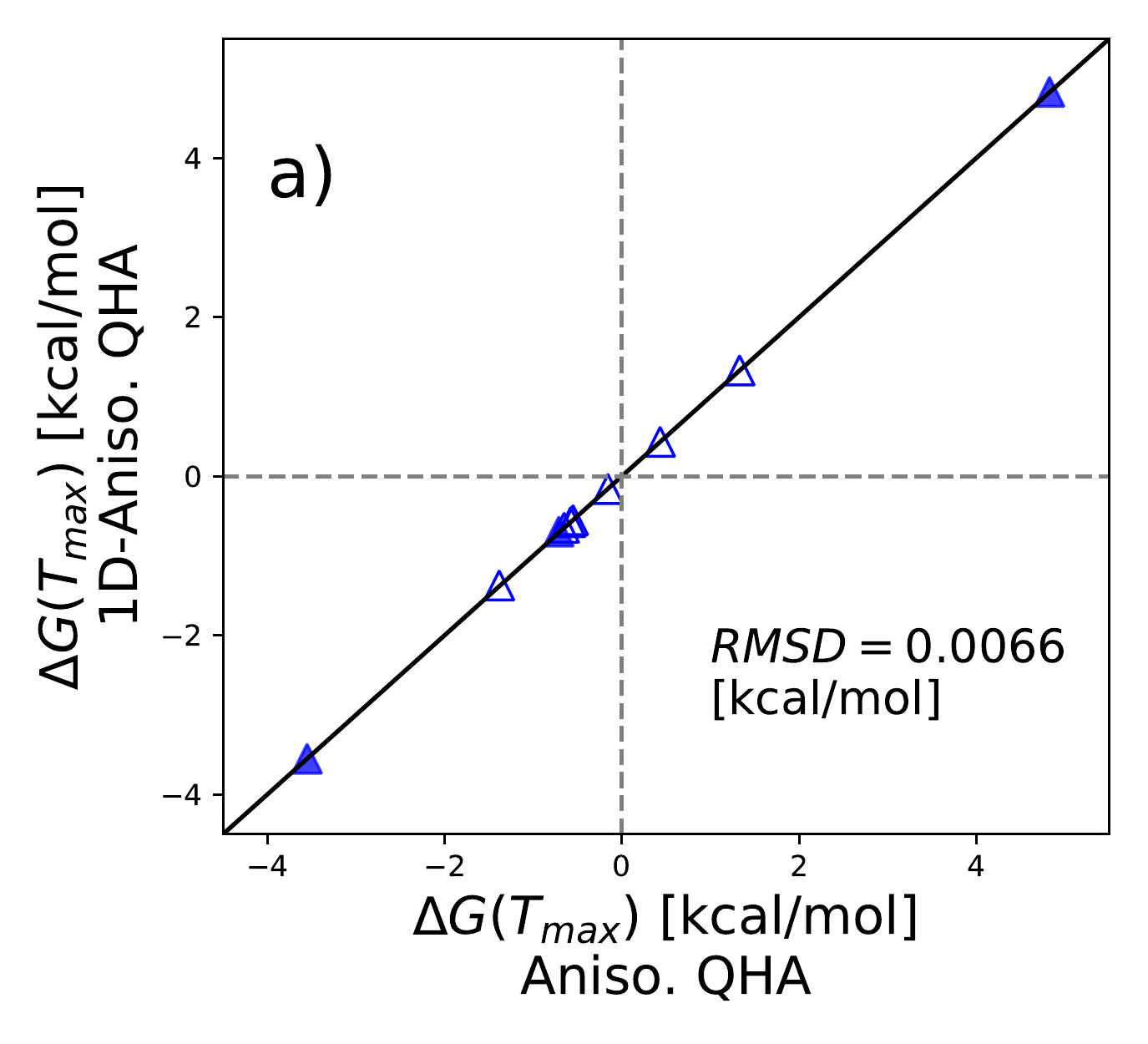}
  \includegraphics[width=8cm]{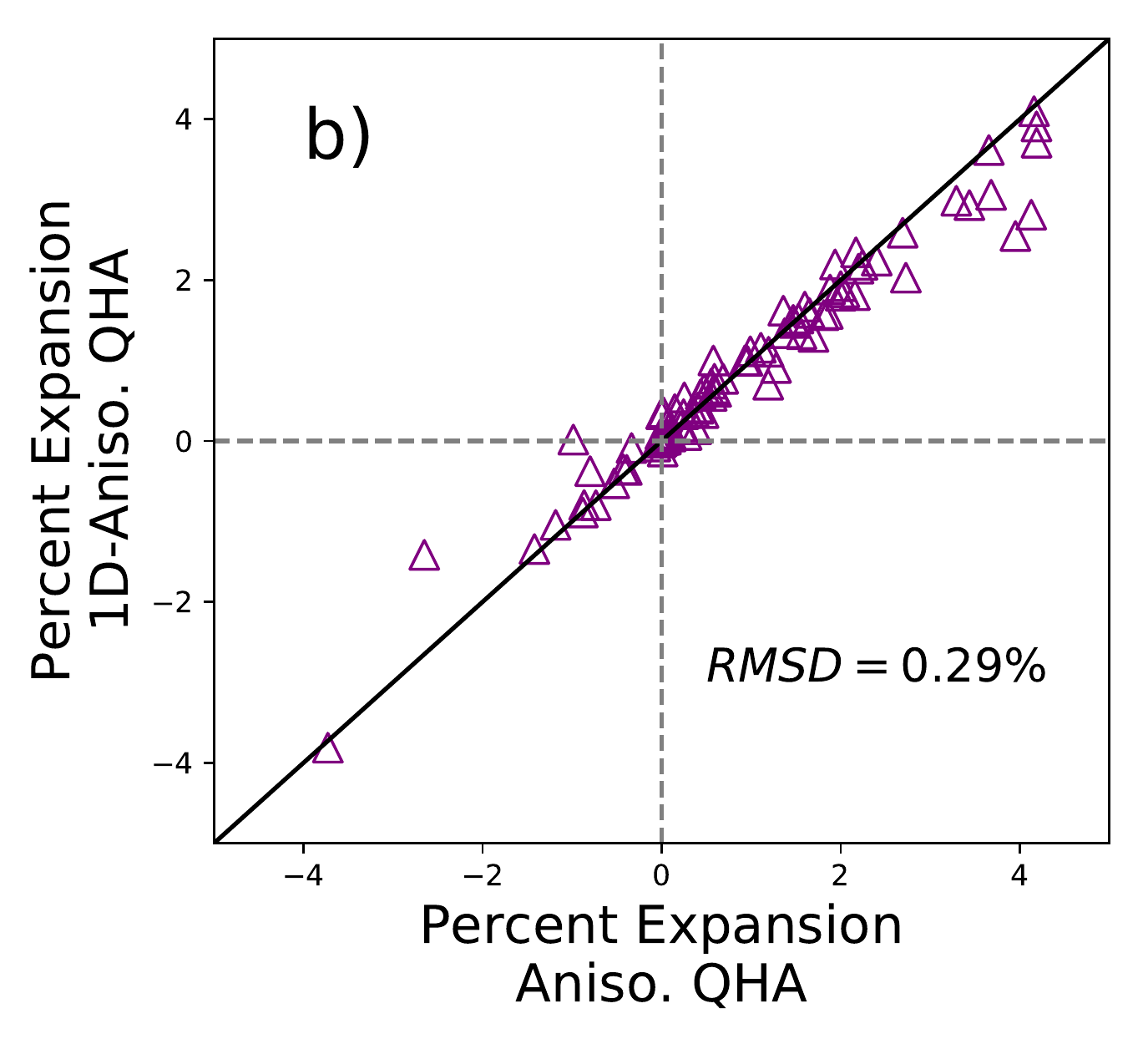}
  \end{center}
  \caption{Comparison of the results of 1D- and fully-anisotropic QHA applied to the 
polymorphic unit cells. In all plots, the filled-in markers are the molecules with multiple 
molecular conformers (tolbutamide, chlorpropamide, and aripiprazole). a) Polymorph 
free energy differences ($\Delta 
G(T_{max})$) computed with 1D and full anisotropic approaches are nearly equal.
The black line is plotted as $y=x$ and the data has a RMSD = 0.0066 
kcal/mol relative to that line. b) The percent expansion from 0 K to $T_{max}$ 
for the three lattice vectors and angles are plotted. The RMSD in percent 
expansion between 1D and anisotropic QHA is 0.29\%. 
  \label{figure:1D_vs_6D}}
\end{figure*}

    For the 10 molecules in this study, the RMSD in polymorph free energy 
difference between 1D-anisotropic and anisotropic QHA is 0.0066 kcal/mol, 
which is well below physically meaningful sensitivity thresholds. 
Figure~\ref{figure:1D_vs_6D}a shows the polymorph free energy differences 
computed with 1D-QHA versus those computed with anisotropic QHA at 
$T_{max}$. The free energies of most polymorphic pairs vary by less than 0.010 
kcal/mol between 1D-QHA and anisotropic QHA. The only exception is 
aripiprazole, where the two QHA methods for determining $\Delta G$ between 
polymorphs differ by 0.016 kcal/mol. The error in aripiprazole comes primarily 
from form X. The error in the individual free energies of polymorphs between 
1D and fully anisotropic QHA is $<0.006$ kcal/mol (excluding aripiprazole 
form X), whereas the error in isotropic expansion is between 0.01--0.21 
kcal/mol (Figure~S\ref{figure:1D_vs_6D_G}). Similarly, the high temperature 
lattice geometries of the two approaches have an RMSD of 0.29\% for the 
expansion relative to lattice minimum structure.

    For the molecules studied, 1D-anisotropic QHA is thus a sufficiently accurate approach 
to model anisotropic expansion for most purposes, and is much more efficient.
The 1D approach produces polymorph free energy differences within 0.02 kcal/mol 
of the fully anisotropic approach at about 10\% of the computational cost. More 
importantly, the 1D approach has greater numerical stability than full 
anisotropic expansion, allowing us obtain results at higher temperatures. The 
close numerical agreement between the 1D and fully anisotropic approach up to 
$T_{max}$ shows that the 1D approach is a realistic constraint for these molecules. 
We will therefore use the 1D approach for all further evaluations for anisotropic QHA.

\subsection{The Cell Size Has a Non-negligible Effect on Polymorph Stability}
\begin{figure}[H]
  \begin{center}
  \includegraphics[width=8cm]{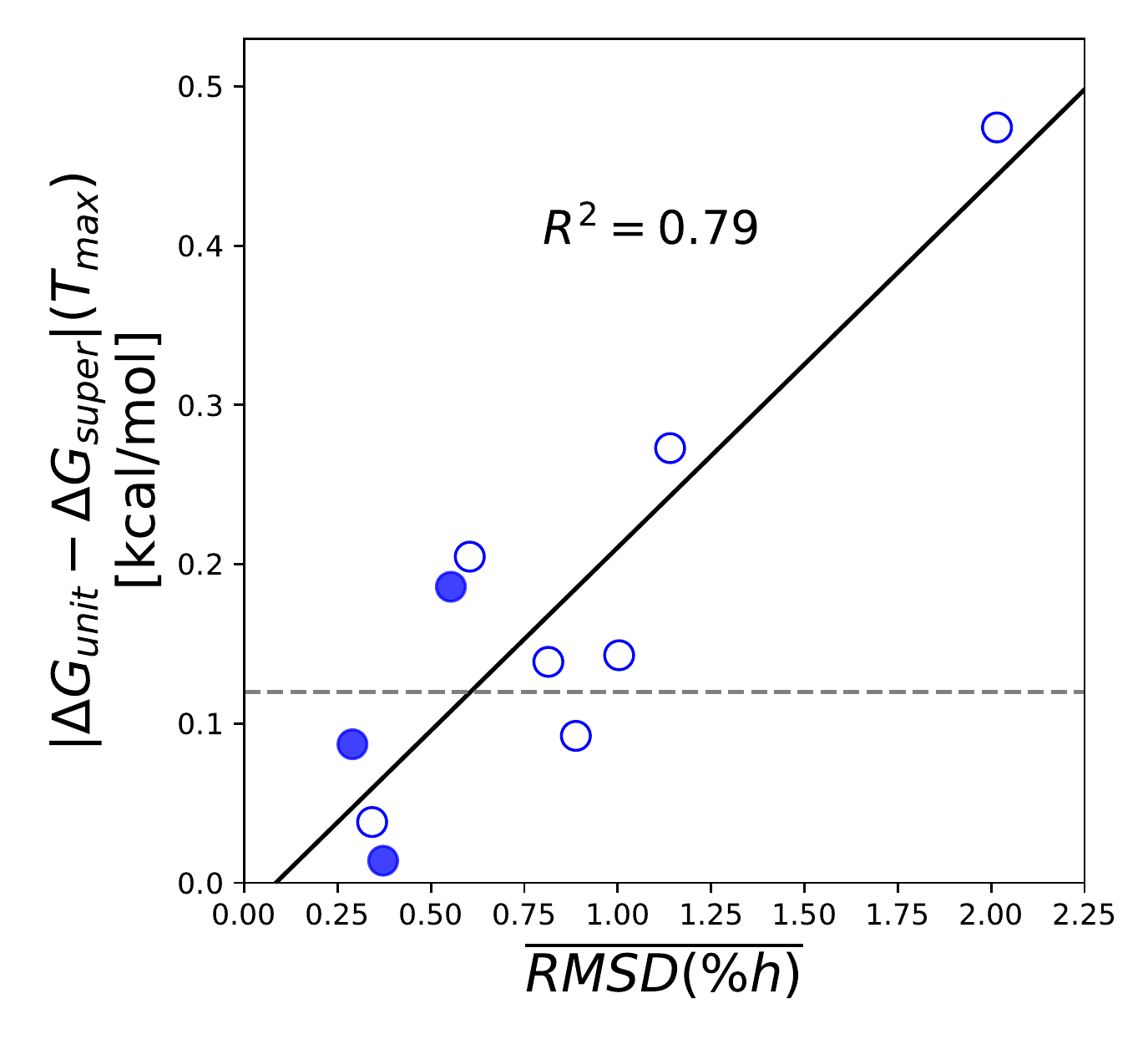}
  \end{center}
  \caption{Differences between 1D-QHA in unit and supercells are shown for 
polymorph free energy differences against the average RMSD in percent expansion 
at $T_{max}$. There is a moderate ($R^{2} = 0.79$) correlation between the 
difference in unit and supercell percent expansion and the finite size error in free energy.
The grey line is the 0.12 kcal/mol threshold, where values below the line have a 
90\% estimated confidence that re-ranking in crystal stability would not 
occur.~\protect{\cite{nyman_static_2015}} In all plots, the filled-in markers 
are the molecules with multiple molecular conformers (tolbutamide, chlorpropamide, 
and aripiprazole).
  \label{figure:U_vs_S}}
\end{figure}

    We observe differences in the quasi-harmonic free energy differences 
between unit cells and supercells ranging from 0.04--0.5 kcal/mol and are 
loosely correlated to differences in the high temperature cell geometry. In our 
previous study we assumed that an energy contribution less than 0.12 kcal/mol would 
give us 90\% confidence that a re-ranking in crystal stability would not 
occur,~\cite{dybeck_capturing_2017} a number based on a previous much larger 
study of lattice energies versus quasi-harmonic free 
energies.~\cite{nyman_static_2015} Six of the ten polymorphic pairs in Figure 
\ref{figure:U_vs_S} exceed that cutoff, showing that finite size errors are 
sufficiently large to affect the stability ranking of polymorphs. All unit cells
were quenched from the experimental crystals prior to running QHA and the 
supercells were constructed from the quenched unit cells. These results used 
the 1D-anisotropic approach for QHA.

   The error in the free energy differences between unit and supercells is correlated to the difference in expansion 
of the lattice parameters between unit and supercells. In Figure~\ref{figure:U_vs_S} we plot the 
deviations in the polymorph free energy differences versus the average RMSD of 
the percent expansions in both crystals. We found that there was a moderate 
linear correlation ($R^{2} = 0.79$) between percent expansion and deviations in 
the free energy due to finite size effects. Tolbutamide and chlorpropamide 
(filled-in markers) still fall under the cutoff suggesting, though with a 
limited data set, that there is not necessarily significant correlation between 
molecular flexibility and finite size error. The largest error (0.47 kcal/mol) 
is for paracetamol where  the supercell is only 6 times the size of the unit 
cell. This is the smallest difference between unit and super cell used in this 
study, with the largest being 24 times. Furthermore, we looked at the 
correlation between finite size errors and the ratio of unit and super cell and 
find that there is no correlation ($R^{2}=0.109$ in Figure 
S\ref{figure:size_correlation}). The cell dimensions of supercells relative to 
unit cells can be found in Table S\ref{table:systems}.

\subsection{Peforming QHA with a Lattice Minima Consistent with MD Ensembles}
   The lattice minimum structure quenched from experiment does not always 
coincide with the low temperature structures found from MD simulation of the 
same model, raising the question of what it means to compare QHA to MD. For some 
of the polymorphs studied here, we have found that the crystal supercells 
clearly reorganized into different configurational ensembles when heated up to 
300 K using MD and can be quenched to a number of lattice 
minimum.~\cite{dybeck_capturing_2017,dybeck_exploring_2019} This suggests than 
QHA can be performed either independent or dependent of MD equilibration. QHA 
independent of MD equilibration, using a minimized experimental structure, has 
the potential risk of occupying a different basin than the MD simulation does, 
leading to potentially large differences in both lattice energy and 
configurational entropy between the approaches. In contrast, if QHA is run from 
a thermally annealed structure of the MD simulations, we have a better chance of 
matching the high-temperature MD ensemble with QHA, as the lattice energy and 
vibrational modes will be more nearly the same, giving a purer test of whether 
the high temperature behavior is indeed quasi-harmonic.

\begin{figure}[H]
  \begin{center}
  \includegraphics[height=6cm]{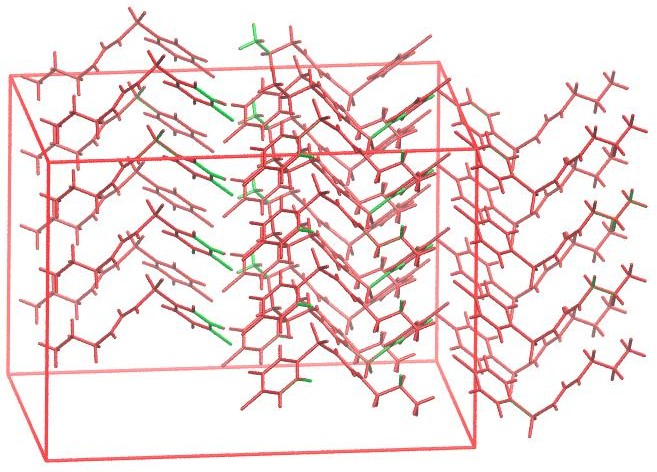}
  \end{center}
  \caption{Lattice minimum of chlorpropamide form I from quenching the experimental structure (green), the structure used in our previous study (blue),~\cite{dybeck_capturing_2017} quenched from a 10 K replica exchange simulation (red). The RMSD between all structures is $<$ 0.0001 \r{A}.
  \label{figure:restruc_bedmig1}}
\end{figure}

\begin{figure}[H]
  \begin{center}
  \includegraphics[height=5.5cm]{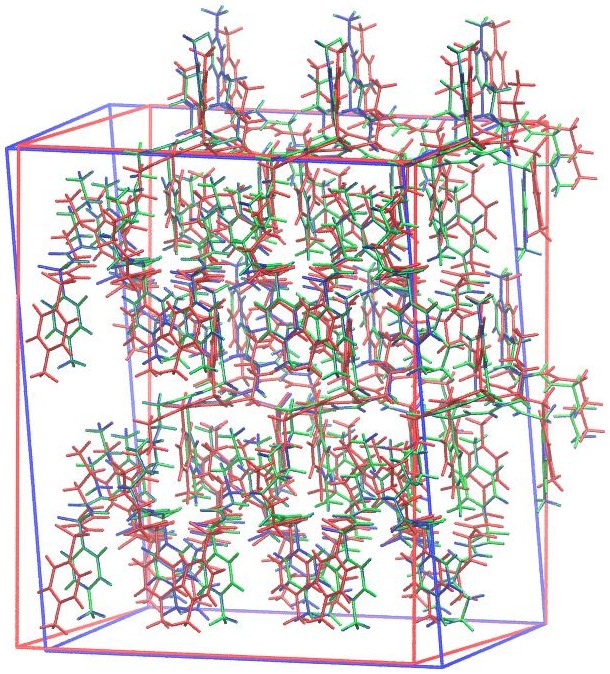}
  \end{center}
  \caption{Lattice minimum of tolbutamide form I from quenching the experimental structure (green), the structure used in our previous study (blue),~\cite{dybeck_capturing_2017} quenched from a 10 K replica exchange simulation (red). The RMSD between the structure quenched from experiment and from our previous paper relative to the minima found from REMD are  1.35 and 1.35 \r{A}, respectively.
  \label{figure:restruc_zzzpus2}}
\end{figure}

    To maximize the chance that we can identify the free energy minimum at low 
temperature, we performed temperature replica exchange molecular dynamics for the 
polymorphs of tolbutamide, chlorpropamide, and aripiprazole. For these six polymorphs,
five frames from the equilibrated 10 K REMD simulation were quenched in order to find
a lattice minimum for QHA consistent with the MD ensemble. All frames from the MD 
trajectory quenched to minima that had their potential energy within 0.02 
kcal/mol, lattice vectors within 0.2 \r{A}, and angles within 
1.0\textdegree~from one another. Differences in minima arose from 
conformations of the alkyl tails in both polymorphs of tolbutamide and 
chlorpropamide form V. In Table~\ref{table:restruc_U} we report the potential 
energy of the polymorphs when quenched from experiment, the lattice minimum used 
in our 2017 paper,~\cite{dybeck_capturing_2017}  and the lowest energy structure 
quenched from REMD, relative to minimum found using REMD. Quenching from REMD 
produces the lowest lattice energy structure for all polymorphs. Overlaid 
structures of restructured crystals are provided in Figures 
S\ref{figure:restruc_bedmig1}--S\ref{figure:restruc_zzzpus2}.

\begin{center}
\begin{table*}
\caption{Potential Energy of Lattice Minimum Relative to 10 K REMD Quenched Structure
\label{table:restruc_U}}
\begin{tabular}{ |l|c|c|c|c|c|c| }
\hline
 & \multicolumn{2}{|c|}{Tolbutamide} & \multicolumn{2}{|c|}{Chlorpropamide} & \multicolumn{2}{|c|}{Aripiprazole} \\
 \hline
 & I & II & I & V & I & X \\
 \hline 
$\Delta U_{\mathrm{Exp.}}$\textsuperscript{\emph{a}} & 1.837 & 0.250 & 0.000 & 3.095 & 1.989 & 0.302 \\

$\Delta U_{\mathrm{Previous}}$ \textsuperscript{\emph{b}} & 0.442 & 0.250 & 0.000 & 0.294 & 0.048 & 0.000 \\

$\Delta U_{\mathrm{REMD}}$ & 0.000 & 0.000 & 0.000 & 0.000 & 0.000 & 0.000 \\
\hline
\end{tabular}

\textsuperscript{\emph{a}} Potential energy differences of the lattice minimum quenched from the experimental structure and quenched from a 10 K REMD simulation;
\textsuperscript{\emph{b}} Potential energy difference of the lattice minimum quenched from the restructured crystal found in our previous study~\protect{\cite{dybeck_capturing_2017}} and quenched from a 10 K REMD simulation.
\end{table*}
\end{center}

    When the lowest lattice energy minimum is chosen, the error between 
isotropic QHA and MD is reduced by 0.083 kcal/mol from the error previously 
reported.~\cite{dybeck_capturing_2017} The error previously reported for these 
three systems was 0.083 -- 0.397 kcal/mol, but with the new minima is 0.014 -- 
0.247 kcal/mol. In Table ~\ref{table:restruc_G} the error of QHA for polymorph 
free energy differences relative to MD using replica exchange is shown at 
$T_{max}$ for the minimum found from our previous study and the minimum quenched 
from the 10 K REMD simulation. We exclude the lattice minimum quenched from the 
experimental structure since the lattice energy differences are so large. For 
isotropic QHA, the minimum quenched from REMD reduces the error relative to MD 
by 0.15 kcal/mol for tolbutamide and chlorpropamide. We see an increase in the 
error for aripiprazole, which is minimal compared to the reduction in error for 
the other two sets of polymorphs.

    When using the lower lattice energy minimum, anisotropic QHA determines 
the stability of chlorpropamide and aripiprazole within 0.05 kcal/mol relative to MD. 
The error in QHA for tolbutamide, 0.206 kcal/mol, is still large, but is half the 
magnitude of what was previously reported.~\cite{dybeck_capturing_2017} Despite 
the new minimum of aripiprazole producing higher error in isotropic QHA, using
anisotropic expansion reduced the error to 0.045 kcal/mol, implying that an 
isotropic constraint is inappropriate. The small increase in error when switching 
to anisotropic treatments seen in chlorpropamide for anisotropic QHA is most 
likely due to cancellation of error between polymorphs. For these polymorphs, all 
further analysis of QHA will be shown for the restructured crystals found from 
quenching low-temperatures replicas of REMD simulations.
\begin{center}
\begin{table*}
\caption{Error of QHA Relative to MD at $T_{max}$ Previously Reported~\cite{dybeck_capturing_2017} and Using a Lattice Minimum Quenched from REMD
\label{table:restruc_G}}
\begin{tabular}{ |l|c|c|c|c|c|c| }
\hline
 & \multicolumn{2}{|c|}{Tolbutamide} & \multicolumn{2}{|c|}{Chlorpropamide} & \multicolumn{2}{|c|}{Aripiprazole} \\
 \hline
QHA & Iso. & Aniso. & Iso. & Aniso. & Iso. & Aniso. \\
 \hline 
$\delta(\Delta G_{\mathrm{Prev.}})$\textsuperscript{\emph{c}}
& 0.397 & -- & 0.159 & -- & 0.083 & -- \\
$\delta(\Delta G_{\mathrm{REMD}})$\textsuperscript{\emph{d}}
& 0.247 & 0.206 & 0.014 & 0.033 & 0.128 & 0.045 \\
\hline
\end{tabular}

\textsuperscript{\emph{c}} Error between QHA and MD using the lattice minimum in our previous study~\cite{dybeck_capturing_2017};
\textsuperscript{\emph{d}} Error between QHA and MD using the lattice minimum quenched from the 10 K replica of REMD.
\end{table*}
\end{center}

\subsection{Comparison of thermal expansion and thermodynamics using QHA and MD}
\begin{figure*}
      \begin{center}
      \includegraphics[width=8cm]{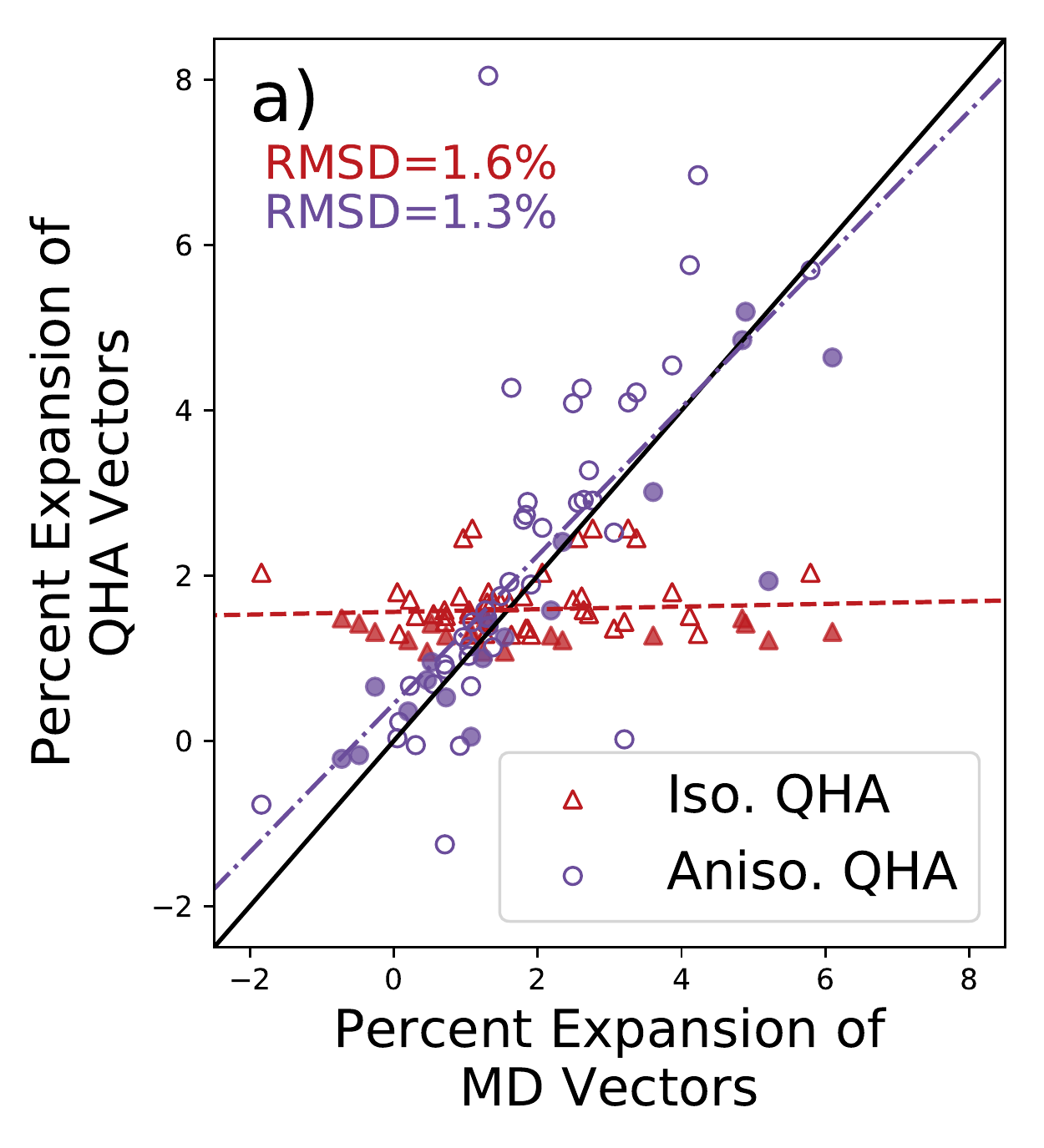} 
      \includegraphics[width=8cm]{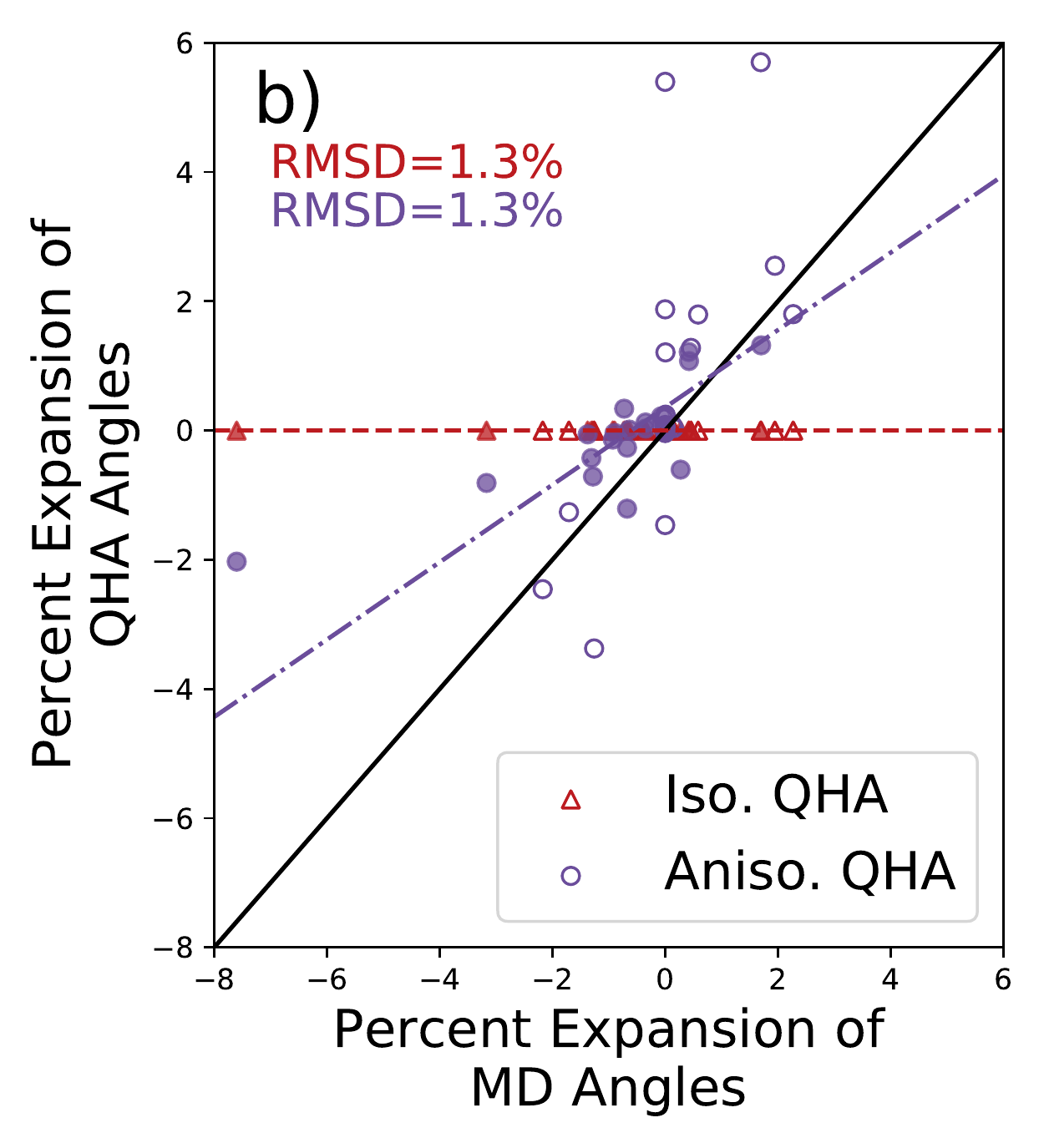}
      \end{center}
      \caption{Scatter plots of the percent expansion of QHA versus MD for the 
lattice a) vectors and b) angles for isotropic and 1D-anisotropic QHA lattice 
parameters. The RMSD to MD is marginally improved for the lattice vectors and 
not improved for the angles when using anisotropic expansion over isotropic 
expansion. A least square fit between MD  anisotropic QHA shows that anisotropic 
expansion provides a better $y=x$ fit (black line) to MD than isotropic 
expansion. The slopes for anisotropic and isotropic expansion are a) 
0.89$\pm$0.22 and 0.02$\pm$0.06\ and b) 0.60$\pm$0.14 and 0.0 (exact), 
respectively. In all plots, the filled-in markers are the molecules with 
significant internal conformational freedom (tolbutamide, chlorpropamide, and 
aripiprazole).
     \label{figure:MD_expansion}}
    \end{figure*}

    Anisotropic QHA allows the crystal lattice to relax into geometries that are 
more similar to MD, but there are still persistent errors in the high 
temperature geometries computed with QHA. In Figure~\ref{figure:MD_expansion}a 
and Figure \ref{figure:MD_expansion}b we show the percent expansion of QHA 
lattice vectors and angles against MD, respectively. The black line represents a 
perfect agreement between the QHA method and MD, while the dashed lines are 
least squared fits between the two QHA methods and MD. For the lattice vectors, 
there is marginal improvement in the RMSD when using anisotropic expansion 
(1.3\%) over isotropic QHA (1.6\%). There is no improvement to RMSD for the 
lattice angles (anisotropic 1.3\%, isotropic 1.3\%). Despite there being 
little-to-no reduction in the RMSD's, the least square fit between anisotropic 
QHA and MD is similar to the $y=x$ line for both lattice vectors and angles. 
This contrasts with isotropic QHA, which is essentially independent of MD box 
vectors, which is expected for the fixed isotropic angles. Quantitatively, the 
slopes for anisotropic and isotropic expansion are 0.89$\pm$0.22 and 
0.02$\pm$0.06 for the box vectors ($a$, $b$, and $c$) and 0.60$\pm$0.14 and 0.0 
(exactly) for the box angles ($\alpha$, $\beta$ and $\gamma$), indicating that 
while anisotropic expansion may over or undershoot the changes, the directions 
of change are generally being properly predicted.

\begin{figure*}
      \begin{center}
      \includegraphics[width=8cm]{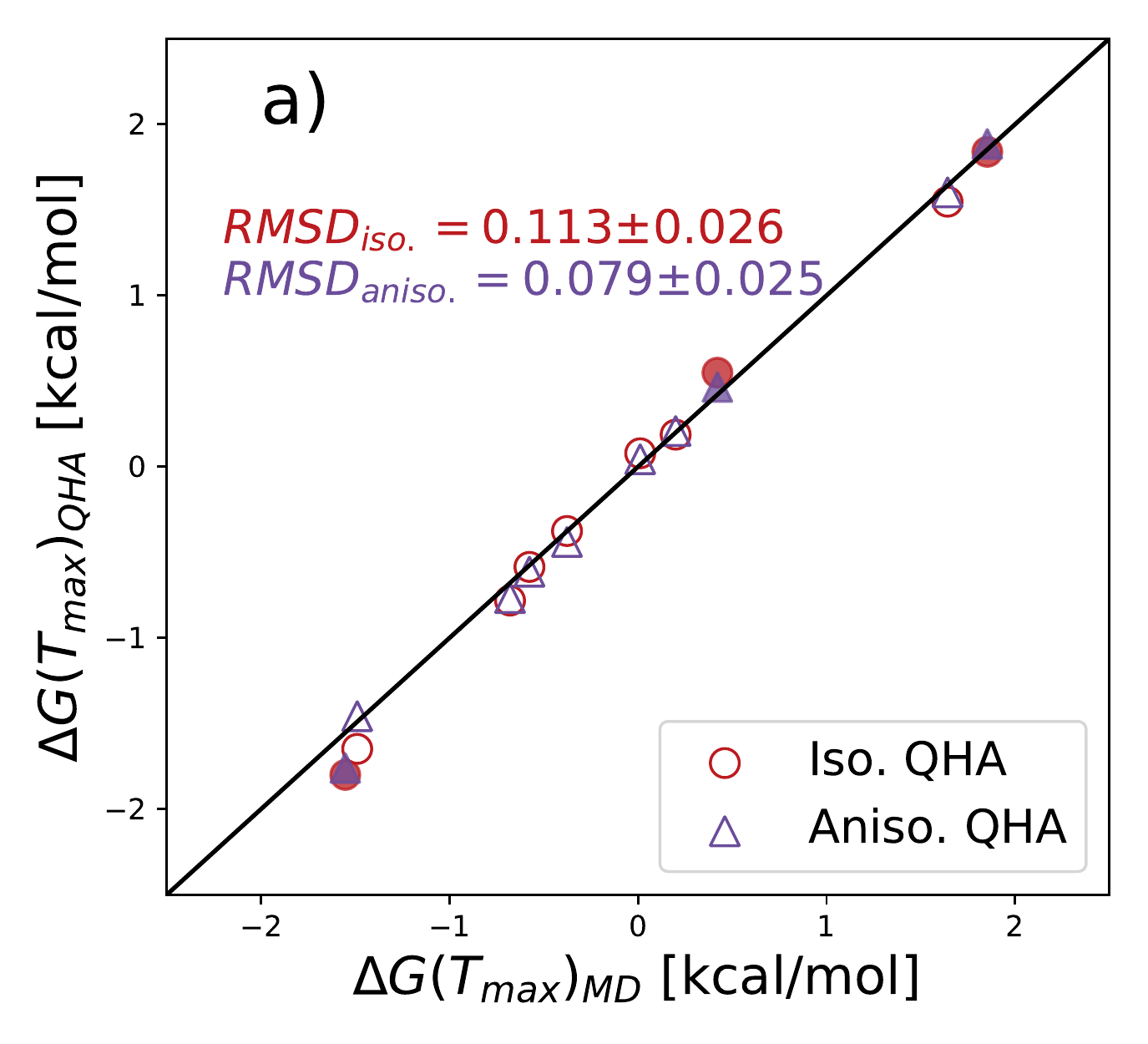}  
      \includegraphics[width=8cm]{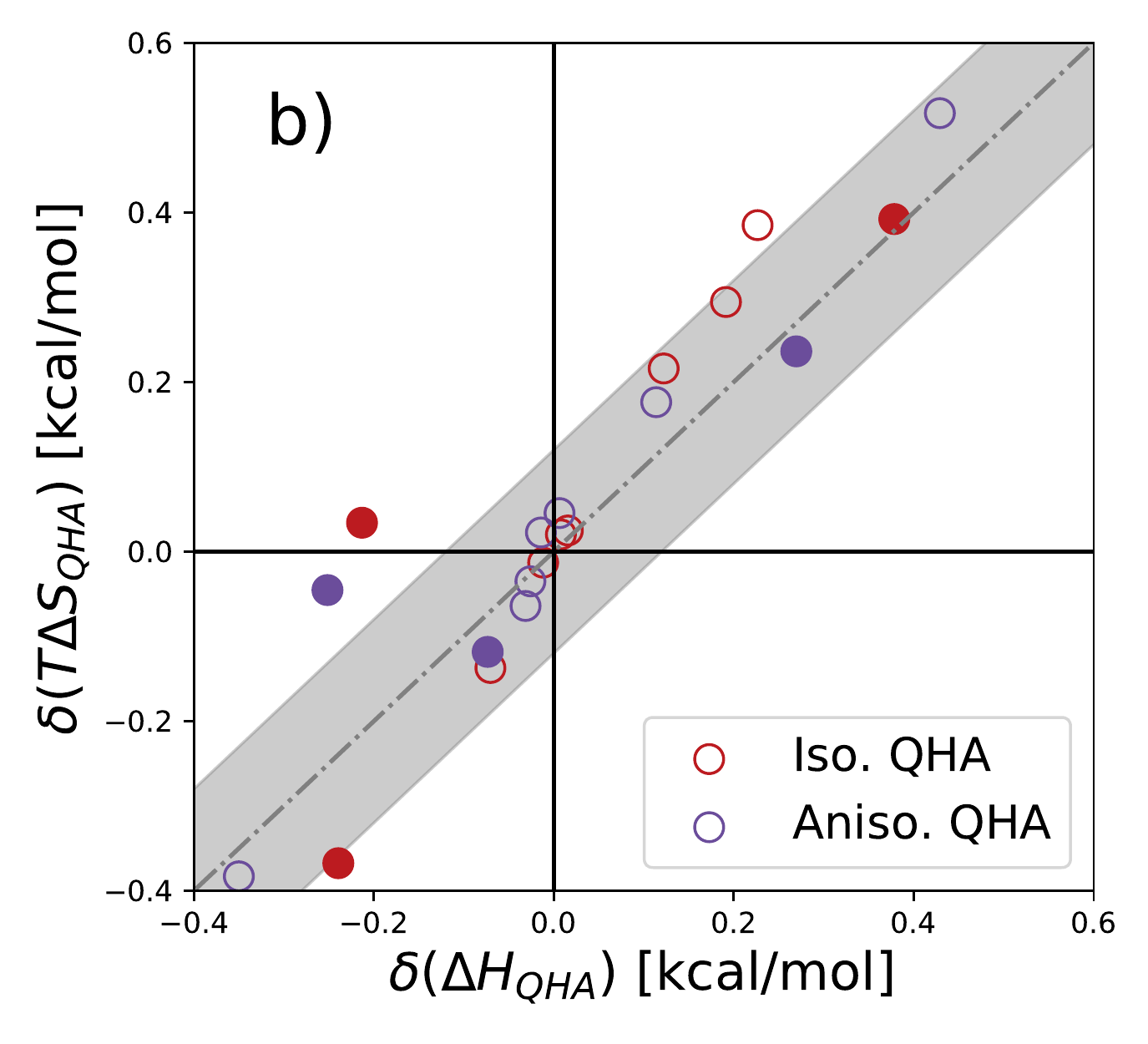}
      \end{center}
      \caption{a) Gibbs free energy differences between polymorphs for QHA relative to 
MD and b) entropy--enthalpy compensation in errors of QHA relative of MD. 
a) Overall, there is a slight improvement of RMSD in the Gibbs free energy when 
using anisotropic expansion over isotropic expansion. This improvement is only 
slightly larger than the bootstrapped error in the RMSD. b) Grey shaded area indicates 
free energy differences $<0.12$ kcal/mol between QHA and MD, which for most of the graph indicates
significant entropy--enthalpy compensation. The only molecule to fall outside 
of these bounds for anisotropic QHA is tolbutamide, which has the largest error 
in QHA relative to MD. In all plots, the filled-in markers are the 
molecules with multiple molecular conformers (tolbutamide, chlorpropamide, and aripiprazole).
      \label{figure:MDvQHA_energy}}
    \end{figure*}

\begin{figure}[H]
  \begin{center}
  \includegraphics[width=16cm]{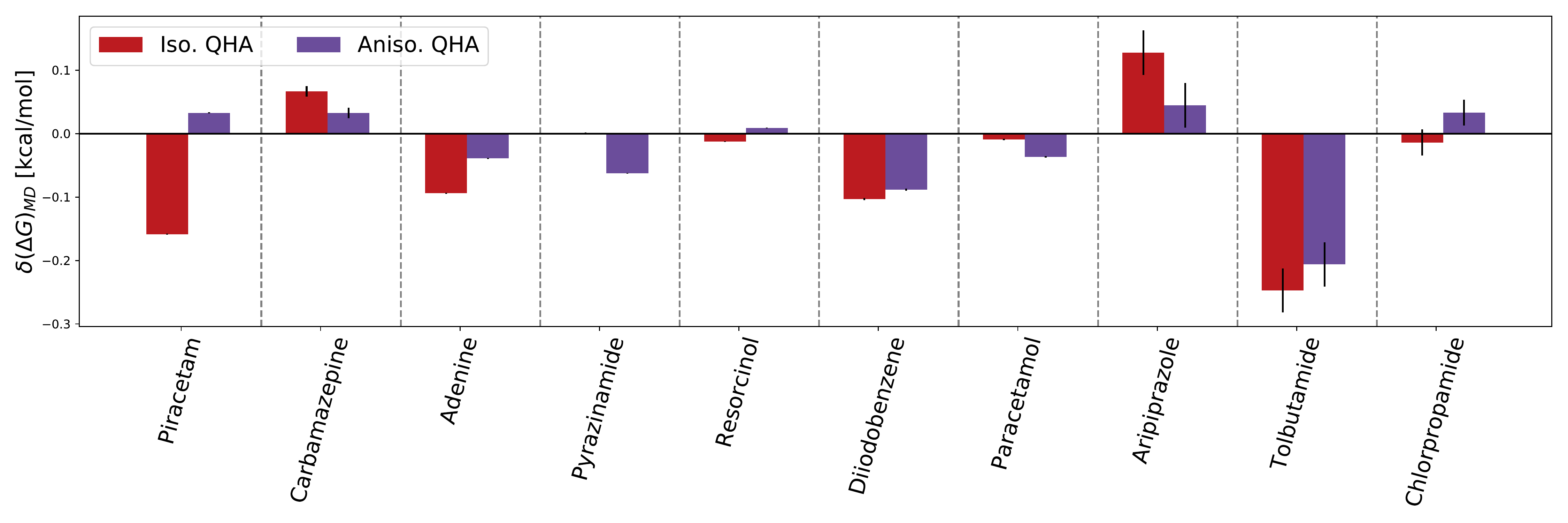}
  \end{center}
  \caption{Deviations in the free energy differences from MD at 300 K (except for tolbutamide at 200 K, and chlorpropamide at 250 K). We report results for isotropic QHA and 1D-anisotropic QHA relative to MD. Overall, there is no evidence that a more complex thermal expansion model reduces the error between QHA and MD.
  \label{figure:dG_MD}}
\end{figure}

    Anisotropic expansion provides minimal improvement in the error of polymorph 
free energy differences between QHA and MD. In Figure 
\ref{figure:MDvQHA_energy}a we plot the Gibbs free energy differences of both 
QHA methods against the results for MD at $T_{max}$. The RMSD of $\Delta G$ for 
isotropic and anisotropic QHA relative to MD are 0.113 and 0.079 kcal/mol, 
respectively. There is only a 0.034 kcal/mol improvement to the RMSD when using 
an anisotropic thermal expansion model, which is comparable to the bootstrapped 
error in the RMSD. Anisotropic QHA in fact has a larger deviation from MD than 
isotropic QHA for 3 of the 10 molecules, though the deviation from MD is small. 
Those molecules and their deviation of anisotropic QHA from MD ($\delta(\Delta 
G_{1D})_{MD}$) are pyrazinamide (0.037 kcal/mol), resorcinol (0.062 kcal/mol), 
and chlorpropamide (0.033 kcal/mol). These results demonstrate that the 
deviations in the free energy differences between MD and QHA are due to 
something other than an insufficiently accurate thermal expansion model. Summary 
results for each molecule is provided in the Figure S\ref{figure:dG_MD}.

\begin{figure}[H]
  \begin{center}
  \includegraphics[width=4cm]{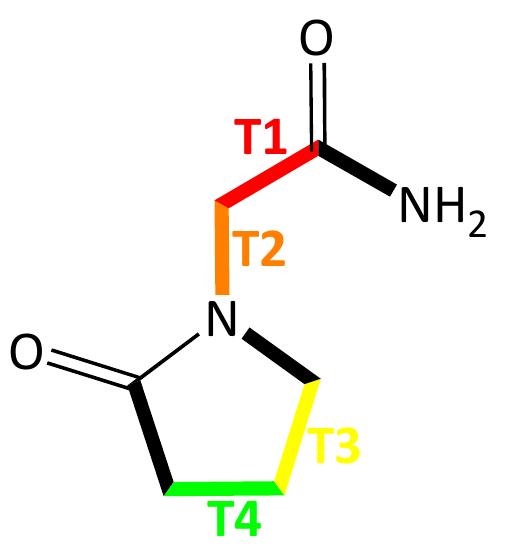}
  \includegraphics[width=6cm]{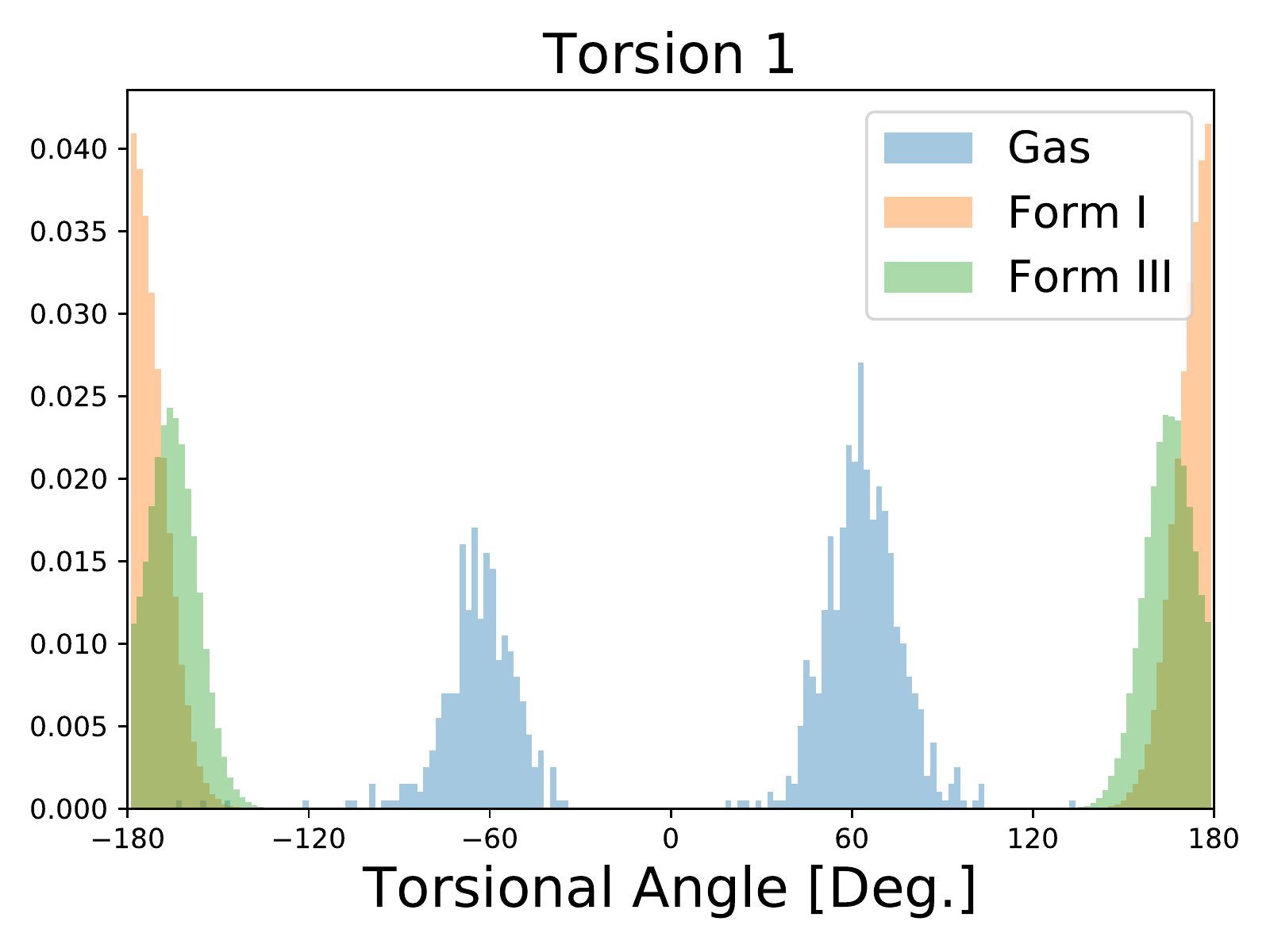}
  \includegraphics[width=6cm]{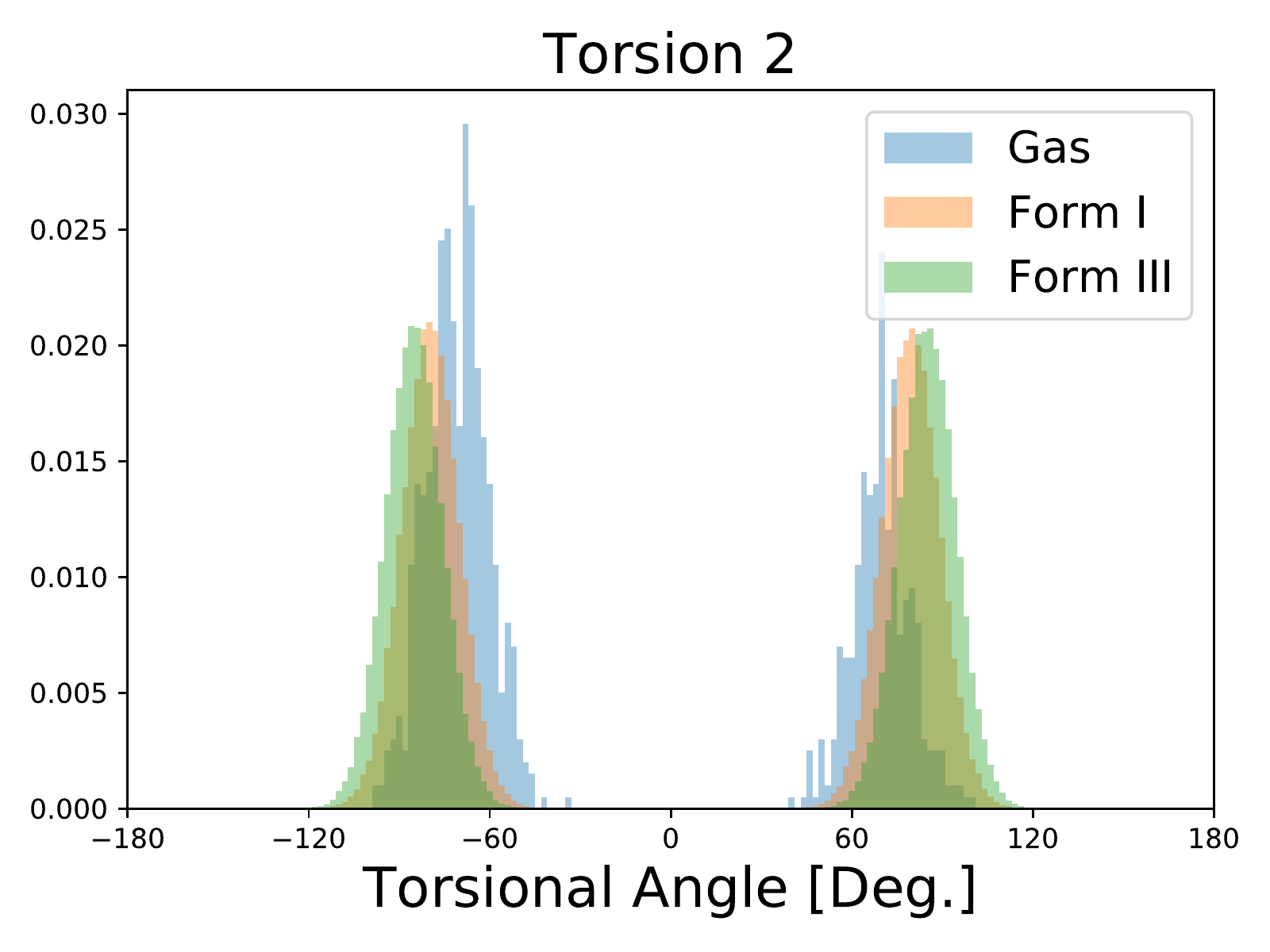}
  \includegraphics[width=6cm]{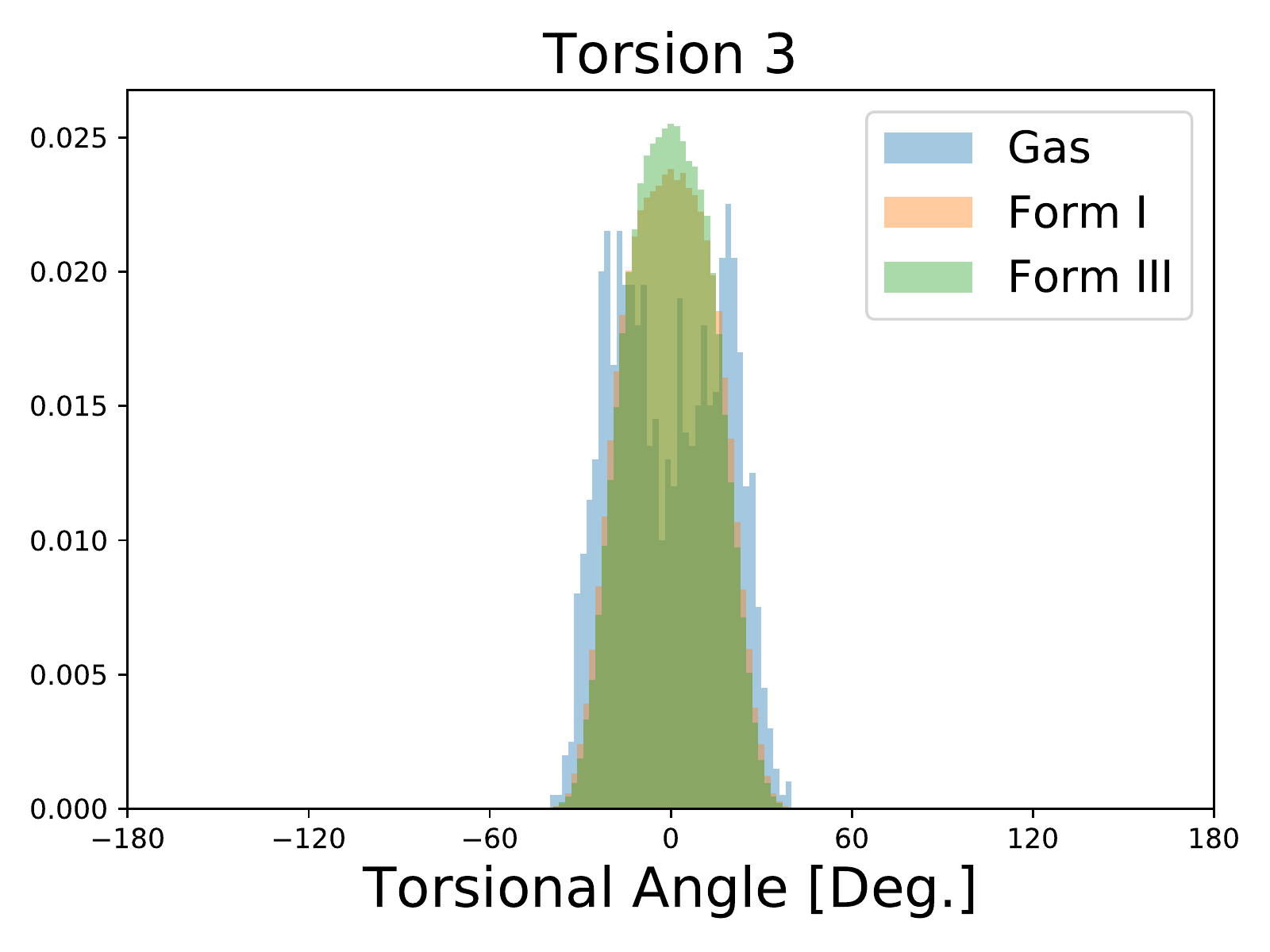}
  \includegraphics[width=6cm]{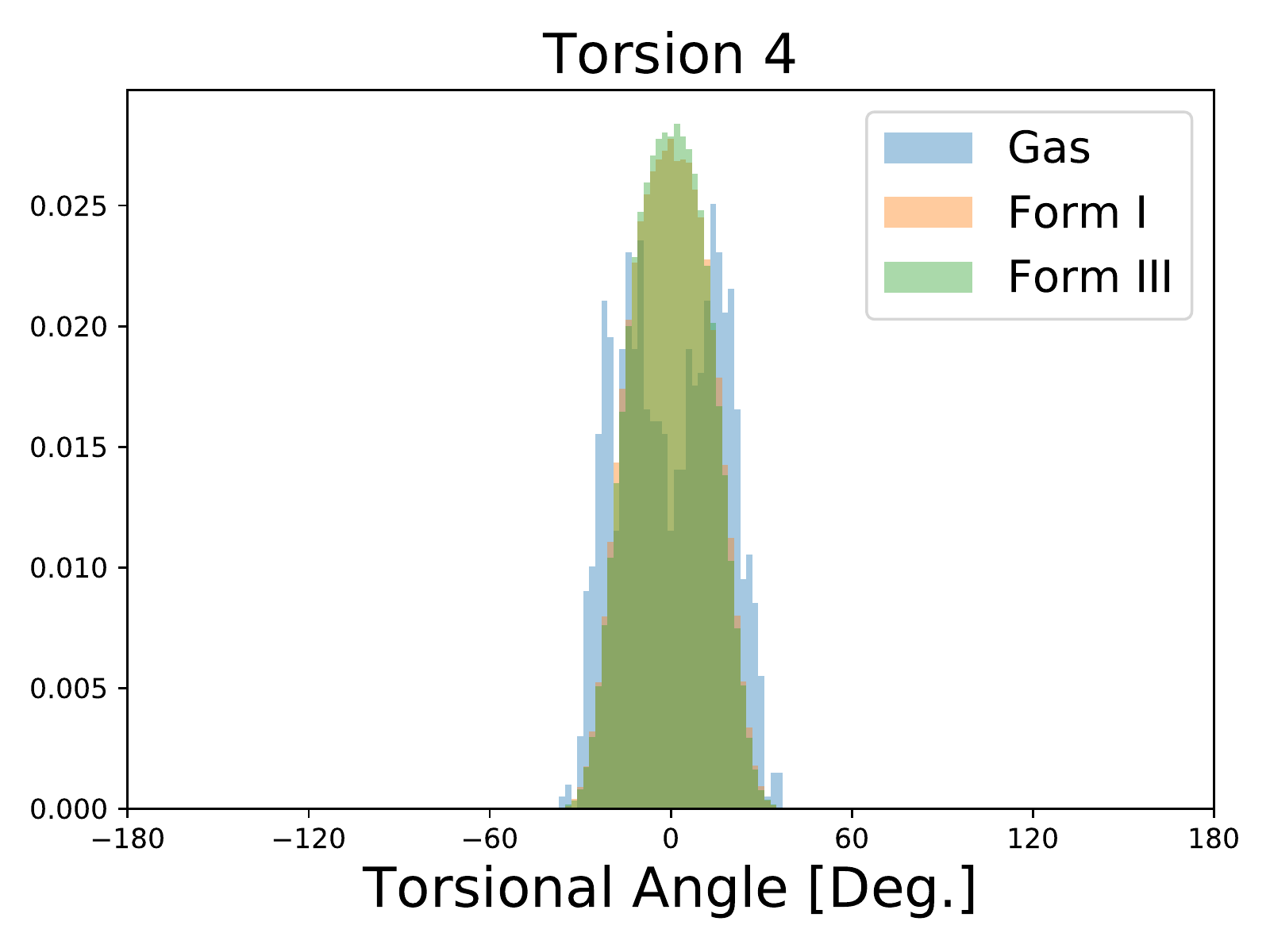}
  \end{center}
  \caption{Torsional distributions of piracetam at 300 K for the molecule in gas and both polymorphs.
  \label{figure:bismev_torsion}}
\end{figure}

\begin{figure}[H]
  \begin{center}
  \includegraphics[width=6cm]{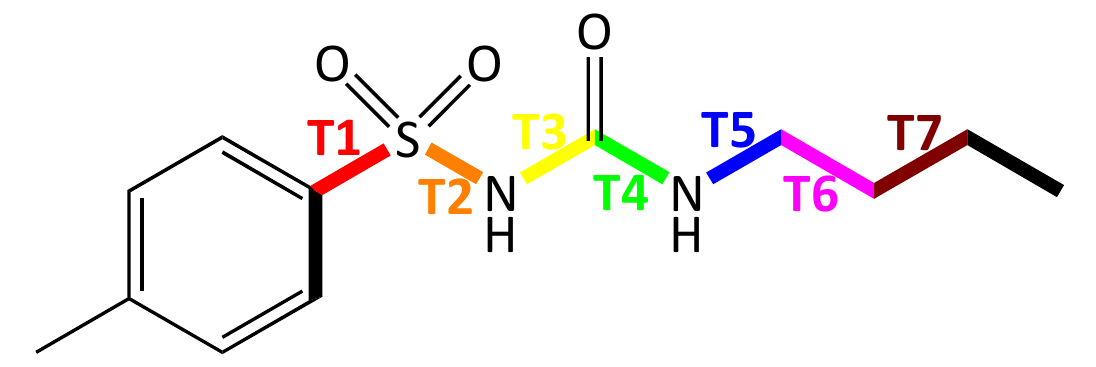}
  \includegraphics[width=6cm]{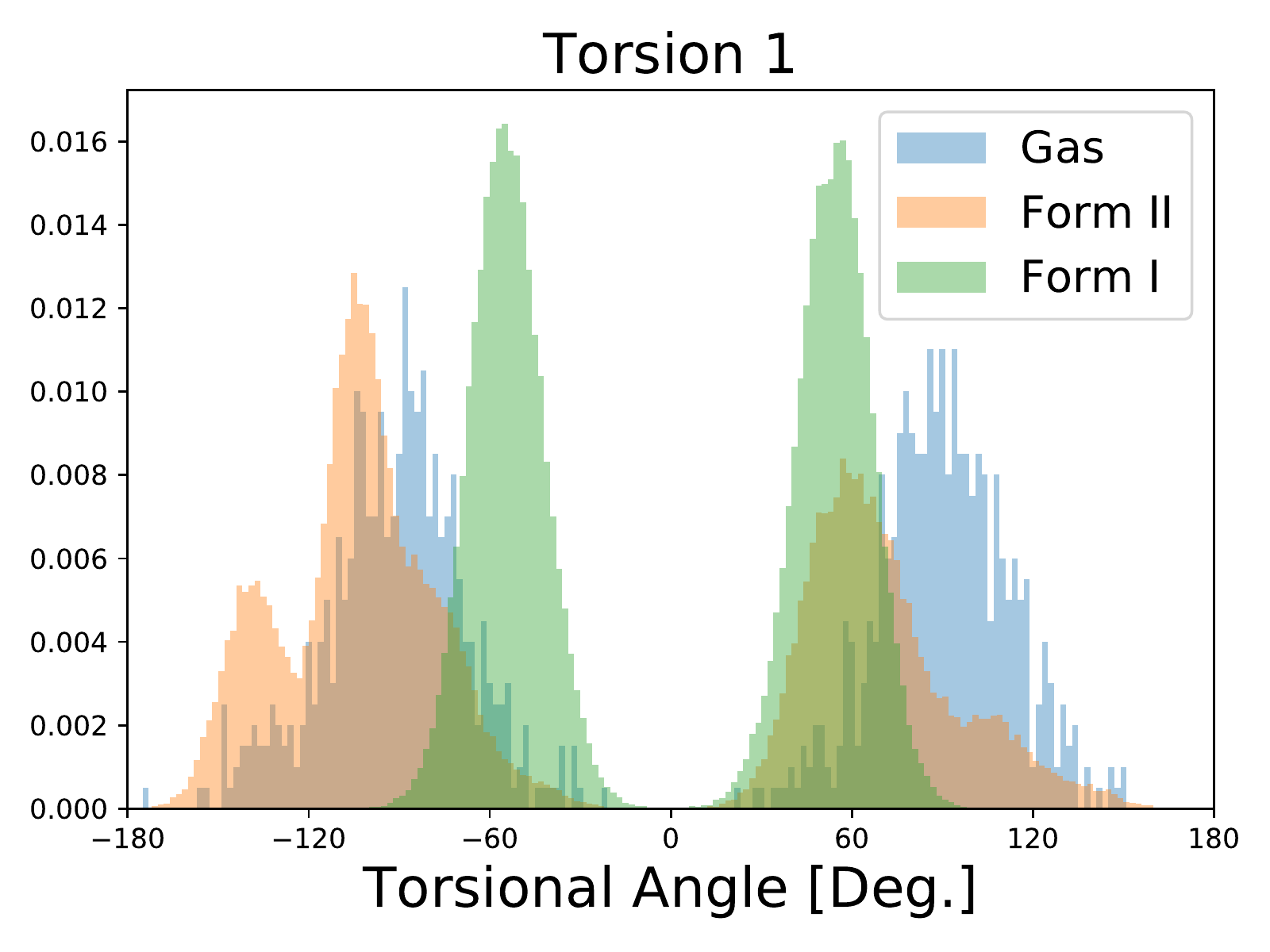}
  \includegraphics[width=6cm]{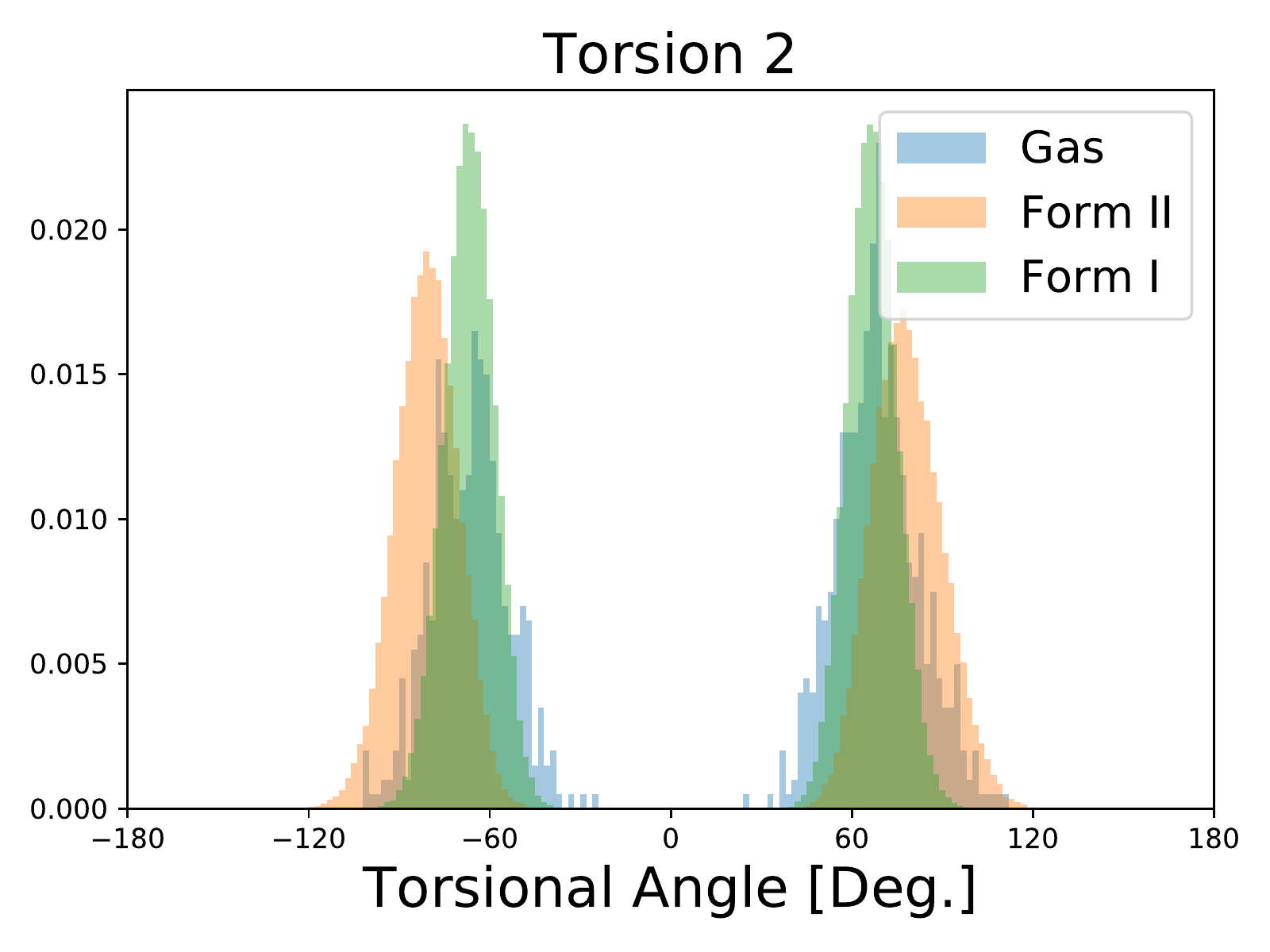}
  \includegraphics[width=6cm]{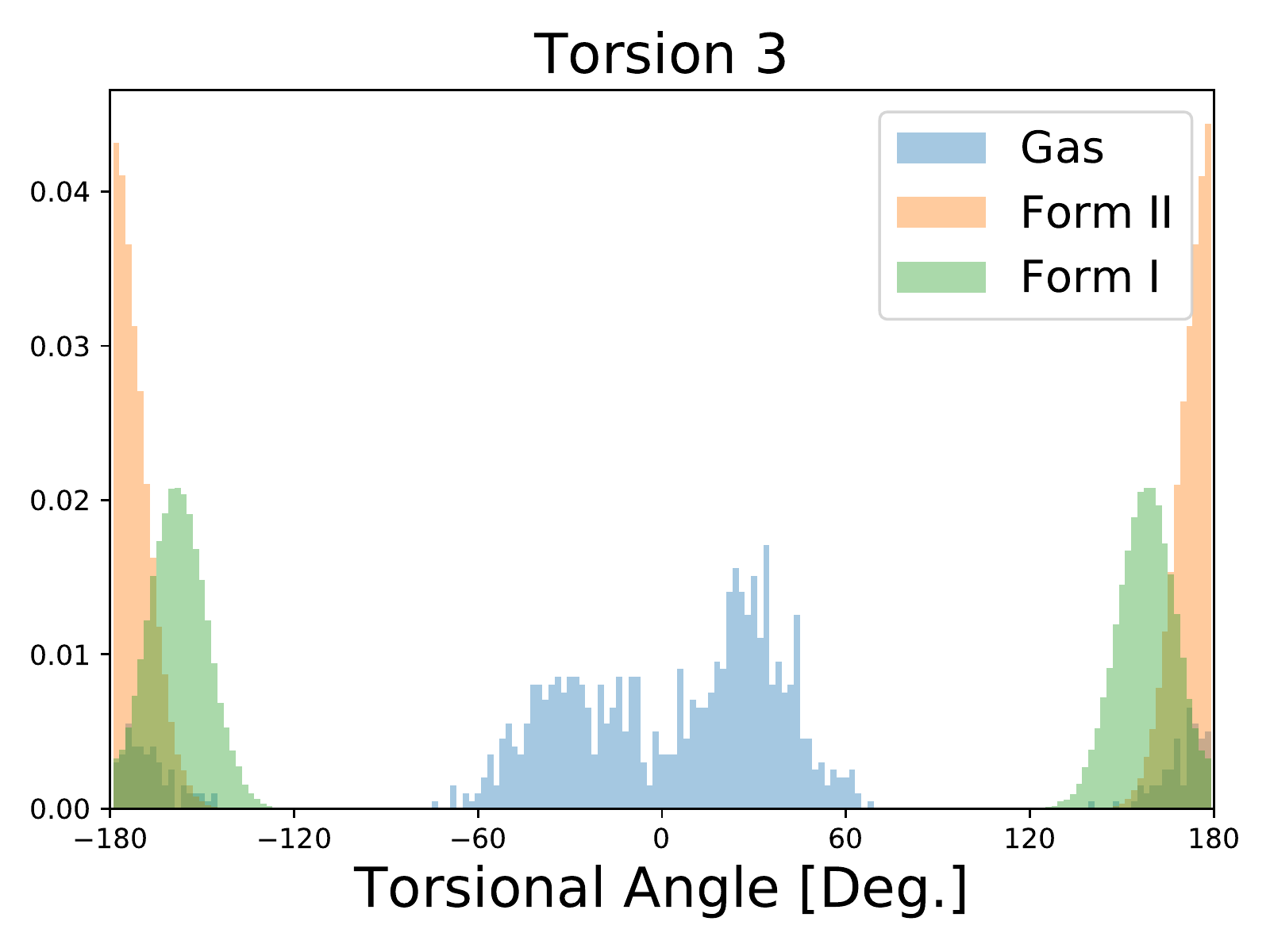}
  \includegraphics[width=6cm]{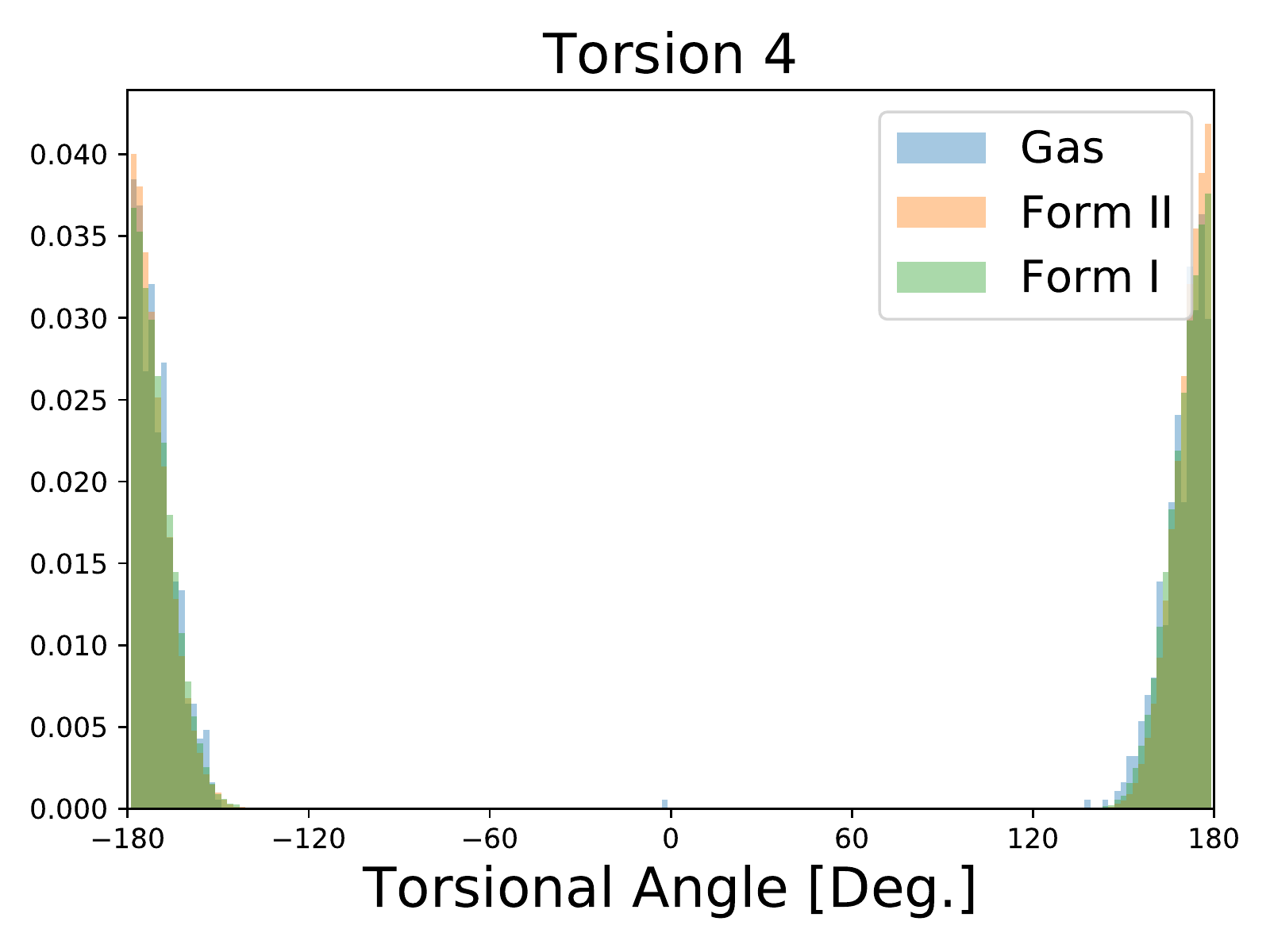}
  \includegraphics[width=6cm]{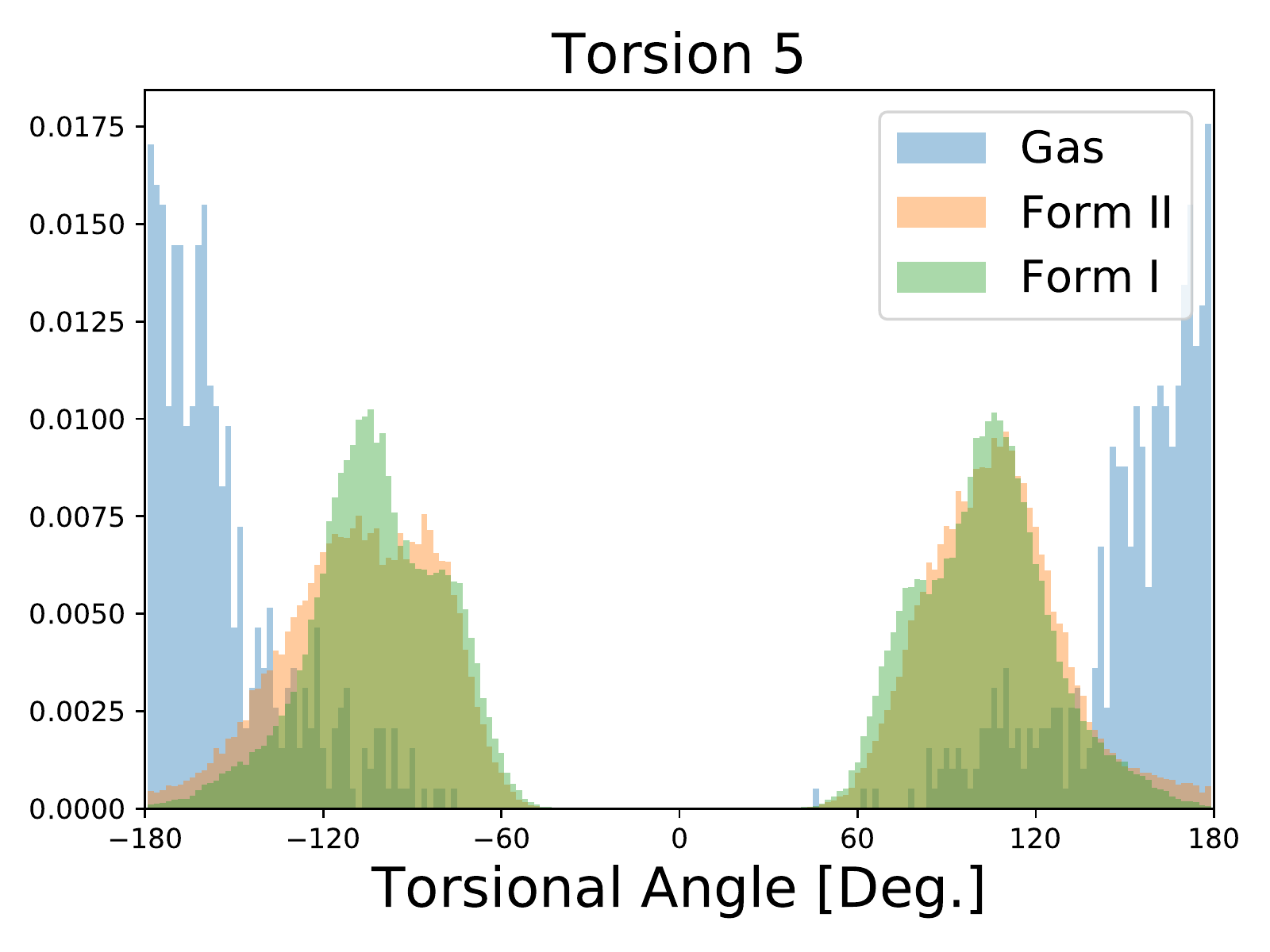}
  \includegraphics[width=6cm]{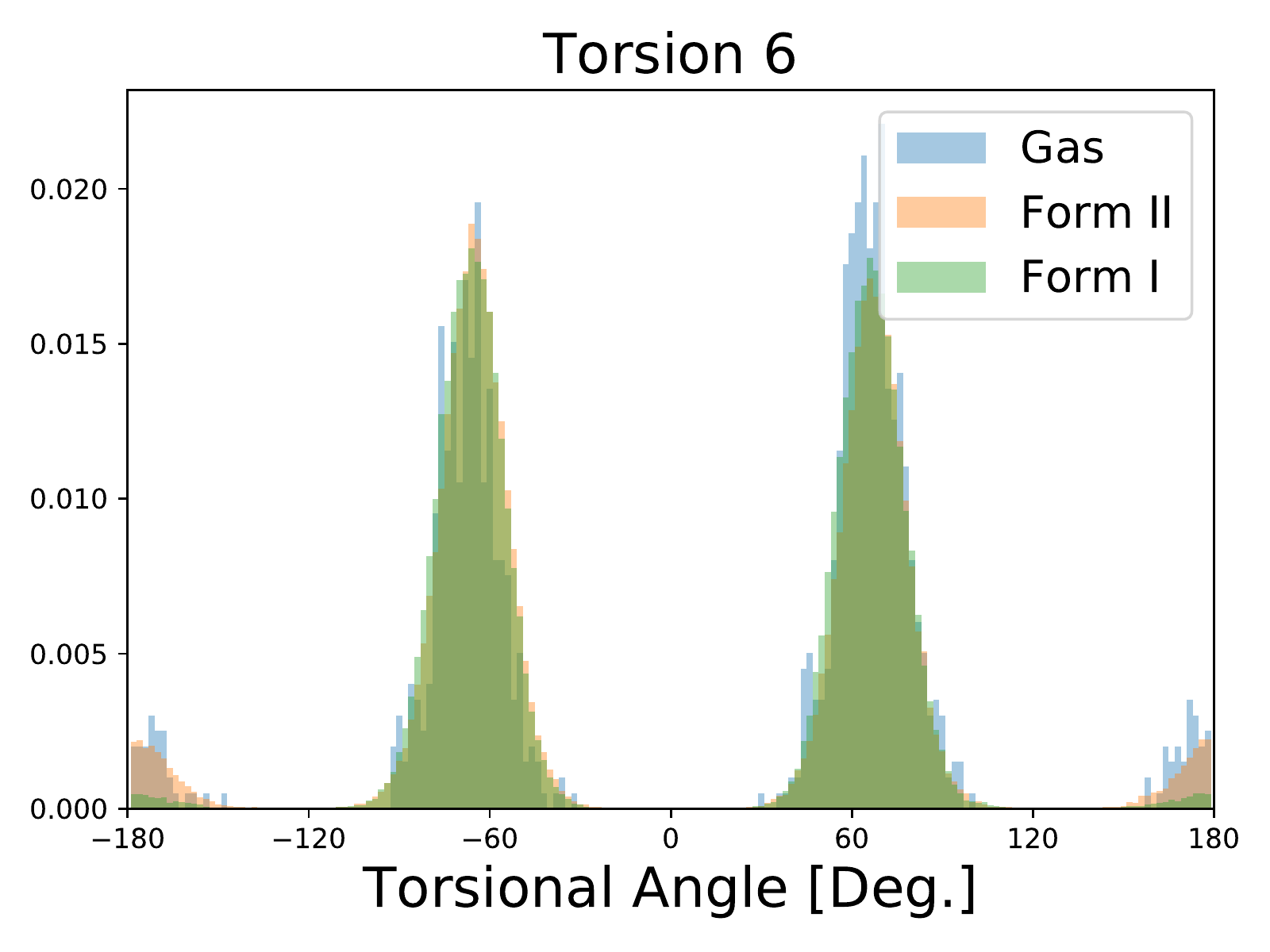}
  \includegraphics[width=6cm]{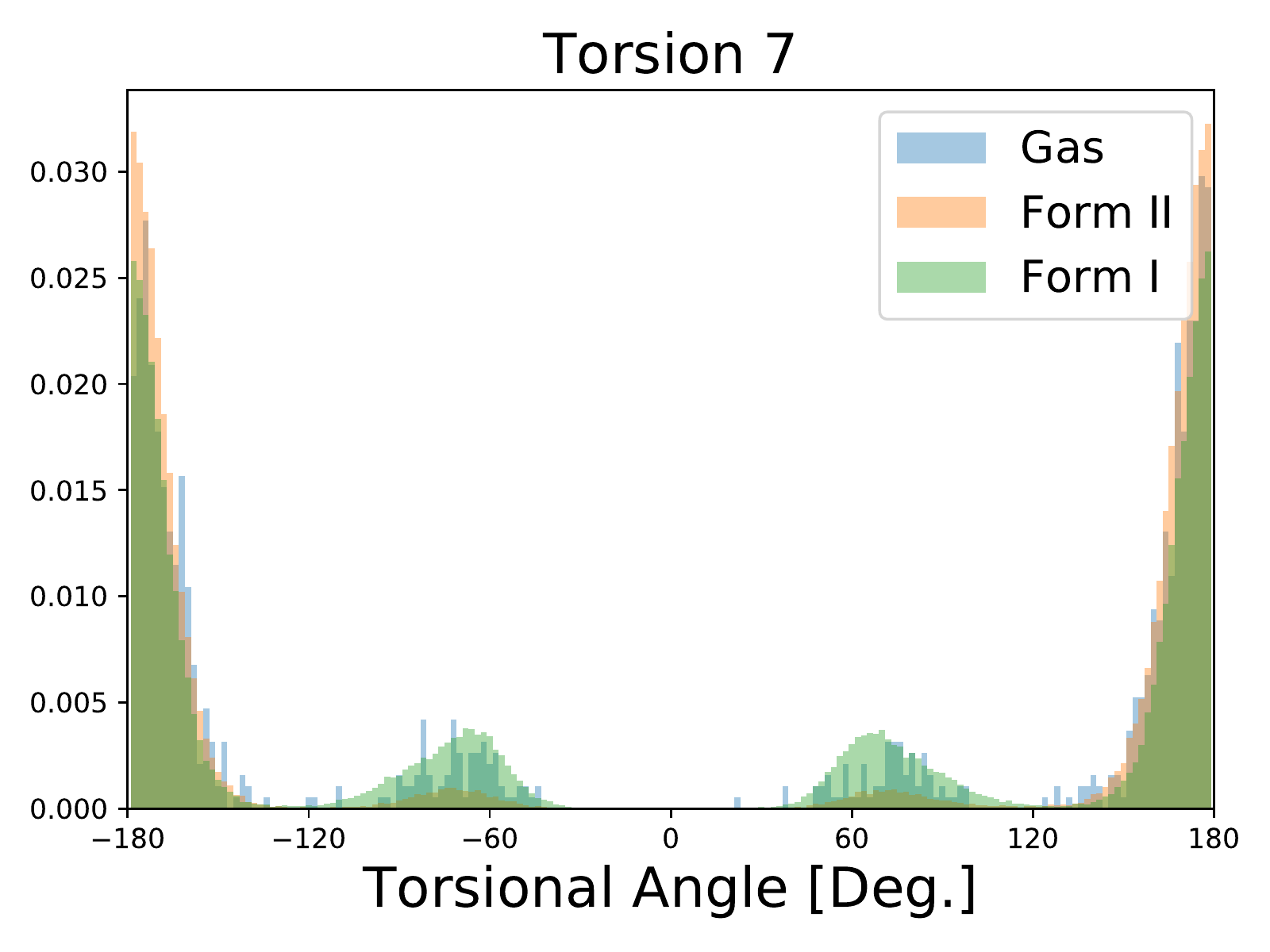}
  
  \end{center}
  \caption{Torsional distributions of tolbutamide at 300 K for the molecule in gas and both polymorphs.
  \label{figure:zzzpus_torsion}}
\end{figure}

The difference between QHA and MD is minimized when a lattice minimum
corresponding the sampled MD lattice parameters and anisotropic QHA is used.
The largest errors for anisotropic QHA is seen for tolbutamide 
($\delta(\Delta G_{1D})_{MD}=$ 0.206 kcal/mol), paracetamol (0.088 kcal/mol), 
and resorcinol (0.062 kcal/mol) with all other molecules $<$ 0.045 kcal/mol. 
Under the condition that a more consistent minimum with MD is used, we no longer 
find that the error in QHA trends with molecular size and 
flexibility.~\cite{dybeck_capturing_2017} However, molecular size and 
flexibility are correlated with a greater number of alternative lattice 
minima.~\cite{dybeck_exploring_2019} We also see no correlation between the 
error in QHA and the number of hydrogen bonds in the crystal ($R^{2} < 0.1$ in 
Figure S\ref{figure:hbonds_0K} and Figure S\ref{figure:hbonds_300K}). 
Tolbutamide form II has a Z$^{\prime}$ of 4 and all other polymorphs are $\leq$ 
2, while this is not statistically significant it could explain why the error in 
QHA is an outlier. We also compared the gas and crystal phase torsions, but 
found no differences that distinguish tolbutamide from the rest of the 
molecules (Torsions shown in figures 
S\ref{figure:bismev_torsion}--S\ref{figure:zzzpus_torsion}). 

    We see entropy--enthalpy compensation in the differences between anisotropic QHA and 
MD of up to 0.12 kcal/mol for all molecules except for tolbutamide. Despite the 
entropy and enthalpy for QHA having errors up to 0.5 kcal/mol from MD, the 
cancellation of those errors allows QHA to give free energy differences within 
0.09 kcal/mol of the MD-calculated free energy differences for all systems except tolbutamide. In 
Figure~\ref{figure:MDvQHA_energy}b the error in the entropic differences of 
polymorphs is plotted versus the error in enthalpic differences of QHA relative 
to MD. We define a set of polymorphs demonstrating entropy--enthalpy compensation 
(gray shaded area) if the differences between errors in entropy and enthalpy 
(i.e. the error in the free energy difference)
is $<0.12$ kcal/mol, giving us 90\% confidence that a re-ranking in crystal stability 
would not occur due to this error.~\cite{dybeck_capturing_2017} 
Anharmonic motions may be important for computing the enthalpy and entropy differences
of polymorphs, but for most of the systems studied here the error is minimal for
computing the free energy. The only molecule that falls 
outside of this region for anisotropic QHA is tolbutamide. Out of the 10 
molecules, 8 have larger entropic error in QHA than enthalpic error, which is 
most likely due to anharmonic motions. The only two sets of polymorphs that have 
larger enthalpic error in QHA are tolbutamide and chlorpropamide, which are 
further explained in the next section.

\subsection{Molecular Conformational Ensembles Can Still Exist Low Temperatures}
\begin{figure*}
      \begin{center}
      \includegraphics[width=16cm]{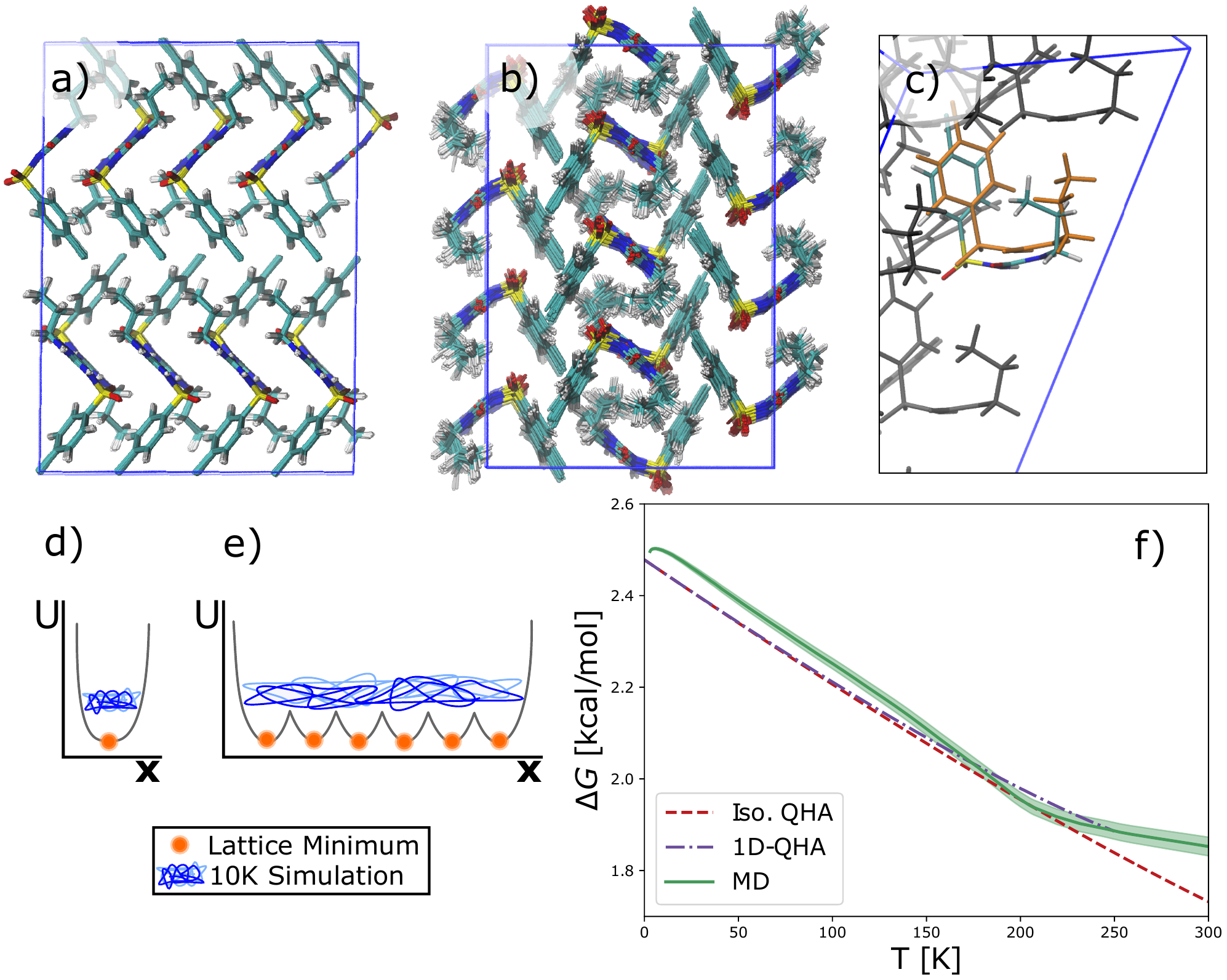}
      \end{center}
      \caption{Snapshots of the 10 K ensemble for chlorpropamide form a) I and b) V taken from 
the last 5 ns of the replica exchange simulation, with 0.1 ns between each 
frame. Even at low temperatures, this crystal structure samples a surprisingly 
diverse conformational ensemble, with the greatest variability in the 
fluctuations around the alkyl tail. The RMSD of the entire 10 K simulation 
ensemble relative to the lattice minimum for form I is 0.165 $\pm$ 0.02 \r{A} 
and form V is 0.691 $\pm$ 0.21 \r{A}. c) Two frames of chlorpropamide form V 
that are 150 ps apart, showing that the position of the terminal carbon is 
sampling two torsional conformations at 10 K. Graphical representation of a 
crystal with d) a single configurational minima and e) multiple 
configurational minima drawn on a potential energy ($U$) against the 
configurational space ($\boldsymbol{x}$). f) Gibbs free energy difference of 
chlorpropamide form V relative to form I.
      \label{figure:bedmig_conformation}}
    \end{figure*}

    Even at low temperatures, some crystals have substantial conformational degeneracies 
allowing quenched structures to settle into lattice minima that break the 
symmetry of the supercell. In figure~\ref{figure:bedmig_conformation}b we show 
frames of the trajectory of chlorpropamide, with each frame taken 0.1 ns apart 
for the last 5 ns of the 10 K ensemble produced during replica exchange. From 
this low temperature ensemble, molecules are able to sample a number of 
configurations due to the degrees of freedom accessible to the alkyl tail, and 
the low energy differences between different alkyl rotamer configurations. When 
quenched, the tails fall into mostly independent conformations, breaking the 
crystal symmetry.

    The thermodynamic accessibility of many configurations at low temperatures 
is a result of the existence of many configurationally diverse lattice minima 
with similar low energies. In Figure~\ref{figure:bedmig_conformation}b the RMSF 
of form V is 0.69 $\pm$ 0.21 \r{A} at 10 K, which is four times greater than 
form I where the RMSF is 0.17 $\pm$ 0.02 \r{A} 
(Figure~\ref{figure:bedmig_conformation}a). The more configurationally diverse behavior of form 
V helps to explain why quenching the MD simulations leads to so many different 
lattice minima. Figure~\ref{figure:bedmig_conformation}d provides a graphical 
representation of the single lattice minimum and small configurational spaced 
covered at 10 K for form I, which contrasts with 
Figure~\ref{figure:bedmig_conformation}e representation of form V.

\begin{figure*}
  \begin{center}
  \includegraphics[width=12cm]{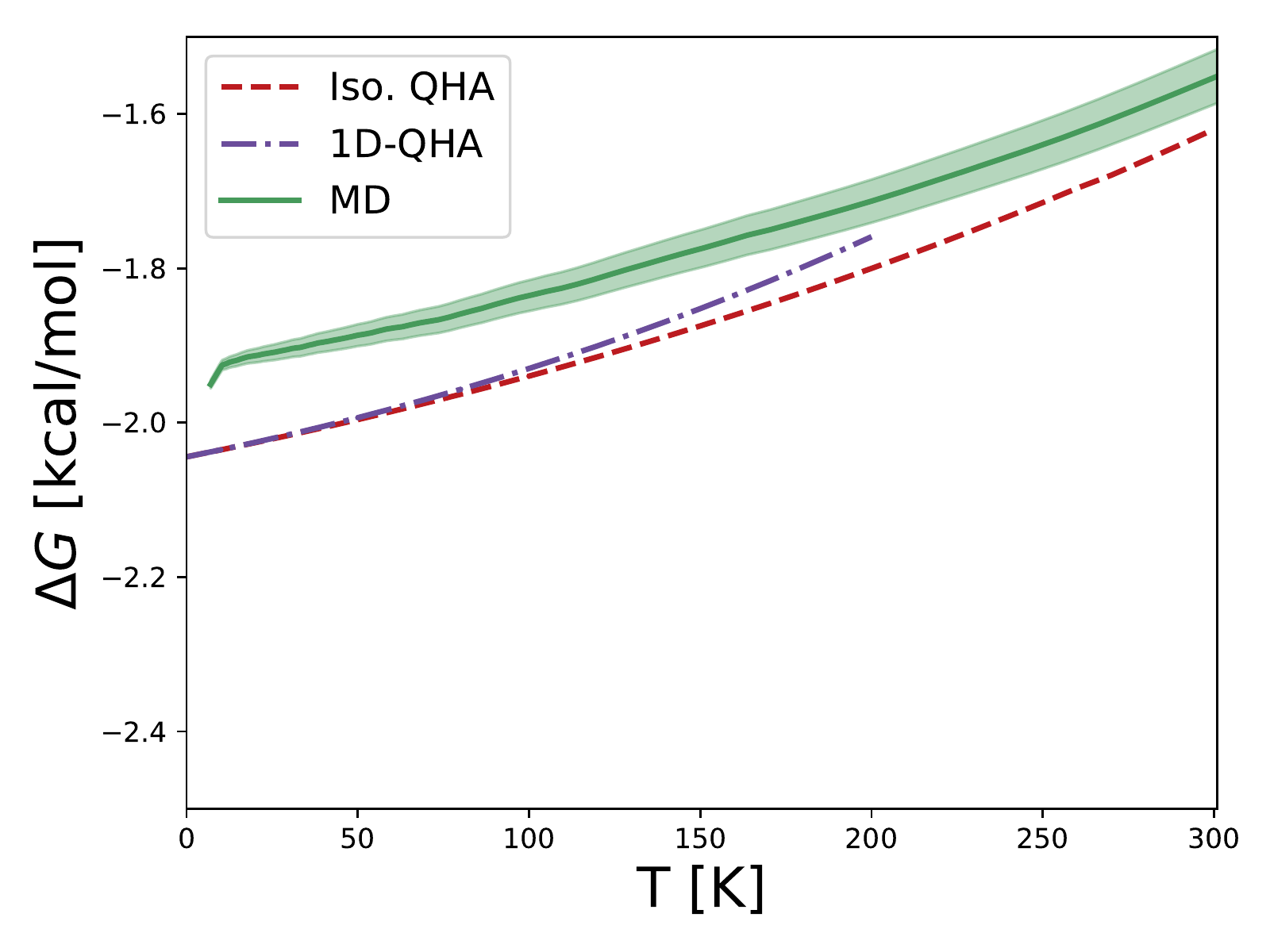}
  \end{center}
  \caption{Polymorph free energy differences of tolbutamide form II relative to form I-restructured.
   \label{figure:zzzpus_dG}}
\end{figure*}

    The number of accessible low-energy minima causes a rapid deviation of QHA 
of up to 0.06 kcal/mol in the free energy difference relative to MD for 
chlorpropamide as low as 10 K. Figure~\ref{figure:bedmig_conformation}f shows the 
polymorph free energy differences for chlorpropamide form V relative to form I 
for QHA and MD. Based on the $G(T)$ plot from MD, between 0 and 10, the 
polymorph free energy difference for MD quickly changes, causing QHA to diverge 
by 0.06 kcal/mol at 10 K. While QHA can accurately model form I at low 
temperatures, it cannot capture the behavior of the alkyl tails in form V, which 
is why we see the immediate divergence of the two methods as $T$ increases from 
zero. At temperatures greater than 10 K the deviation of the free energy from MD does not grow 
because of the entropy--enthalpy compensation, but we cannot necessarily conclude that this would be 
the case for other crystals that could have alternate configurational minimum. 
Both polymorphs of tolbutamide also sample a large number of conformations at low 
temperatures, causing a similar divergent behavior in $\Delta G$ for QHA and MD 
at low temperatures (Figure S\ref{figure:zzzpus_dG}). 

\begin{figure*}
  \begin{center}
  \includegraphics[width=8cm]{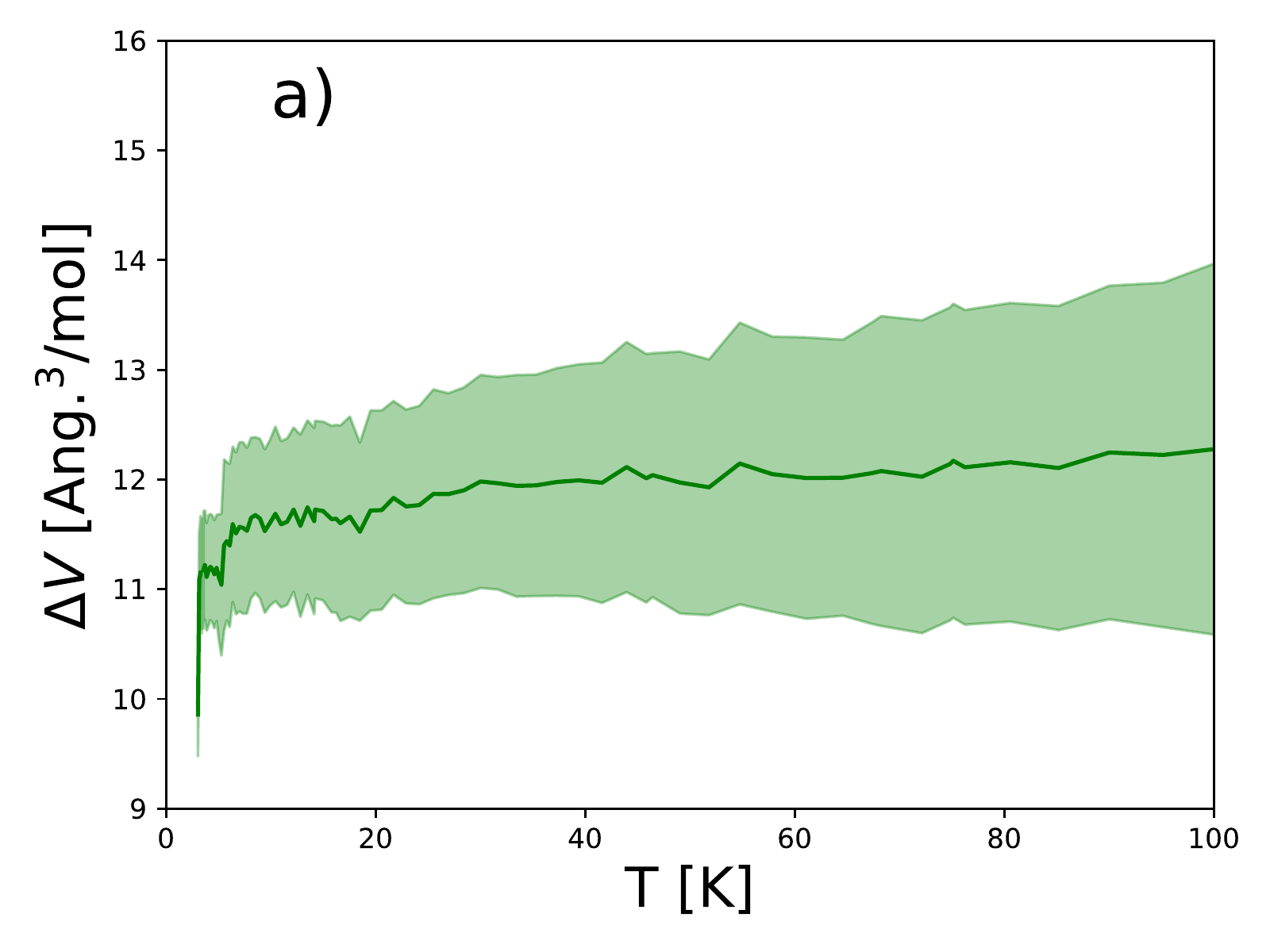}
  \includegraphics[width=8cm]{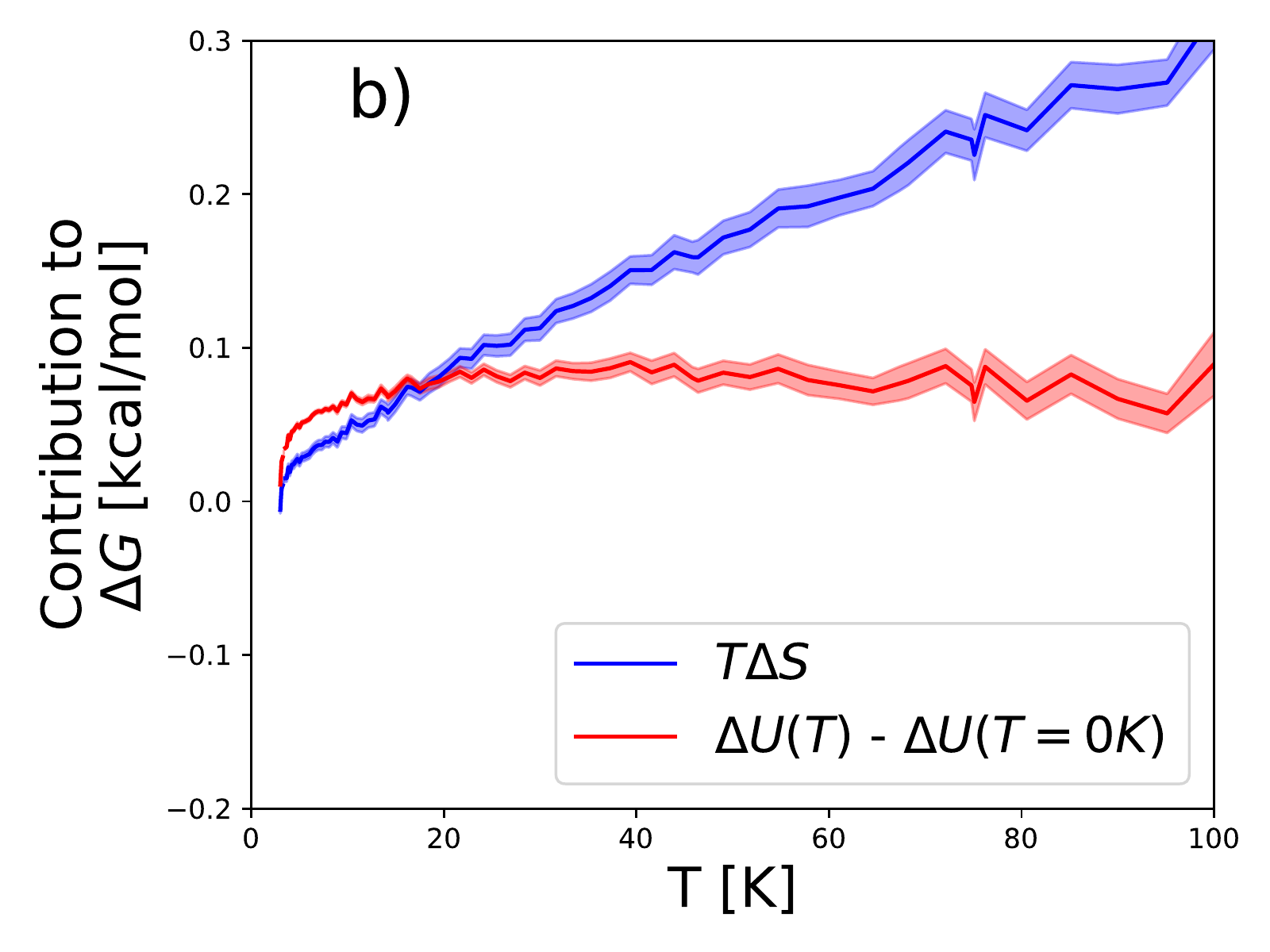}
  \end{center}
  \caption{(a) Volume differences of chlorpropamide form V relative to form I using MD. Initially, form V expands much faster than form I. Above 10 K, the difference in volume between the forms is almost constant. (b) The entropic and enthalpic contributions to the Gibbs free energy difference of chlorpropamide form V relative to I. Below 10 K, the potential energy is the dominating free energy contribution making form V less favorable, explained by the sudden expansion of form V seen at low temperatures.
    \label{figure:dVSU_bedmig}}
\end{figure*}

   The concavity in the free energy of chlorpropamide at low temperatures is due 
to dominating enthalpic changes close to 0 K, which quickly diminish relative to 
the entropic contributions as the temperature increases. Surprisingly, although 
form V has a larger low T configurational ensemble, it initially becomes less 
stable as a function of T relative to the more constrained form I. Up to 10 K, 
form V expands faster than form I (Figure \ref{figure:dVSU_bedmig}a) causing an 
unfavorable enthalpic contribution to form V relative to form I 
(Figure~\ref{figure:dVSU_bedmig}b). Since $T\Delta S$ is small at low 
temperatures, the changes in the enthalpic difference dominates the free energy 
differences. At temperatures greater than 10 K the difference in volume and 
enthalpy between polymorphs is constant. The initial expansion of form V allows 
the alkyl tails to have more conformational space than form I and by 20 K 
the entropic contributions in form V are large enough to switch the stability of the polymorphs.

%

\section{Conclusions}
    Although we have studied a relatively small number of enantiotropic 
polymorphs in this study, we find there here are persistent shortcomings of QHA 
that highlight the importance of anharmonic configurational sampling in 
determining accurate free energy differences of crystal polymorphs. The 
shortcomings identified in this study are: 1) that including anisotropic 
expansion provides minimal improvement in the error of polymorph free energy 
differences between QHA and MD, 2) the number of accessible low-energy minima 
causes a rapid deviation of QHA relative to MD for temperatures as low as 10 K, 
3) some crystals reorganize into clearly different structures when heated up to 
300 K using MD, and 4) polymorph restructuring prevents QHA from being run 
independently of MD because the minimum energy configuration is likely to be 
inaccessible from an arbitrary low-energy starting point.

   We found that the OPLS-AA point charge potential incorrectly ranked 4 of the 
10 enthalpic differences of polymorphs relative to experiment, but correctly 
ranked the entropic differences of all 10 molecules. Despite the lattice 
geometries for MD differing moderately from experiment, the expansion of vectors 
and orthogonal angles for MD generally agree with experiment. The RMSD of both 
enthalpic and entropic differences between MD and experiment were 4.0 and 2.0 
kcal/mol, respectively, which is roughly 2 orders of magnitude larger than 
errors seen of QHA relative to MD if properly performed.

    Our one-dimensional approach to model anisotropic thermal expansion predicts 
essentially identical high temperature lattice geometries and polymorph free 
energy differences as the full anisotropic QHA for the 10 polymorph pairs 
studied. Specifically, the computed polymorph free energy differences are within 
0.02 kcal/mol at $T_{max}$ for our 1D- and fully-anisotropic QHA methods for the 
10 molecules studied. The lattice geometries of the two approaches remained 
essentially the same, with an RMSD of 0.29\% for the expansion relative to 
lattice minimum structure. Additionally, 1D-anisotropic QHA is able to find a 
free energy minimum up to 300 K for all but one crystal unit cell, whereas the 
fully anisotropic approach fails to reach those desired temperatures for 50\% of 
the crystals, due to the larger likelihood of a rearrangement into an alternate 
minimum when multiple derivatives are used, resulting in numerical instability 
of the approach.

    Finite size effects do not always cancel out for polymorph free
energy differences; instead, differences between unit cells and supercells 
can be up to 0.5 kcal/mol and therefore must be accounted for when attempting to accurately rank
crystal stability. We found that this error is correlated with the error in the lattice 
geometries ($R^{2} = $ 0.79), meaning they are reduced but not eliminated if the 
high temperature lattice geometries are the same between unit and supercells. 

    Anisotropic QHA only marginally improves the error in free energy 
differences from MD for the polymorph pairs studied. Using anisotropic QHA 
reduces the RMSD of $\Delta G$ relative to MD by 0.034 kcal/mol, which is only 
slightly larger than the standard error in the RMSD. The errors in QHA with 
respect to MD are due primarily to anharmonic motions, because using an 
anisotropic thermal expansion model does not improve the deviations between QHA 
and MD. Differences between anisotropic QHA and MD do not appear strongly 
correlated with molecular flexibility. Tolbutamide with both a large deviation 
and high flexibility is an exception, and this difference could also be due to 
the large Z$^{\prime}$ value of 4 for form II. The free energy difference in 
tolbutamide had an error of 0.21 kcal/mol between anisotropic QHA and MD, 
whereas all other molecules deviated $<0.09$ kcal/mol.

    The degree of flexibility of the molecules can result in a more significant 
chance of crystal restructuring upon annealing with replica exchange. Once a lower
energy lattice minimum was found, the error in isotropic QHA was on average 0.08 kcal/mol lower than 
previously reported.~\cite{dybeck_capturing_2017}. Anisotropic QHA from the new minimum was able to reduce
the free energy differences between chlorpropamide and aripiprazole to $<0.05$
kcal/mol. We were consistently able to find lower lattice minimum than those quenched directly
from experiment for more flexible molecules by running replica exchange molecular dynamics and quenching
frames from the low temperature replica.

    The polymorphs of conformationally flexible molecules exhibit differences in
$\Delta G$ of QHA relative to MD on the magnitude of 0.06 kcal/mol even as low as 10 K, 
despite the 0 K  lattice energies being within 0.01 kcal/mol at 0 K, showing that 
agreement between QHA and MD cannot be assumed even at very low temperatures. These 
differences are due to either one or both of the polymorphs expanding quickly at 
low temperatures, which the provides enough free volume to access a diverse 
number of conformations around the alkyl tail. We note that the temperature at 
which a crystal can access these conformations is structure dependent, as seen 
with the different polymorphs of chlorpropamide.

\section*{Acknowledgments}
    The authors thank Eric Dybeck for producing majority of the MD results used 
from our previous paper. Final results for QHA were performed on the Extreme 
Science and Engineering Discovery Environment (XSEDE), which is supported by 
National Science Foundation grant number ACI-1548562. Specifically, it used the 
Bridges system, which is supported by NSF award number ACI-1445606, at the 
Pittsburgh Supercomputing Center (PSC). This work was also supported financially by 
NSF through the grant CBET-1351635.

\newpage
\newpage

\newpage
\newpage

\end{singlespace}
\end{multicols}

\begin{multicols}{2}
\maketitle

\begin{singlespace}
\section{Systems Studied} \label{section:systems_studied}
    In Table \ref{table:systems} we provide the systems studied, their polymorphs names/numberings, and corresponding Cambridge Structure Database
reference code (CSD Refcode), and the dimensions of the unit and supercells relative to the symmetric cell pulled from the
database. The supercell sizes are the same as our previous study and the unit cells are the smaller cell size that assures 
that both crystals have the same number of molecules.
\end{singlespace}
\end{multicols}
\singlespacing
      \begin{longtable}{l|l|l}
      \hline \hline
      \textbf{Piracetam} & Form I & Form III \\
      CSD Refcode & BISMEV03 & BISMEV02 \\
      Space Group & P2$_{1}$/n & P2$_{1}$/n \\
      Z / Z$^\prime$ & 4/1 & 4/1 \\
      Unit cell Dimensions & $ 1\times 1\times 1$ & $ 1\times 1\times 1$ \\
      Supercell Dimensions & $ 4\times 2\times 3$ & $ 2\times 3\times 4$ \\
      \hline \hline
      
      \textbf{Carbamazepine} & Form I & Form III \\
      CSD Refcode & CBMZPN11 & CBMZPN02 \\
      Space Group & P$\bar{1}$ & P2$_{1}$/n \\
      Z / Z$^\prime$ & 8/2 & 4/1 \\
      Unit cell Dimensions & $ 1\times 1\times 1$ & $ 2\times 1\times 1$ \\
      Supercell Dimensions & $ 4\times 2\times 1$ & $ 4\times 2\times 2$ \\
      \hline \hline
      
      \textbf{Adenine} & Form I & Form II \\
      CSD Refcode & KOBFUD  & KOBFUD01 \\
      Space Group & P2$_{1}$/c & Fdd2 \\
      Z / Z$^\prime$ & 8/2 & 16/1 \\
      Unit cell Dimensions & $ 1\times 1\times 2$ & $ 1\times 1\times 1$ \\
      Supercell Dimensions & $ 3\times 1\times 4$ & $ 3\times 1\times 2$ \\
      \hline \hline

      \textbf{Pyrazinamide} & Form $\gamma$ & Form $\delta$ \\
      CSD Refcode & PYRZIN20 & PYRZIN16 \\
      Space Group & Pc & P$\bar{1}$ \\
      Z / Z$^\prime$ & 2/1 & 2/1 \\
      Unit cell Dimensions & $ 1\times 2\times 1$ & $ 2\times 1\times 1$ \\
      Supercell Dimensions & $ 4\times 6\times 2$ & $ 4\times 3\times 4$ \\
      \hline \hline

      \textbf{Resorcinol} & Form $\alpha$ & Form $\beta$ \\
      CSD Refcode & RESORA03 & RESORA08 \\
      Space Group & Pna2$_{1}$ & Pna2$_{1}$ \\
      Z / Z$^\prime$ & 4/1 & 4/1 \\
      Unit cell Dimensions & $ 1\times 1\times 1$ & $ 1\times 1\times 1$ \\
      Supercell Dimensions & $ 2\times 2\times 4$ & $ 2\times 2\times 4$ \\
      \hline \hline

      \textbf{1,4-Diiodobenzene} & Form $\alpha$ & Form $\beta$ \\
      CSD Refcode & ZZZPRO03 & ZZZPRO04 \\
      Space Group & Pbca & Pccn \\
      Z / Z$^\prime$ & 4/1 & 4/1 \\
      Unit cell Dimensions & $ 1\times 1\times 1$ & $ 1\times 1\times 1$ \\
      Supercell Dimensions & $ 2\times 3\times 3$ & $ 2\times 3\times 3$ \\
      \hline \hline

      \textbf{Paracetamol} & Form I & Form II \\
      CSD Refcode & HXACAN01 & HXACAN \\
      Space Group & P2$_{1}$/a & Pcab \\
      Z / Z$^\prime$ & 4/1 & 8/1 \\
      Unit cell Dimensions & $ 1\times 2\times 1$ & $ 1\times 1\times 1$ \\
      Supercell Dimensions & $ 2\times 2\times 3$ & $ 2\times 1\times 3$ \\
      \hline \hline

      \textbf{Aripiprazole} & Form I & Form X \\
      CSD Refcode & MELFIT01 & MELFIT05 \\
      Space Group & P2$_{1}$ & P2$_{1}$ \\
      Z / Z$^\prime$ & 2/1 & 2/1 \\
      Unit cell Dimensions & $ 1\times 1\times 1$ & $ 1\times 1\times 1$ \\
      Supercell Dimensions & $ 3\times 4\times 2$ & $ 3\times 4\times 2$ \\
      \hline \hline

      \textbf{Tolbutamide} & Form I & Form II \\
      CSD Refcode & ZZZPUS04 & ZZZPUS05 \\
      Space Group & Pna2$_{1}$ & Pc \\
      Z / Z$^\prime$ & 4/1 & 8/4 \\
      Unit cell Dimensions & $ 1\times 2\times 1$ & $ 1\times 1\times 1$ \\
      Supercell Dimensions & $ 2\times 3\times 2$ & $ 3\times 2\times 1$ \\
      \hline \hline

      \textbf{Chlorpropamide} & Form I & Form V \\
      CSD Refcode & BEDMIG & BEDMIG04 \\
      Space Group & P2$_{1}$2$_{1}$2$_{1}$ & Pna2$_{1}$ \\
      Z / Z$^\prime$ & 4/1 & 4/1 \\
      Unit cell Dimensions & $ 1\times 1\times 1$ & $ 1\times 1\times 1$ \\
      Supercell Dimensions & $ 1\times 4\times 2$ & $ 1\times 4\times 2$ \\
      \hline \hline
      \caption{Table of systems studied in the main paper. We have included the polymorph names, CSD reference codes, and
               the cell dimensions relative to the symmetric unit cell pulled down from the database.
      \label{table:systems}}
      \end{longtable}
\begin{multicols}{2}
\section{1D vs. Full Anisotropic QHA} \label{section:1v6D_supporting}
   In Table \ref{table:1D_6D_dG} we report the free energy differences of the unit cell of each system using 1D and
fully anisotropic QHA at the maximum temperature that all methods could achieve $T_{max}$. We also report the deviation
between the two methods ($\delta (|\Delta G(T_{max})|)$) in the far right column, which shows that both methods are within
0.02 kcal/mol of each other for all systems except for chlorpropamide.
    \begin{table*}
      \begin{tabular}{c|c|c|c|c}
                        &           & \multicolumn{2}{c}{$\Delta G(T_{max})$ [kcal/mol]} & $\delta (|\Delta G(T_{max})|)$ \\
      System            & $T_{max}$ [K] & 1D-QHA & Aniso.-QHA &  \\ \hline \hline
      Piracetam         & 255       & -1.374 & -1.384         & 0.010 \\
      Carbamazepine     & 300       & -0.160 & -0.157         & 0.003 \\
      Adenine           & 300       &  1.330 &  1.328         & 0.002 \\
      Pyrazinamide      & 180       & -0.552 & -0.551         & 0.001 \\
      Resorcinol        & 300       & -0.578 & -0.578         & 0.000 \\
      1,4-Diiodobenzene & 210       &  0.431 &  0.431         & 0.000 \\
      Paracetamol       & 150       & -0.648 & -0.652         & 0.004 \\
      Aripiprazole      & 180       & -0.696 & -0.712         & 0.016 \\
      Tolbutamide       & 100       & -3.592 & -3.593         & 0.001 \\
      Chlorpropamide    & 150       &  4.830 &  4.824         & 0.006 \\
      \end{tabular}
      \caption{Free energy differences for 1D and fully anisotropic QHA at $T_{max}$ for the unit cells. $T_{max}$ is
               the maximum temperature that both crystals and both methods can reach while still being at a free energy
               minimum. For all systems there is $<$ 0.02 kcal/mol difference in $\Delta G$ with aripiprazole having the
               largest difference of 0.016 kcal/mol.
      \label{table:1D_6D_dG}}
    \end{table*}

\begin{figure*}
      \begin{center}
      \includegraphics[width=8cm]{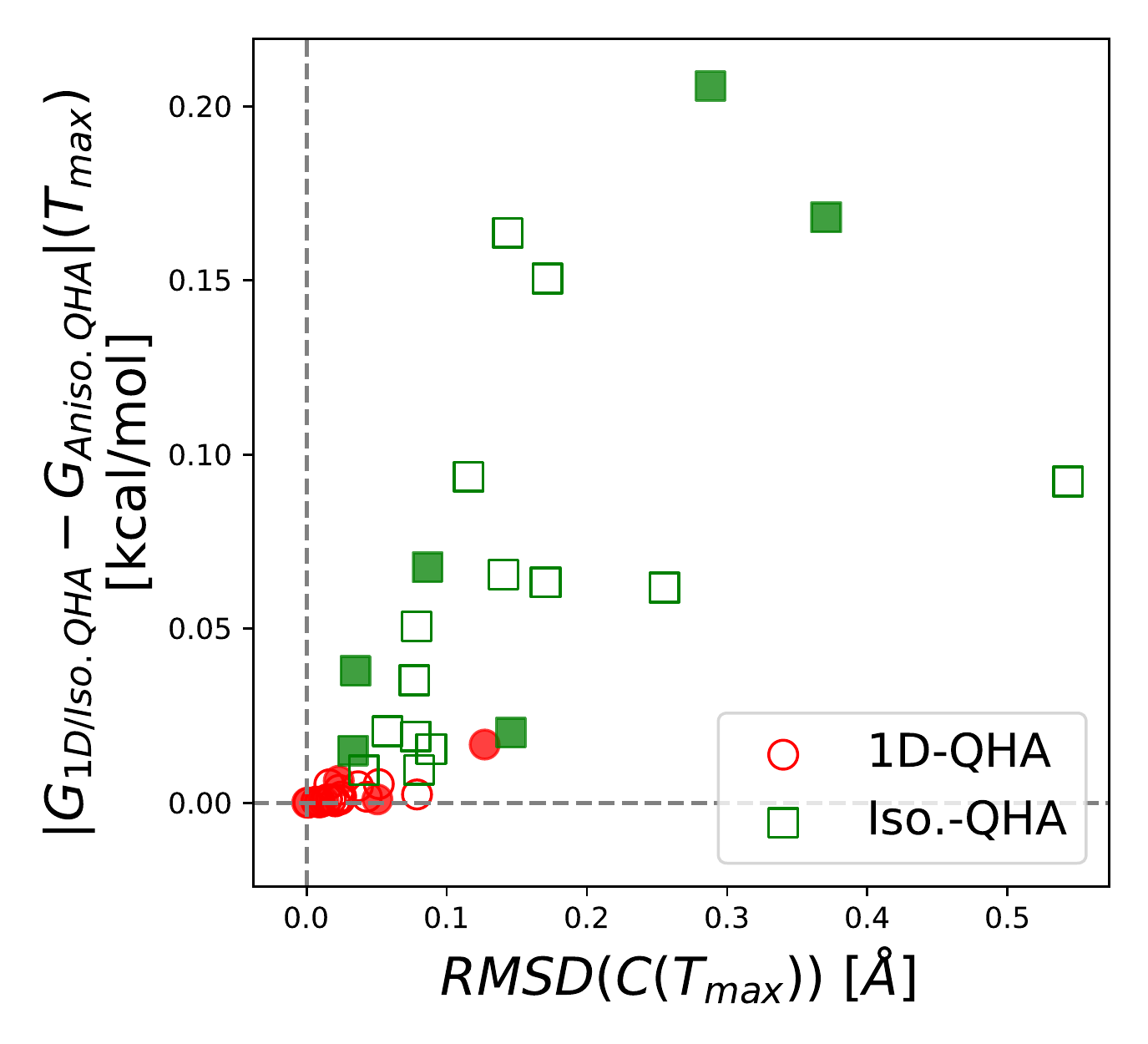}
      \end{center}
      \caption{Deviations in the free energy of the constrained QHA (isotropic QHA and 1D-QHA) 
from anisotropic QHA versus the box vector RMSD from the unconstrained QHA structure. All values are 
reported at $T_{max}$. For individual free energies the 1D-variant deviates $<$ 0.02 kcal/mol, which
is 10 times more precise than isotropic QHA that deviates up to 0.21 kcal/mol from the fully anisotropic
approach. 
Filled in markers are the systems with multiple molecular conformers (tolbutamide, chlorpropamide, and aripiprazole). 
      \label{figure:1D_vs_6D_G}}
    \end{figure*}

    Differences in $\Delta G$ of the anisotropic methods are correlated with the 
errors in the predicted high temperature lattice  geometries.
Figure~\ref{figure:1D_vs_6D_G} has the absolute free energy deviations between 
1D- and fully-anisotropic QHA versus the root-mean-square deviation (RMSD) between 
$T_{max}$ lattice geometries for individual polymorph. The RMSD is computed as:
    \begin{eqnarray}
      RMSD = \sqrt{\frac{1}{6} \sum_{i=1}^{6} (C_{i}^{1D}(T_{max}) - C_{i}^{Full}(T_{max}))}
    \end{eqnarray}
where the RMSD is taken between the six lattice parameters of the upper triangular 
crystal tensor form. From 
Figure~\ref{figure:1D_vs_6D}c, we can see that as the RMSD between the 1D- and 
fully-anisotropic $T_{max}$ structure becomes larger, so does the error in the 
computed free energy of the polymorph. The error in thermal expansion and free 
energy for isotropic QHA is reduced by using the 1D-QHA variant. The differences
in the free energy are a direct result of the crystal expanding to a slightly 
different geometry, but we can see that this has little effect on the free energy
differences.

    The 1D-variant reduces the error of the isotropic QHA up to 90\% for the systems studied relative
to the fully anisotropic approach. The deviations between 1D-QHA and fully ansiotropic QHA are $<$ 0.02
kcal/mol, whereas the error for isotropic QHA is as large as 0.21 kcal/mol for individual polymorphs 
(Figure ~\ref{figure:1D_vs_6D_G}). The largest deviations in free energy for isotropic expansion come from aripiprazole
form X and chlorpropamide form V. The largest deviation in free energy for 1D-QHA (0.0167 kcal/mol) is from aripiprazole
form X.

\section{Unit vs. Supercell 1D-QHA}
   In Table \ref{table:UvS_dG} we report the free energy differences of the unit and supercell of each system using 1D-QHA 
at the maximum temperature all cell sizes could achieve $T_{max}$. We also report the deviation between the two cell sizes
($\delta (|\Delta G(T_{max})|)$) in the far right column. We see that there are significant finite size errors for all crystals
that range between 0.03 -- 0.50 kcal/mol.
    \begin{table*}
      \begin{tabular}{c|c|c|c|c}
                        &           & \multicolumn{2}{c}{$\Delta G(T_{max})$ [kcal/mol]} & $\delta (|\Delta G(T_{max})|)$ \\
      System            & $T_{max}$ [K] & Unit & Super &  \\ \hline \hline
      Piracetam         & 300       & -1.364 & -1.457         & 0.093 \\
      Carbamazepine     & 300       & -0.160 &  0.044         & 0.204 \\
      Adenine           & 300       &  1.330 &  1.603         & 0.273 \\
      Pyrazinamide      & 300       & -0.756 & -0.613         & 0.143 \\
      Resorcinol        & 300       & -0.578 & -0.439         & 0.139 \\
      1,4-Diiodobenzene & 300       &  0.427 &  0.389         & 0.038 \\
      Paracetamol       & 300       & -1.241 & -0.767         & 0.474 \\
      Aripiprazole      & 300       & -0.764 & -0.950         & 0.186 \\
      Tolbutamide       & 180       & -3.602 & -3.489         & 0.113 \\
      Chlorpropamide    &  50       &  5.374 &  5.287         & 0.087 \\
      \end{tabular}
      \caption{Free energy differences for unit and supercells at $T_{max}$ using 1D-QHA are reported. $T_{max}$ is
               the maximum temperature that both crystals and cell sizes can reach while still being at a free energy
               minimum. 
               We find that size errors on the magnitude of 0.03 -- 0.5 kcal/mol for polymorph free energy differences.
      \label{table:UvS_dG}}
    \end{table*}

\begin{figure}[H]
  \begin{center}
  \includegraphics[width=8cm]{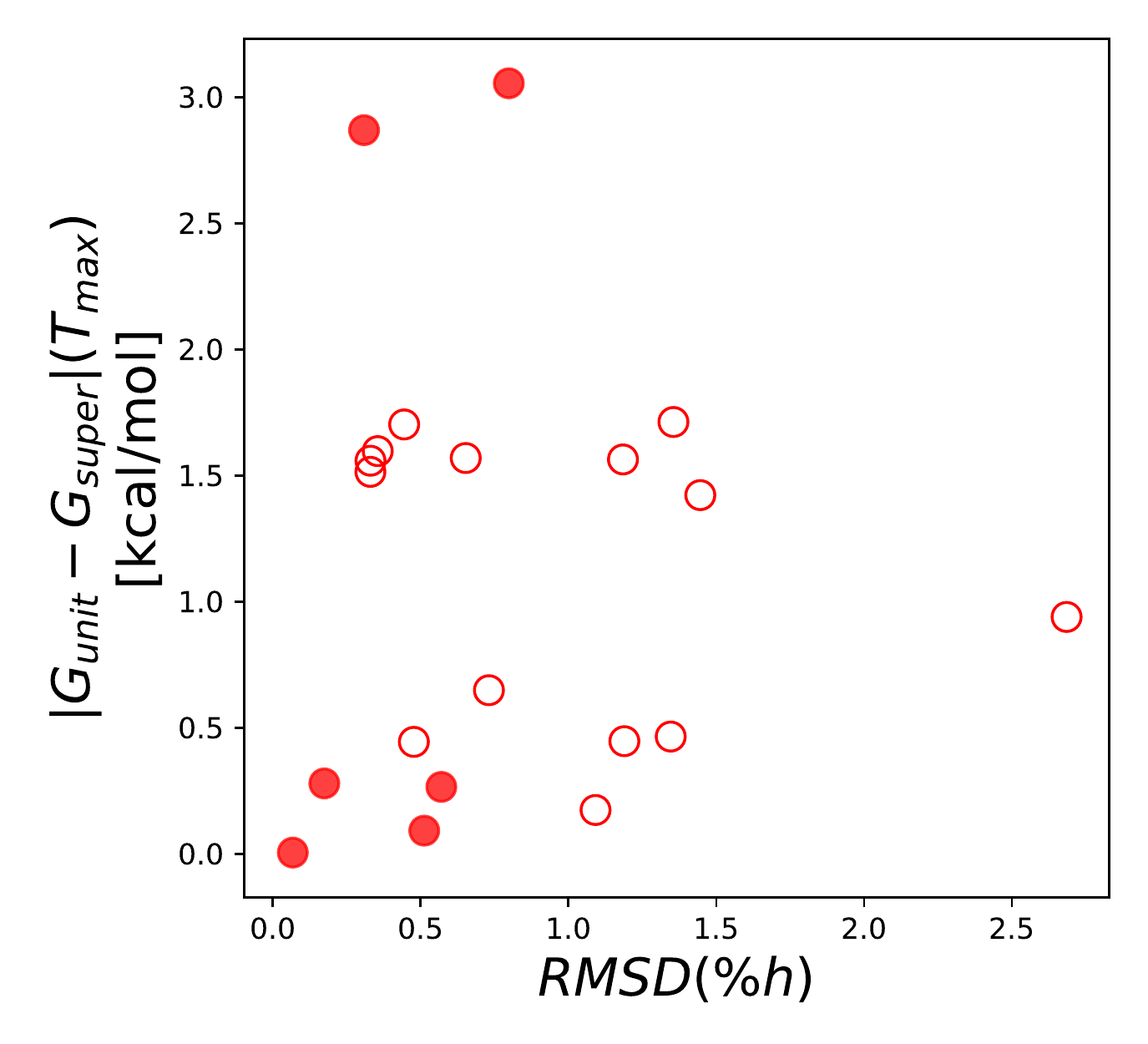}
  \end{center}
  \caption{The deviation in the free energy between unit and super cell is shown relative to the RMSD in the percent expansion. For individual polymorphs there is no correlation between error in the free energy and the error in the high temperature lattice geometries. The filled in markers are the systems with multiple molecular conformers (tolbutamide, chlorpropamide, and aripiprazole)
  \label{figure:U_vs_S_G}}
\end{figure}

    For individual polymorphs, there is no trend between the differences in 
free energy and the RMSD in the percent expansion up to $T_{max}$ for unit and 
supercells; the trends only appear for polymorph differences. 
Figure \ref{figure:U_vs_S_G} compares the absolute deviation in free energy with 
the RMSD in the percent expansion for unit and supercells. The error in 
free energy between unit and super cells is between 0 -- 3 kcal/mol and does 
not correlate with the cells percent expansion at $T_{max}$.

\begin{figure}[H]
  \begin{center}
  \includegraphics[width=8cm]{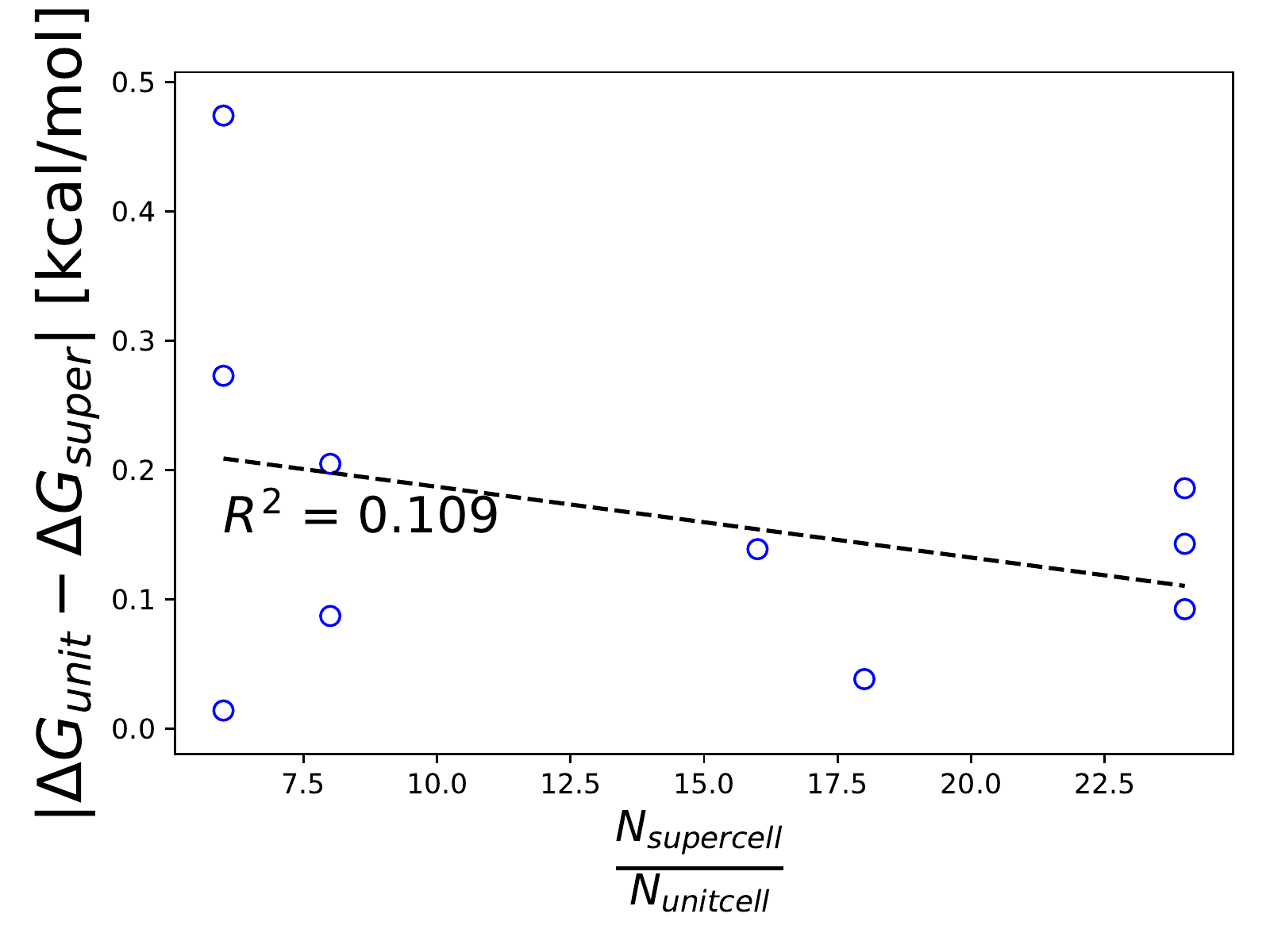}
  \end{center}
  \caption{Plotting the finite size errors at $T_{max}$ versus the ratio of number of molecules between the super- and unit cell.
  \label{figure:size_correlation}}
\end{figure}

    The finite size errors do not correlate with the ratio of number of molecules
in the unit and supercells. We have super cells that have 6 -- 24 times the number
of molecules in the unit cells. We would expect the systems with similar number of
molecules in the unit cell and supercell to have smaller deviations in the free
energy, but that is not the case. In Figure~\ref{figure:size_correlation} the 
free energy deviation between the unit and supercell at 300 K is plotted against
the ratio of molecules in the super cell relative to the unit cell. The greatest error
is seen for paracetamol, which is one of the systems with the most similar unit and super 
cells. There is no correlation between the finite size error and ratio of number of 
molecules ($R^{2} = 0.109$).

\pagebreak
\section{Re-structuring}
\begin{figure}[H]
  \begin{center}
  \includegraphics[height=6cm]{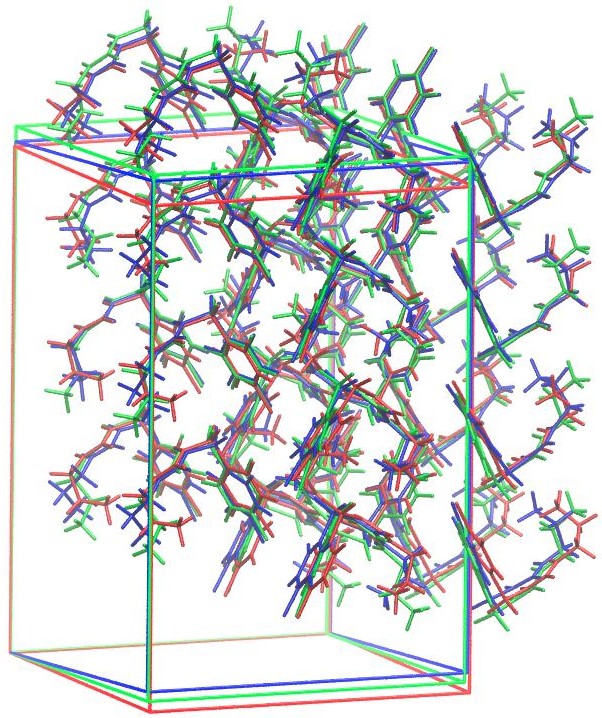}
  \end{center}
  \caption{Lattice minimum of chlorpropamide form V from quenching the experimental structure (green), the structure used in our previous study (blue),~\cite{dybeck_capturing_2017} quenched from a 10 K replica exchange simulation (red). The RMSD between the structure quenched from experiment and from our previous paper relative to the minima found from REMD are  1.26 and 1.07 \r{A}, respectively.
  \label{figure:restruc_bedmig5}}
\end{figure}
\begin{figure}[H]
  \begin{center}
  \includegraphics[height=6cm]{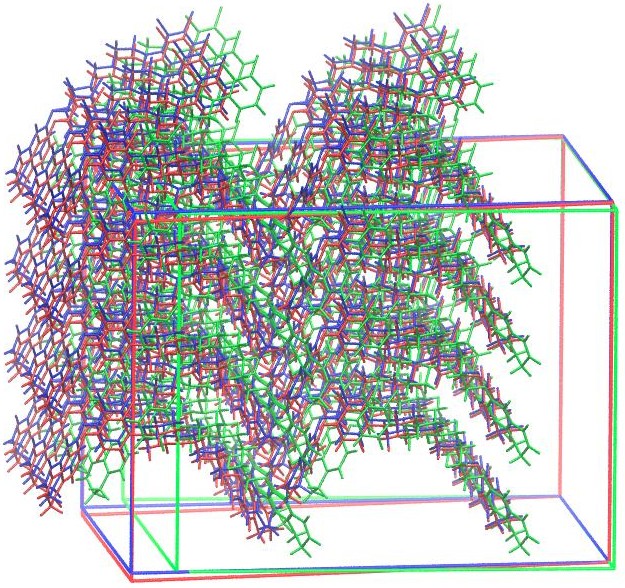}
  \end{center}
  \caption{Lattice minimum of aripiprazole form I from quenching the experimental structure (green), the structure used in our previous study (blue),~\cite{dybeck_capturing_2017} quenched from a 10 K replica exchange simulation (red). The RMSD between the structure from our previous paper relative to the minima found from REMD is 0.36 \r{A}.
  \label{figure:restruc_melfit1}}
\end{figure}
\begin{figure}[H]
  \begin{center}
  \includegraphics[height=6cm]{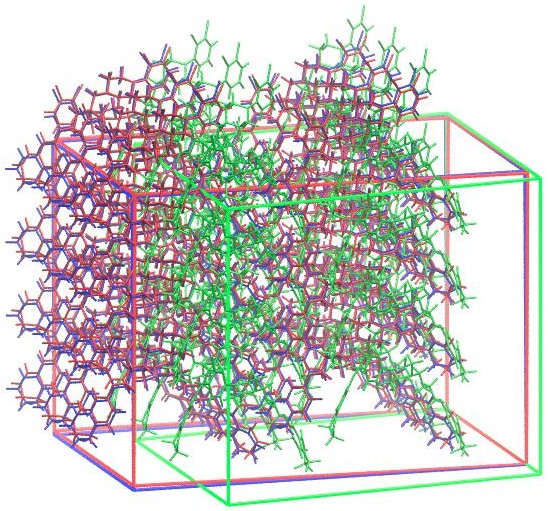}
  \end{center}
  \caption{Lattice minimum of aripiprazole form X from quenching the experimental structure (green), the structure used in our previous study (blue),~\cite{dybeck_capturing_2017} quenched from a 10 K replica exchange simulation (red). The RMSD between the structure from our previous paper relative to the minima found from REMD is 0.51 \r{A}.
  \label{figure:restruc_melfit5}}
\end{figure}
\begin{figure}[H]
  \begin{center}
  \includegraphics[height=4.5cm]{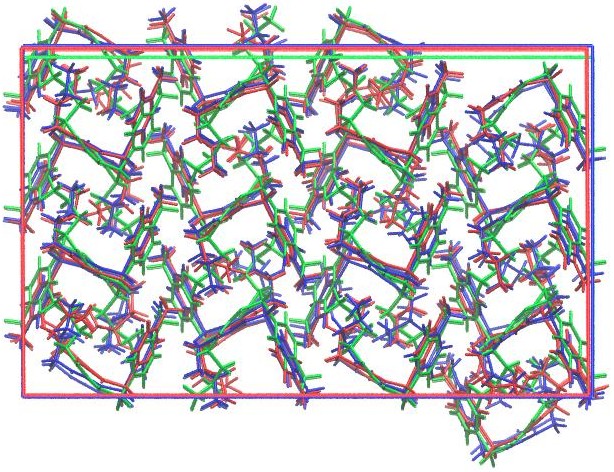}
  \end{center}
  \caption{Lattice minimum of tolbutamide form I from quenching the experimental structure (green), the structure used in our previous study (blue),~\cite{dybeck_capturing_2017} quenched from a 10 K replica exchange simulation (red). The RMSD between the structure quenched from experiment and from our previous paper relative to the minima found from REMD are  1.48 and 1.09 \r{A}, respectively.
  \label{figure:restruc_zzzpus1}}
\end{figure}

\pagebreak

\section{Free Energy Differences and Expansion with Temperature} \label{section:complete_results}

    The original MD data for several systems were run with constraints that prevented the 
angles from changing. While there fluctuations in the angles should have been considered, 
the angles remain 90\textdegree across the entire temperature range. For the following 
systems the MD angles were fixed at 90\textdegree: adenine from II, resorcinol from 
$\alpha$ and $\beta$, diiodobenzene form $\alpha$ and $\beta$-restructured, and paracetamol 
form II. For all crystals where the angles are fixed at 90\textdegree, we ran 1 ns 
simulations at 50, 100, 200, and 300 K with the angles unfixed and verified that average 
value of the angle remains at 90\textdegree.
\end{multicols}

\subsection{Piracetam}
\begin{figure*}
  \begin{center}
  \includegraphics[width=16cm]{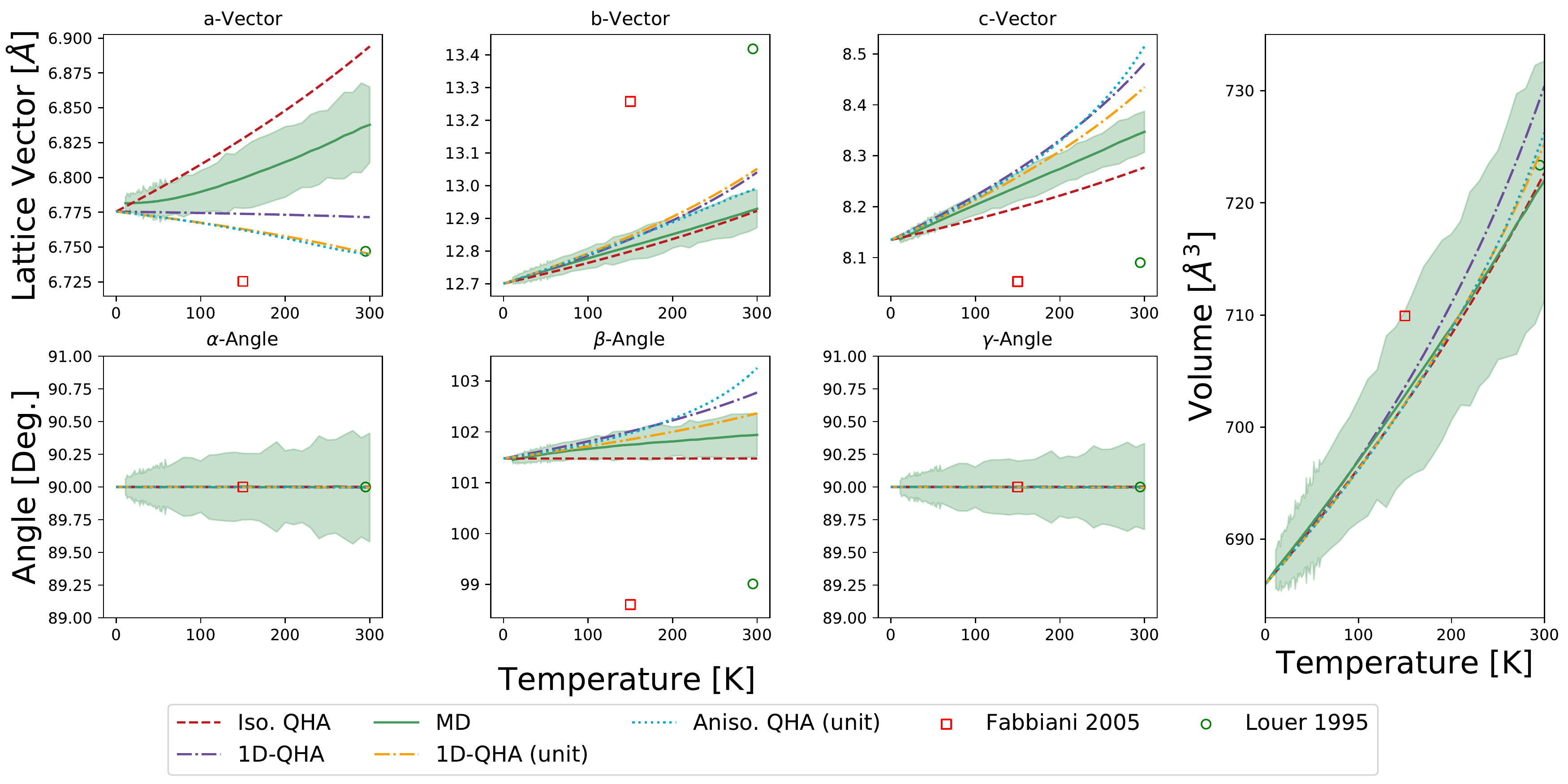}
  \end{center}
  \caption{Piracetam form I percent expansion from 0 K with experimental results.~\cite{fabbiani_exploration_2005,louer_structure_1995}}
\end{figure*}
\begin{figure*}
  \begin{center}
  \includegraphics[width=16cm]{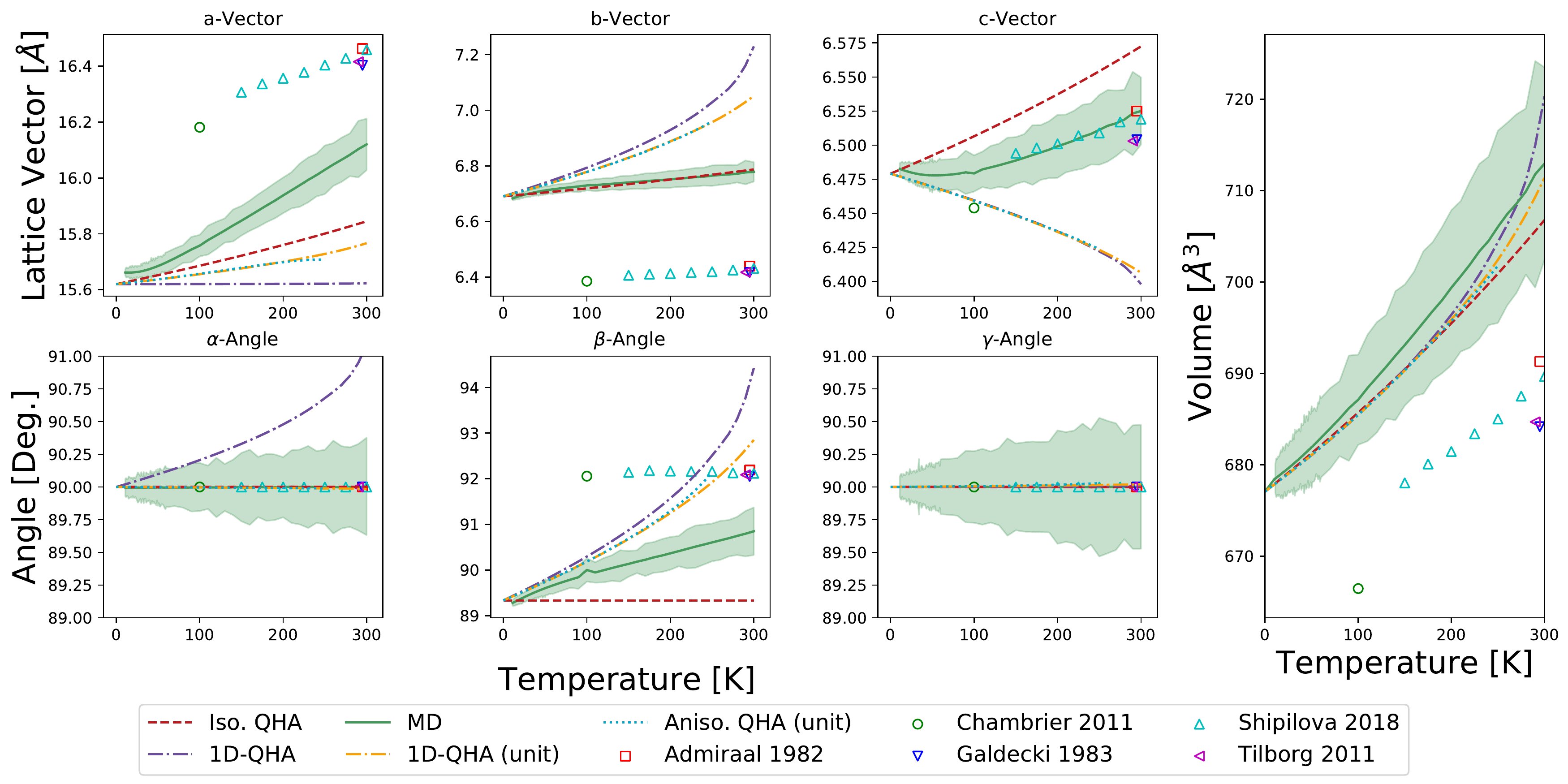}
  \end{center}
  \caption{Piracetam form III percent expansion from 0 with experimental results.~\cite{admiraal_structures_1982,chambrier_electron_2011,galdecki_1983,shipilova_x-ray_2018,tilborg_structural_2011}}
\end{figure*}
%

\clearpage
\subsection{Carbamazepine}
\begin{figure*}
  \begin{center}
  \includegraphics[width=12cm]{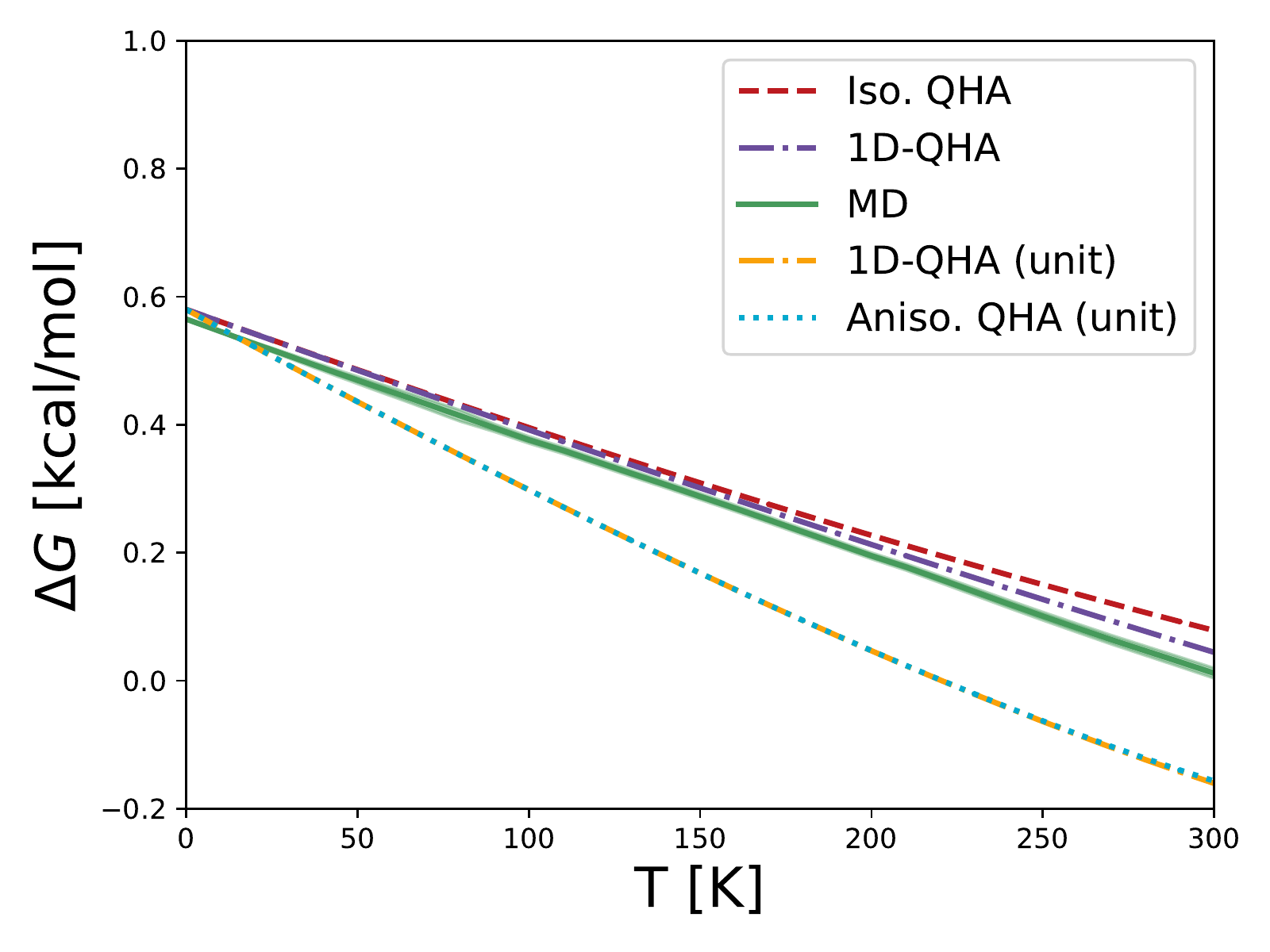}
  \end{center}
  \caption{Polymorph free energy differences of carbamazepine form I relative to form III}
\end{figure*}
\begin{figure*}
  \begin{center}
  \includegraphics[width=16cm]{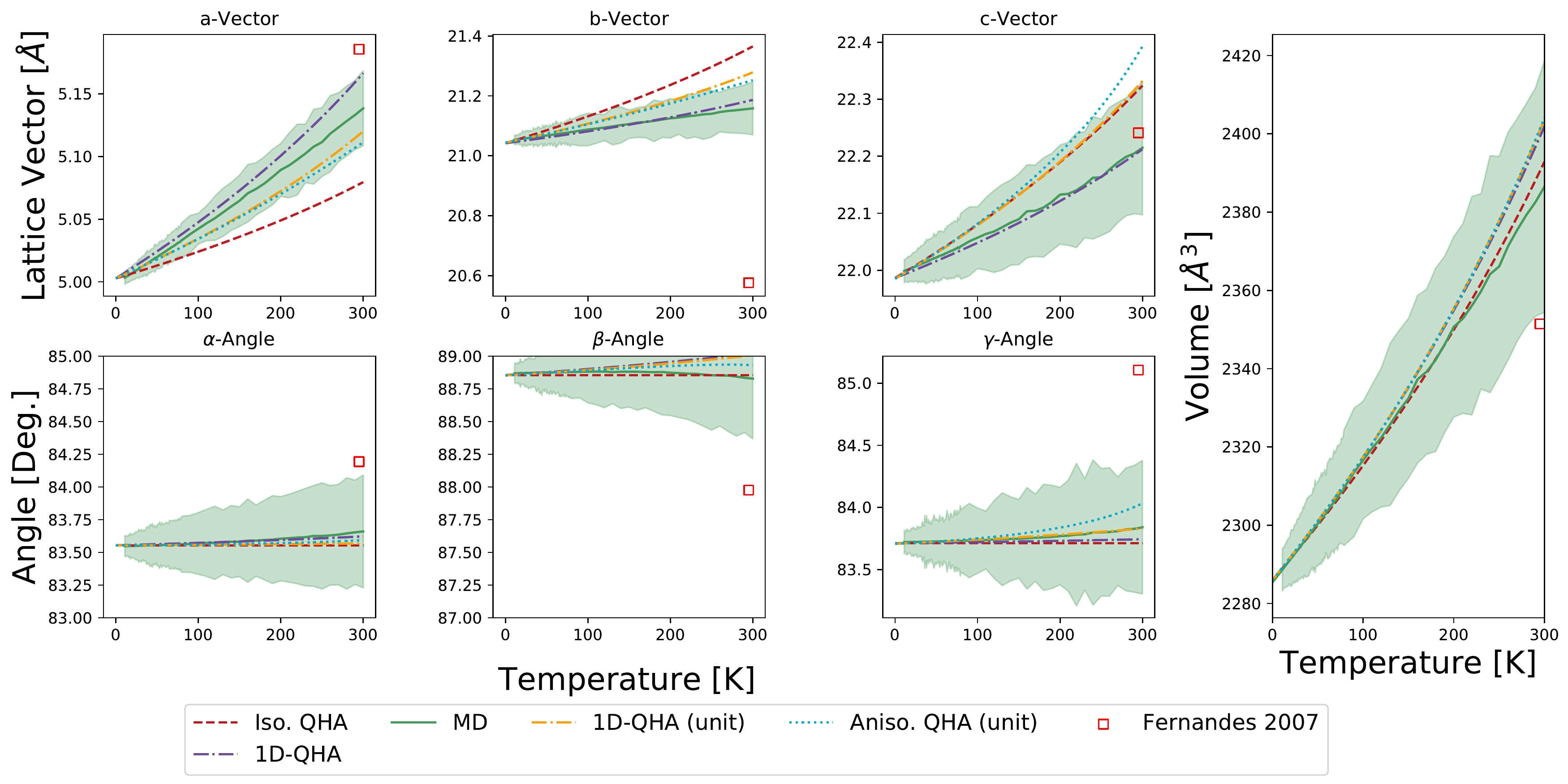}
  \end{center}
  \caption{Carbamazepine form I percent expansion from 0 K with experimental results.~\cite{fernandes_solving_2007}}
\end{figure*}
\begin{figure*}
  \begin{center}
  \includegraphics[width=16cm]{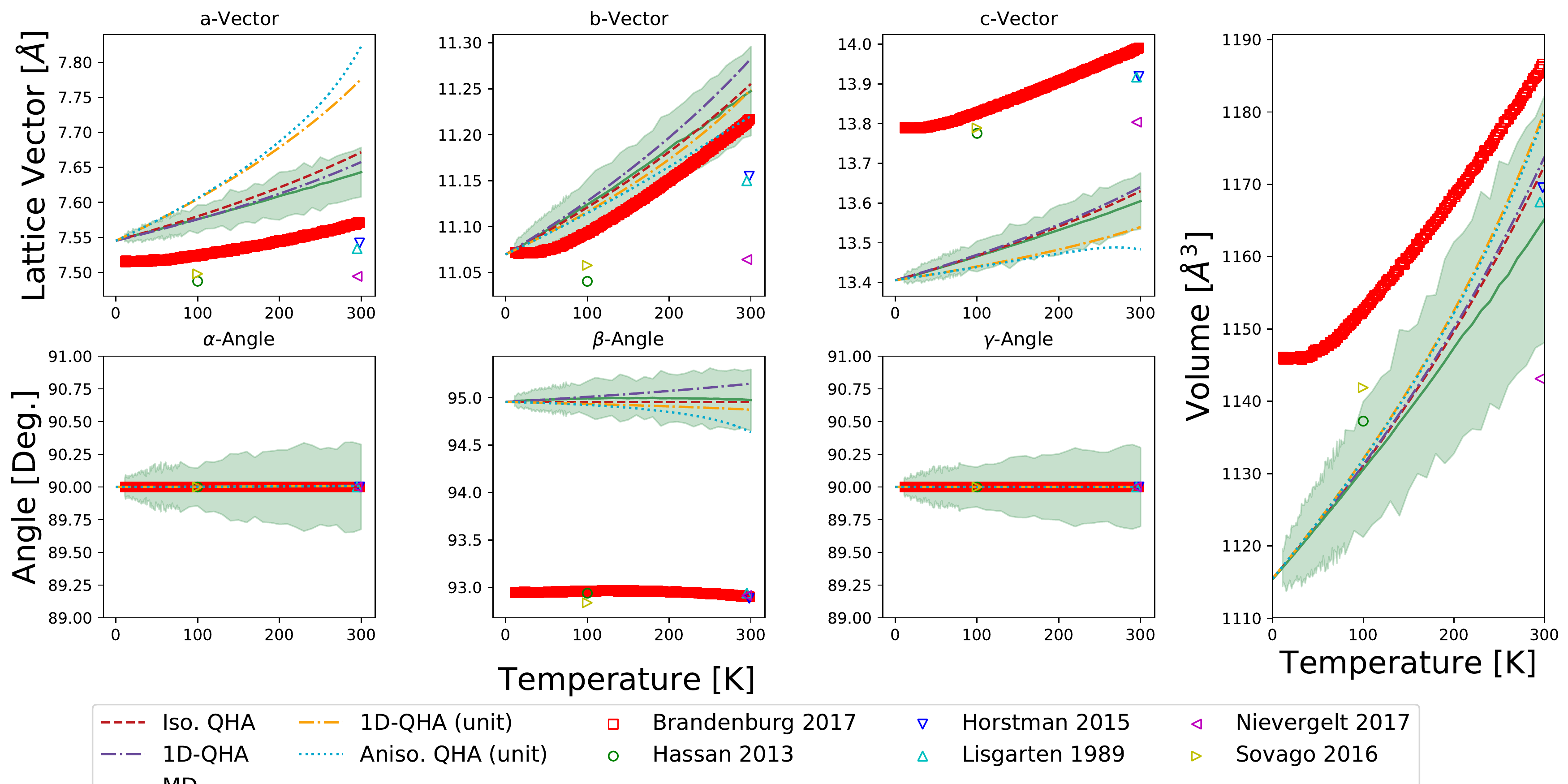}
  \end{center}
  \caption{Carbamazepine form III percent expansion from 0 K with experimental results.~\cite{brandenburg_thermal_2017,horstman_crystallization_2015,lisgarten_crystal_1989,nievergelt_growing_2016,sovago_electron_2016,el_hassan_electron_2013}}
\end{figure*}
%

\clearpage
\subsection{Adenine}
\begin{figure*}
  \begin{center}
  \includegraphics[width=12cm]{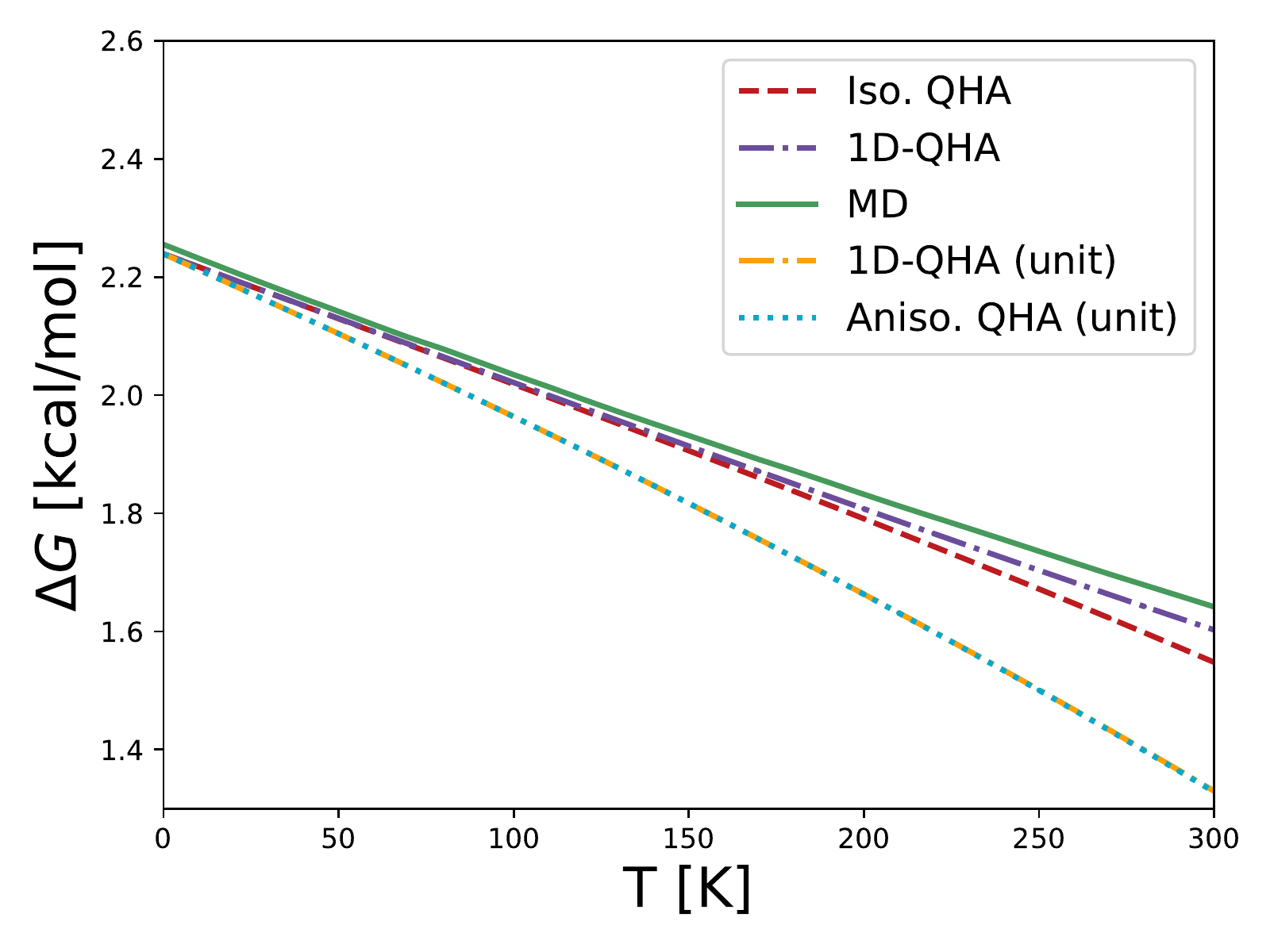}
  \end{center}
  \caption{Polymorph free energy differences of adenine form II relative to form I}
\end{figure*}
\begin{figure*}
  \begin{center}
  \includegraphics[width=16cm]{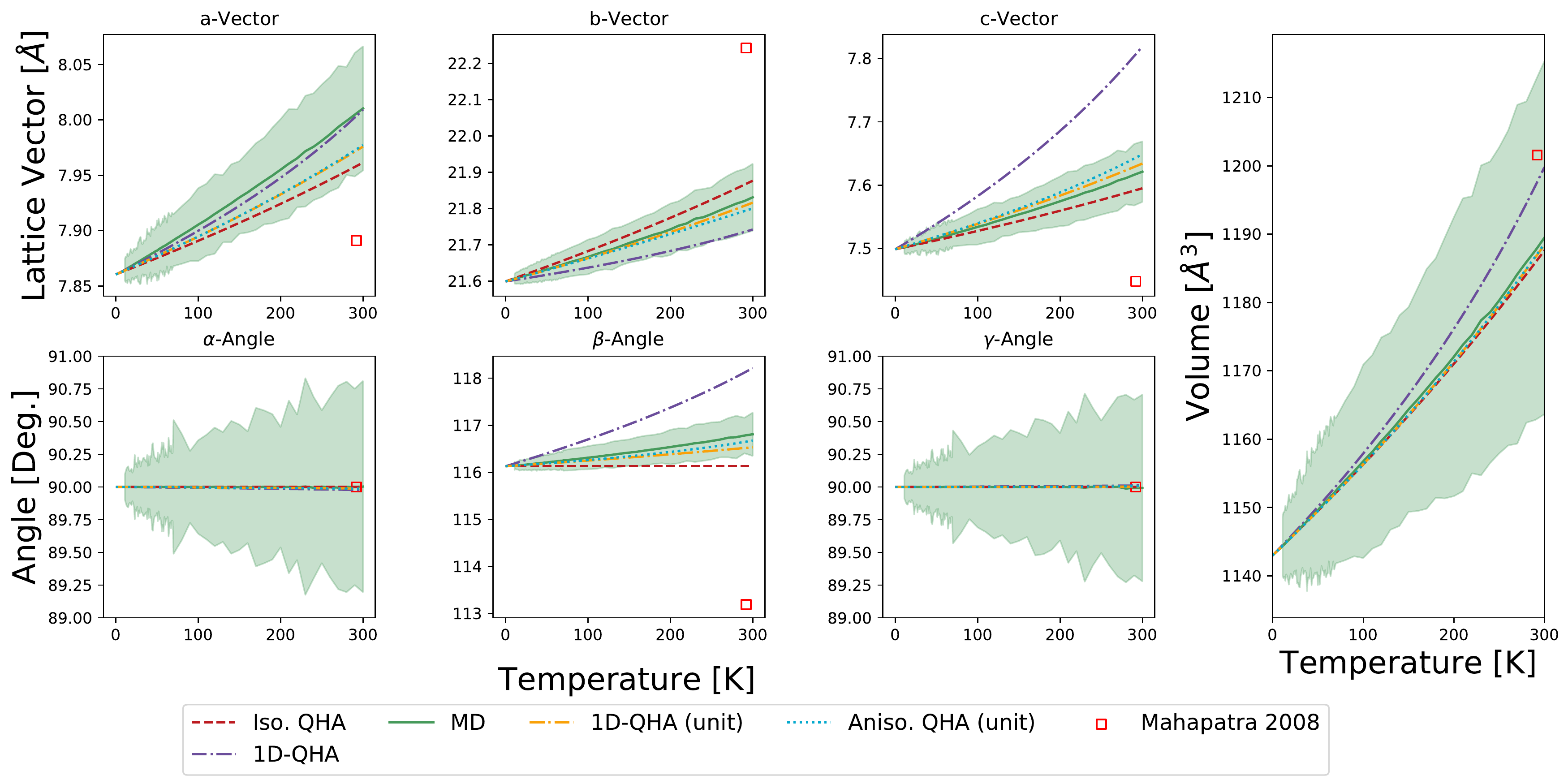}
  \end{center}
  \caption{Adenine from I percent expansion from 0 K with experimental results.~\cite{mahapatra_anhydrous_2008}}
\end{figure*}
\begin{figure*}
  \begin{center}
  \includegraphics[width=16cm]{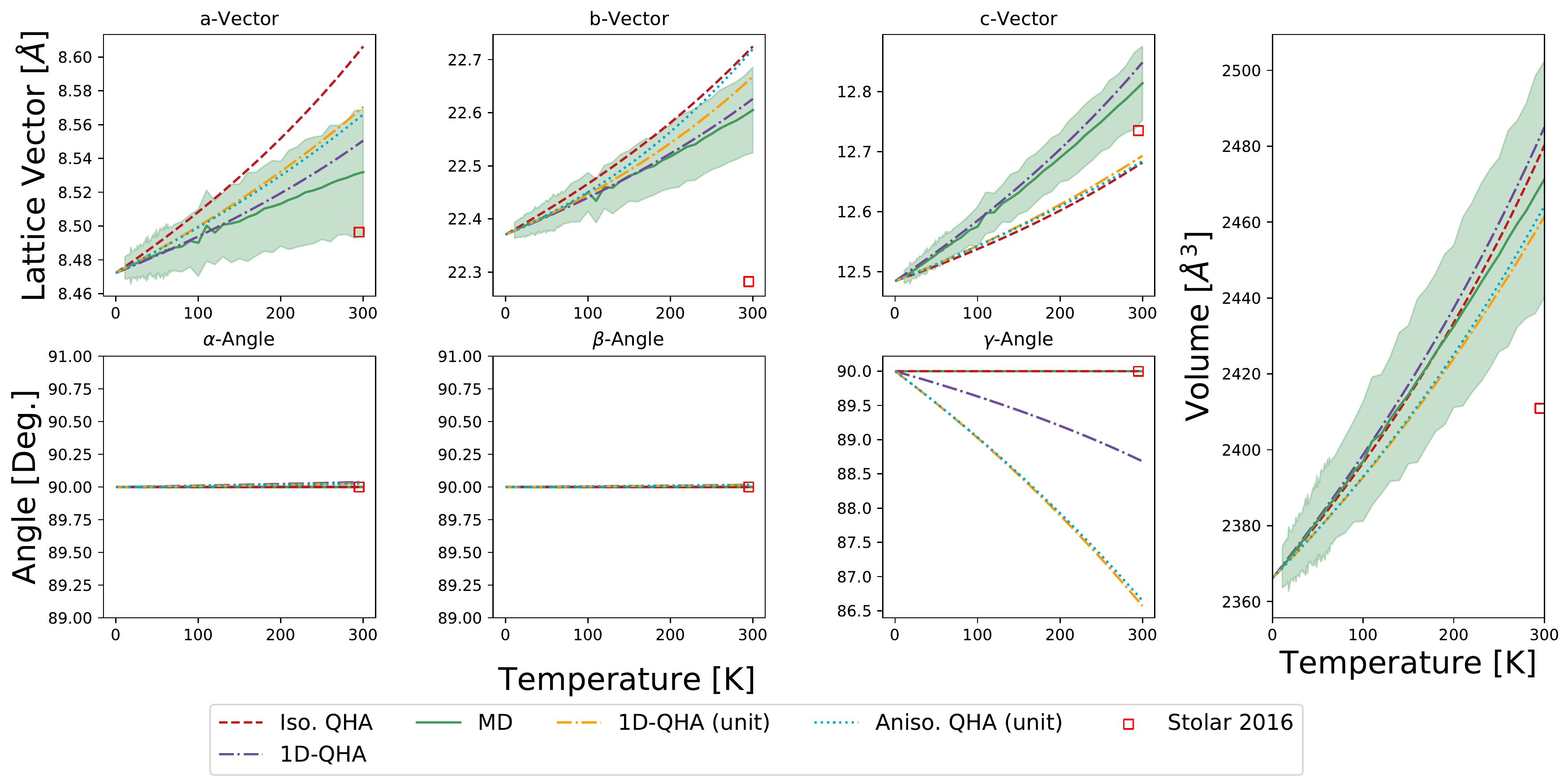}
  \end{center}
  \caption{Adenine form II percent expansion from 0 K with experimental results.~\cite{stolar_solid-state_2016}}
\end{figure*}
%
\clearpage
\subsection{Pyrazinamide}
\begin{figure*}
  \begin{center}
  \includegraphics[width=12cm]{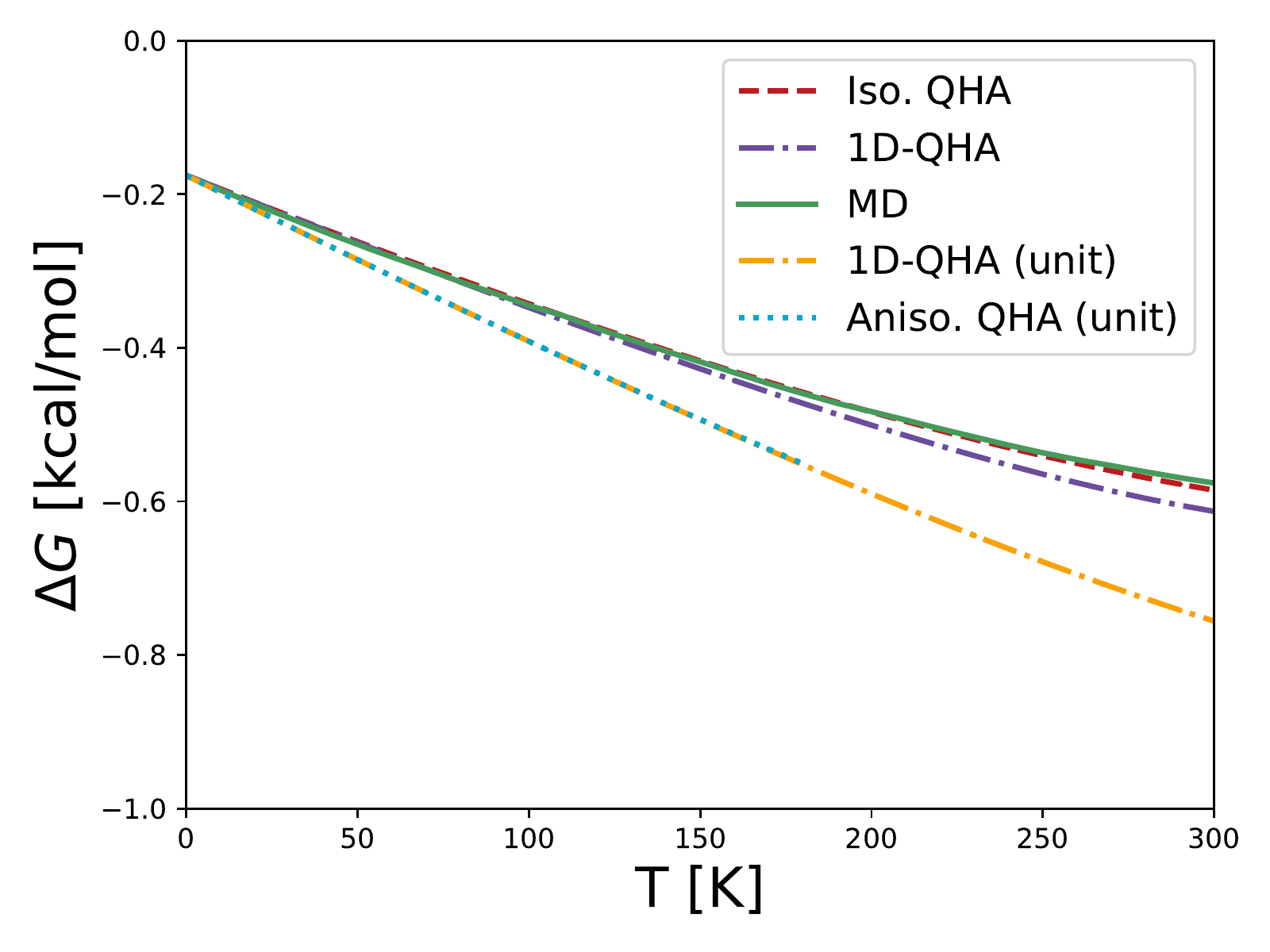}
  \end{center}
  \caption{Polymorph free energy differences of pyrazinamide form $\gamma$ relative to form $\delta$}
\end{figure*}
\begin{figure*}
  \begin{center}
  \includegraphics[width=16cm]{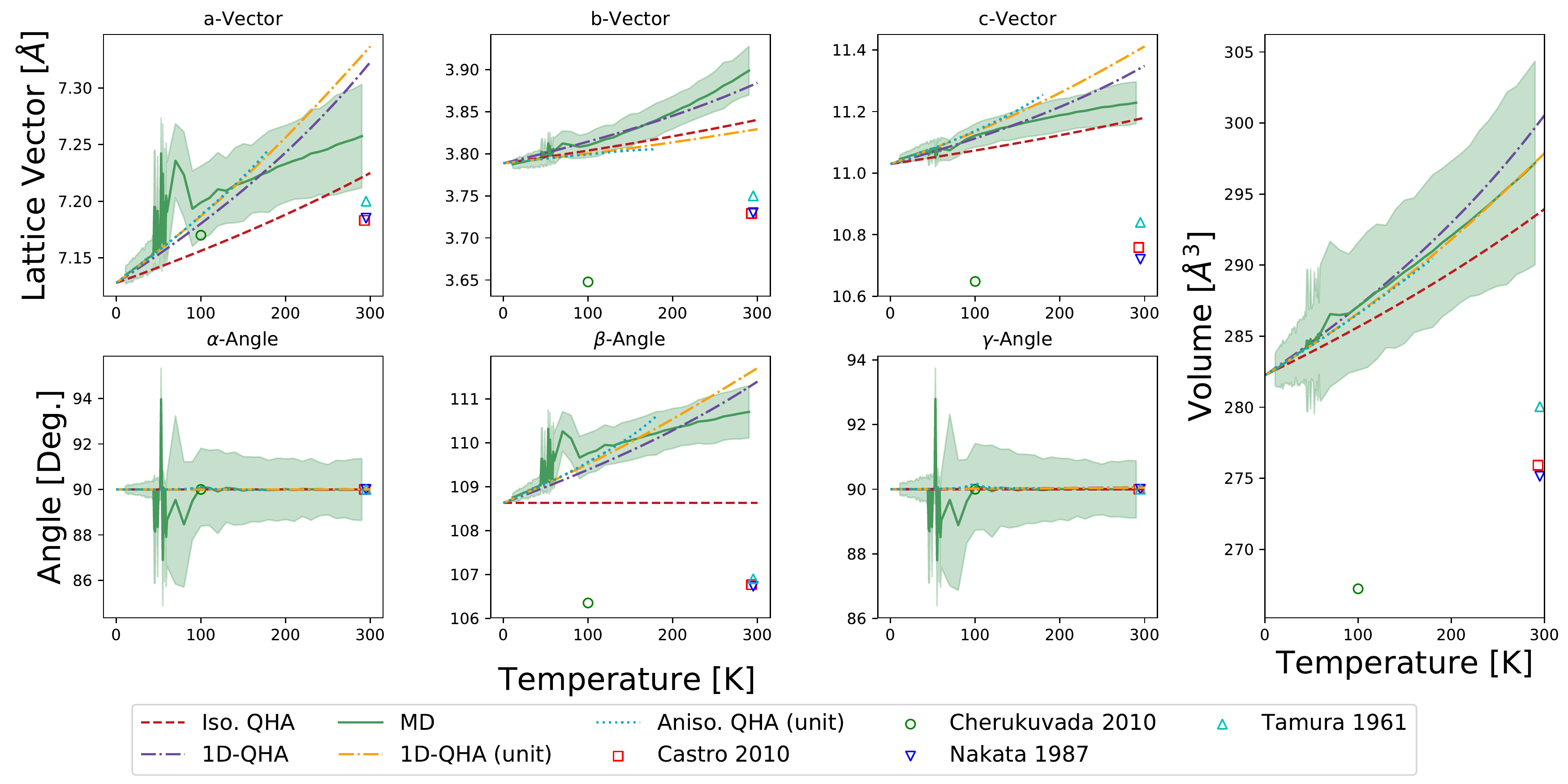}
  \end{center}
  \caption{Pyrazinamide form $\gamma$ percent expansion from 0 K with experimental results.~\cite{nangia_2005,ro_crystal_1972}}
\end{figure*}
\begin{figure*}
  \begin{center}
  \includegraphics[width=16cm]{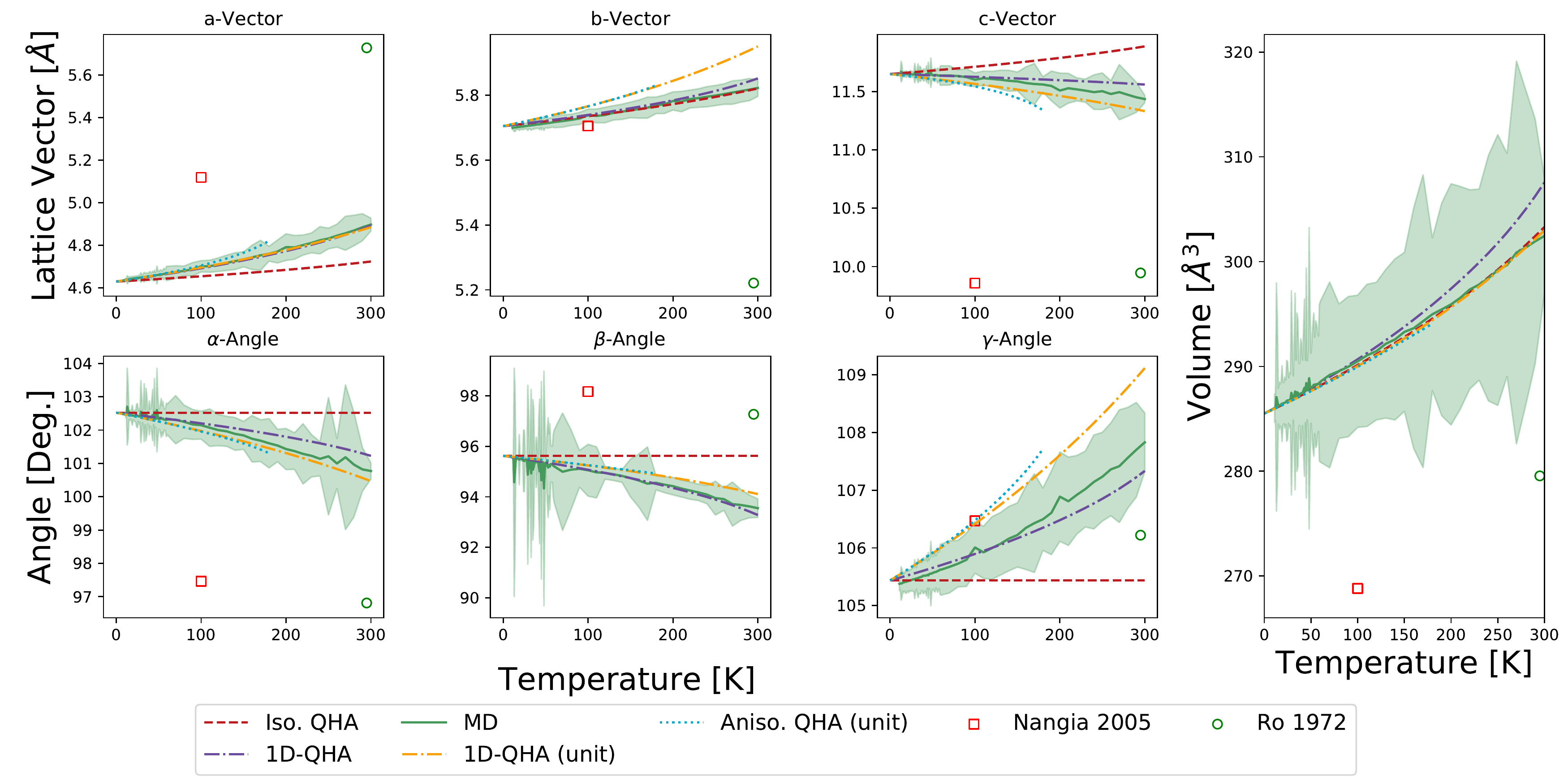}
  \end{center}
  \caption{Pyrazinamide form $\delta$ percent expansion from 0 K with experimental results.~\cite{castro_new_2010,cherukuvada_pyrazinamide_2010,nakata_1987,tamura_crystallographic_1961}}
\end{figure*}
%
\clearpage
\subsection{Resorcinol}
\begin{figure*}
  \begin{center}
  \includegraphics[width=12cm]{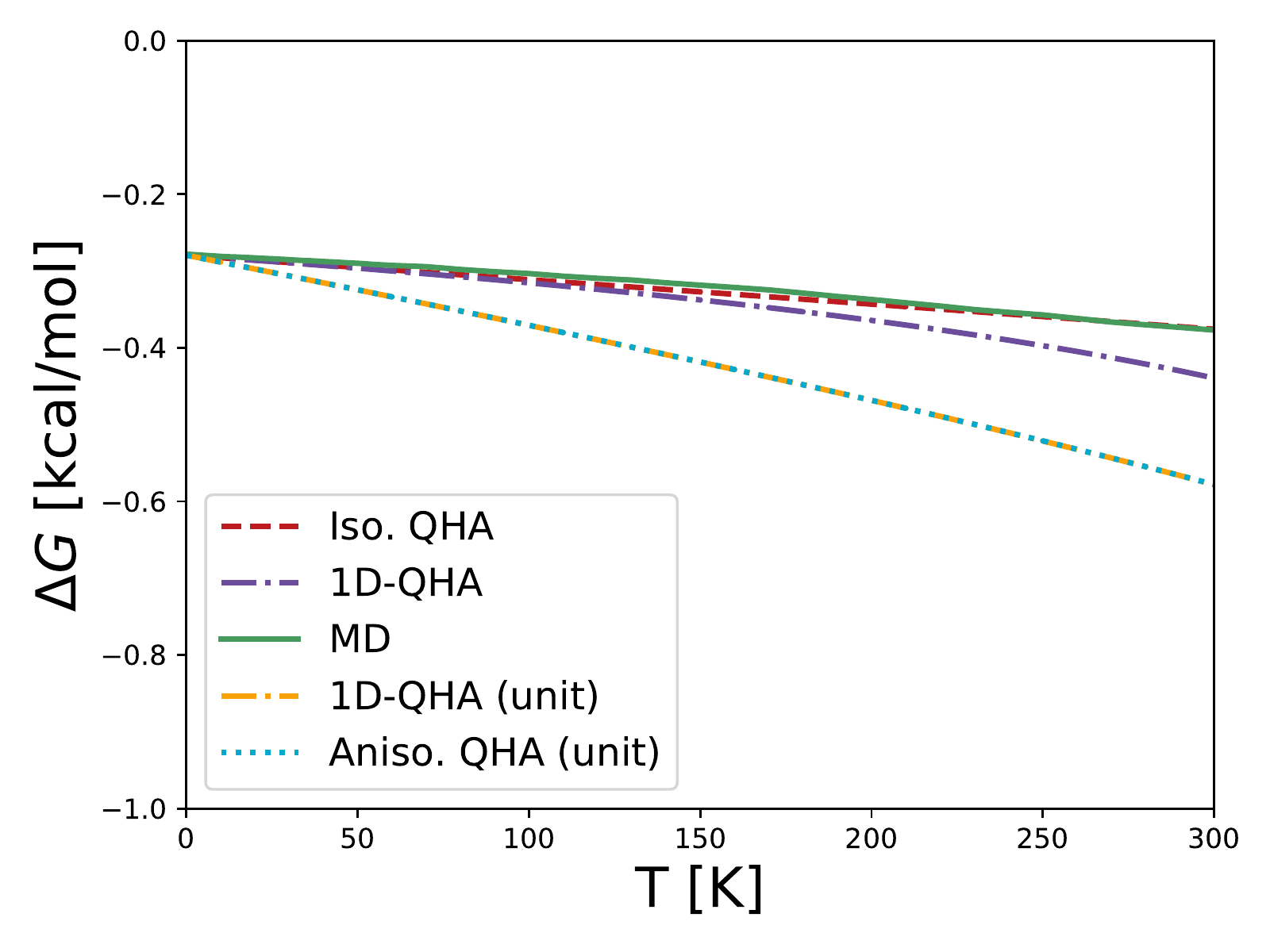}
  \end{center}
  \caption{Polymorph free energy differences of carbamazepine form $\beta$ relative to form $\alpha$}
\end{figure*}
\begin{figure*}
  \begin{center}
  \includegraphics[width=16cm]{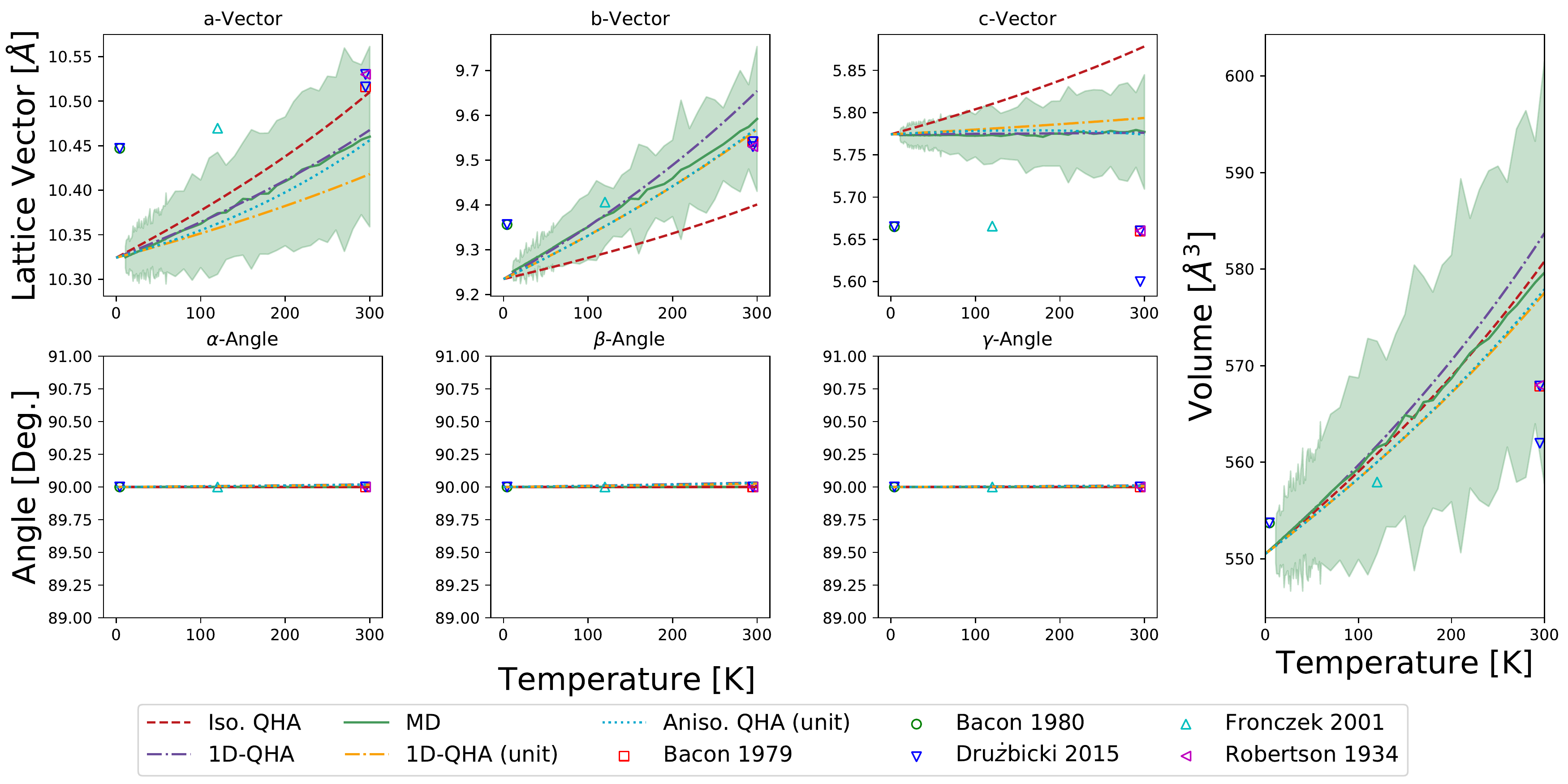}
  \end{center}
  \caption{Resorcinol form $\alpha$ percent expansion from 0 K with experimental results.~\cite{bacon_neutron_1979,bacon_neutron_1980,druzbicki_polymorphism_2015,fronczek_2001,robertson_j._monteath_new_1938}}
\end{figure*}
\begin{figure*}
  \begin{center}
  \includegraphics[width=16cm]{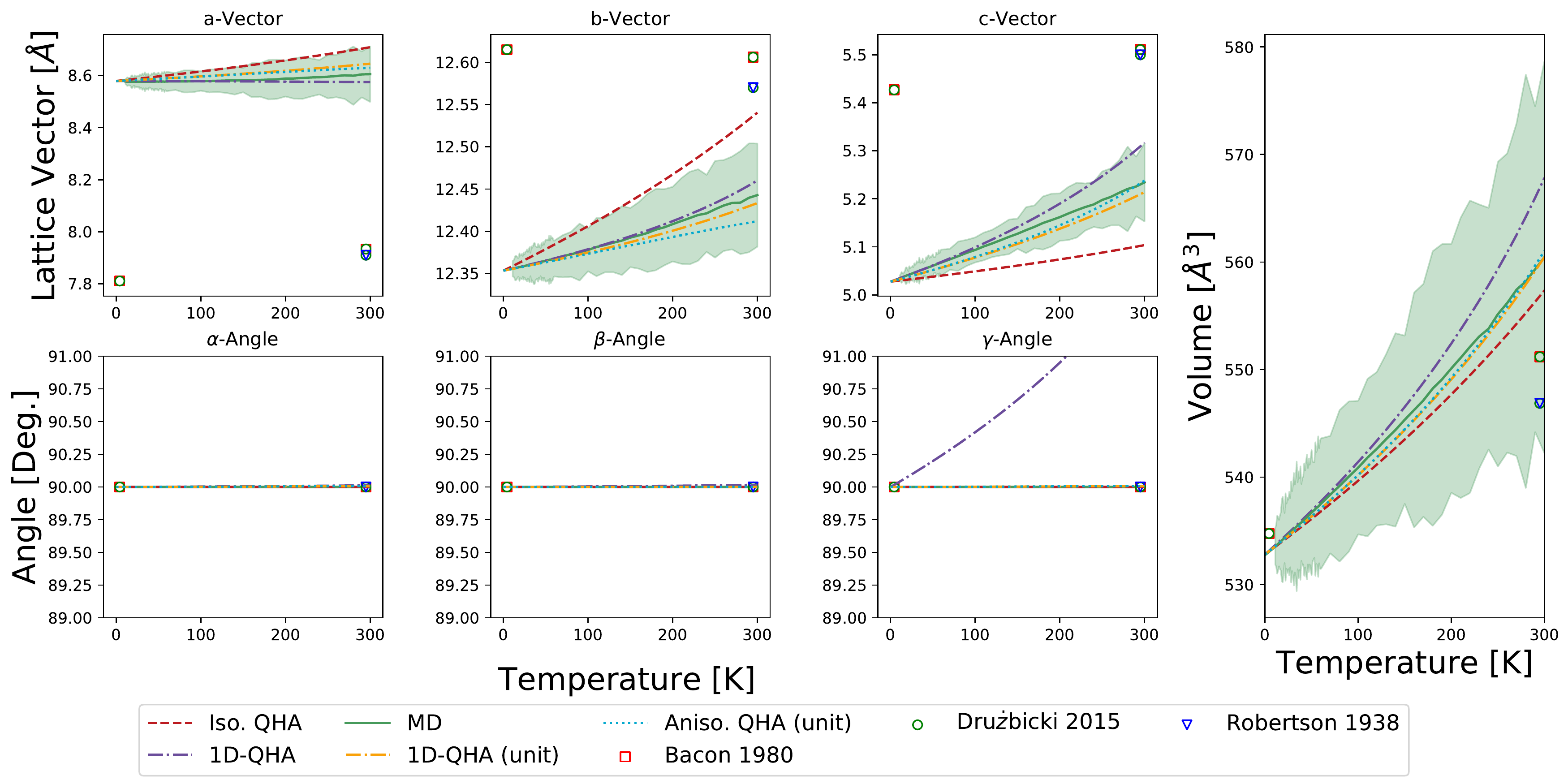}
  \end{center}
  \caption{Resorcinol form $\beta$ percent expansion from 0 K with experimental results.~\cite{bacon_neutron_1980,druzbicki_polymorphism_2015,robertson_j._monteath_new_1938}}
\end{figure*}
%
\clearpage
\subsection{1,4-Diiodobenzene}
\begin{multicols}{2}
   For MD, form $\beta$ was found to restructure at high temperature to form $\beta$-restructured. We show the results for form $\beta$ using QHA only, where the lattice minimum was found by directly quenching the experimental crystal structure. The results for form $\beta$ were used for the comparison of the 1D variant and fully anisotropic QHA, as well as the error with cell size. Form $\beta$-restructured was used for the comparison to MD.
\end{multicols}
\begin{figure*}
  \begin{center}
  \includegraphics[width=12cm]{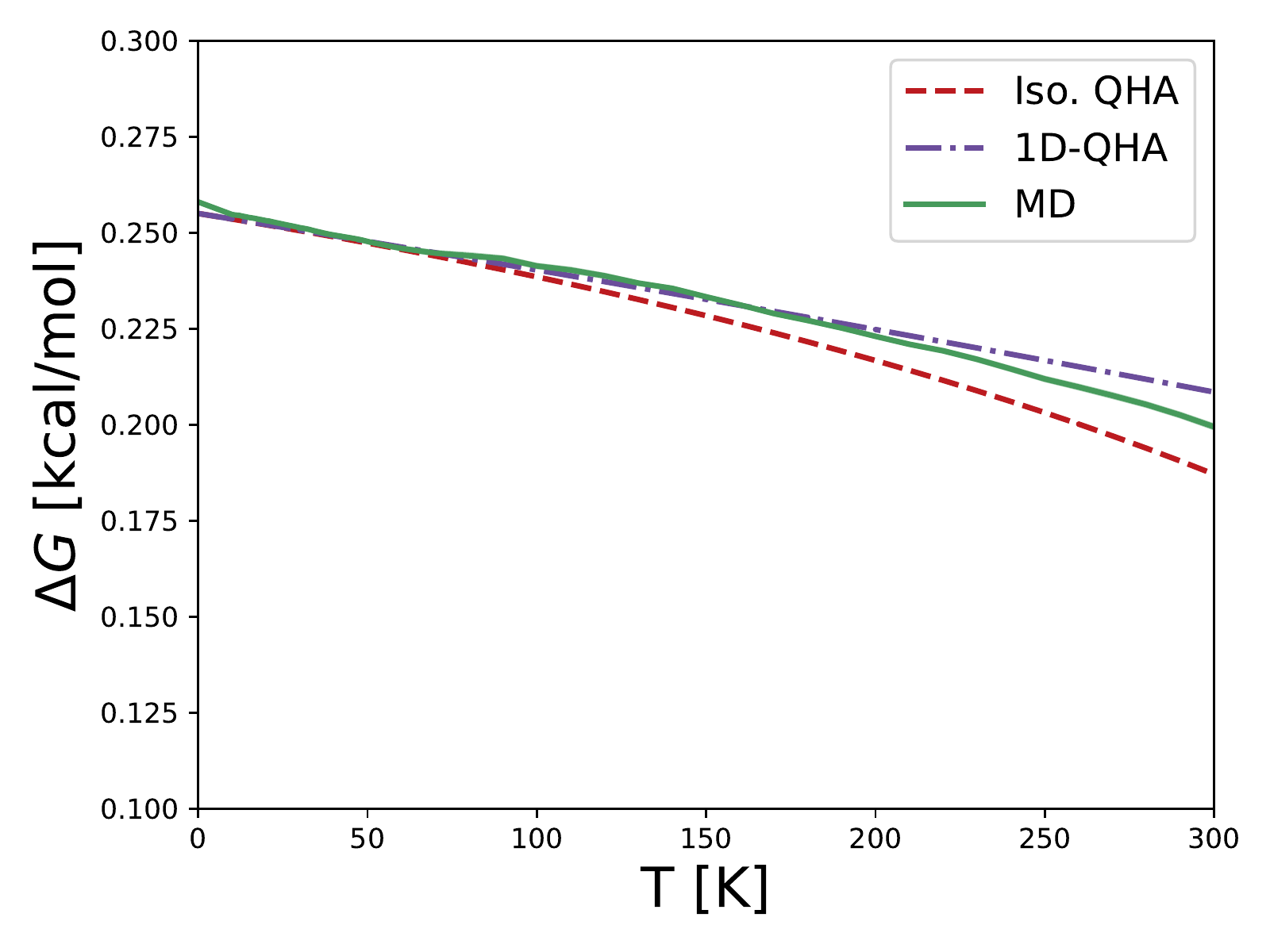}
  \end{center}
  \caption{Polymorph free energy differences of diiodobenzene form $\beta$-restructured relative to form $\alpha$}
\end{figure*}
\begin{figure*}
  \begin{center}
  \includegraphics[width=16cm]{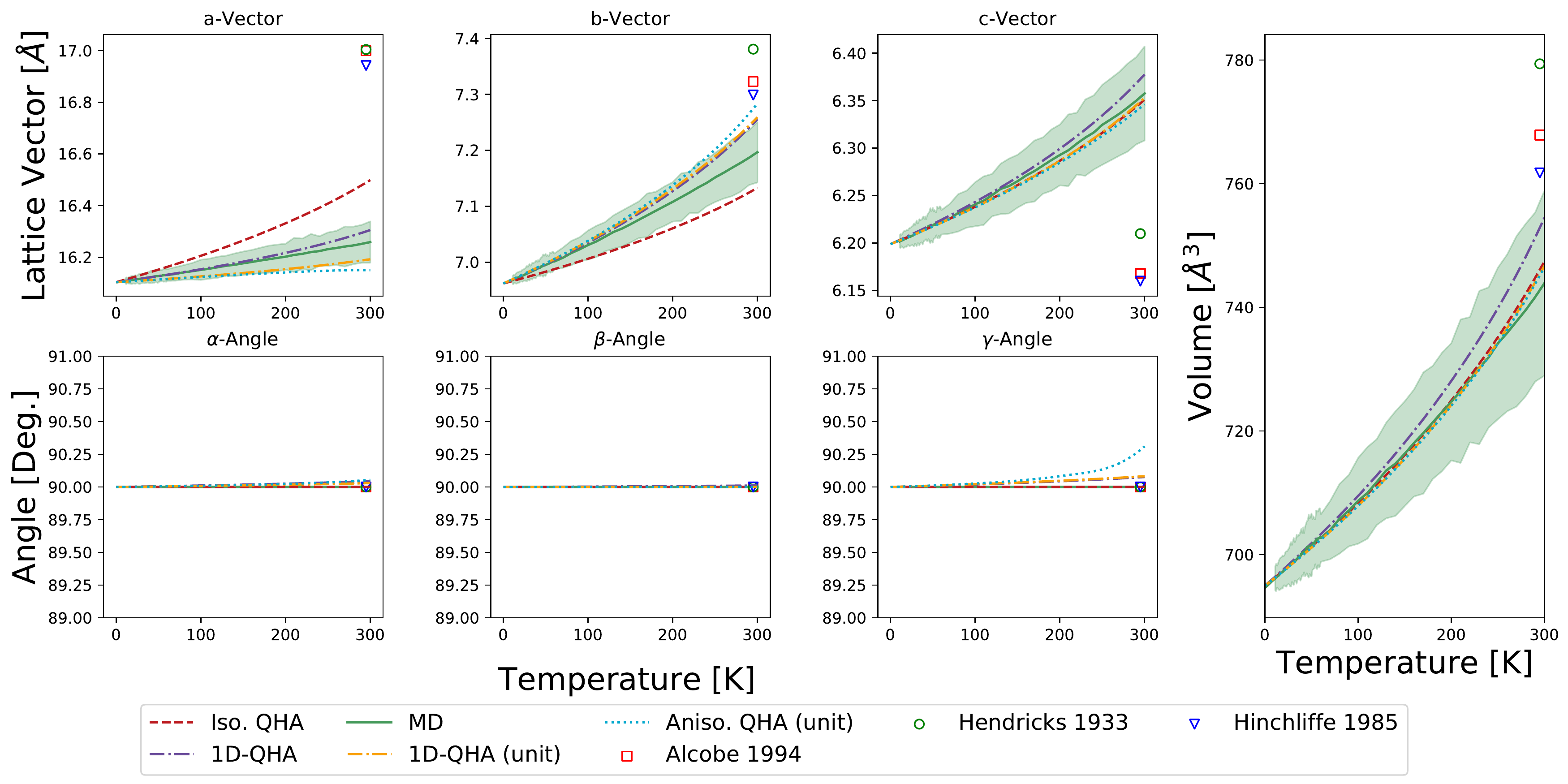}
  \end{center}
  \caption{Diiodobenzene form $\alpha$ percent expansion from 0 K with experimental results.~\cite{alcobe_temperature-dependent_1994,hendricks_xray_1933,hinchliffe_structure_1985}}
\end{figure*}
\begin{figure*}
  \begin{center}
  \includegraphics[width=16cm]{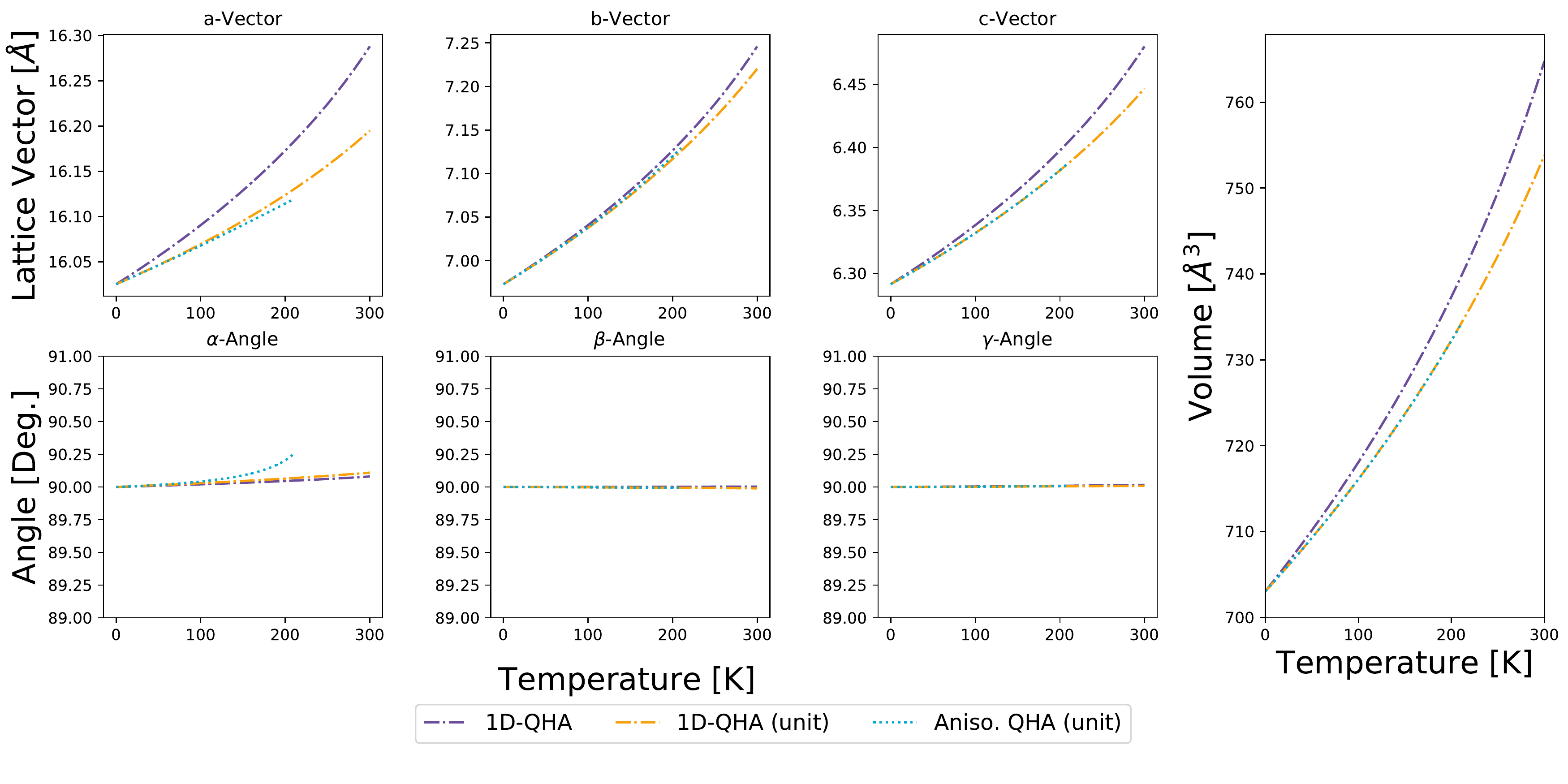}
  \end{center}
  \caption{Diiodobenzene form $\beta$ percent expansion from 0 K with experimental results.~\cite{}. MD is not shown for the original structure of $\beta$ because the system restructured at high temperature.}
\end{figure*}
\begin{figure*}
  \begin{center}
  \includegraphics[width=16cm]{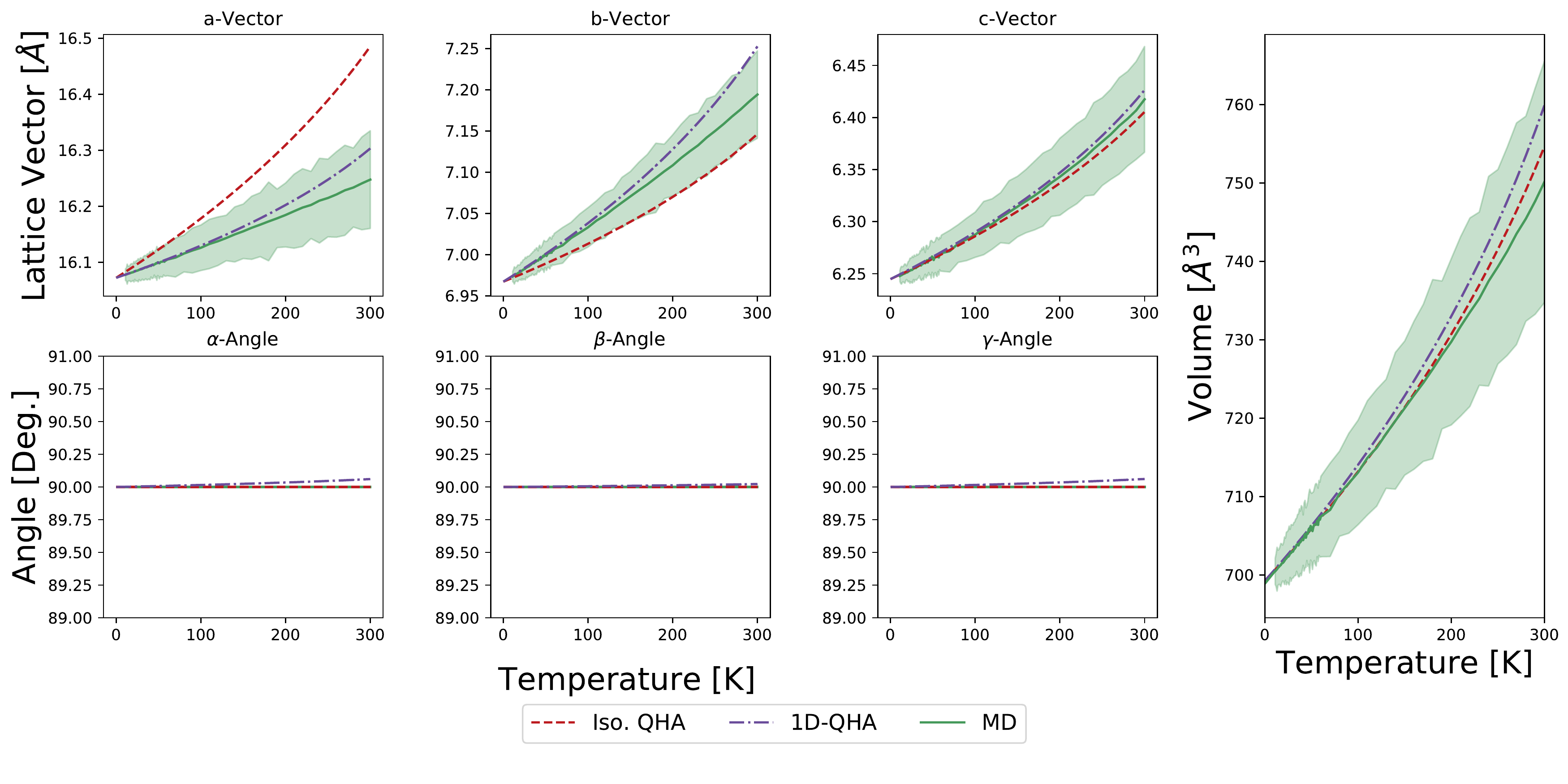}
  \end{center}
  \caption{Diiodobenzene form $\beta$-restructured percent expansion from 0 K with experimental results.~\cite{}}
\end{figure*}
%
\clearpage
\subsection{Paracetamol}
\begin{figure*}
  \begin{center}
  \includegraphics[width=12cm]{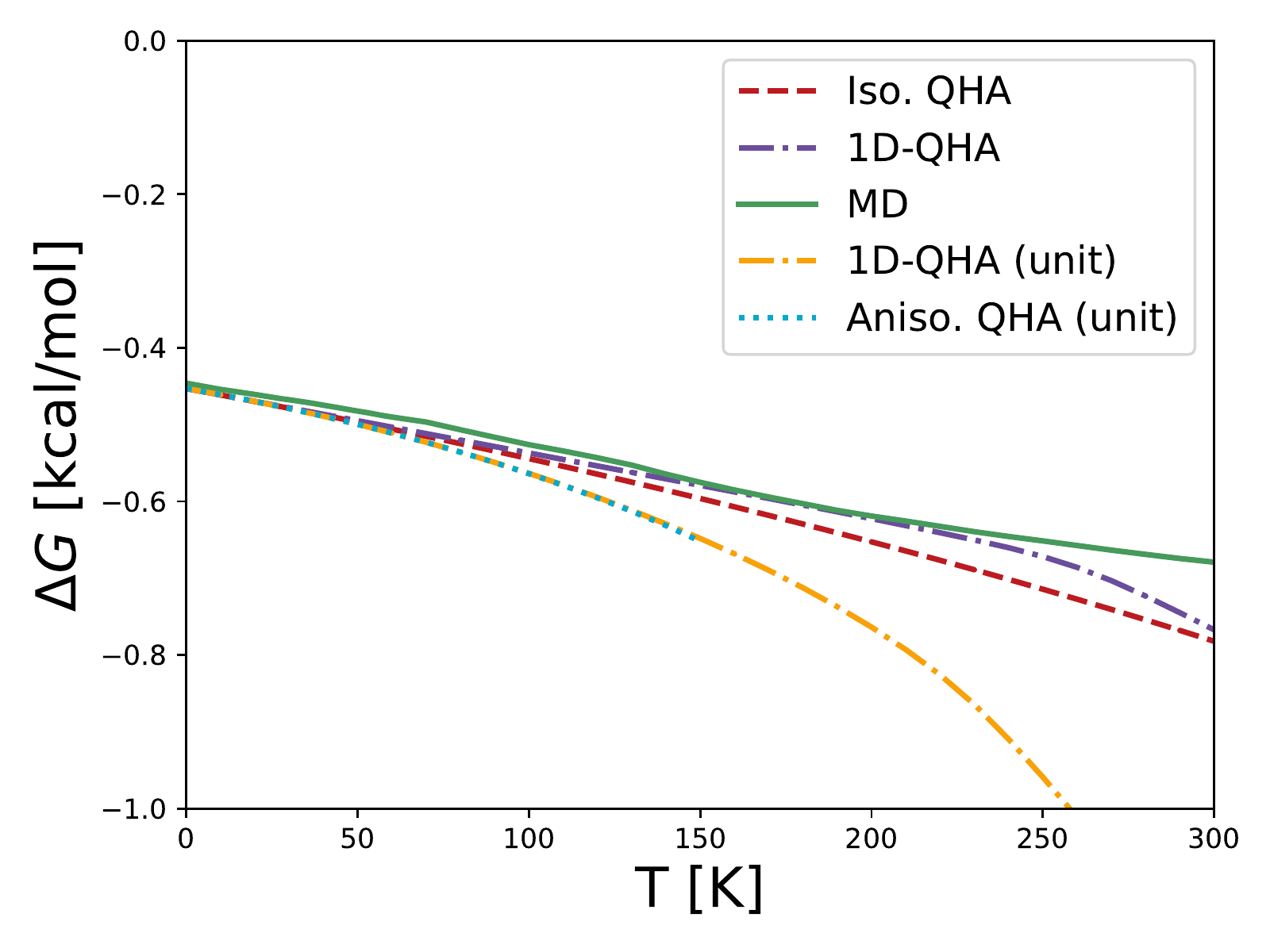}
  \end{center}
  \caption{Polymorph free energy differences of paracetamol form I relative to form II}
\end{figure*}
\begin{figure*}
  \begin{center}
  \includegraphics[width=16cm]{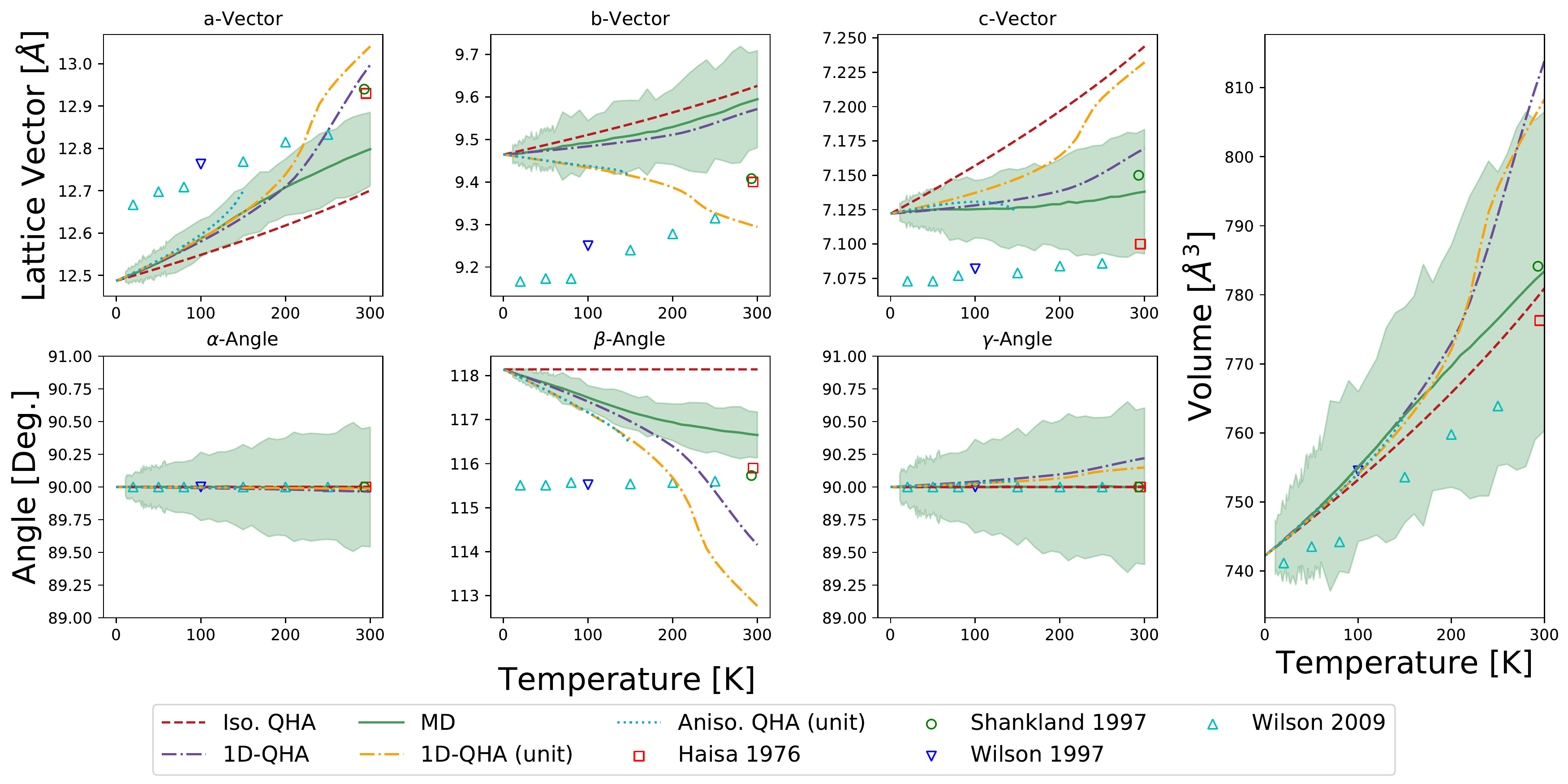}
  \end{center}
  \caption{Paracetamol form I percent expansion from 0 K with experimental results.~\cite{haisa_monoclinic_1976,wilson_variable_2009,wilson_single-crystal_1997,wilson_neutron_1997}}
\end{figure*}
\begin{figure*}
  \begin{center}
  \includegraphics[width=16cm]{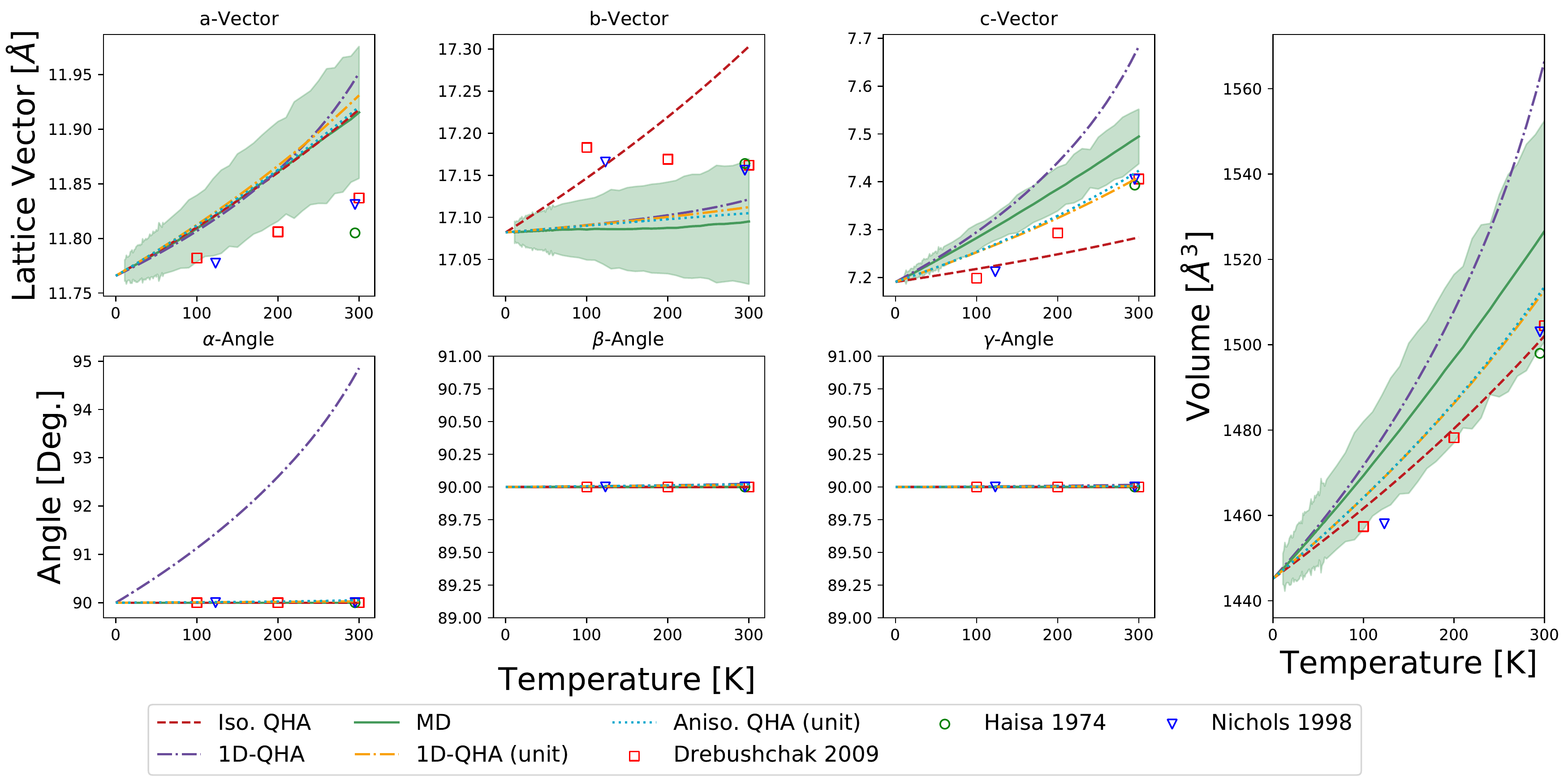}
  \end{center}
  \caption{Paracetamol form II percent expansion from 0 K with experimental results.~\cite{haisa_orthorhombic_1974,drebushchak_variable_2009,nichols_physicochemical_1998}}
\end{figure*}
%
\clearpage
\subsection{Aripiprazole}
\begin{multicols}{2}
    For MD, both crystal structures were found to restructure at high temperature to form. We show the results for for I and X using QHA only, where the lattice minimum was found by directly quenching the experimental crystal structure. These results were used for the comparison of the 1D variant and fully anisotropic QHA, as well as the error with cell size. Form I-restructured and form X-restructured were used for the comparison to MD.
\end{multicols}
\begin{figure*}
  \begin{center}
  \includegraphics[width=16cm]{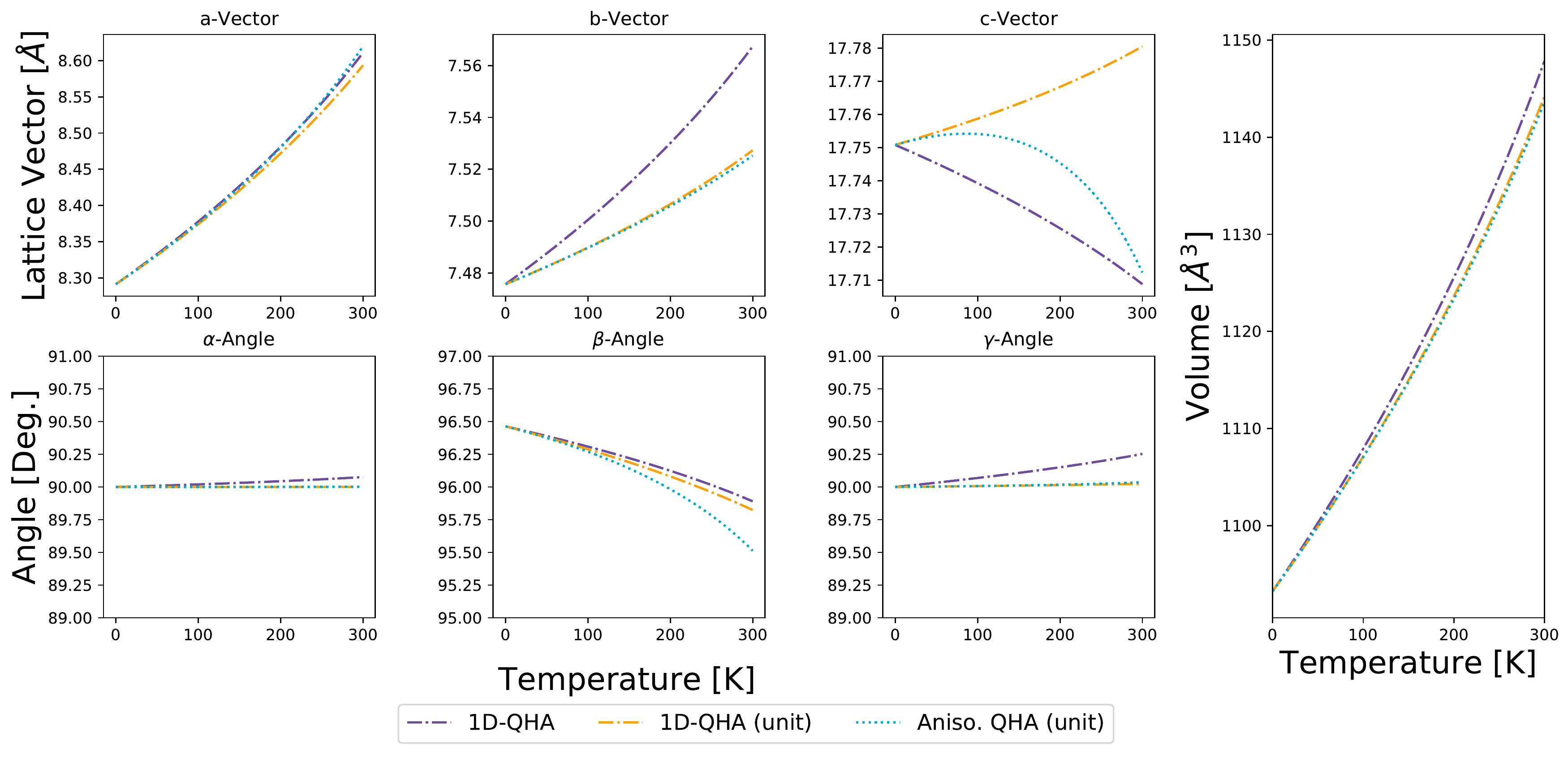}
  \end{center}
  \caption{Aripiprazole form I percent expansion from 0 K. MD is not shown for the original structure of I because the system restructured at high temperature.}
\end{figure*}
\begin{figure*}
  \begin{center}
  \includegraphics[width=16cm]{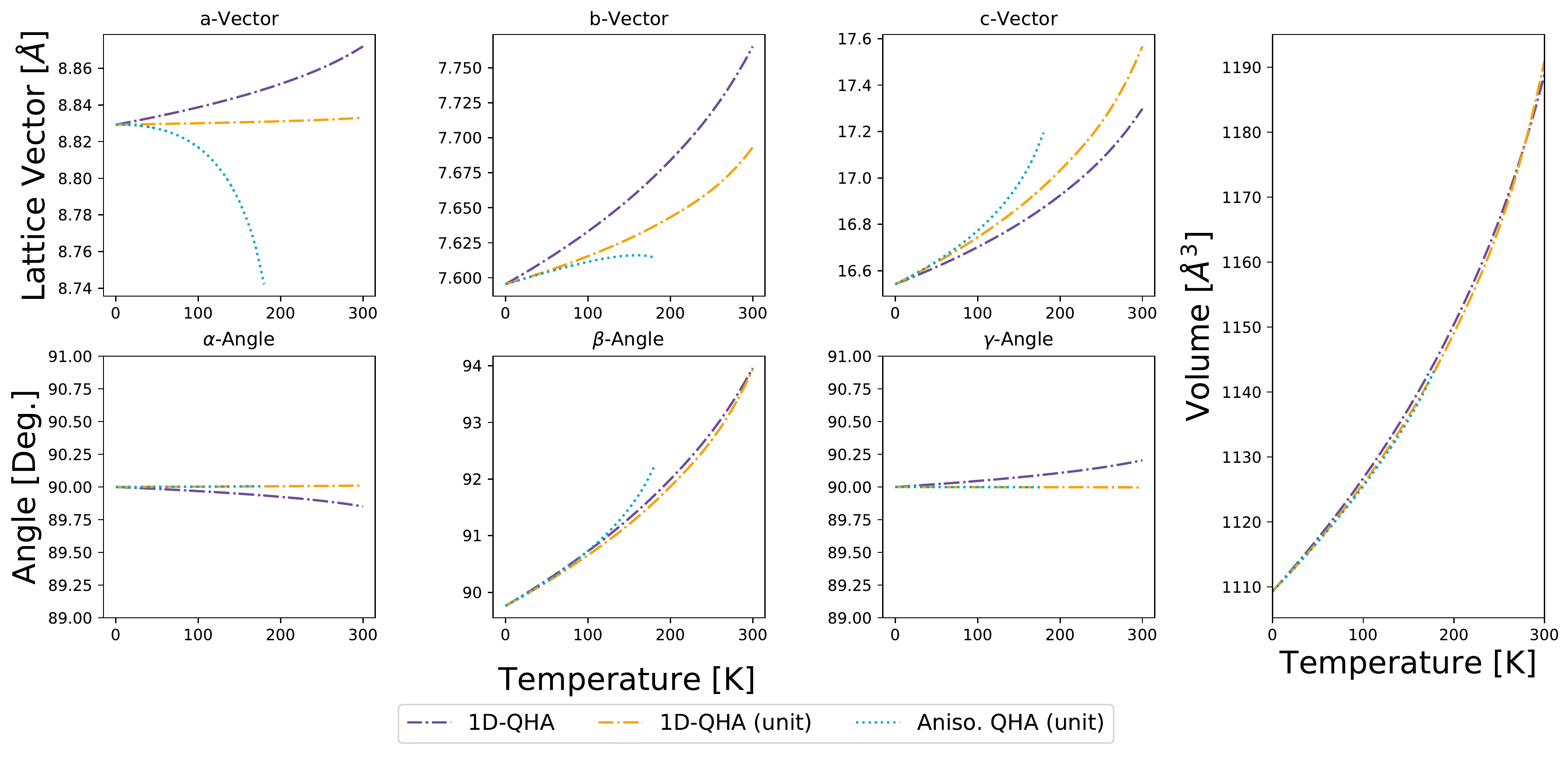}
  \end{center}
  \caption{Aripiprazole form X percent expansion from 0 K. MD is not shown for the original structure of X because the system restructured at high temperature.}
\end{figure*}
\begin{figure*}
  \begin{center}
  \includegraphics[width=12cm]{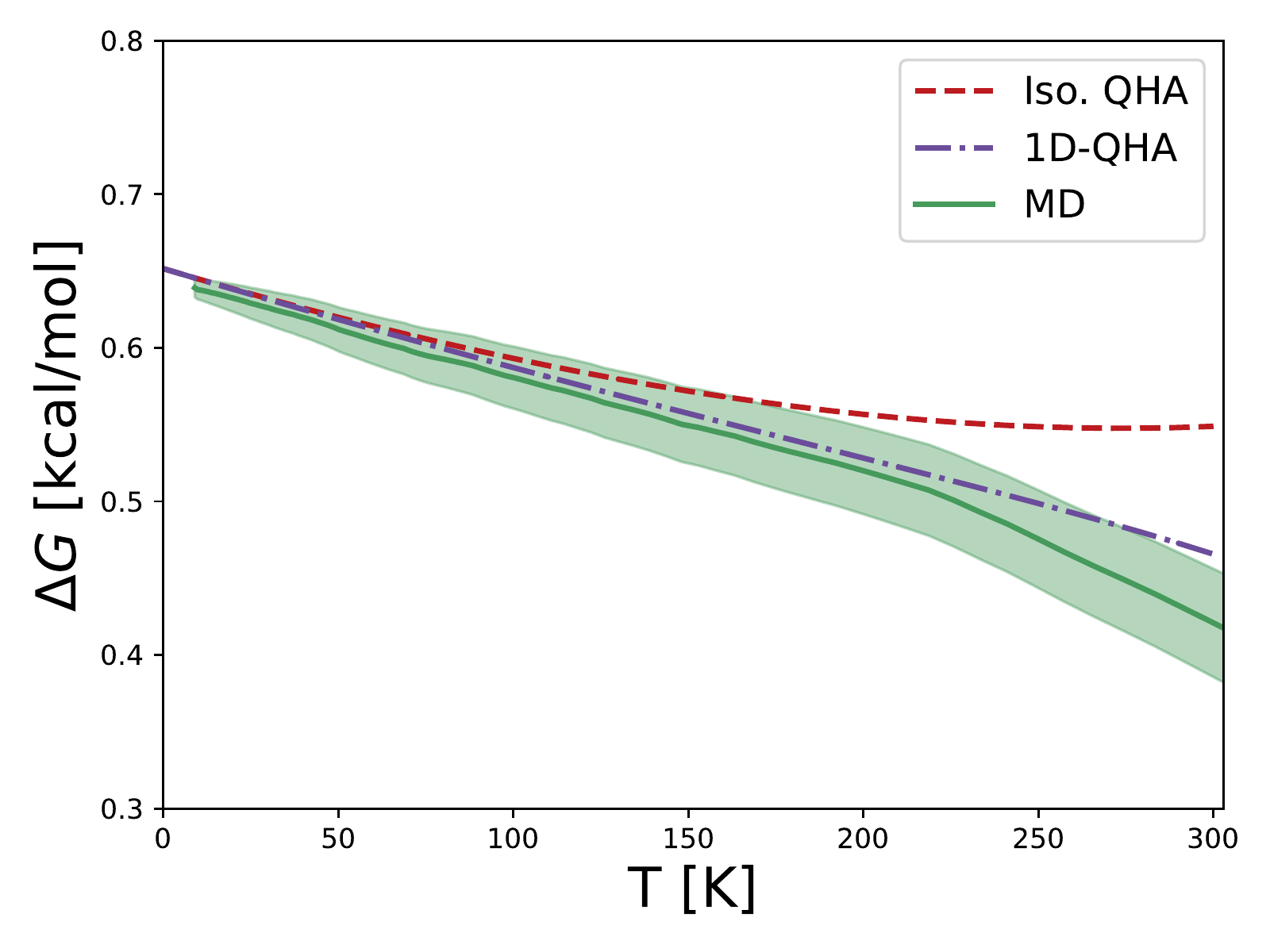}
  \end{center}
  \caption{Polymorph free energy differences of aripiprazole form X-restructured relative to form I-restructured}
\end{figure*}
\begin{figure*}
  \begin{center}
  \includegraphics[width=16cm]{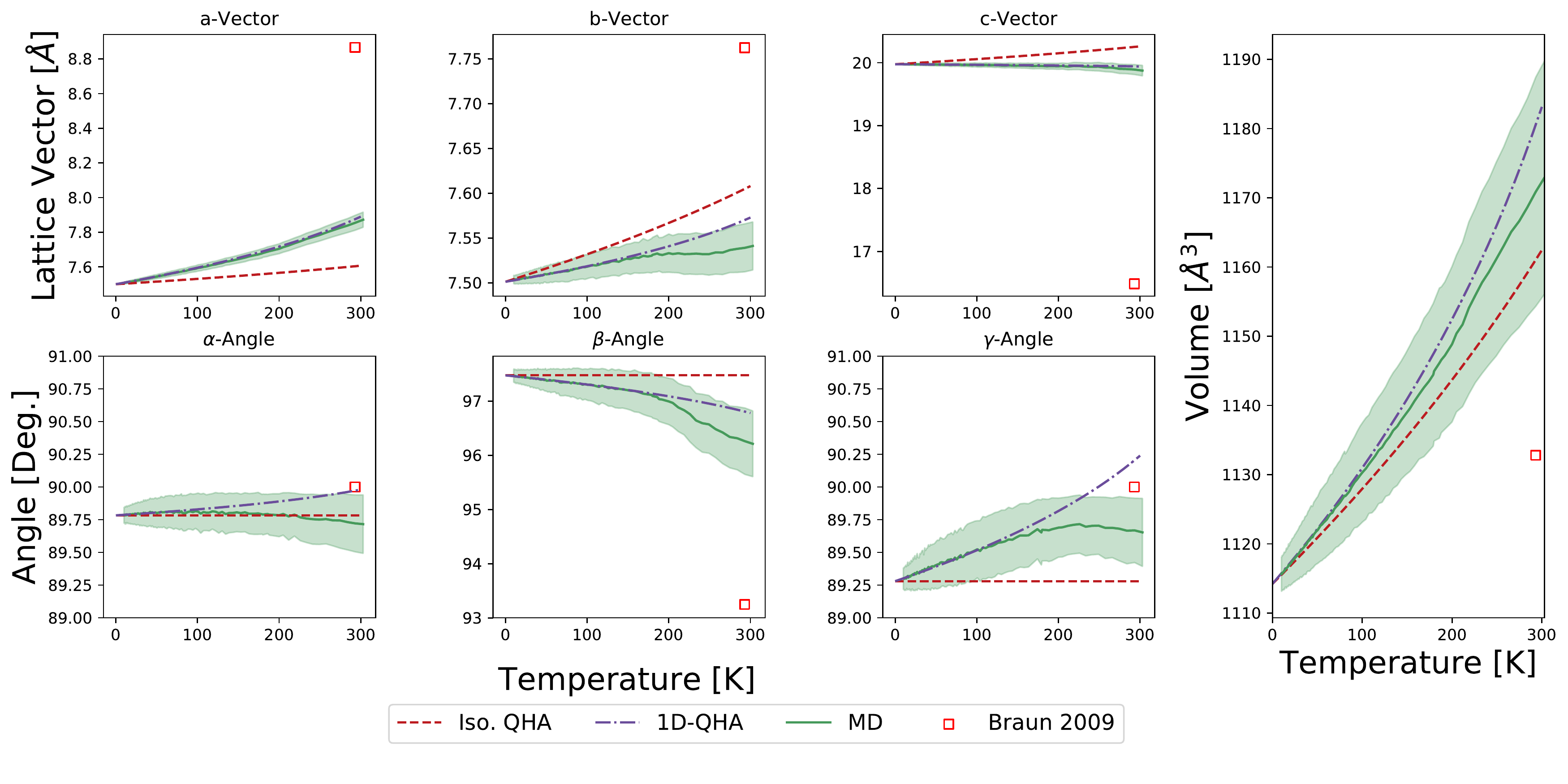}
  \end{center}
  \caption{Aripiprazole form X-restructured percent expansion from 0 K with experimental results.~\cite{braun_conformational_2009}}
\end{figure*}
\begin{figure*}
  \begin{center}
  \includegraphics[width=16cm]{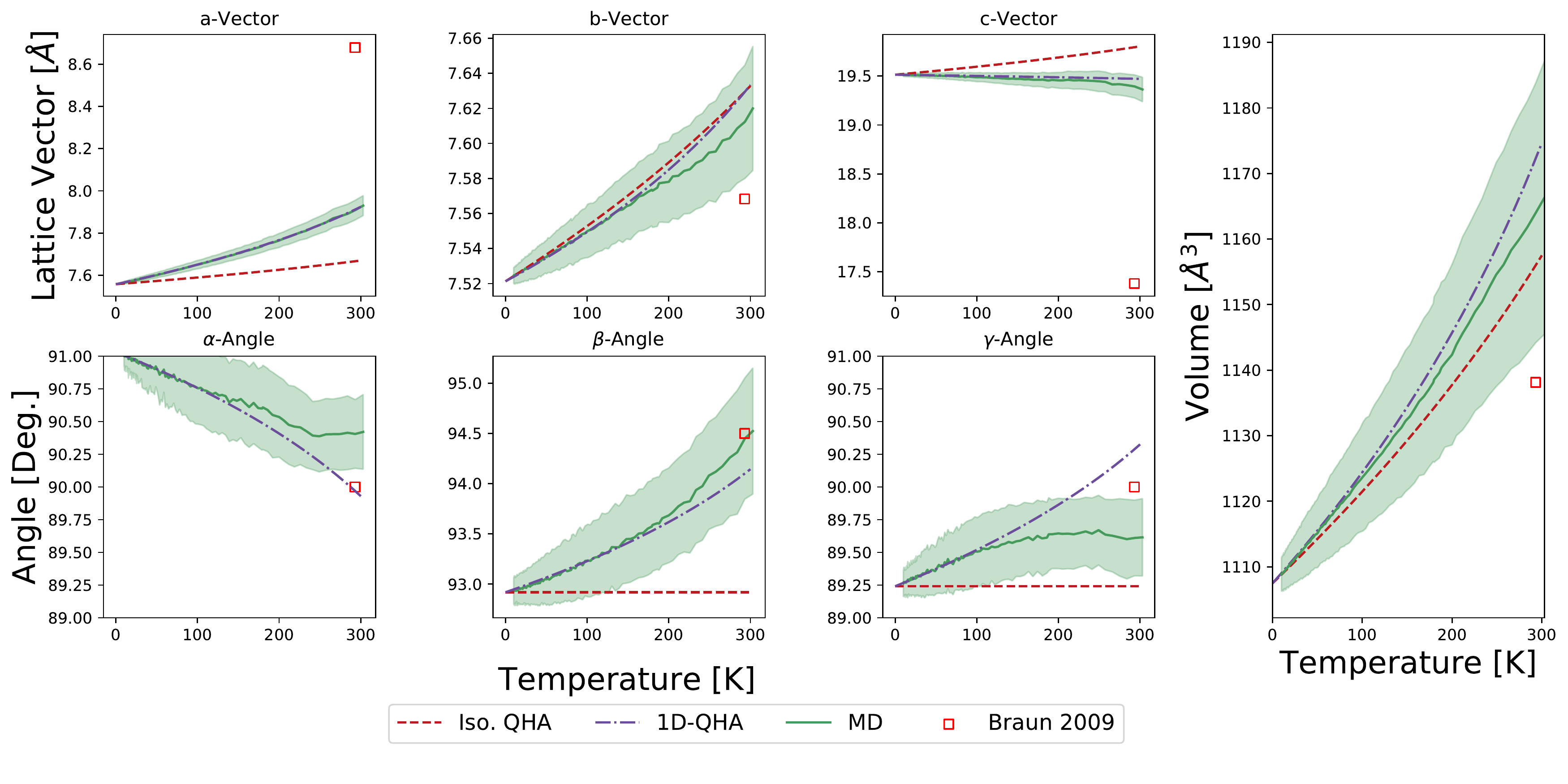}
  \end{center}
  \caption{Aripiprazole form I-restructured percent expansion from 0 K with experimental results.~\cite{braun_conformational_2009}}
\end{figure*}
%

\clearpage
\subsection{Tolbutamide}
\begin{multicols}{2}
For MD, form I was found to restructure at high temperature to form I-restructured. We show the results for form I using QHA only, where the lattice minimum was found by directly quenching the experimental crystal structure. The results for form I were used for the comparison of the 1D variant and fully anisotropic QHA, as well as the error with cell size. Form I-restructured was used for the comparison to MD.
\end{multicols}
\begin{figure*}
  \begin{center}
  \includegraphics[width=16cm]{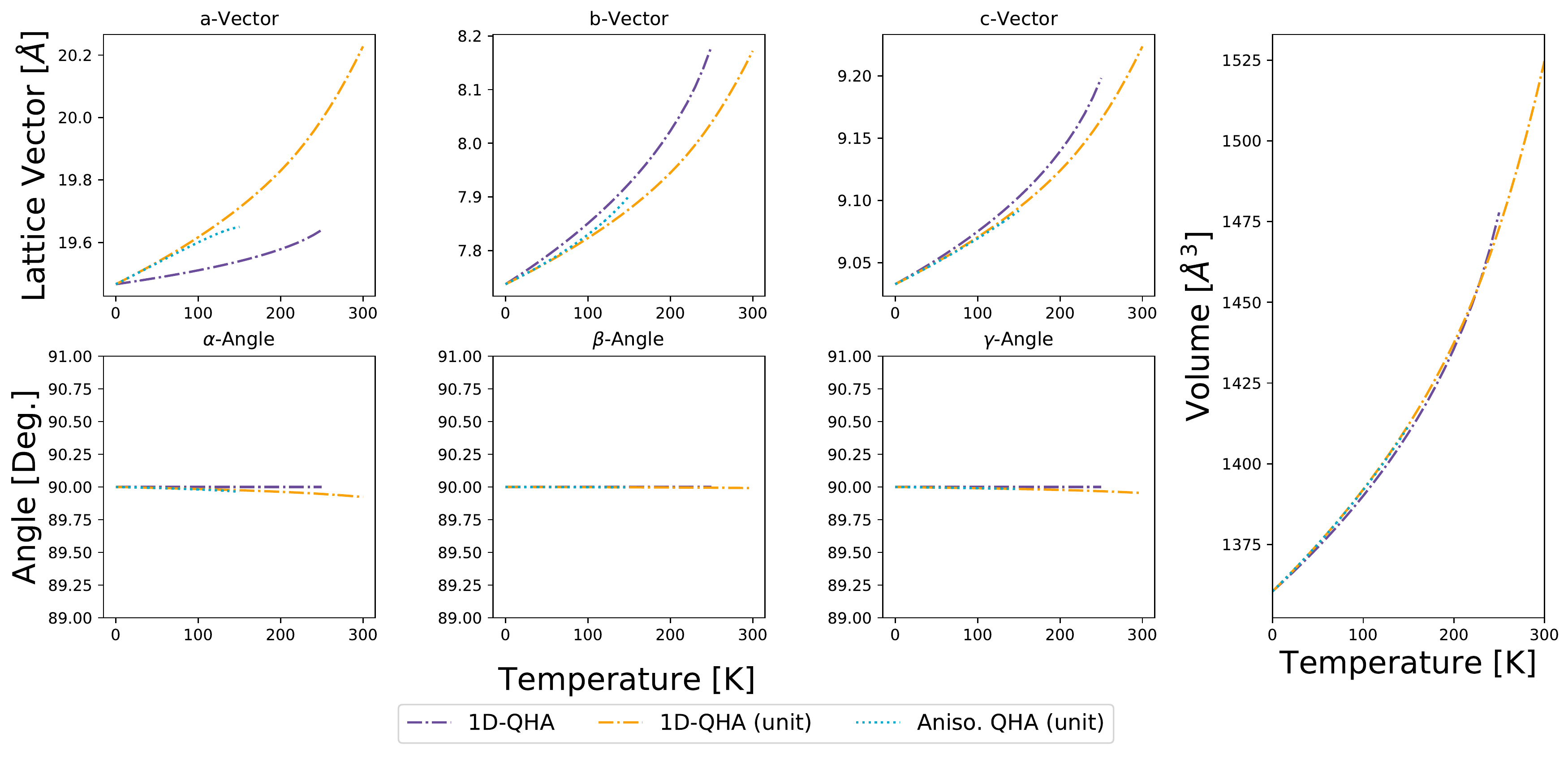}
  \end{center}
  \caption{Tolbutamide form I percent expansion from 0 K.MD is not shown for the original structure of I because the system restructured at high temperature.}
\end{figure*}

\begin{figure*}
  \begin{center}
  \includegraphics[width=16cm]{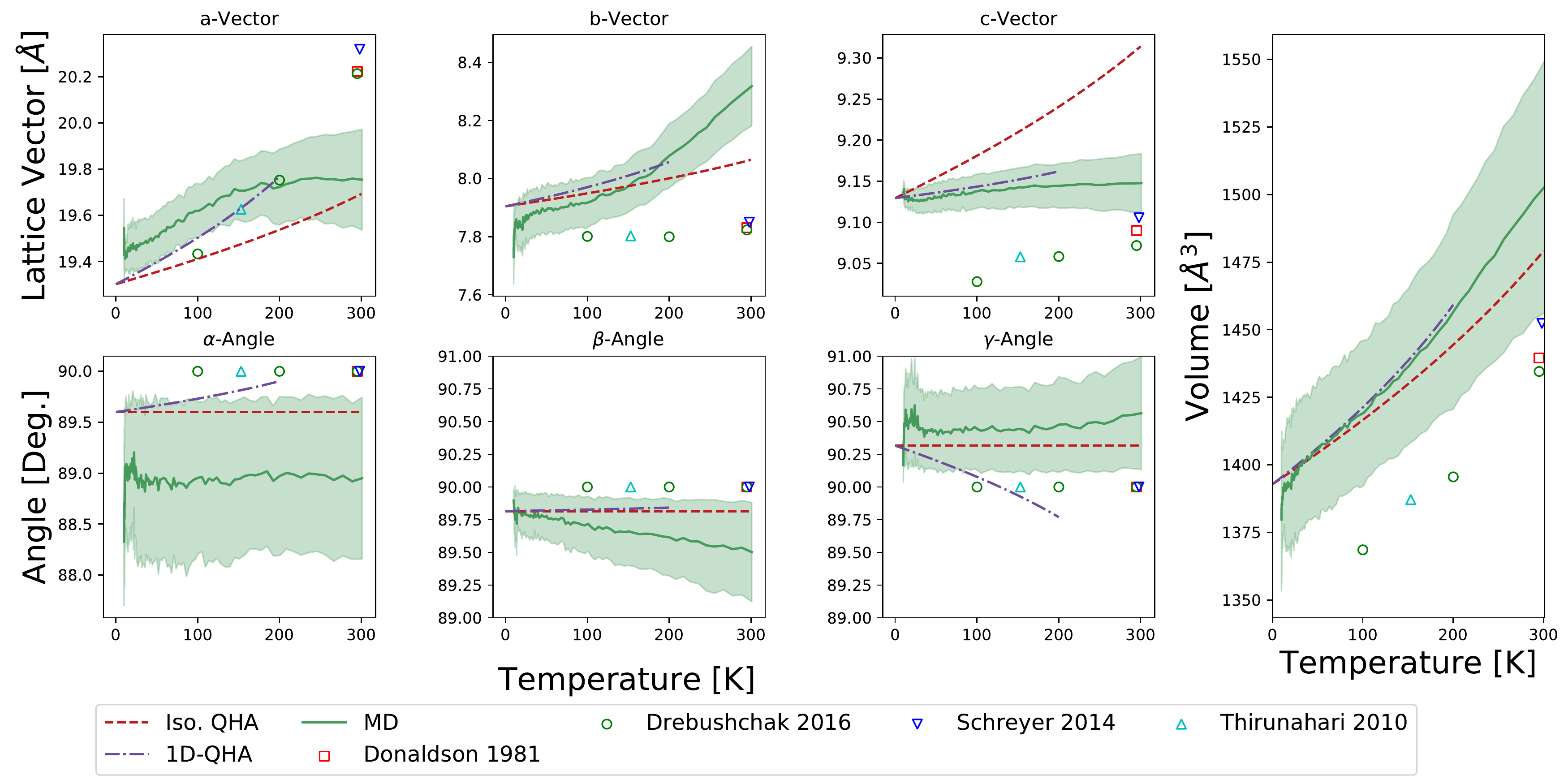}
  \end{center}
  \caption{Tolbutamide form I-restructured percent expansion from 0 K with experimental results.~\cite{donaldson_structure_1981,drebushchak_single-crystal_2016,schreyer_simultaneous_2014,thirunahari_conformational_2010}}
\end{figure*}
\begin{figure*}
  \begin{center}
  \includegraphics[width=16cm]{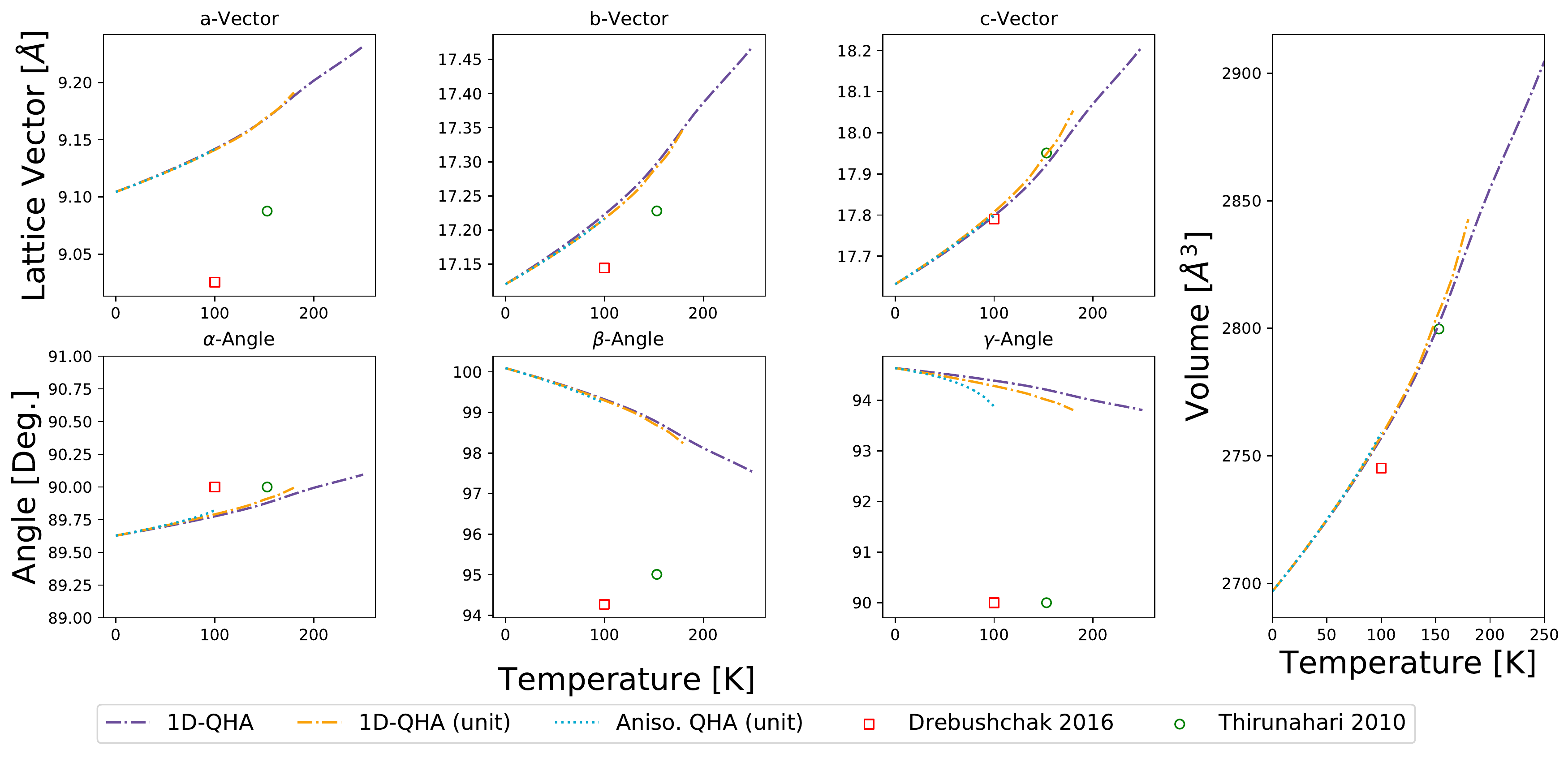}
  \end{center}
  \caption{Tolbutamide form II percent expansion from 0 K with experimental results.~\cite{drebushchak_single-crystal_2016,kimura_characterization_1999,thirunahari_conformational_2010}}
\end{figure*}
\begin{figure*}
  \begin{center}
  \includegraphics[width=16cm]{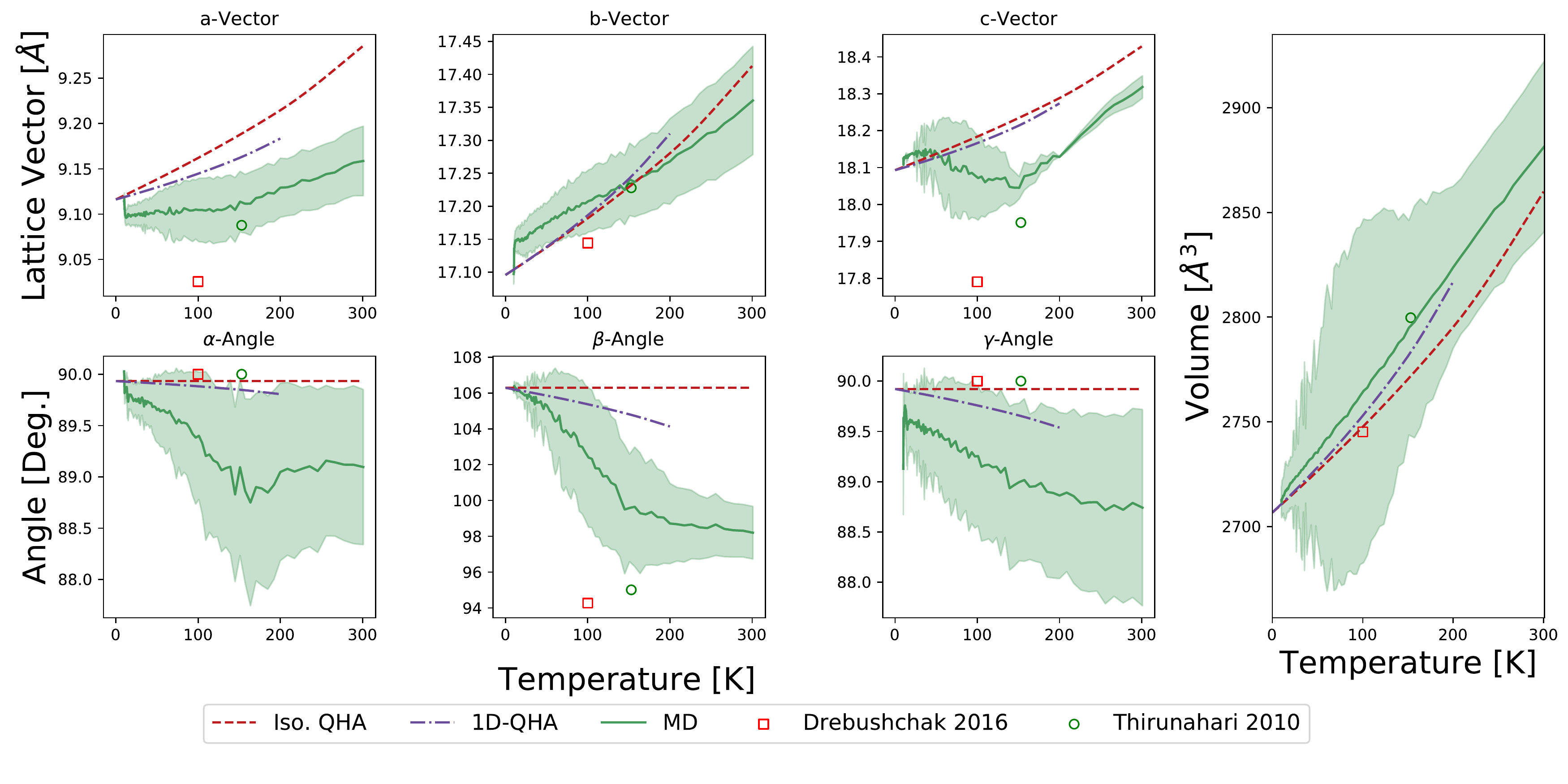}
  \end{center}
  \caption{Tolbutamide form II-restructured percent expansion from 0 K with experimental results.~\cite{drebushchak_single-crystal_2016,kimura_characterization_1999,thirunahari_conformational_2010}}
\end{figure*}
%

\clearpage
\subsection{Chlorpropamide}
\begin{multicols}{2}
For MD, form V was found to restructure at high temperature to form V-restructured. We show the results for form V using QHA only, where the lattice minimum was found by directly quenching the experimental crystal structure. The results for form V were used for the comparison of the 1D variant and fully anisotropic QHA, as well as the error with cell size. Form V-restructured was used for the comparison to MD.
\end{multicols}
\begin{figure*}
  \begin{center}
  \includegraphics[width=12cm]{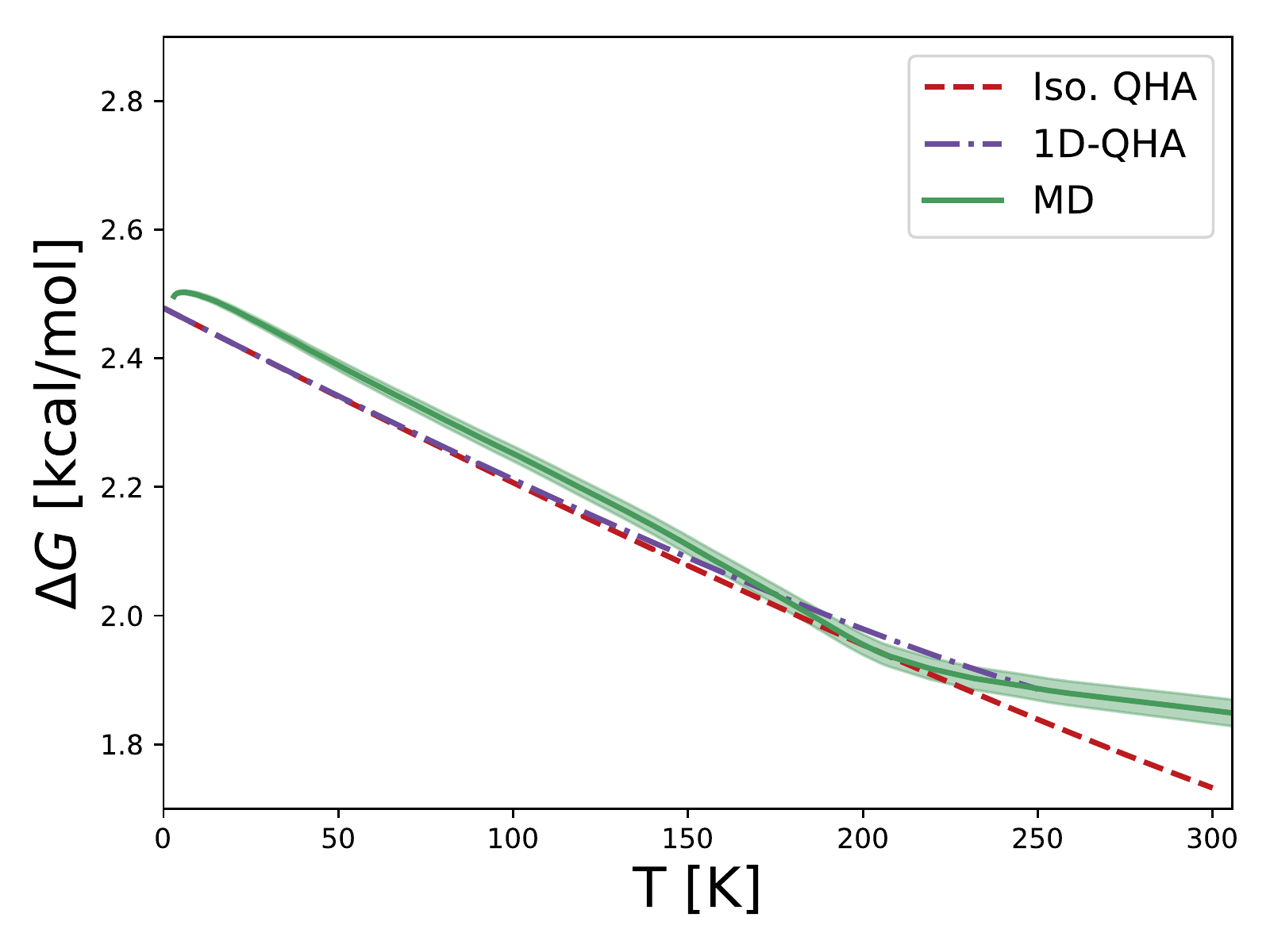}
  \end{center}
  \caption{Polymorph free energy differences of chlorpropamide form V-restructured relative to form I-Restructured}
\end{figure*}
\begin{figure*}
  \begin{center}
  \includegraphics[width=16cm]{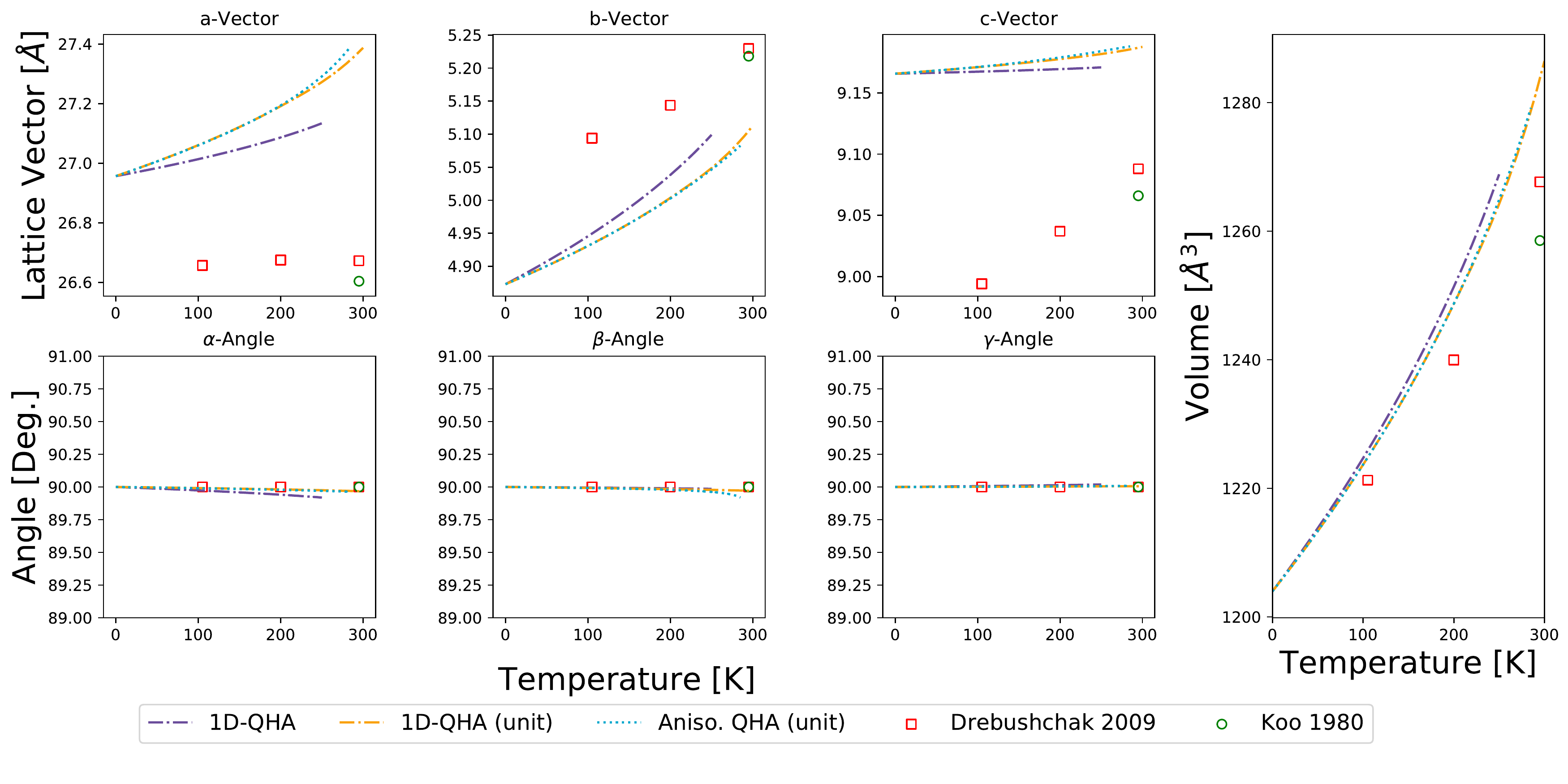}
  \end{center}
  \caption{Chlorpropamide form I percent expansion from 0 K with experimental results.~\cite{drebushchak_conformational_2009,koo_crystal_1980} MD is not shown here because this is QHA using the minima directly quenched from the experimental structure, which does not correspond
           with the 0 K minima found with MD.}
\end{figure*}
\begin{figure*}
  \begin{center}
  \includegraphics[width=16cm]{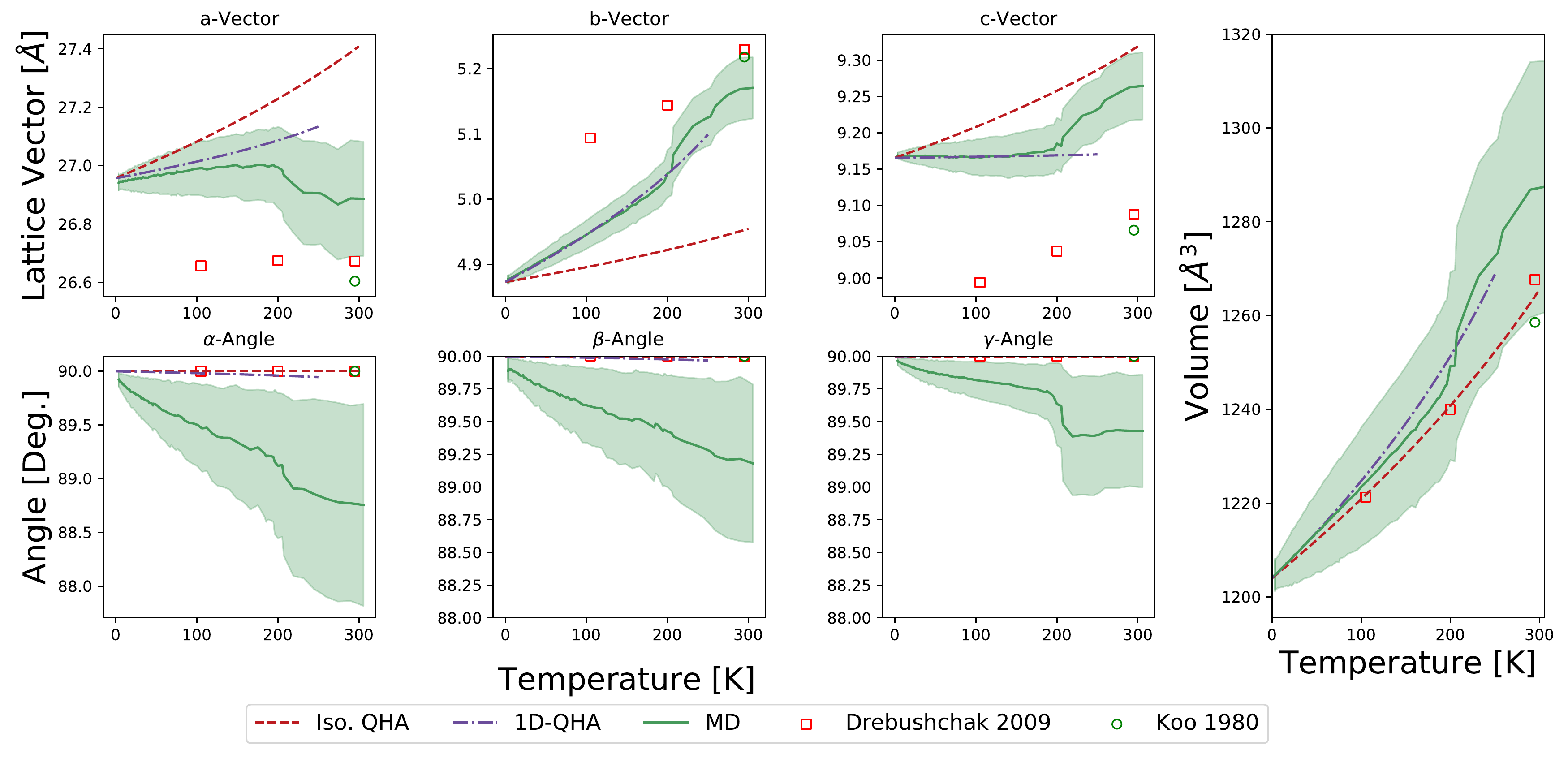}
  \end{center}
  \caption{Chlorpropamide form I-Restructured percent expansion from 0 K with experimental results.~\cite{drebushchak_conformational_2009,koo_crystal_1980}}
\end{figure*}

\begin{figure*}
  \begin{center}
  \includegraphics[width=16cm]{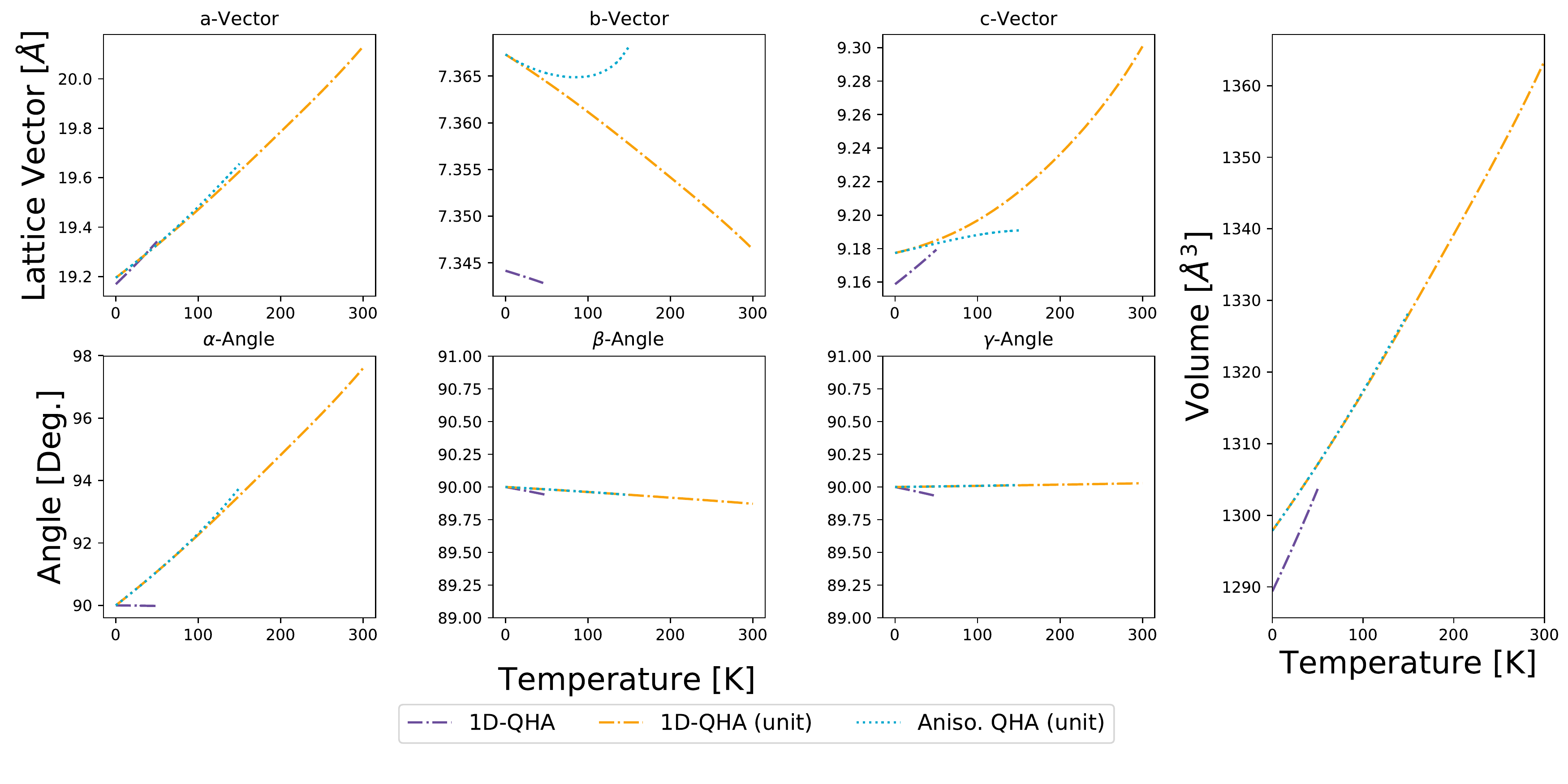}
  \end{center}
  \caption{Chlorpropamide form V percent expansion from 0 K. MD is not shown here because this is QHA using the minima directly quenched from the experimental structure, which does not correspond
           with the 0 K minima found with MD.}
\end{figure*}

\section{Summary of Results}
\begin{figure}[H]
  \begin{center}
  \includegraphics[width=16cm]{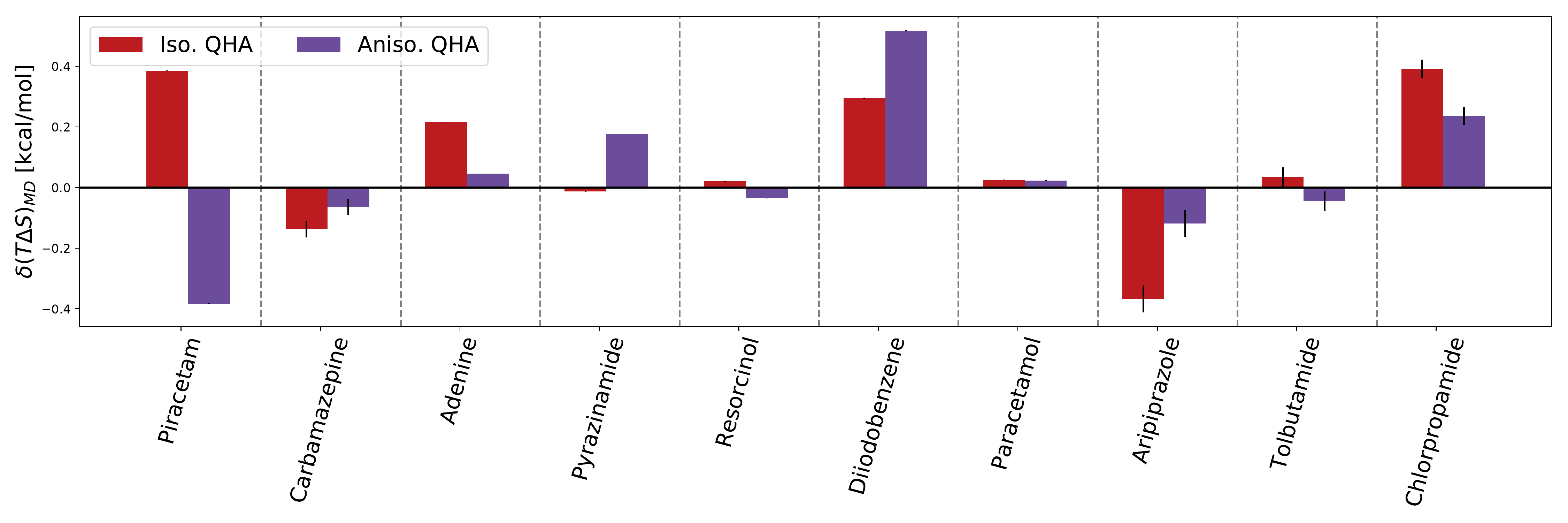}
  \end{center}
  \caption{Deviations in the entropic contribution to the Gibbs free energy differences from MD at 300 K (except for tolbutamide at 200 K, and chlorpropamide at 250 K). We report results for isotropic QHA and 1D-anisotropic QHA relative to MD. The entropy differences between methods are compensated by a larger enthalpy difference, resulting in a smaller free energy differences.
  \label{figure:dS_MD}}
\end{figure}
\begin{figure}[H]
  \begin{center}
  \includegraphics[width=16cm]{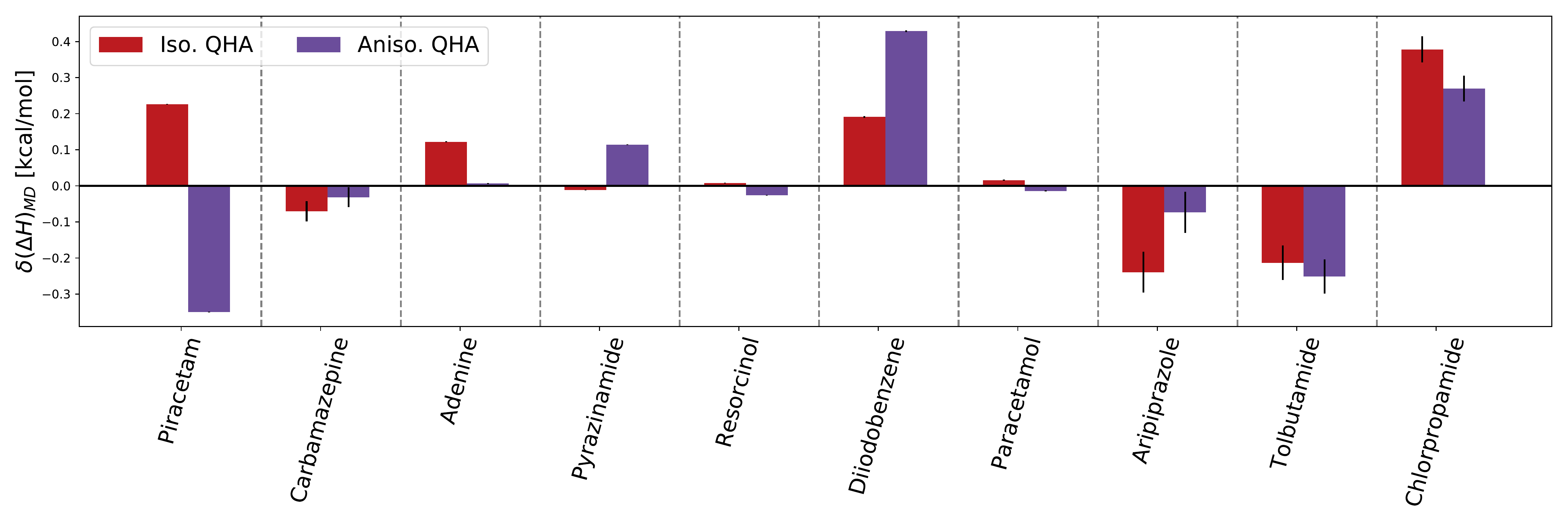}
  \end{center}
  \caption{Deviations in the enthalpic contribution to the Gibbs free energy differences from MD at 300 K (except for tolbutamide at 200 K, and chlorpropamide at 250 K). We report results for isotropic QHA and 1D-anisotropic QHA relative to MD. The enthalpy differences between methods are compensated by a larger enthalpy difference, resulting in a smaller free energy differences.
  \label{figure:dH_MD}}
\end{figure}

\clearpage
\section{Replica Exchange}

    \begin{longtable}{c|c|p{8cm}}
      System            & Simulation Time [ns] & Temperatures Sampled [K]  \\ \hline \hline
      Aripiprazole      & 20 & 10.0, 10.2, 10.4, 10.6, 10.8, 11.0, 11.2, 11.5, 11.8, 12.1, 12.4, 12.7, 13.0, 13.3, 13.6, 13.9, 14.2, 14.6, 15.0, 15.4, 15.8, 16.2, 16.6, 17.0, 17.4, 17.9, 18.4, 18.9, 19.4, 19.9, 20.4, 21.0, 21.6, 22.2, 22.8, 23.4, 24.1, 24.8, 25.5, 26.2, 27.0, 27.8, 28.6, 29.4, 30.3, 31.2, 32.1, 33.1, 34.1, 35.1, 36.2, 37.3, 38.4, 39.6, 40.8, 42.0, 43.3, 44.6, 46.0, 47.4, 48.9, 50.4, 52.0, 53.6, 55.3, 57.0, 58.8, 60.7, 62.6, 64.6, 66.6, 68.7, 70.9, 73.2, 75.5, 77.9, 80.4, 81.1, 83.0, 85.7, 88.5, 91.4, 94.4, 97.5, 100.7, 104.0, 107.4, 110.9, 114.5, 118.2, 122.1, 126.1, 130.2, 134.5, 138.9, 143.5, 148.2, 153.1, 158.1, 163.3, 168.7, 174.3, 179.1, 180.0, 185.9, 188.1, 192.0, 200.0, 204.9, 211.7, 218.7, 225.9, 233.4, 241.1, 249.1, 257.4, 265.9, 274.7, 283.8, 293.2, 302.9, 313.0, 323.4, 334.2, 345.3, 350.0 \\ \hline
      Tolbutamide       & 20 & 10.0, 10.30, 10.6, 10.9, 11.2, 11.5, 11.8, 12.1, 12.5, 12.9, 13.3, 13.7, 14.1, 14.5, 15.0, 15.5, 16.0, 16.5, 17.0, 17.6, 18.2, 18.8, 19.4, 20.1, 20.8, 21.5, 22.3, 23.1, 23.9, 24.8, 25.7, 26.6, 27.6, 28.6, 29.7, 30.8, 31.9, 33.1, 34.3, 35.6, 36.9, 38.3, 39.8, 41.3, 42.9, 44.6, 46.3, 48.1, 50.0, 52.0, 54.0, 56.1, 58.3, 60.6, 63.0, 65.5, 68.1, 70.8, 73.6, 76.6, 79.7, 82.9, 86.2, 89.7, 93.3, 97.1, 101.0, 105.1, 109.4, 113.9, 118.6, 123.4, 128.5, 133.8, 139.3, 145.0, 151.0, 157.2, 163.7, 170.5, 177.5, 184.8, 192.4, 200.0, 208.7, 217.3, 226.3, 235.7, 245.5, 255.7, 266.3, 277.4, 288.9, 300.9, 313.4, 326.5, 340.1, 350.0 \\ \hline
      Chlorpropamide    & 30 & 3.00, 3.12, 3.25, 3.38, 3.52, 3.67, 3.83, 4.00, 4.18, 4.37, 4.57, 4.78, 5.00, 5.24, 5.49, 5.75, 6.03, 6.32, 6.63, 6.96, 7.31, 7.68, 8.07, 8.48, 8.92, 9.38, 9.87, 10.39, 10.94, 11.52, 12.13, 12.78, 13.46, 14.10, 14.18, 14.94, 15.75, 16.20, 16.60, 17.50, 18.46, 19.47, 20.54, 21.67, 22.87, 24.14, 25.48, 26.9, 28.4, 29.98, 31.65, 33.42, 35.29, 37.27, 39.36, 41.58, 43.92, 45.80, 46.40, 49.02, 51.79, 54.72, 57.82, 61.10, 64.57, 67.30, 68.24, 72.12, 74.80, 75.10, 76.22, 80.56, 85.15, 90.00, 95.13, 100.56, 106.30, 112.37, 118.79, 125.58, 132.76, 140.36, 148.39, 156.89, 160.90, 165.88, 175.38, 183.60, 185.43, 187.90, 194.40, 196.06, 200.00, 205.30, 207.30, 219.19, 231.76, 245.06, 253.20, 259.13, 274.01, 289.74, 305.70, 306.38, 323.98, 332.20, 333.50, 342.59, 362.27, 382.20, 383.09, 390.70, 400.00 \\ \hline
      \caption{Systems that replica exchange we run on, the amount of time run for, and the temperatures used for exchange. Chlorpropamide was run to lower temperatures to validate our findings at low temperatures and therefore required 10 ns longer of simulation time for these low temperatures to converge conformationally.
      \label{table:REMD}}
    \end{longtable}
\begin{multicols}{2}
\section{Comparison of Gas \& Crystal Torsions}
    For all molecules that contain 1 or more free rotating torsions, we compared the torsional distributions in the gas state with the crystals at 300 K. The gas states were run with REMD from 300 to 1000 K with about 0.2 probability exchange between replica.
\end{multicols}
\begin{figure}[H]
  \begin{center}
  \includegraphics[width=4cm]{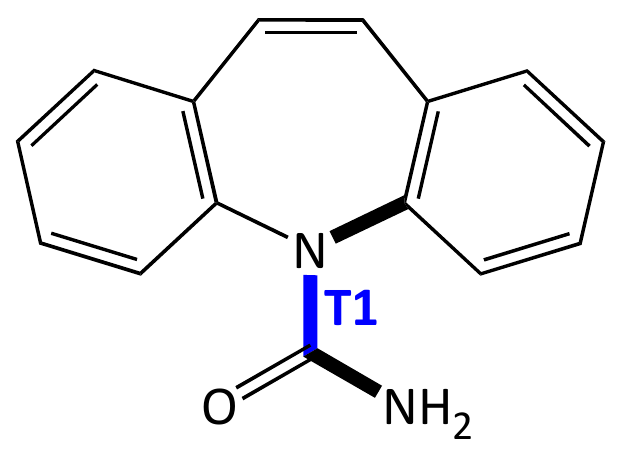}
  \includegraphics[width=6cm]{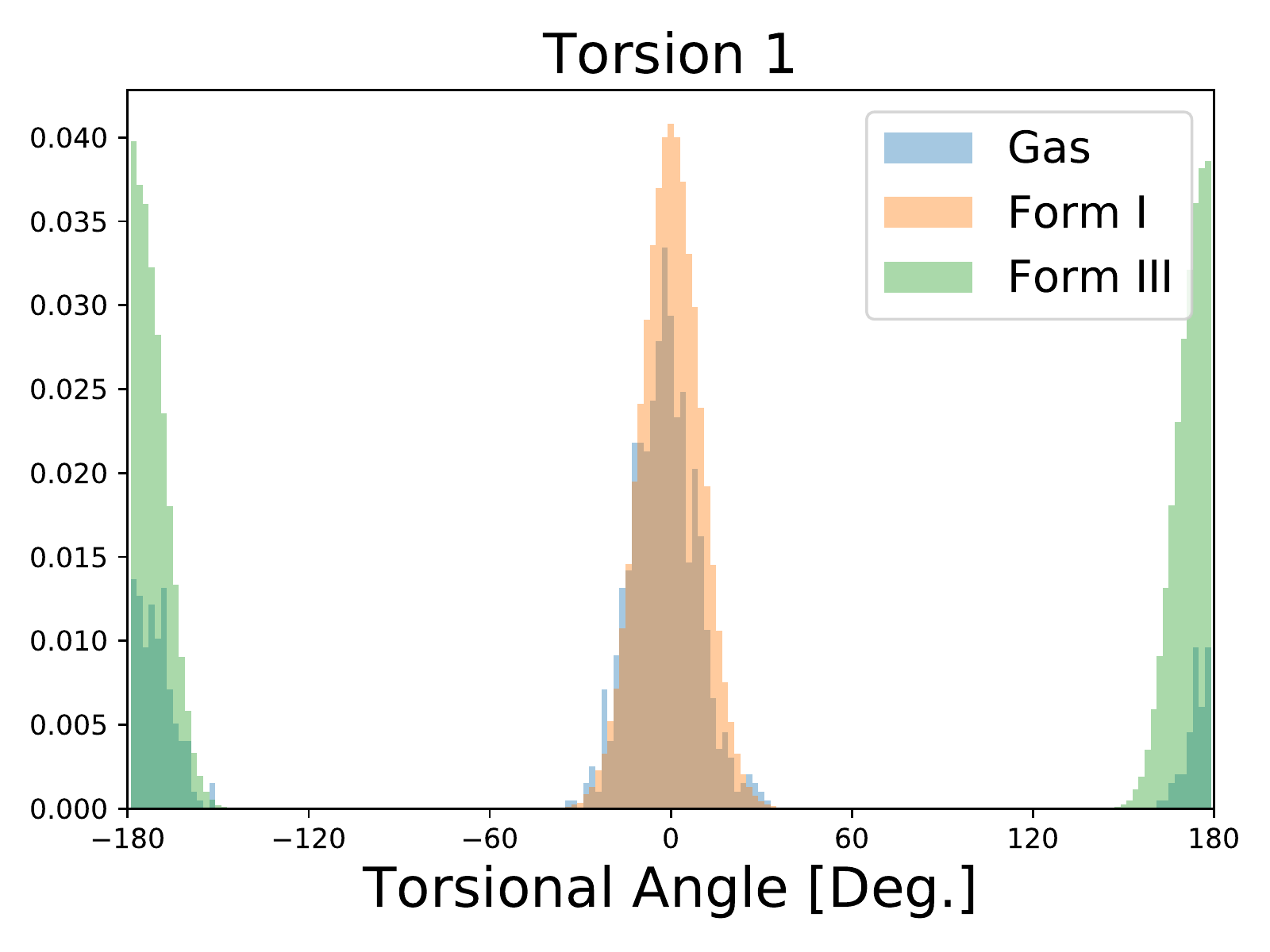}
  \end{center}
  \caption{Torsional distributions of carbamazepine at 300 K for the molecule in gas and both polymorphs.
  \label{figure:cbmzpn_torsion}}
\end{figure}
\begin{figure}[H]
  \begin{center}
  \includegraphics[width=4cm]{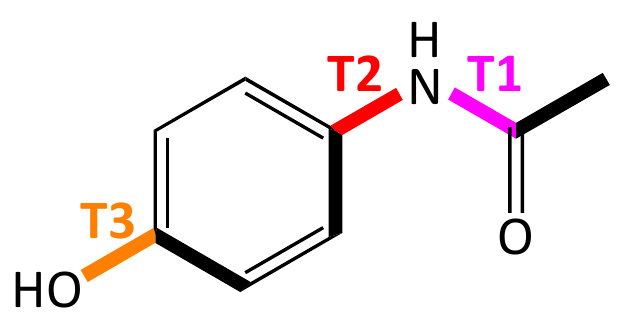}
  \includegraphics[width=6cm]{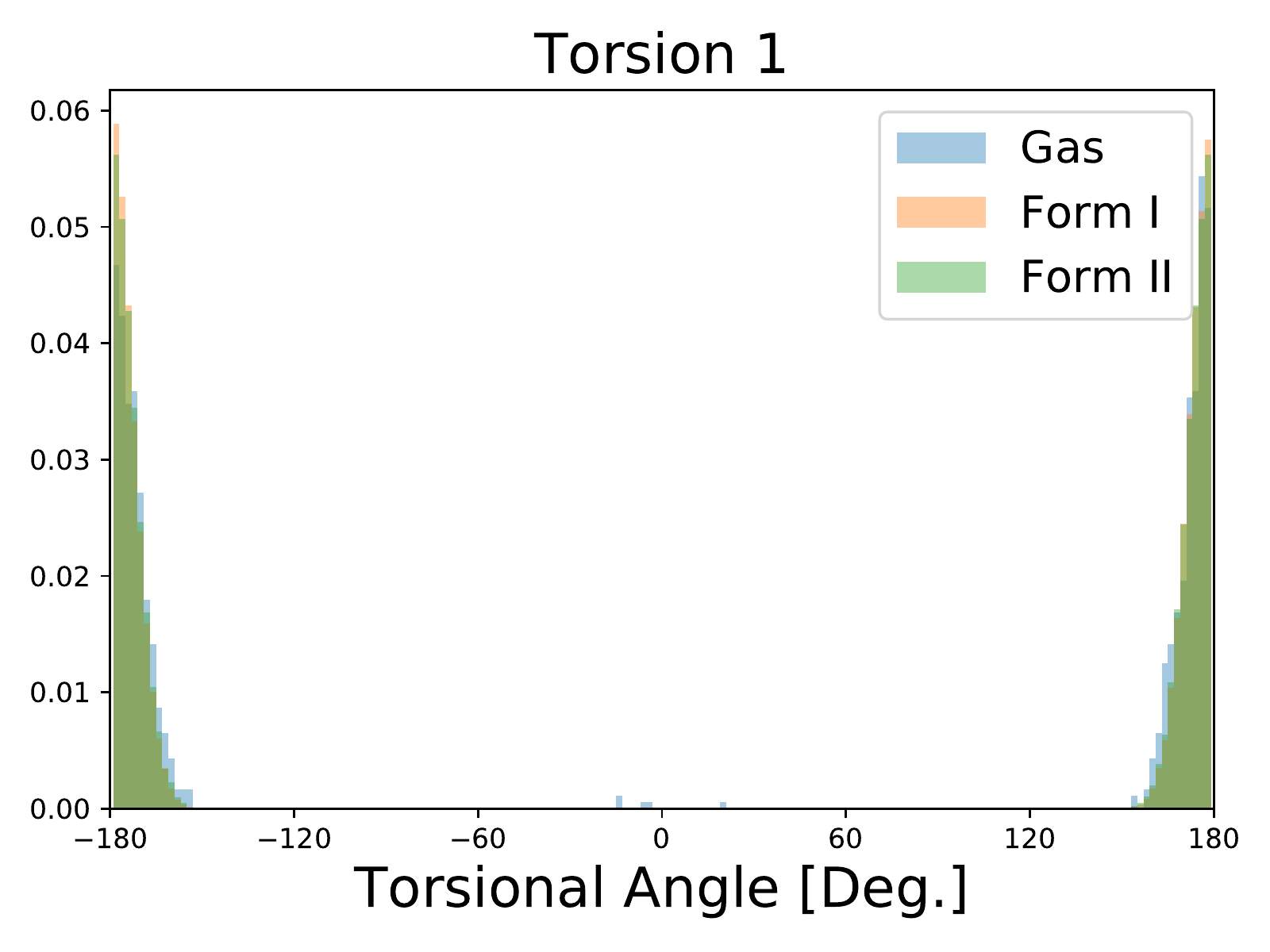}
  \includegraphics[width=6cm]{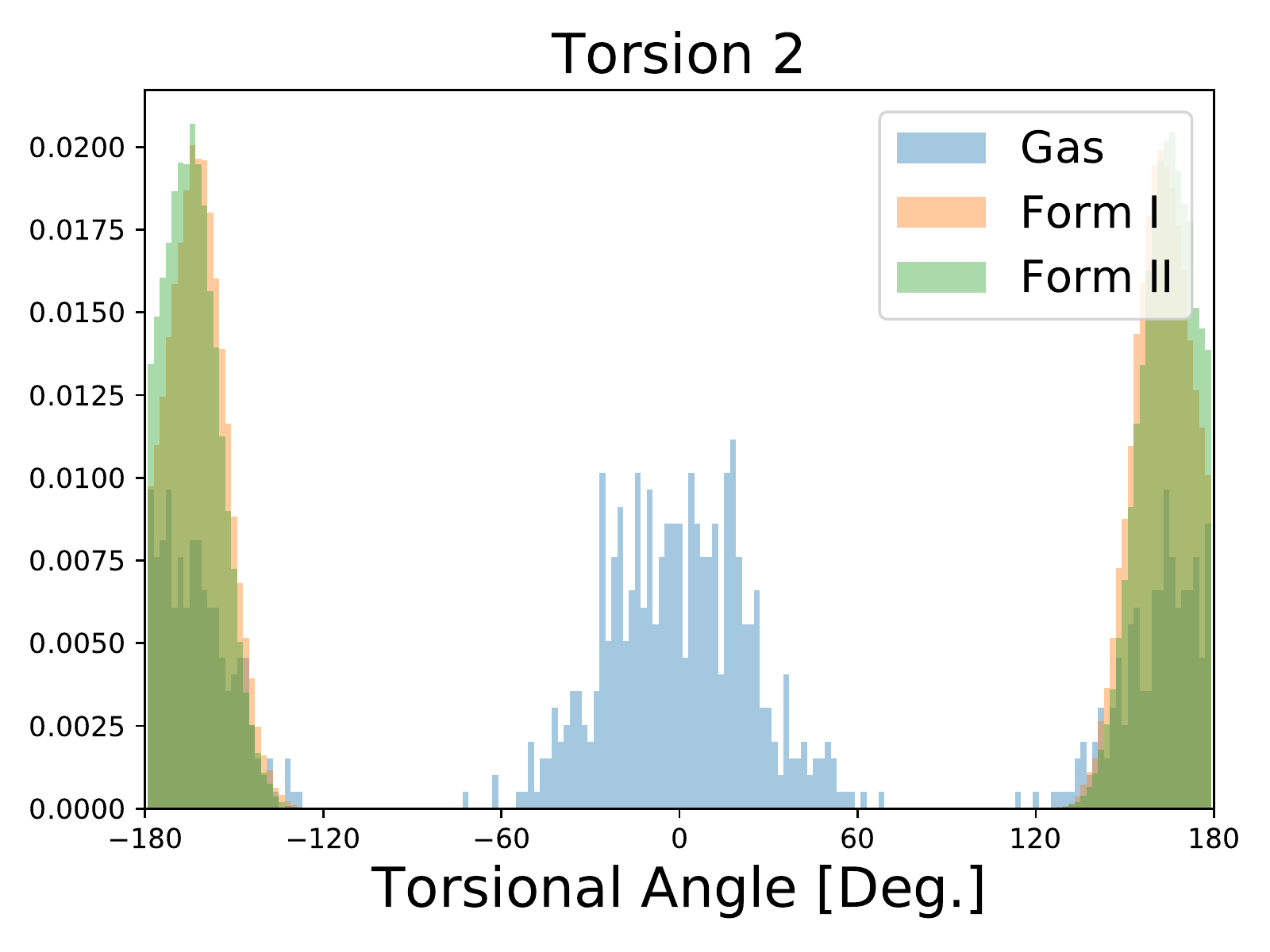}
  \includegraphics[width=6cm]{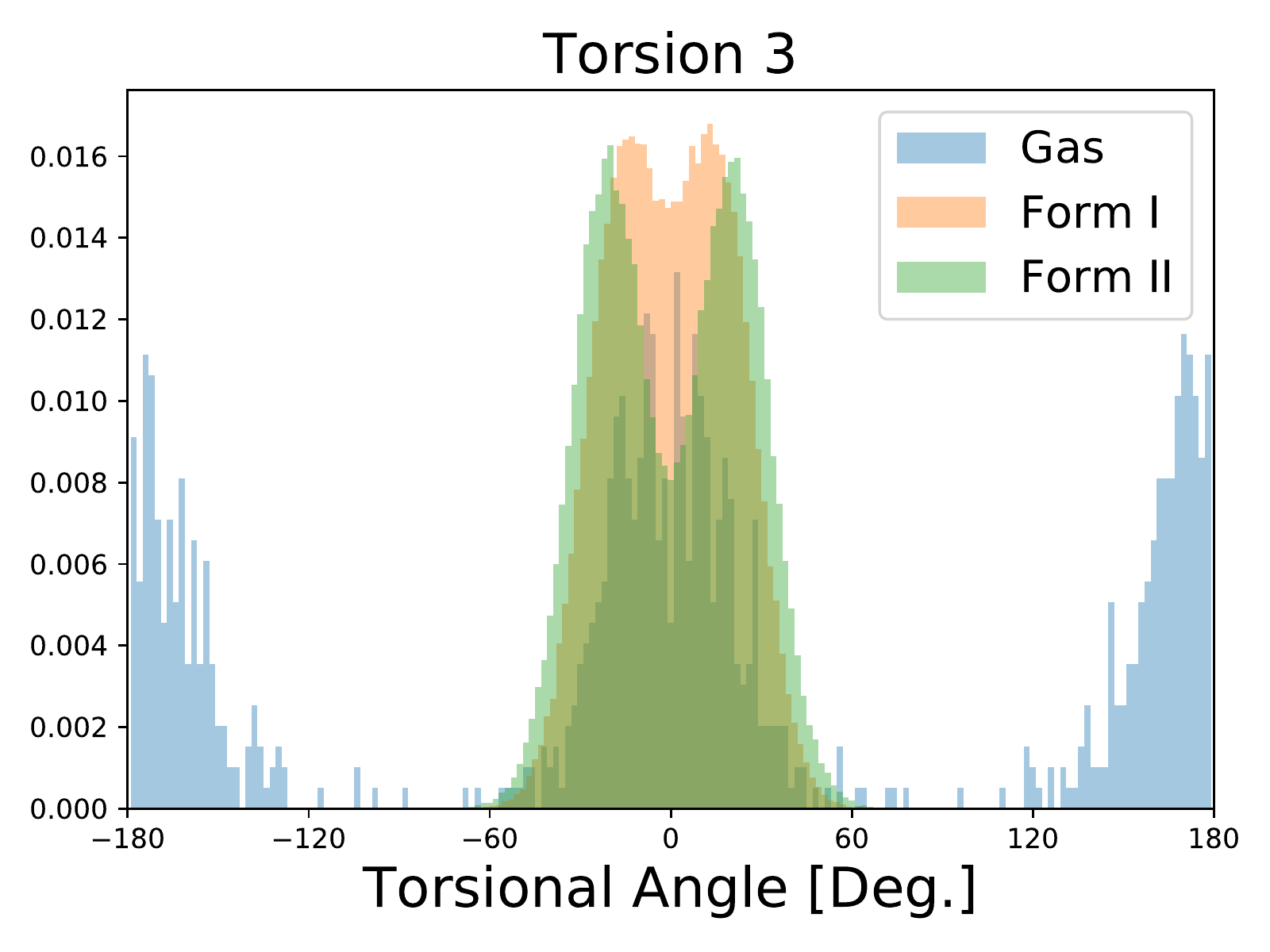}
  \end{center}
  \caption{Torsional distributions of paracetamol at 300 K for the molecule in gas and both polymorphs.
  \label{figure:hxacan_torsion}}
\end{figure}
\begin{figure}[H]
  \begin{center}
  \includegraphics[width=6cm]{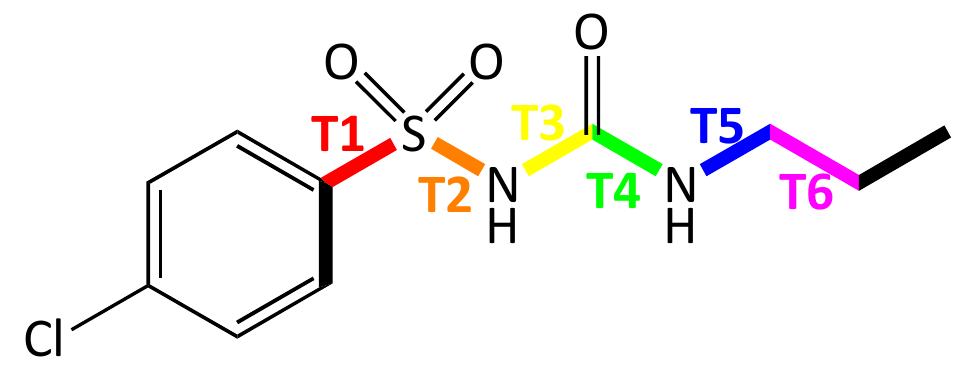}
  \includegraphics[width=6cm]{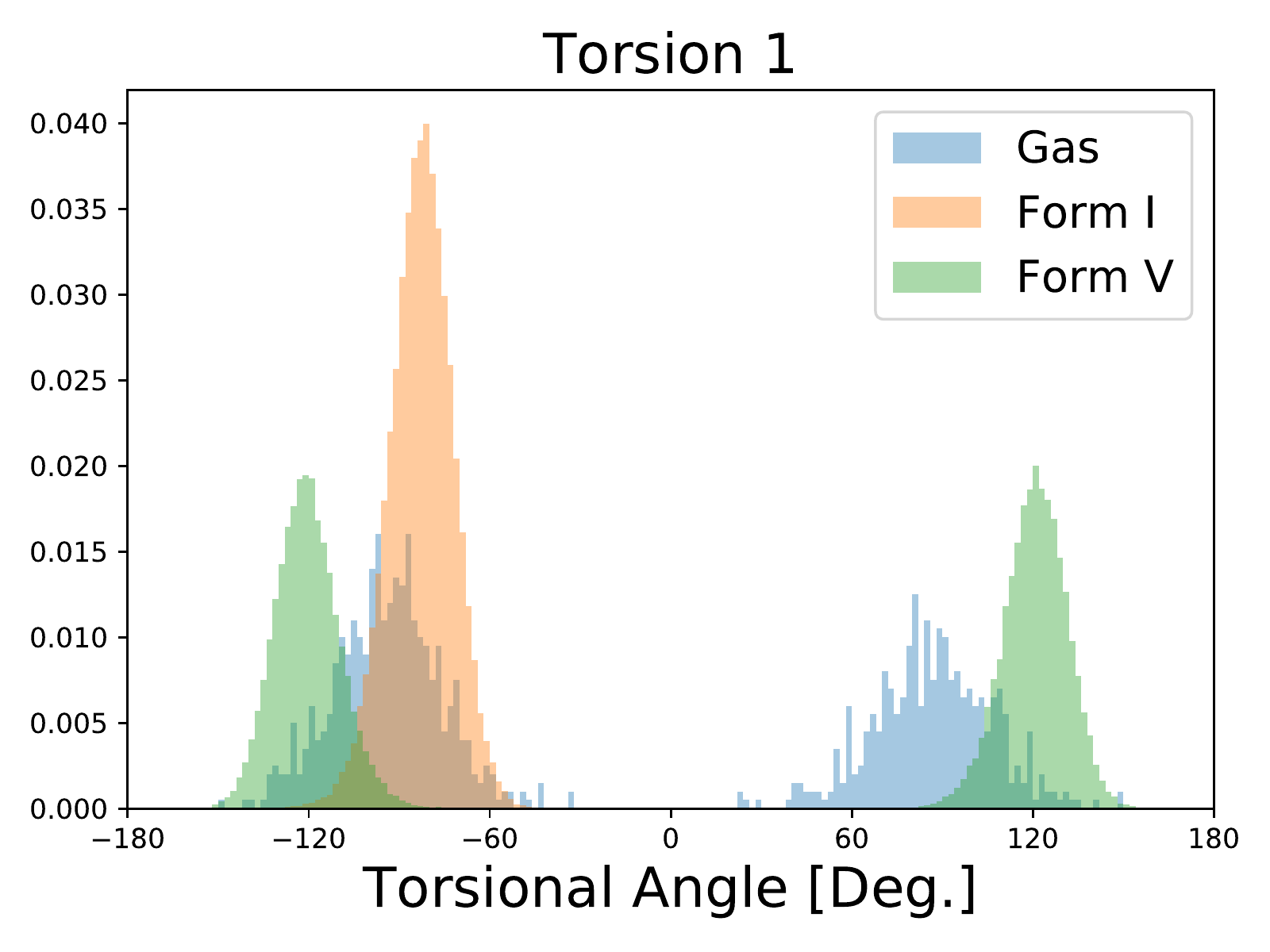}
  \includegraphics[width=6cm]{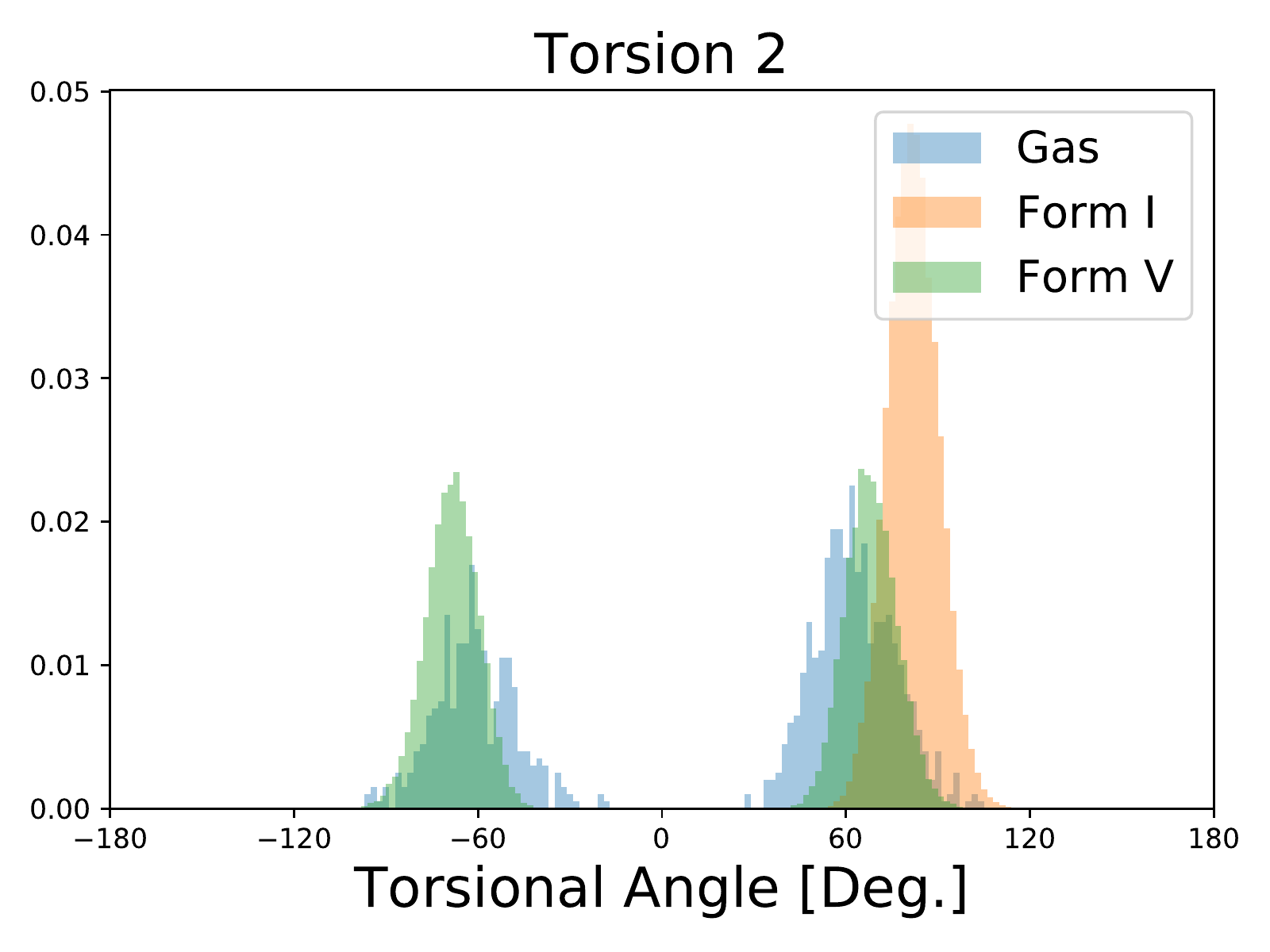}
  \includegraphics[width=6cm]{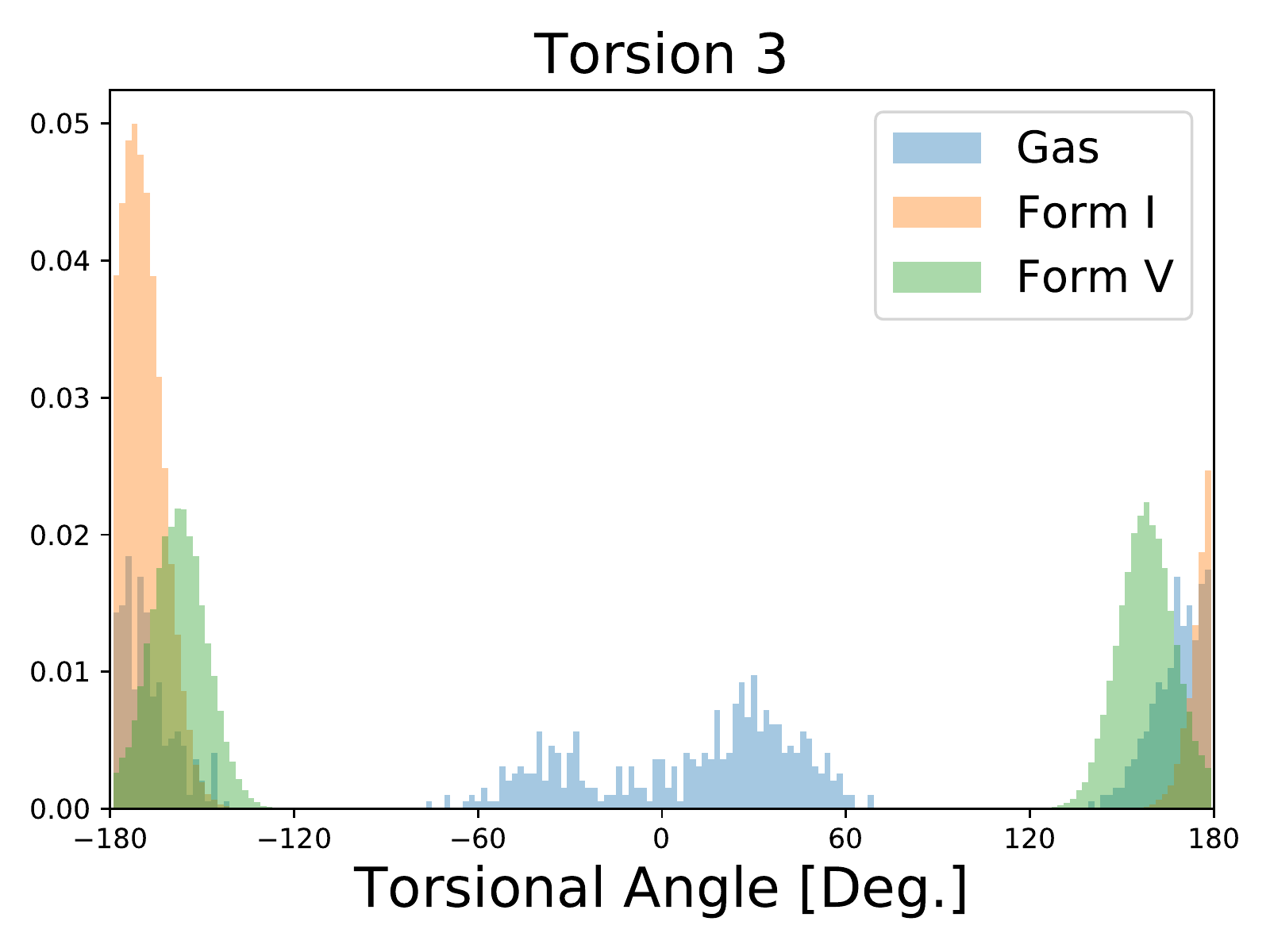}
  \includegraphics[width=6cm]{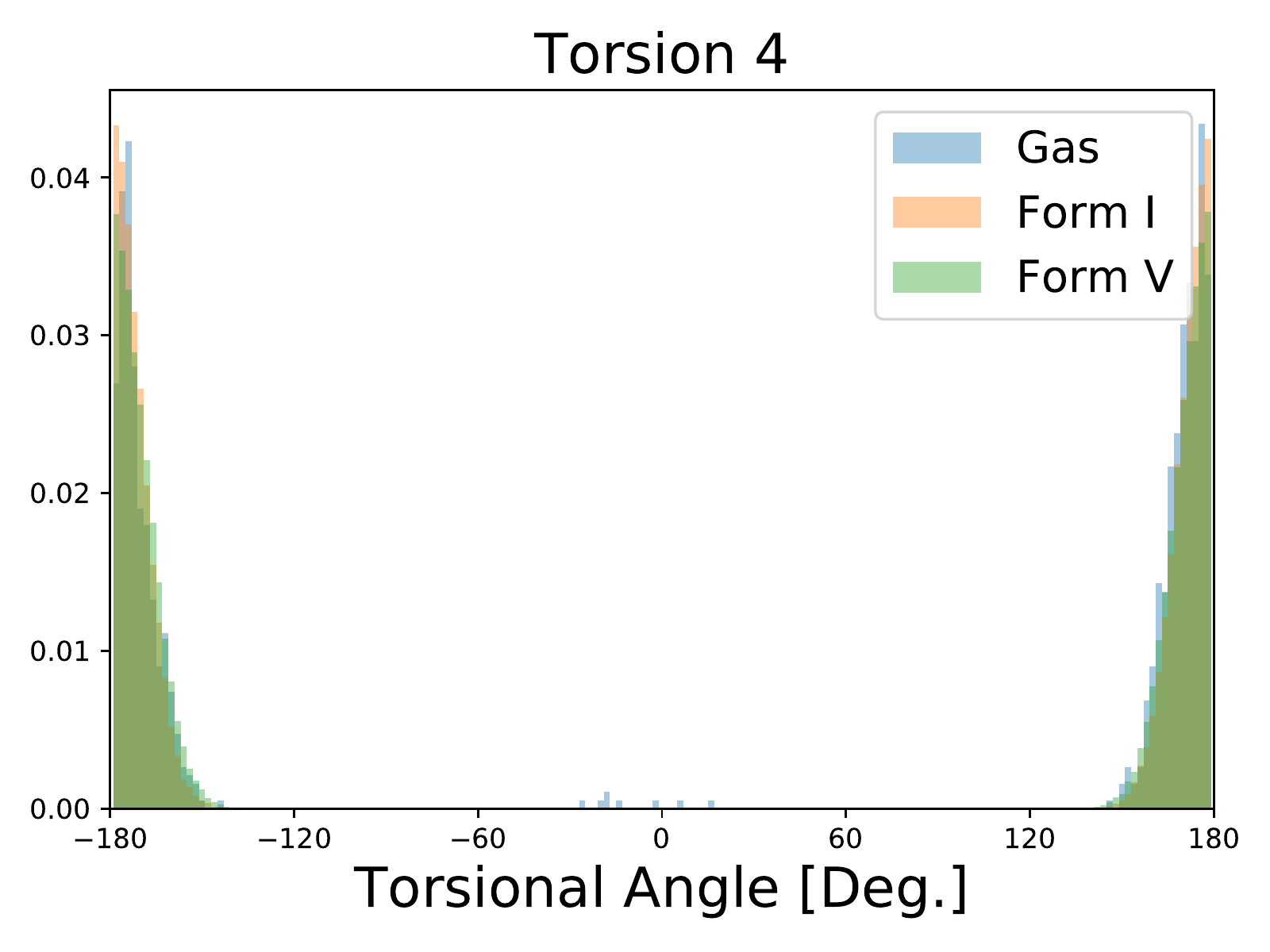}
  \includegraphics[width=6cm]{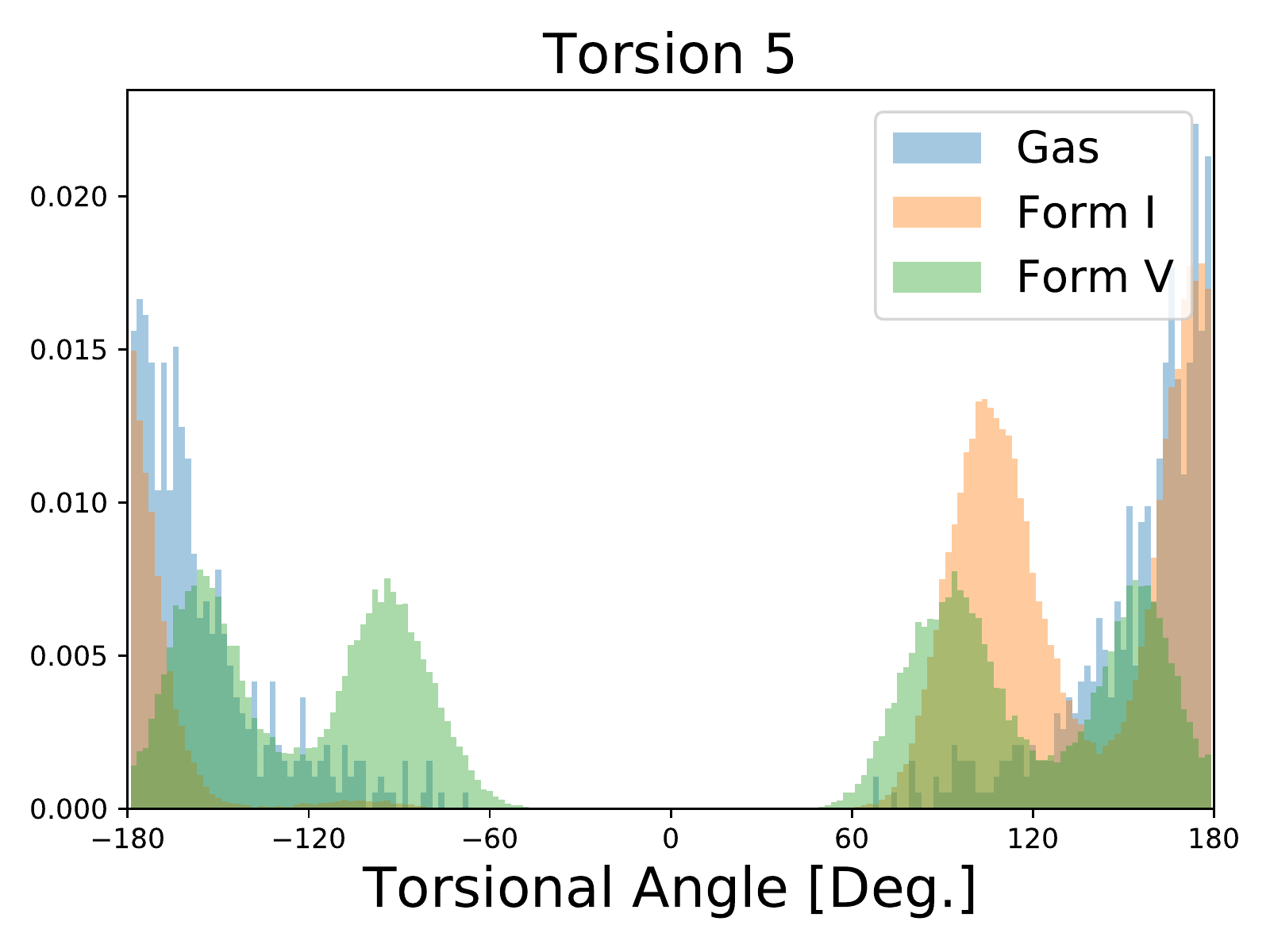}
  \includegraphics[width=6cm]{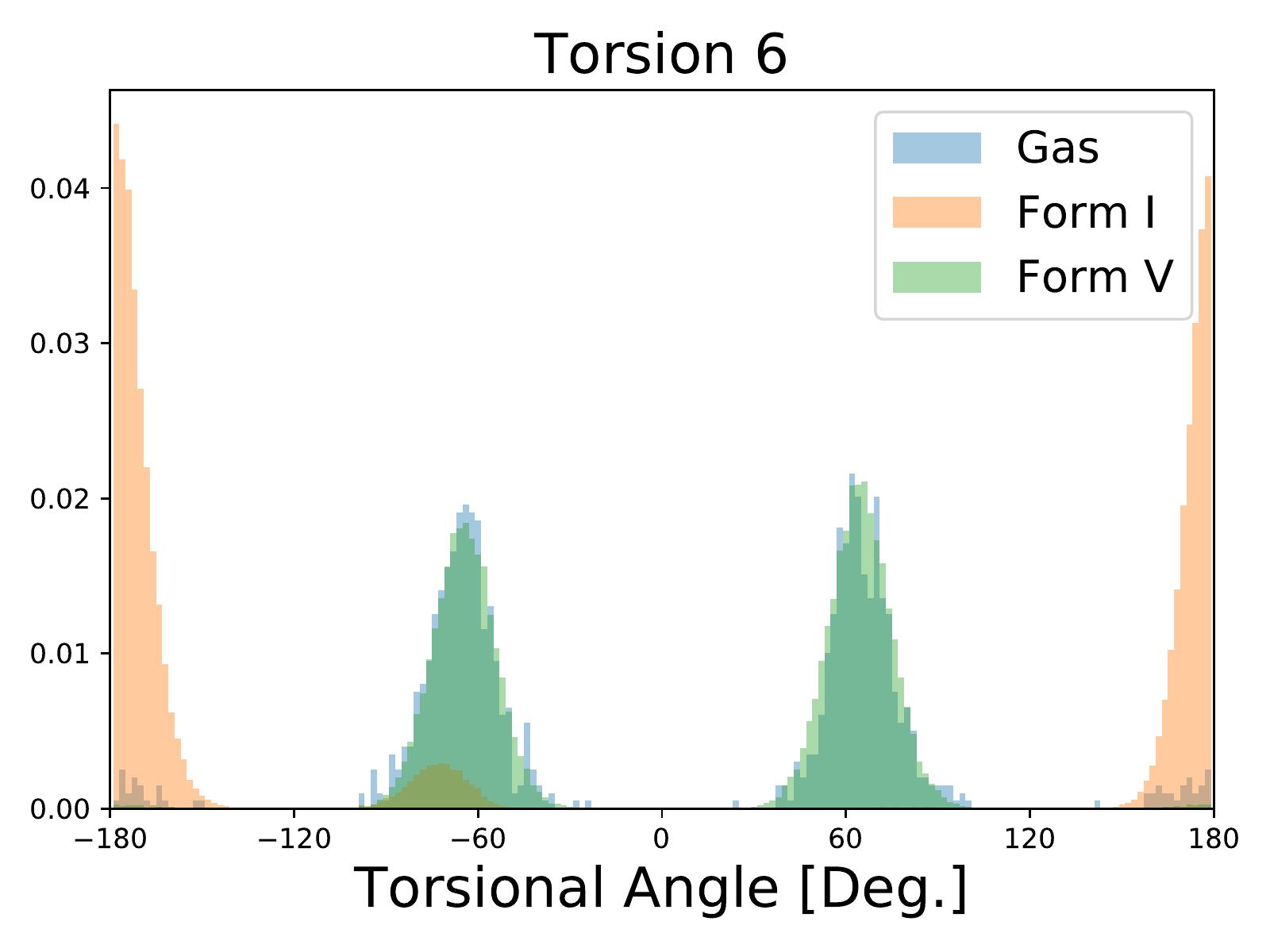}
  \end{center}
  \caption{Torsional distributions of chlorpropamide at 300 K for the molecule in gas and both polymorphs.
  \label{figure:bedmig1_torsion}}
\end{figure}
\begin{figure}[H]
  \begin{center}
  \includegraphics[width=6cm]{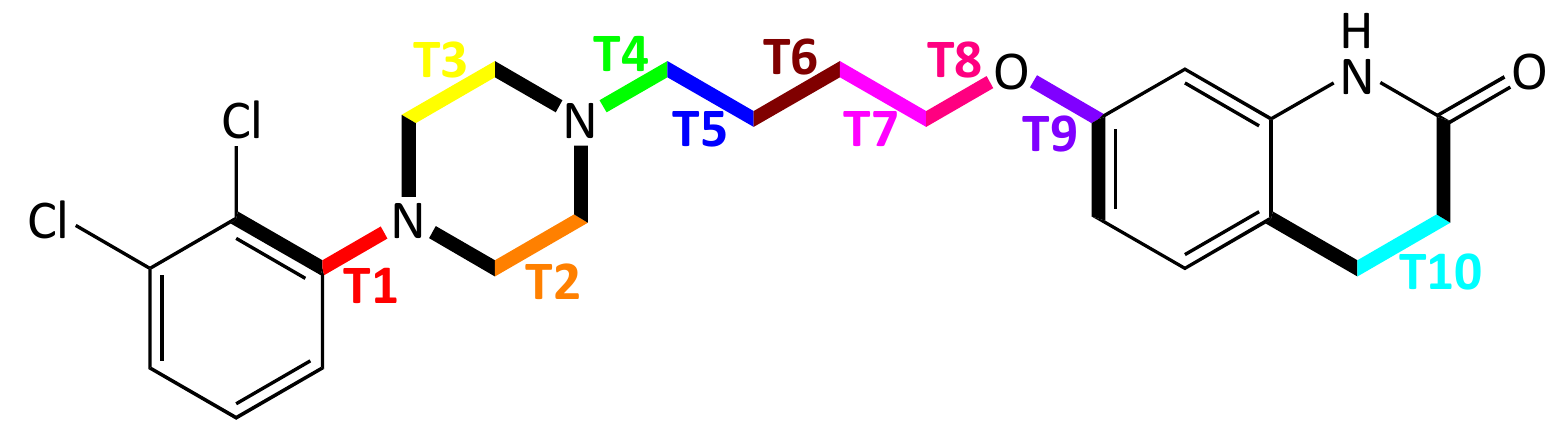}
  \includegraphics[width=6cm]{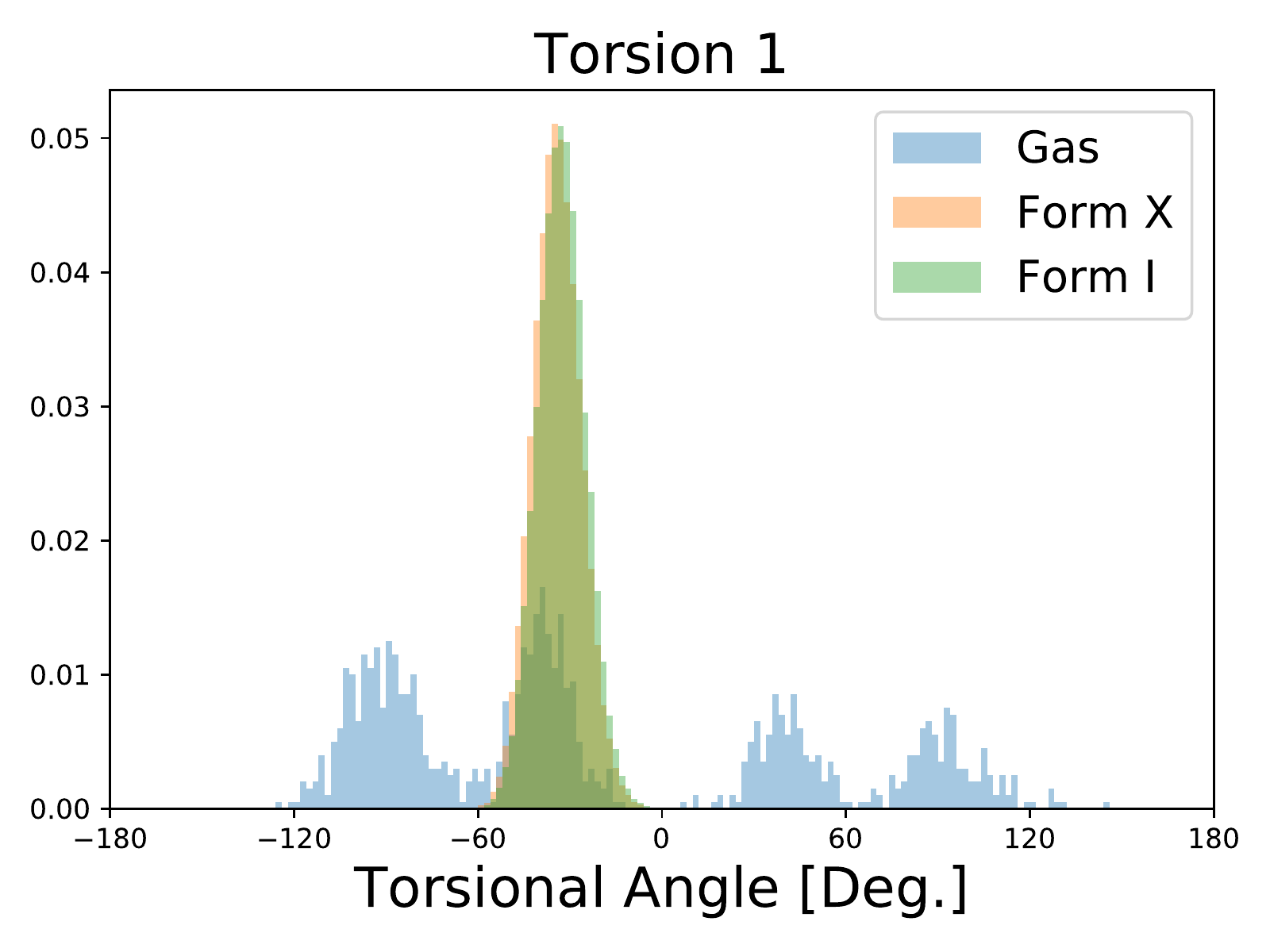}
  \includegraphics[width=6cm]{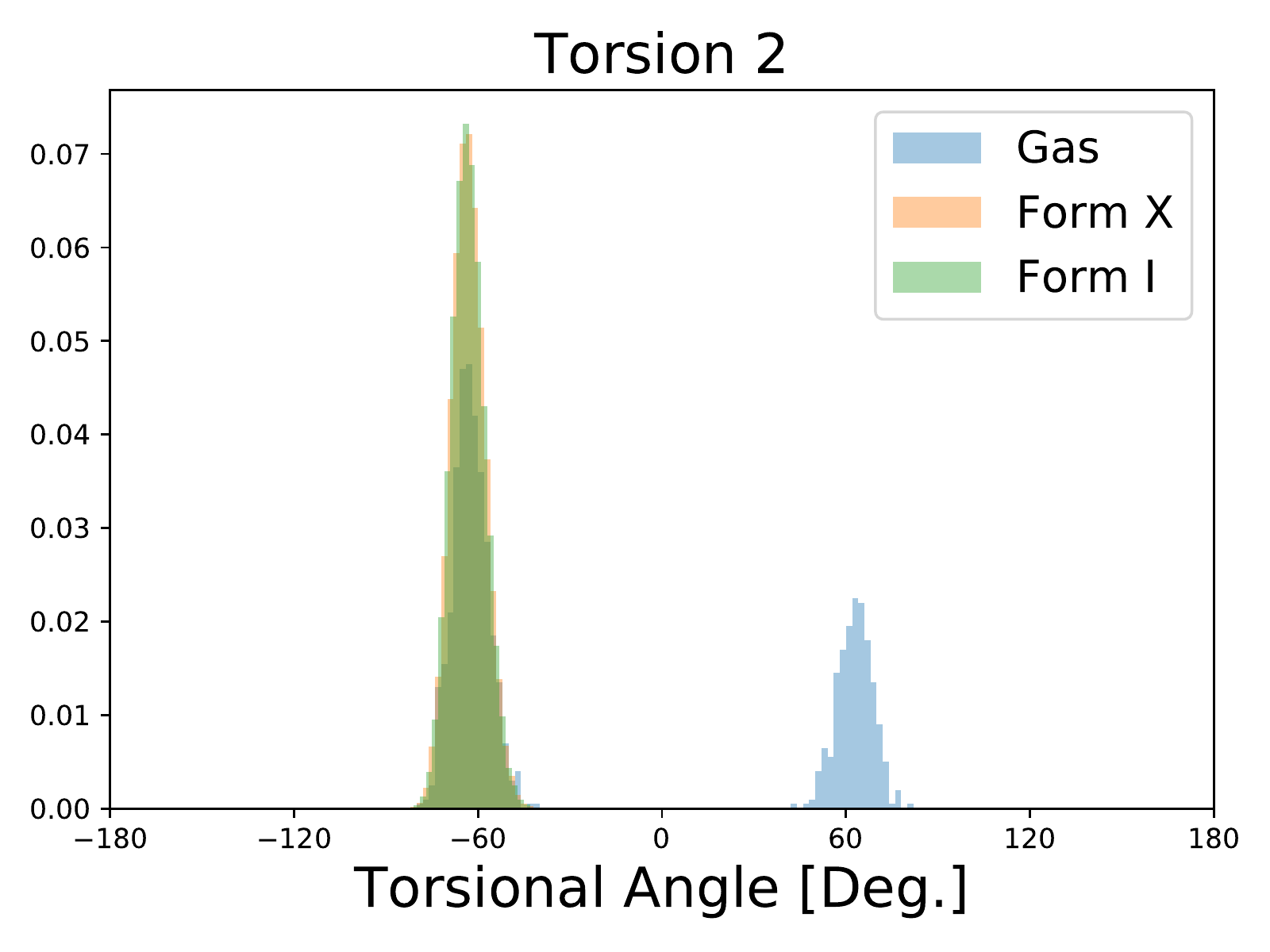}
  \includegraphics[width=6cm]{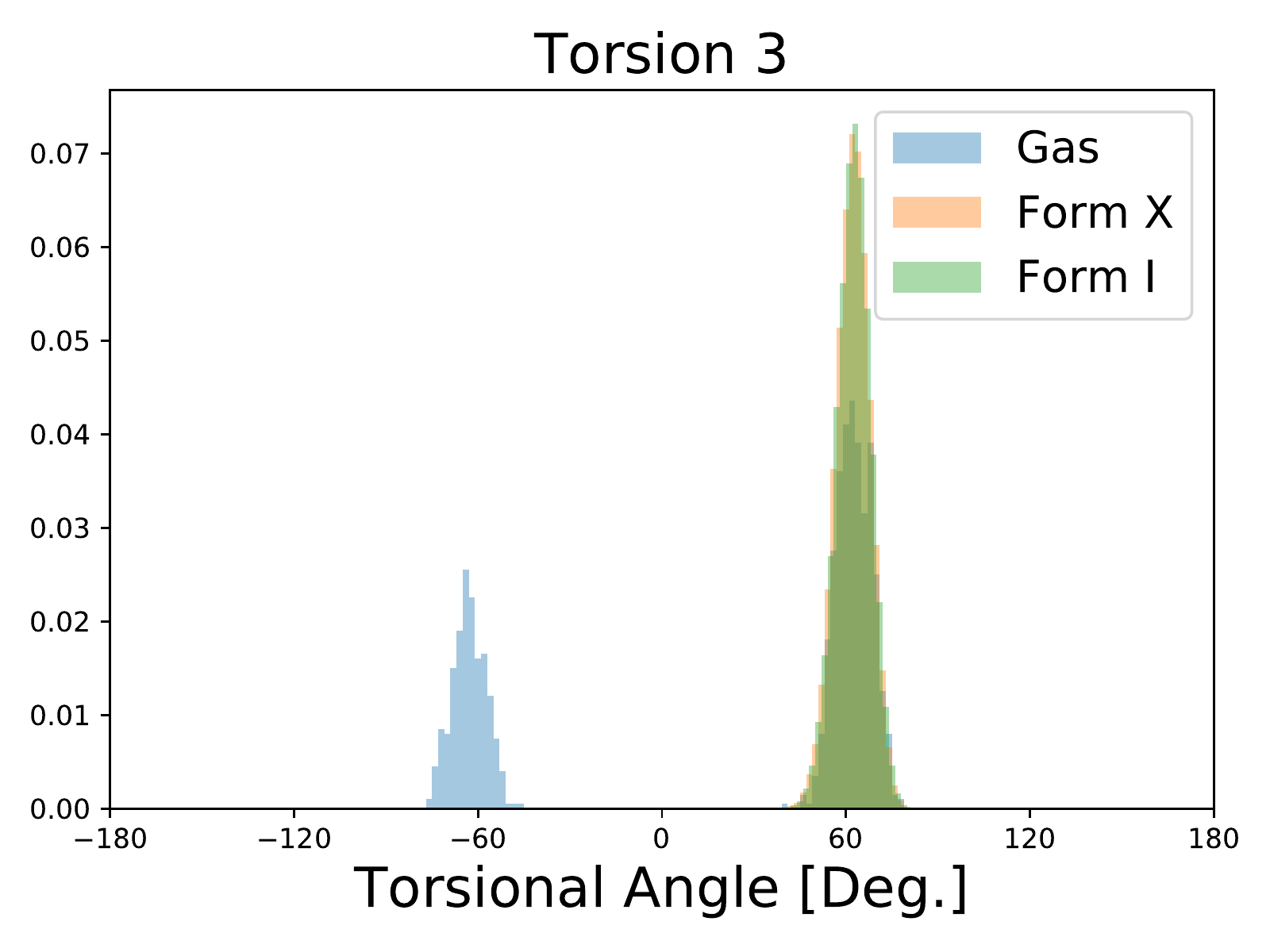}
  \includegraphics[width=6cm]{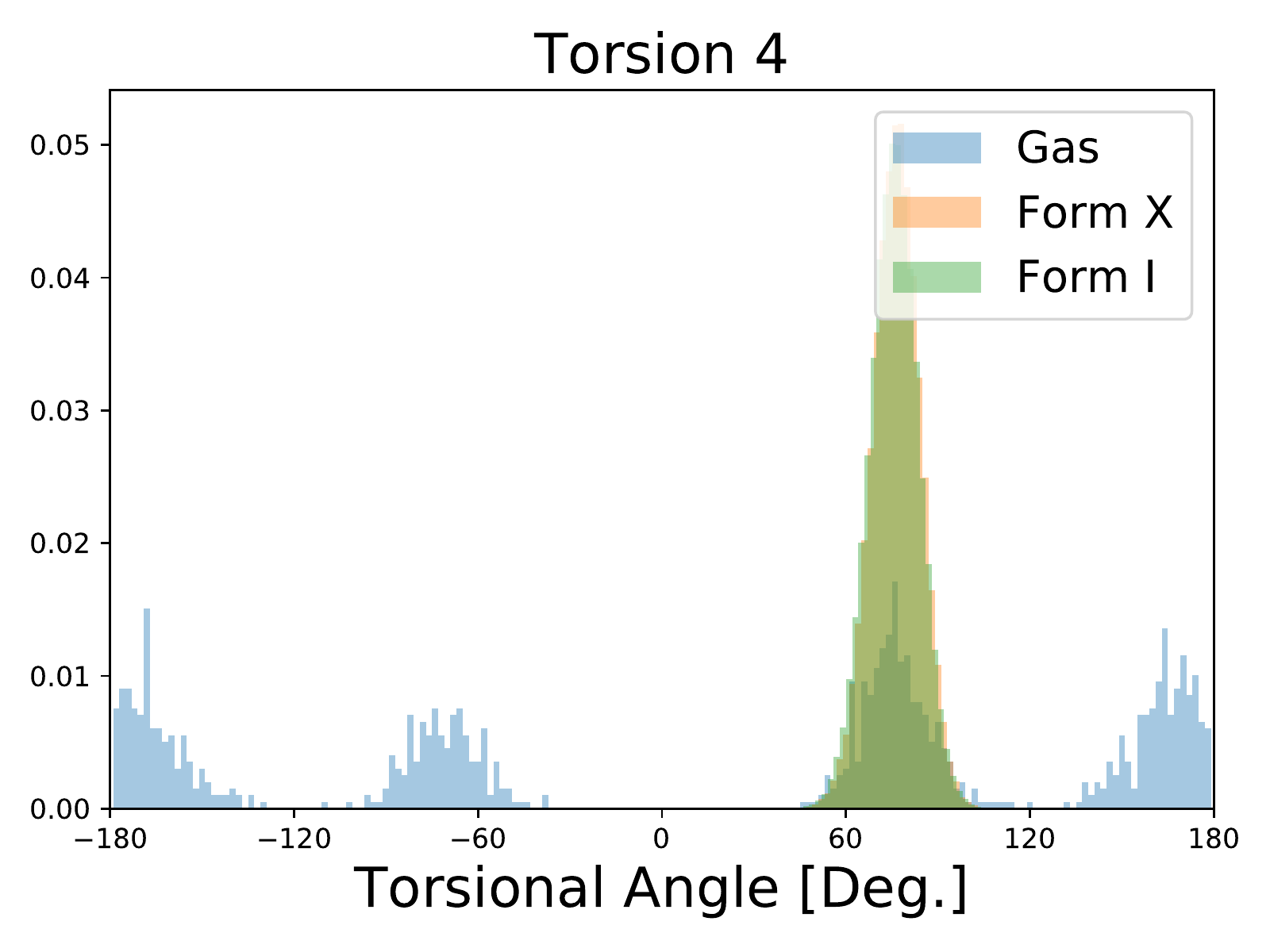}
  \includegraphics[width=6cm]{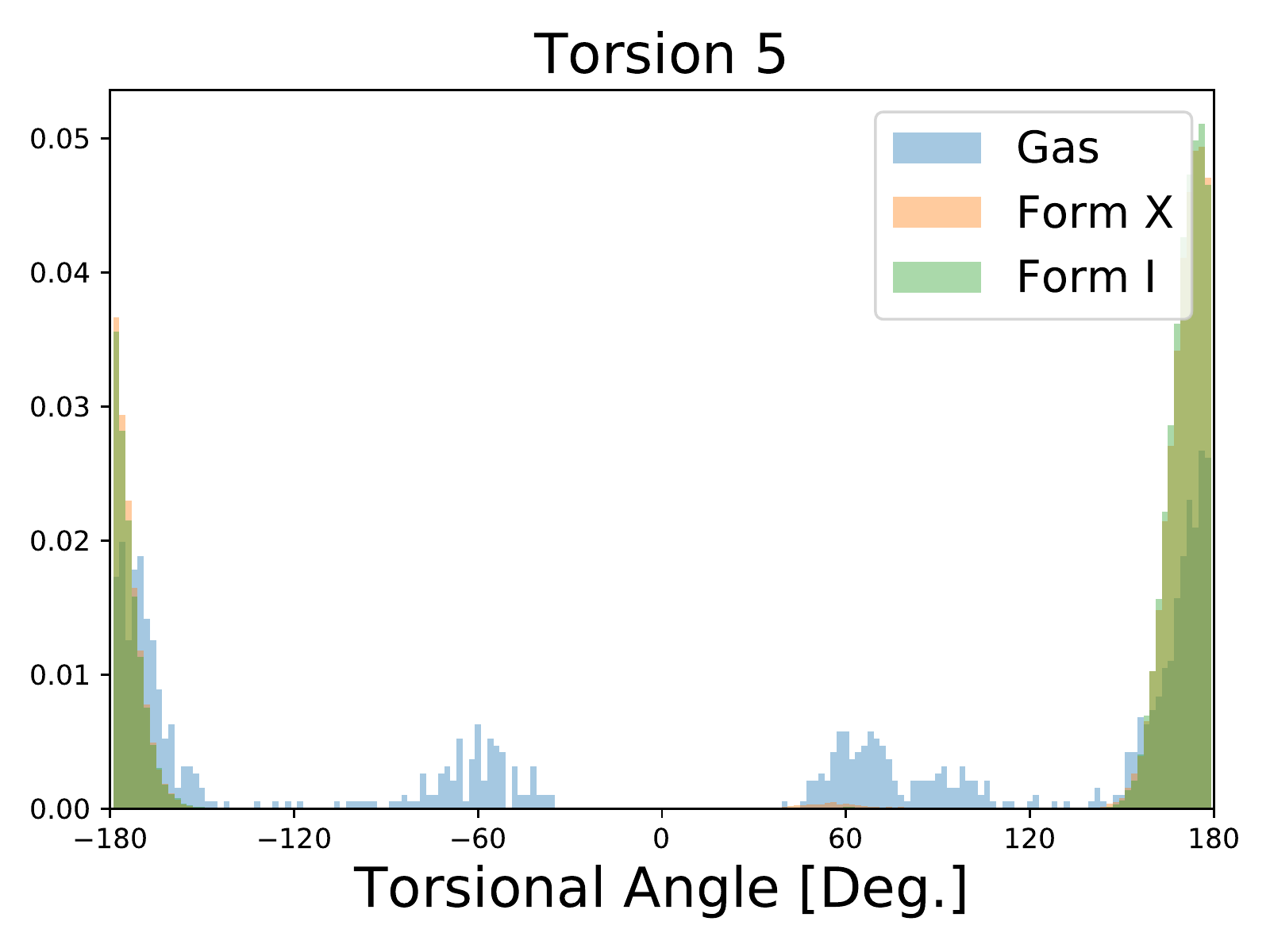}
  \end{center}
  \caption{Torsional distributions of aripiprazole at 300 K for the molecule in gas and both polymorphs for torsions 1 -- 5.
  \label{figure:bedmig2_torsion}}
\end{figure}
\begin{figure}[H]
  \begin{center}
  \includegraphics[width=6cm]{melfit_torsion.pdf}
  \includegraphics[width=6cm]{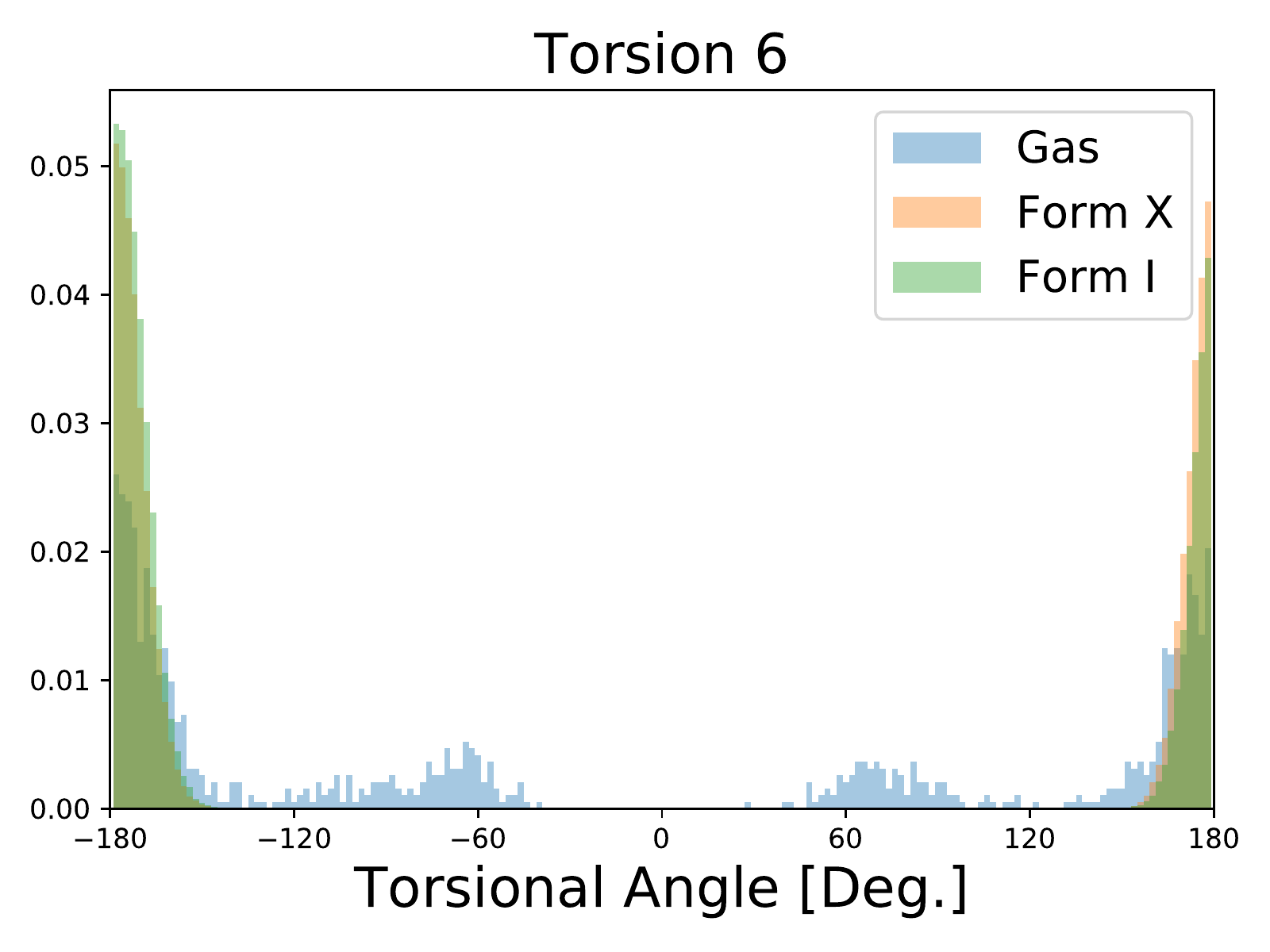}
  \includegraphics[width=6cm]{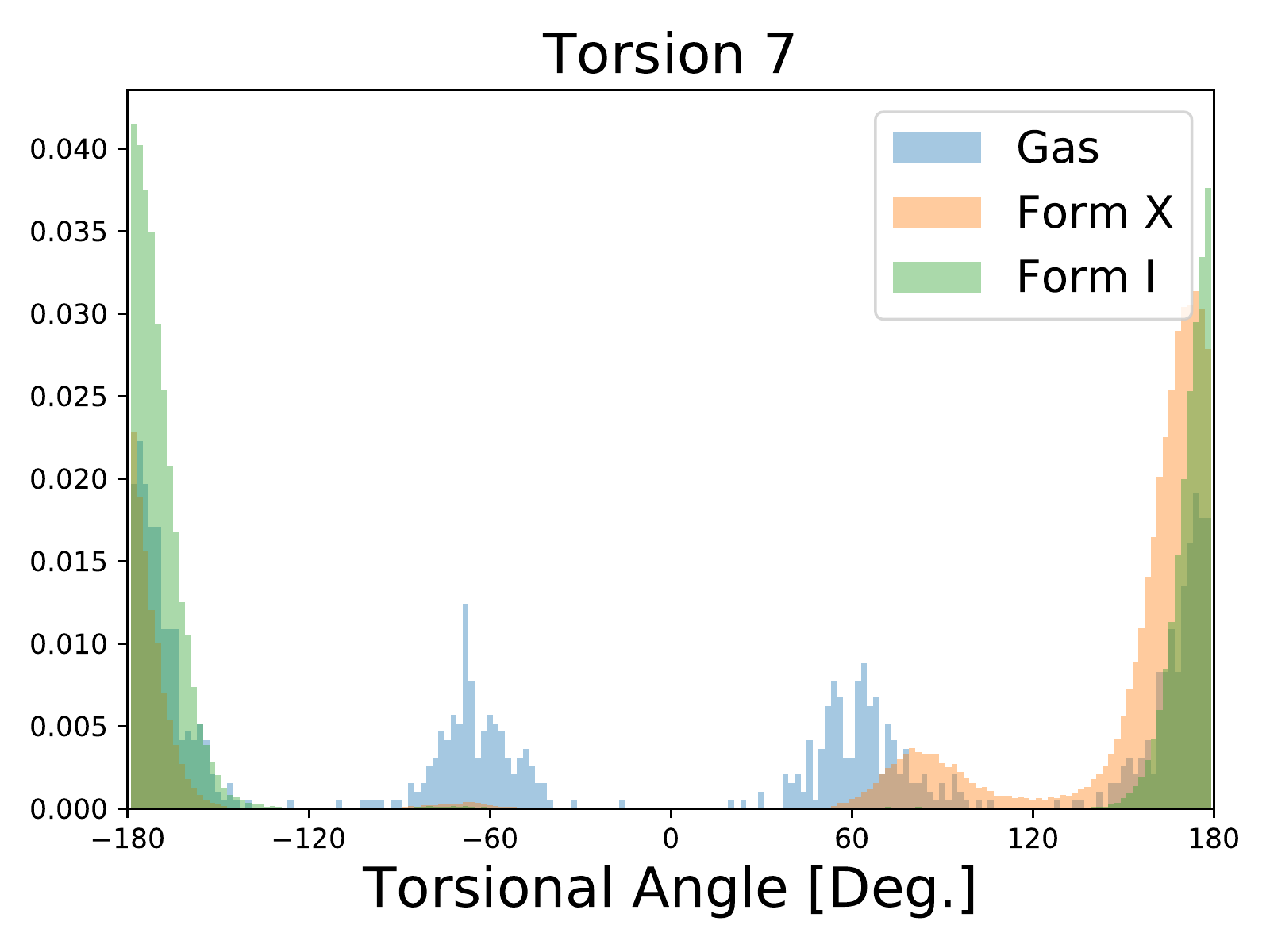}
  \includegraphics[width=6cm]{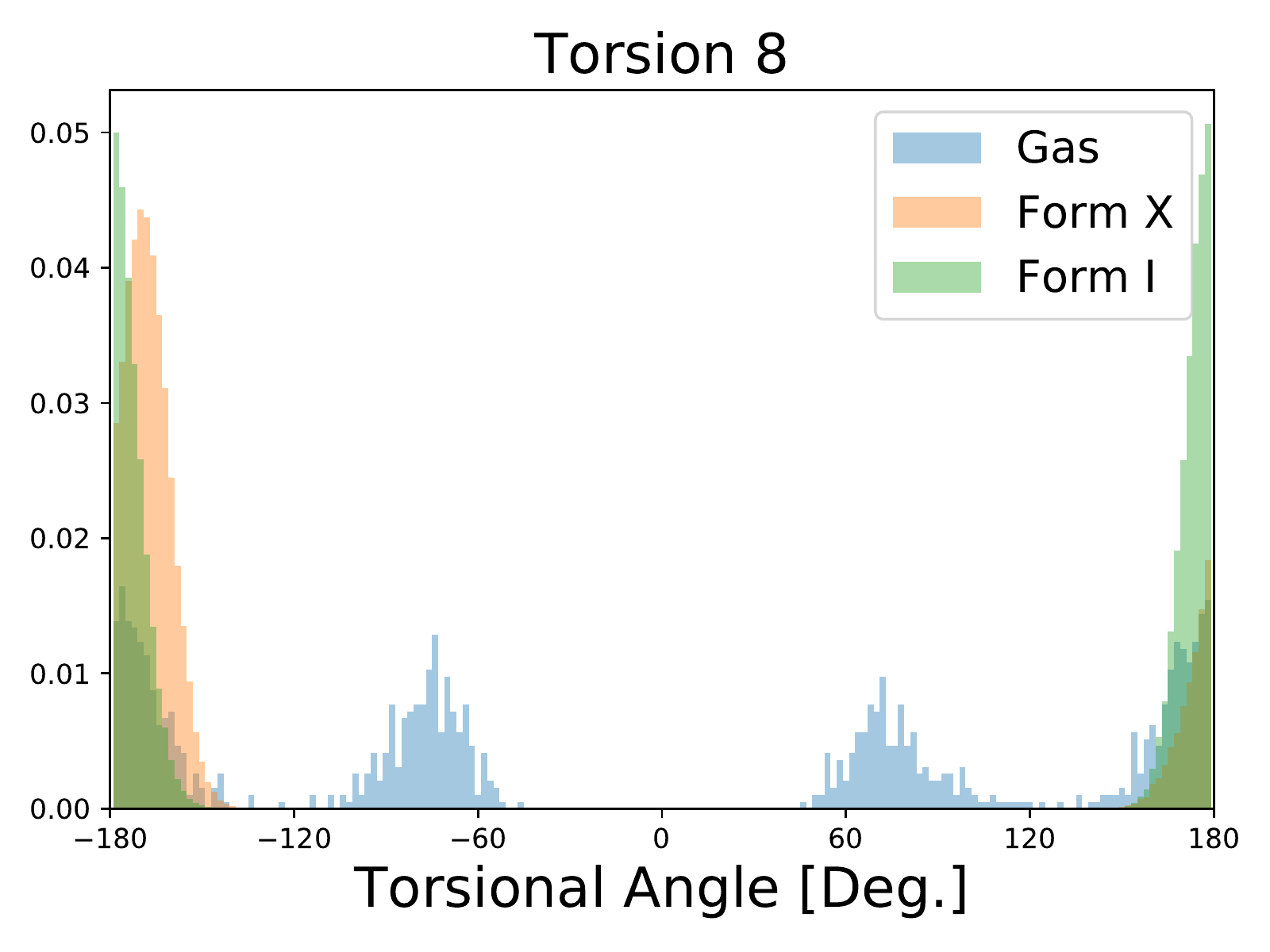}
  \includegraphics[width=6cm]{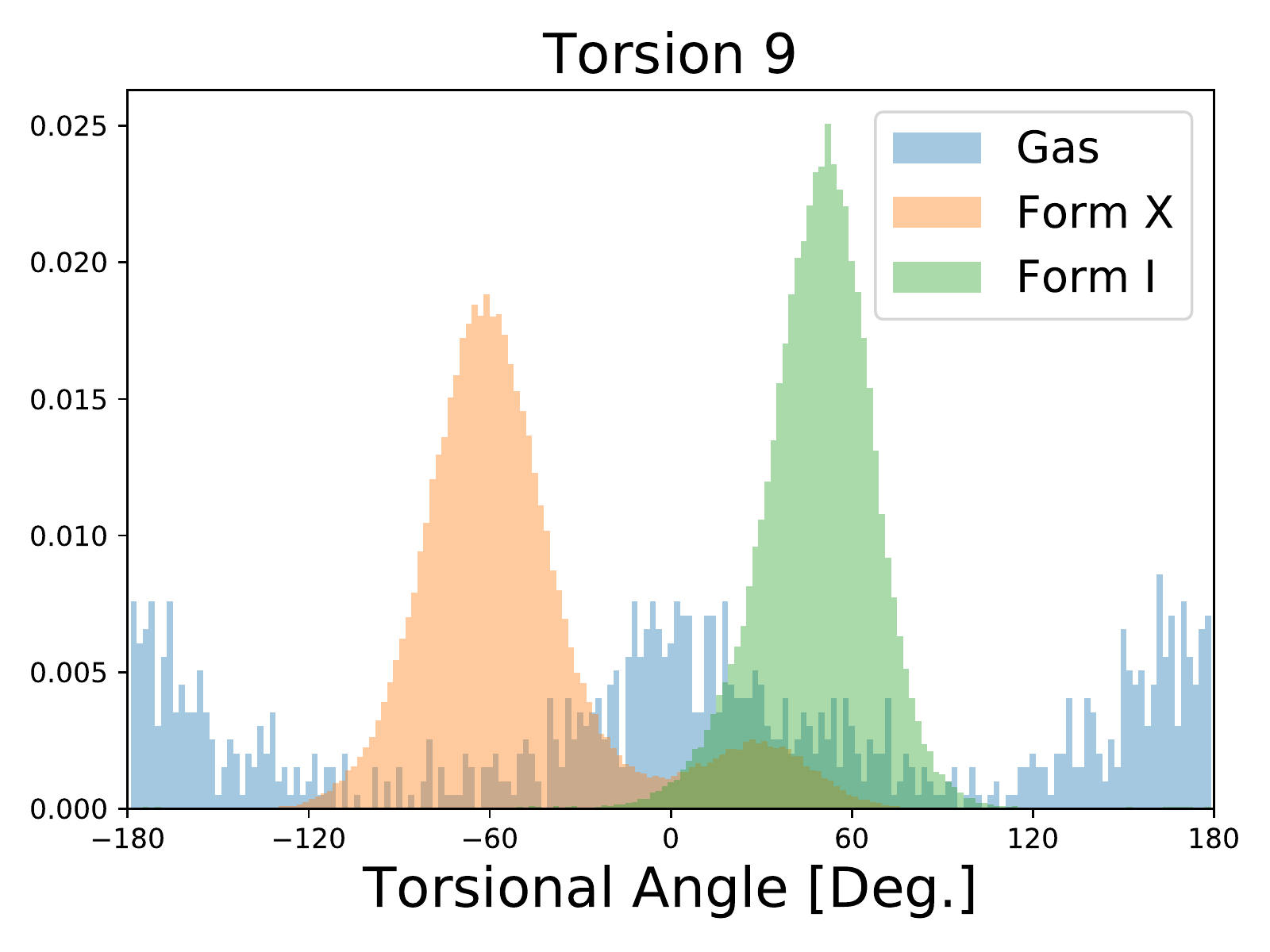}
  \includegraphics[width=6cm]{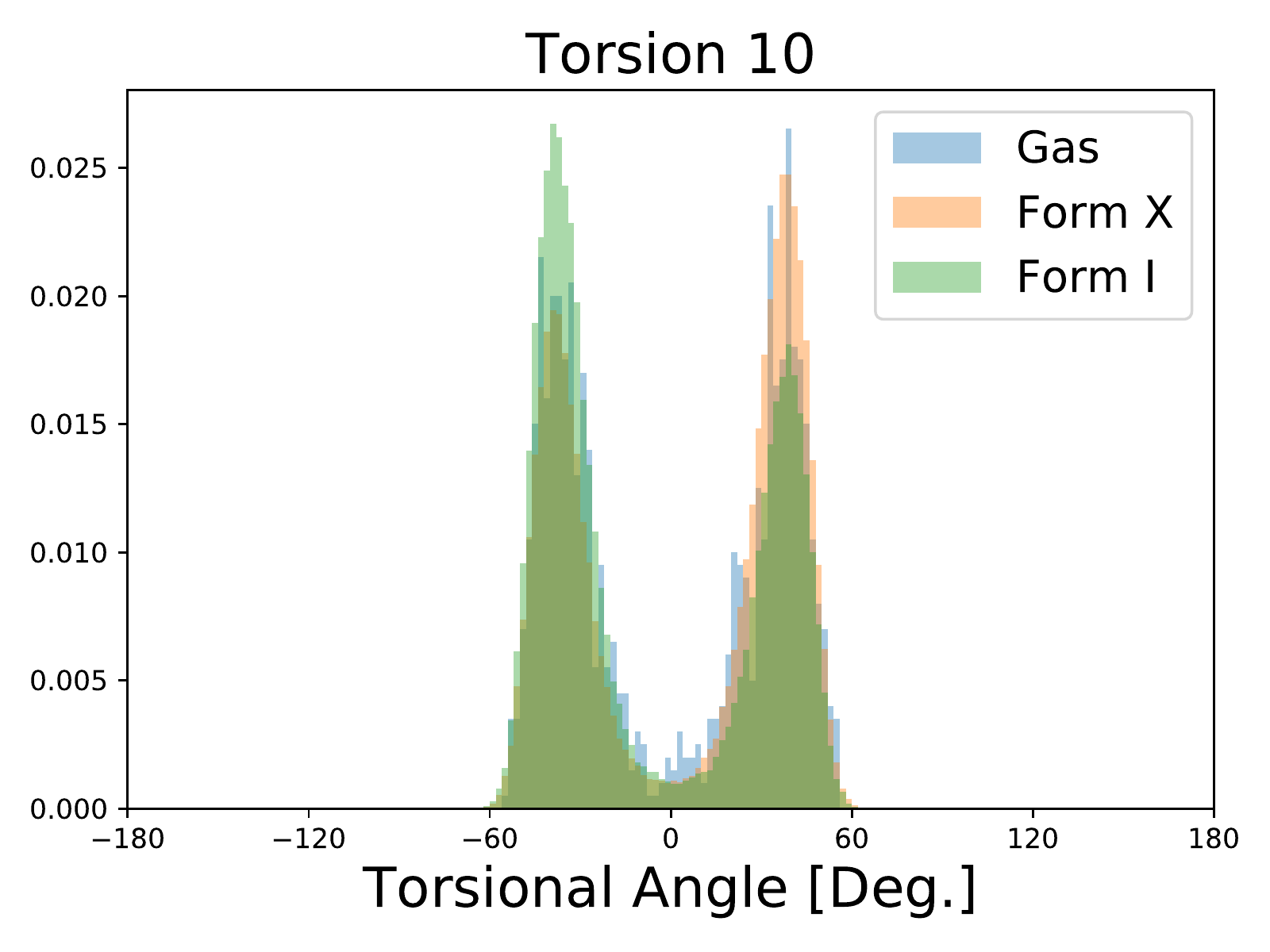}
  \end{center}
  \caption{Torsional distributions of aripiprazole at 300 K for the molecule in gas and both polymorphs for torsions 6 -- 10.
  \label{figure:melfit_torsion}}
\end{figure}

\begin{multicols}{2}
\section{Error in QHA due to Hydrogen Bonds}
\begin{figure}[H]
  \begin{center}
  \includegraphics[width=8cm]{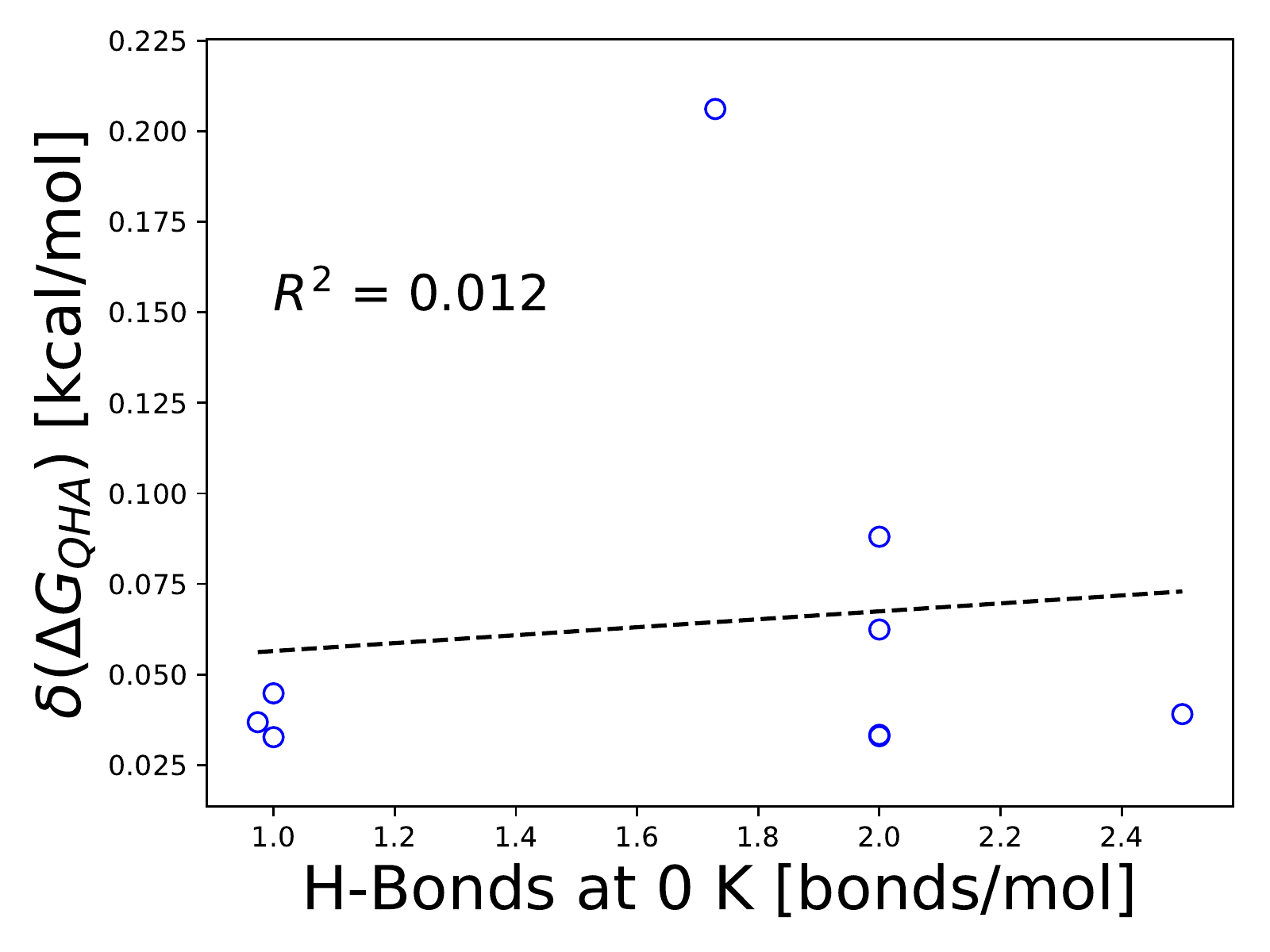}
  \end{center}
  \caption{Error in the polymorph free energy differences for 1D-QHA from MD versus the
           number of hydrogen bonds in the lattice minimum structure. There is no correlation
           between QHA error and 0 K hydrogen bonds.
  \label{figure:hbonds_0K}}
\end{figure}

\begin{figure}[H]
  \begin{center}
  \includegraphics[width=8cm]{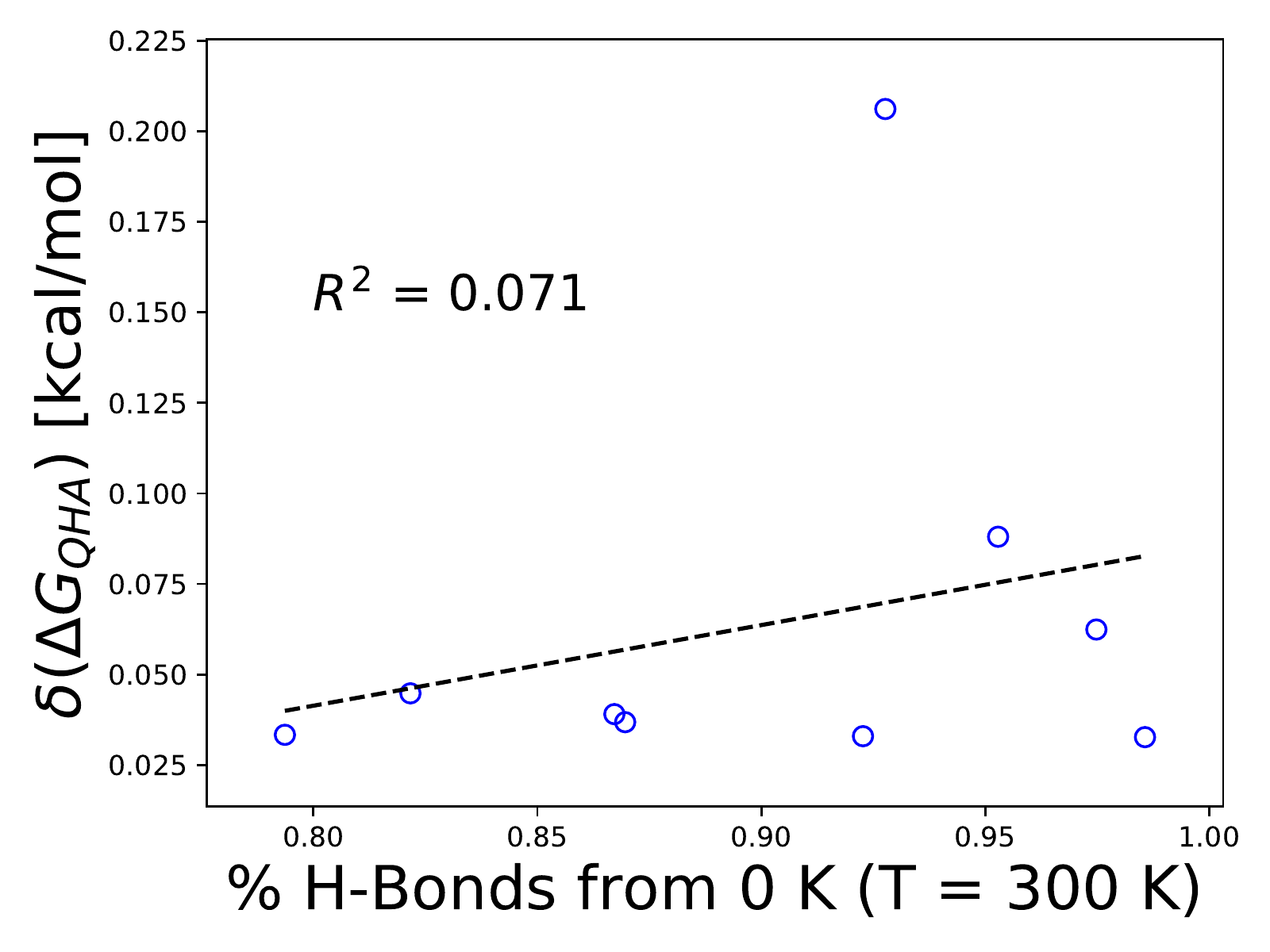}
  \end{center}
  \caption{Error in the polymorph free energy differences for 1D-QHA from MD versus the
           percent of hydrogen bonds remaining at 300 K relative to the lattice minimum. 
           There is no correlation between QHA error and 300 K hydrogen bonds.
  \label{figure:hbonds_300K}}
\end{figure}

We found that there was little to no correlation between the hydrogen bonds and the error
in anisotropic QHA relative to MD. We determine the number of hydrogen bonds on a per molecule
basis for the lattice minimum (Figure ~\ref{figure:hbonds_0K}) and the percent of hydrogen
bonds remaining at 300 K (Figure ~\ref{figure:hbonds_300K}) for MD for all systems to see if there was 
any correlation between error in 1D-QHA from MD ($\delta(\Delta G_{QHA})$) and the number
of hydrogen bonds. Despite having no hydrogen bonds present, diiodobenzne has the lowest
error in the QHA free energy difference and we therefore left it out of both 
Figure ~\ref{figure:hbonds_0K} and ~\ref{figure:hbonds_300K}. In both cases, a linear
fit shows the $R^{2}$ value is less than 0.1 showing that there is little to no correlation.

\end{multicols}


\newpage
\newpage

\begin{singlespace}
\begin{multicols}{2}
\bibliography{../citations}
\end{multicols}
\end{singlespace}

\end{document}